%% file: LRG.tex
\documentclass[twocolumn]{aastex61hack}

\usepackage{natbib}
\newcommand{\HI}{\ensuremath{\mbox{\ion{H}{1}}}}

\newcommand{\MgII}{\ensuremath{\mbox{\ion{Mg}{2}}}}
\newcommand{\FeII}{\ensuremath{\mbox{\ion{Fe}{2}}}}
\newcommand{\CIII}{\ensuremath{\mbox{\ion{C}{3}}}}

\newcommand{\OII}{\ensuremath{[\mbox{\ion{O}{2}}]}}
\newcommand{\OIII}{\ensuremath{[\mbox{\ion{O}{3}}]}}
\newcommand{\OVI}{\ensuremath{\mbox{\ion{O}{6}}}}

\newcommand{\NHI}{\ensuremath{N(\mbox{\ion{H}{1}})}\relax}
\newcommand{\logNHI}{\ensuremath{\log N(\mbox{\ion{H}{1}})}\relax}
\newcommand{\logNCIII}{\ensuremath{\log N(\mbox{\ion{C}{3}})}\relax}
\newcommand{\logNMgII}{\ensuremath{\log N(\mbox{\ion{Mg}{2}})}\relax}
\newcommand{\logNFeII}{\ensuremath{\log N(\mbox{\ion{Fe}{2}})}\relax}
\newcommand{\logNOI}{\ensuremath{\log N(\mbox{\ion{O}{1}})}\relax}

\newcommand{\Rvir}{\ensuremath{R_{\rm vir}}}
\newcommand{\logMstar}{\ensuremath{\log M_\star}}

\newcommand{\nH}{\ensuremath{n_{\rm H}}}
\newcommand{\z}{$z$}
\newcommand{\zlrg}{\ensuremath{z_{\rm LRG}}}

\newcommand{\sdssr}{$r$}

\newcommand{\msun}{\ensuremath{{\rm M}_\odot}}

\newcommand{\msunyr}{\ensuremath{{\rm M}_\odot ~ {\rm yr}^{-1}}}

\newcommand{\column}{cm$^{-2}$}
\newcommand{\percc}{cm$^{-3}$}
\newcommand{\kms}{km\,s$^{-1}$}
\newcommand{\hub}{km\,s$^{-1}$\,Mpc$^{-1}$}

\newcommand{\kcorrect}{{\tt kcorrect}}

\newcommand{\hst}{{\em HST}}

\defcitealias{hm1996}{{\rm HM05}}
\defcitealias{hm2012}{{\rm HM12}}
\defcitealias{prochaska2017}{{\rm P17}}
\defcitealias{chen2018}{{\rm C18}}
\defcitealias{howk2019}{Paper~II}
\defcitealias{tpw2017}{Tumlinson, Peeples, \& Werk 2017}

\shorttitle{The Cool CGM About RDR LRGs} 
\shortauthors{Berg et al.}

\begin{document}

\title{The Red Dead Redemption Survey of Circumgalactic Gas About Massive Galaxies. I. Mass and Metallicity of the Cool Phase}

\correspondingauthor{Michelle A. Berg}
\email{mberg3@nd.edu}

\author[0000-0002-8518-6638]{Michelle A. Berg}
\affiliation{Department of Physics, University of Notre Dame, Notre Dame, IN 46556}

\author[0000-0002-2591-3792]{J. Christopher Howk}
\affiliation{Department of Physics, University of Notre Dame, Notre Dame, IN 46556}

\author[0000-0001-9158-0829]{Nicolas Lehner}
\affiliation{Department of Physics, University of Notre Dame, Notre Dame, IN 46556}

\author[0000-0001-6923-978X]{Christopher B. Wotta}
\affiliation{Department of Physics, University of Notre Dame, Notre Dame, IN 46556}

\author[0000-0002-7893-1054]{John M. O'Meara}
\affiliation{Department of Chemistry and Physics, Saint Michael's College, Colchester, VT, 05439}
\affiliation{W. M. Keck Observatory 65-1120 Mamalahoa Hwy. Kamuela, HI 96743}

\author[0000-0002-5668-0397]{David V. Bowen}
\affiliation{Princeton University Observatory, Princeton, NJ 08544}

\author[0000-0002-1979-2197]{Joseph N. Burchett}
\affiliation{Department of Astronomy and Astrophysics, UCO/Lick Observatory, University of California, Santa Cruz, Santa Cruz, CA 95064}

\author[0000-0003-1455-8788]{Molly S. Peeples}
\affiliation{Space Telescope Science Institute, Baltimore, MD, 21218}
\affiliation{Department of Physics and Astronomy, Johns Hopkins University, Baltimore, MD 21218}

\author[0000-0002-1883-4252]{Nicolas Tejos}
\affiliation{Instituto de F\'{i}sica, Pontificia Universidad Cat\'{o}lica de Valpara\'{i}so, Casilla 4059, Valpara\'{i}so, Chile}


\begin{abstract} 

We present a search for \HI\ in the circumgalactic medium (CGM) of 21 massive ($\langle \logMstar \rangle \sim 11.4$), luminous red galaxies (LRGs) at $z\sim0.5$. Using UV spectroscopy of QSO sightlines projected within 500 kpc ($\sim$\Rvir) of these galaxies, we detect \HI\ absorption in 11/21 sightlines, including two partial Lyman limit systems and two Lyman limit systems. The covering factor of $\logNHI \ge 16.0$ gas within the virial radius of these LRGs is $f_c(\rho \le \Rvir) = 0.27^{+0.11}_{-0.10}$, while for optically-thick gas ($\logNHI \ge 17.2$) it is $f_c(\rho \le \Rvir) = 0.15^{+0.10}_{-0.07}$. Combining this sample of massive galaxies with previous galaxy-selected CGM studies, we find no strong dependence of the \HI\ covering factor on galaxy mass, although star-forming galaxies show marginally higher covering factors. There is no evidence for a critical mass above which dense, cold ($T \sim 10^4$ K) gas is suppressed in the CGM of galaxies (spanning stellar masses $9.5 \la \logMstar \la 11.8$). The metallicity distribution in LRGs is indistinguishable from those found about lower-mass star-forming galaxies, and we find low-metallicity gas with $[{\rm X/H}] \approx -1.8$ (1.5\% solar) and below about massive galaxies. \added{About half the cases show super-solar [\FeII/\MgII] abundances as seen previously in cool gas near massive galaxies.} While the high-metallicity cold gas seen in LRGs could plausibly result from condensation from a corona, the low-metallicity gas is inconsistent with this interpretation. 

\end{abstract}


\section{Introduction} \label{sec:intro}

Many galaxies have experienced some process(es) that ``quenched'' their ability
to transform gas into stars. For more massive ``red-and-dead'' galaxies, this
does not mean that they lack a supply of gas: massive and/or elliptical
galaxies contain significant reservoirs of hot gas
\citep[e.g.,][]{anderson2013,singh2018} and even some cool gas in both their
interstellar medium \citep{serra2012,osullivan2018} and the surrounding circumgalactic medium
\citep[CGM,][]{thom2012}, albeit at a lower mass fraction than typical
star-forming galaxies. Indeed, as these galaxies continue to accrete mass from
their surroundings, they will accrete new gas as well \added{(e.g., \citealt{fakhouri2010,oser2010,hausammann2019})}. Massive galaxies thus
contain significant gas mass, but they are not able to efficiently access that
gas to form new stars at a high rate. This is in large part due to the physical
conditions of their gas: it is predominantly hot, with long cooling times.

Simulations and theoretical work have suggested that the transition to
quiescence is perhaps associated with different modes for gas accretion with
galaxy mass (see the general CGM review by \citetalias{tpw2017} and additional
reviews on accretion in \citealt{fox2017}). Low-mass galaxies \added{($M_\star \la 10^{10.5}$ \msun)} may be able to
accrete cold matter directly from the intergalactic medium (IGM\added{, \citealt{cattaneo2006}}); if that gas
can stay cool as it falls to the center of the halo, it may fuel star
formation relatively directly \citep[e.g.,][]{birnboim2003, keres2005, db2006,
stewart2011}. (In this context ``cold'' or ``cool'' is used to describe gas at
$T < T_{\rm vir}$, although we will typically use it to describe dense,
photoionized gas of order $T \sim 10^4$ K.) However, cool material accreting
onto higher-mass galaxies is thought to encounter strong accretion shocks,
heating the gas to $\sim T_{\rm vir}$
\citep[][]{db2006,ocvirk2008,crain2010,correa2018}. Fueling star formation with
matter accreted in this way requires the gas be able to cool again (likely through a
cooling instability) and survive a fall into the central regions of a galaxy to form stars
\citep[][]{mb2004,mccourt2012,voit2015}. The prevailing picture is that this
``hot-mode accretion'' involving cooling from the hot CGM is relatively
inefficient, especially when confronted by processes that suppress the cooling
such as feedback from supermassive blackholes in these galaxies and ionization
by the ultraviolet background \citep{oppenheimer2013,nelson2015}. It is this
inefficiency that ultimately keeps the star formation rate in massive galaxies
low.

Measuring the properties of cold gas in the CGM about massive galaxies offers a
way to directly assess several of the assumptions inherent in this picture. Cold
gas in massive halos has been traditionally expected to be relatively rare \added{(i.e., low volume filling factors, low covering factors)}. Such
gas could arise from expulsion or stripping of gas from satellites or from
cooling instabilities in the hot CGM itself. Both scenarios would produce
relatively metal-rich cold gas and be found preferentially in the inner CGM,
where the gas is denser and metallicities are higher (the latter important for the
cooling process). Generally, any pristine gas accreted onto massive halos should
be shock-heated and unable to cool until it is mixed with more metal-rich
material \citep{wiersma2009}. Thus, we expect very little in the way of cold,
metal-poor gas in the CGM of massive galaxies, in contrast to what is found more
generally \citep[e.g.,][]{lehner2013, lehner2016, wotta2016, wotta2019}.

There are hints that this picture is not so simple (see \citealt{chen2017} for a recent
review of cold gas in the CGM of massive galaxies). For example,
\cite{thom2012} found no discernible difference between the cold \HI\ content in
the CGM of quiescent galaxies \added{($\log$ sSFR $\le -11$ \added{[yr$^{-1}$]} and $10^{10.5} \le M_\star \le 10^{11.5}$ \msun)} and star-forming galaxies \added{($\log$ sSFR $> -11$ \added{[yr$^{-1}$]} and $10^{9.5} \le M_\star \le 10^{11.1}$ \msun) at $z \sim 0.2$} in the COS-Halos survey
\citep[for a recent update, see][hereafter
\citetalias{prochaska2017}]{prochaska2017}. Thus, it appears that the CGM of
these quiescent, \added{high-mass} galaxies has a significant component of cold \HI\ that is not
being used to fuel star formation \citep{tumlinson2013}. \added{We still lack a thorough understanding of why these quiescent systems are not using this gas to form stars nor of what the origins of this gas are. Thus, further study of the mass-dependence of the cold \HI\ content of galaxies is warranted.}

Several recent studies have probed the cold gas content of massive luminous
red galaxies (LRGs) at $z \la 1$ \citep[e.g.,][]{gauthier2009, lundgren2009,
chen2010, bowen2011, zhu2014, perez-rafols2015, huang2016, chen2018,
smailagic2018, zahedy2019}. \added{LRGs are high-mass ($M_\star \ga 10^{11}$ \msun) quiescent galaxies with old stellar
populations that have been selected via color (to select quiescent, high redshift galaxies) and magnitude (to select massive, $\ga$3L* systems) cuts \citep{eisenstein2001}. They have been passively evolving since $z\sim 1$ \citep{banerji2010}. These characteristics make LRGs particularly interesting to study the physics of the baryon cycle because they extend the mass scale to $>$L* (the COS-Halos survey is $\sim$L*), and their passivity ensures we are characterizing the CGM around galaxies that are at the end of their star-forming life cycle.} Given their importance for cosmological studies
\citep[e.g.,][]{eisenstein2005, xu2013, slepian2017}, there are large samples of
LRGs with both SDSS photometry and spectroscopy
\citep{dawson2013,dawson2016,prakash2016,albareti2017}. \added{Due to these SDSS surveys, the
general characteristics of LRGs are well-understood. (Luminous blue galaxies and blue cluster galaxies would also
be good objects to test accretion predictions, but there currently is not a large,
uniformly-selected sample to draw from.)}

Significant metal-line absorption from the CGM of LRGs has been found in several studies, hinting at a prevalence
of metal-enriched cool gas. \citet{bouche2004} found a high cross-correlation
amplitude between \MgII\ absorbers seen against QSOs and LRGs, implying LRG
halos house cold, metal-enriched gas. This result was confirmed by
\citet{gauthier2009} and \citet{lundgren2009} with larger samples. Using
stacked QSO absorption lines to study weaker \MgII\ absorption,
\citet{zhu2014} and \citet{perez-rafols2015} again found \MgII\ absorbers are
correlated with LRGs out to 10 Mpc \citep[a result confirmed recently
by][]{lan2018}. Thus, there is a clear statistical correlation of metal-enriched
cold gas with LRG halos.

Searches for individual strong \MgII\ absorbers associated with individual LRGs
have yielded high covering factor estimates of cold, metal-enriched gas around
LRGs using absorption lines toward background QSOs \citep{chen2010a, bowen2011,
gc2011, huang2016}. For example, \citet{huang2016} found a covering factor
($f_c$) for impact parameters $\rho \le 120$ kpc of $f_c(\rho \leq 120 {\rm \,
kpc}) > 0.15$, a value that falls to $\sim0.05$ for $\rho \leq 500$ kpc. All of
these surveys indicate an abundance of cool, metal-enriched gas in the CGM of
LRGs. However, they have limited diagnostic power. In order to obtain larger
samples of galaxies with QSOs projected at small impact parameters, these works
utilize low-resolution spectroscopy of \MgII, giving sensitivity only to the
strongest absorbers (typically at the level of $W_r(\mbox{\MgII\ 2796}) \ga 0.3$
\AA). While the presence of strong \MgII\ tells us there is metal-enriched gas
about LRGs, it gives us no information on whether metal-poor gas that may trace
new accretion can survive deep in the halos of these massive galaxies.

In order to address these shortcomings, we study the \HI\ content within $\rho
\la 500$ kpc ($\sim$\Rvir) of a sample of massive LRGs using archival
ultraviolet (UV) spectroscopy from the {\em Hubble Space Telescope} (\hst). The
galaxies targeted by our Red Dead Redemption (RDR) survey all have masses in
excess of the predicted critical masses marking the transition between galaxies
thought to obtain their gas via ``cold-mode'' accretion and those acquiring
their gas via ``hot-mode'' accretion
\citep[e.g., $M_h = 10^{12}$ \msun,][]{birnboim2003,db2006,ocvirk2008,nelson2018}. Here we aim to test
whether LRGs and other massive galaxies show a deficit of cold, dense \HI\
compared with lower-mass, star-forming galaxies and to assess the frequency of gas that may represent recent accretion (assessed through its metallicity) in these halos. During the preparation of our survey, a complementary
work by \citet[][hereafter \citetalias{chen2018}]{chen2018} became available.
Their work focused on cool gas in the inner regions of a sample of massive
galaxy halos (their sightlines probe $\rho \la 160$ kpc while ours extend to the full impact parameter range within LRG halos
$\rho \la 500$ kpc). Their findings of a significant covering factor of
optically-thick \HI\ are in agreement with ours, and we combine the samples in
later sections of this paper.

Our paper is organized as follows. We describe our selection of the RDR sample of LRGs with \hst\ UV spectroscopy in \S~\ref{sec:sample} and summarize our data reduction and spectral analysis in \S~\ref{sec:methods}. We consider the column density and covering factors of \HI\ absorbers with impact parameters about the RDR LRGs in \S~\ref{sec:general}. We measure the metallicity of the cool absorbers to constrain their origins in \S~\ref{sec:metallicity}. We explore the implications of our results in \S~\ref{sec:discussion}. Our main results are summarized in \S~\ref{sec:conclusion}. A companion paper, \citet[][hereafter \citetalias{howk2019}]{howk2019} focuses on the highly-ionized phase of the CGM about the RDR LRGs as traced by \OVI.

Throughout this paper we adopt the cosmology from \citet{planck2016}, notably $H_0 = 67.7$ \hub, $\Omega_m (z=0) = 0.309$, and $\Omega_\Lambda(z=0) = 0.691$. We adopt a naming scheme for objects as follows: target galaxies are labelled as ``LRG'' followed by their SDSS (J2000) coordinate designation; QSO targets are listed by their SDSS coordinate name. We will be studying absorption from partial Lyman limit systems (pLLSs) and Lyman limit systems (LLSs) in this work. We adopt definitions of pLLSs as absorbers with \HI\ column densities $16.0 \le \logNHI < 17.2$; LLSs are defined by $17.2 \le \logNHI < 19.0$. We generally refer to absorption systems with $\logNHI \ge 16.0$ as ``strong \HI\ absorbers.''



\section{Sample Selection and Galaxy Properties} \label{sec:sample}


\subsection{Sample Selection}\label{sec:samselec}

In this work we compile a sample of QSO-LRG pairs for which the QSOs have been
observed in the UV by \hst. We selected the LRGs from the SDSS DR13
\citep{albareti2017} LOWZ and CMASS samples of the BOSS and eBOSS surveys
\citep[][with additional information on the eBOSS target selection in
\citealt{prakash2016}]{dawson2013, dawson2016}. The LOWZ sample targets massive
galaxies in the range $0.15 \la z \la 0.43$, whereas the CMASS sample focuses on
$0.43 \la z \la 0.7$ \citep{dawson2013}. \added{The CMASS sample includes massive blue galaxies, while the LOWZ sample does not.} Hereafter, we refer to the galaxies as
``LRGs," even if the \added{CMASS} sample is broader than the traditional LRG definition of
\citet{eisenstein2001}. This includes the late-type galaxies that may make up
$\sim25\%$ of the sample \citep{masters2011}\added{ and may also be in our final RDR sample}. The most
important characteristic of these SDSS samples is that they select
high-mass galaxies, with a mean stellar mass of $\log M_\star \approx 11.3 \pm 0.5$
\citep[e.g.,][]{maraston2013}.\footnote{Unless otherwise stated, all masses in
this work are given in physical solar masses, \msun, with all cosmological
corrections included. We adopt $h \equiv H_0 / (100\, {\rm km \, s^{-1} \,
Mpc^{-1}}) = 0.677$ throughout.}

The LRGs from which we draw our sample all have spectroscopic redshifts
available. We further restrict our LRG selection to the redshift range $0.26 \le
\zlrg \le 1.25$. The minimum redshift is adopted to place the Lyman
break in the UV spectral range accessible to \hst; the Lyman break is usually key to
accurately assessing the total \HI\ column density, \NHI, in high column density
systems that are not damped. The maximum redshift is chosen so
that a UV selection of QSOs can be made based on GALEX FUV magnitudes (see
below). The typical redshift errors from the SDSS pipeline in this sample are
$\sigma(\zlrg) / (1+\zlrg) \approx 10^{-4}$, which corresponds to $\sim$30 \kms.

We drew QSOs from the DR7 QSO catalog of SDSS \citep{schneider2010} cross-matched with GALEX UV
sources \citep{martin2005}. We selected QSOs projected within 500 kpc of the LRGs,
requiring $0.3 \le z_{\rm em} \le 1.5$ and $0.1 \le z_{\rm em} - \zlrg \le 0.5$.
The low-redshift cut-off simply ensures the QSOs are at redshifts higher than
our target LRGs. The high-redshift cut is made in order to avoid the confusing
influence of the dense Lyman-$\alpha$ forest at higher redshifts. The redshift
separation constraint minimizes the contamination from unrelated absorption that
might contaminate the signature of the LRGs' CGM. We also apply a constraint to
the GALEX photometry of the QSOs, $FUV \le 21.0$ in order to ensure \added{the objects are bright enough to produce} good S/N in
even the low-resolution observations \added{(typical S/N for our observations are 8 for COS/G130M,G160M, 5 for COS/G140L, and 14 for FOS/G190H)}.\footnote{These S/N values are per pixel and describe the spectrum as a whole \citep{stoehr2008}.} These constraints yield $\sim$500 QSO-LRG
pairs projected within $\rho \le 500$ kpc.

We used this sample as the basis of a search of the Mikulski Archive for Space Telescopes (MAST) for \hst\ UV spectra of the background QSOs, searching for data from the Goddard High Resolution Spectrograph, the Faint Object Spectrograph, the Space Telescope Imaging Spectrograph, and the Cosmic Origins Spectrograph. Because we are interested in an unbiased search for \HI\ (in particular) associated with LRGs, we excluded any data from program 14171 (PI: Zhu) that targeted (some) LRGs on the
basis of previously-identified \MgII\ absorption. This search yielded 24 sight
lines with UV spectral coverage of the Lyman series lines at the redshift of the
LRGs. Three of these LRGs were later removed from our sample. In one case, the QSO
was initially targeted on the basis of a strong foreground absorber that resides at the
redshift of the LRG. In the second, the galaxy was an outlier in stellar mass from
the rest of the sample (with $\logMstar < 11$), and the data for the third sightline were low enough resolution that we could not definitively determine if the absorption feature is associated with the LRG.

The remaining 21 QSO-LRG pairs listed in Table~\ref{tab:objinfo} form the basis of our study. The 21 LRGs along these
sightlines lie in the range $0.28 \le \zlrg \le 0.61$. The median sample redshift is $\zlrg = 0.46$. To the best of our
knowledge, based on a reading of the publicly-available program abstracts in
MAST, none of the background QSOs were targeted due to the presence of the
foreground LRGs in our sample. (Several of the QSOs were targeted to study other foreground galaxies, but this does not affect our results.) Even with this selection, three sightlines do not have coverage of the Lyman break due to the choice of grating. These LRGs are all probed by Lyman series lines that give us the ability to place strong constraints on the absorption, so we have included them in our sample.

\added{Compared to \citetalias{chen2018}, our LRG selection is more restrictive. They have used all of SDSS to search for their spectroscopically confirmed LRGs, while we have stayed within the CMASS and LOWZ samples. Their sample has comparable redshift and stellar mass distributions to ours (see \S~\ref{sec:galprop}), but our LRGs have a higher median color $(u-g) = 2.03$ and $(g-i) = 1.20$. We note the $u$-band magnitudes have large uncertainties.}


\subsection{Galaxy Properties}\label{sec:galprop}

\begin{figure*}
    \epsscale{1}
    \plotone{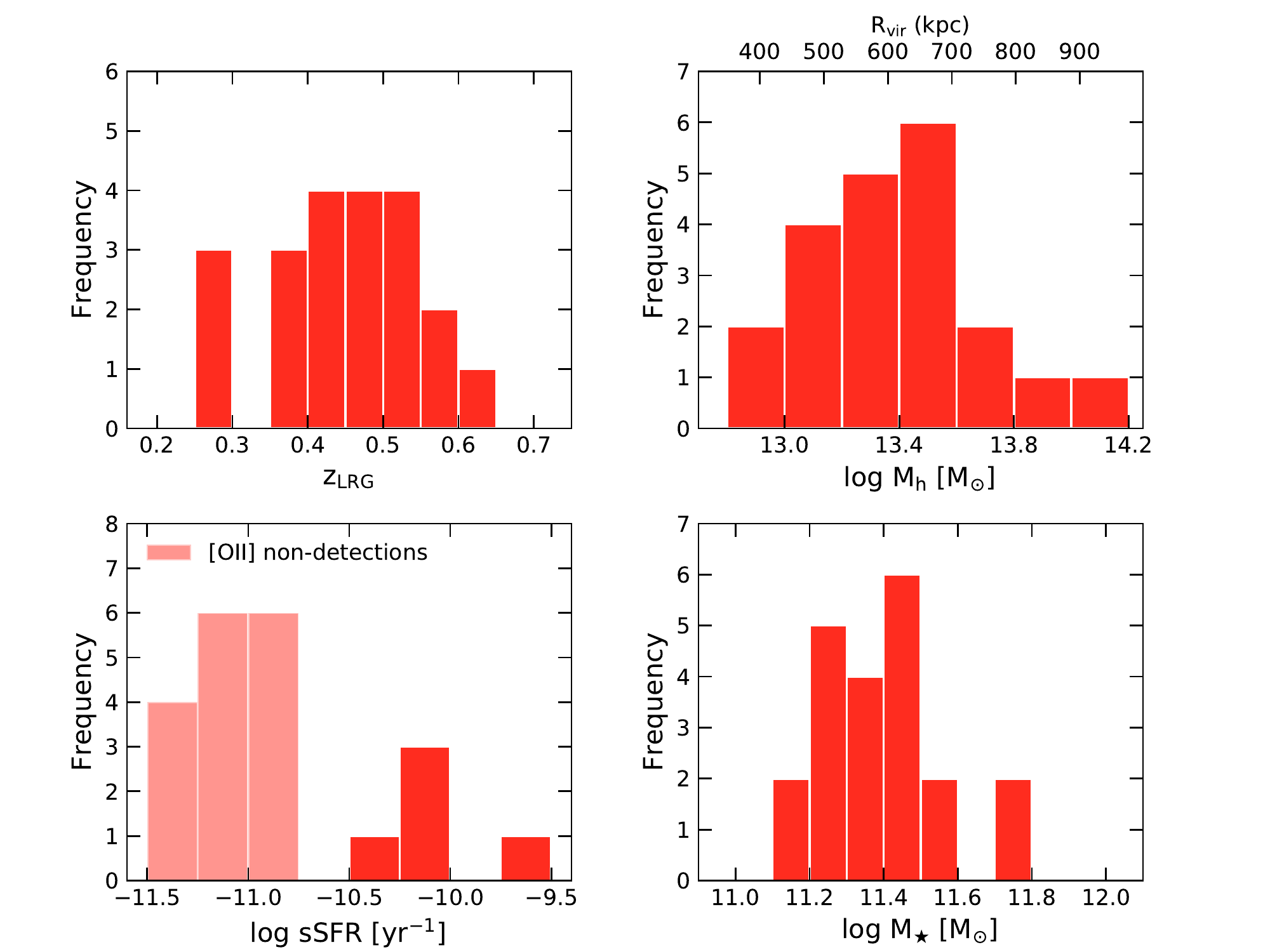}
    \caption{Distributions of \zlrg, $\log M_{\rm h}$, $\log$ sSFR, and $\log M_\star$ for the RDR sample of LRGs. The redshifts come from the eBOSS and BOSS SDSS spectroscopic surveys. We primarily use \OII\ to calculate the SFR, \kcorrect\ to derive $M_\star$ \citep{blanton2007}, and the SMHM relation from \citet{rodriguez-puebla2017} to calculate $M_{\rm h}$. For \OII\ non-detections, we adopt the SFR from the median stack of the LRG spectra showing no detectable \OII\ emission in calculating the sSFRs. The sSFRs should be considered upper limits (see discussion in the text).} \label{fig:LRGdistributions}
\end{figure*}

\begin{figure}
    \epsscale{1.1}
    \plotone{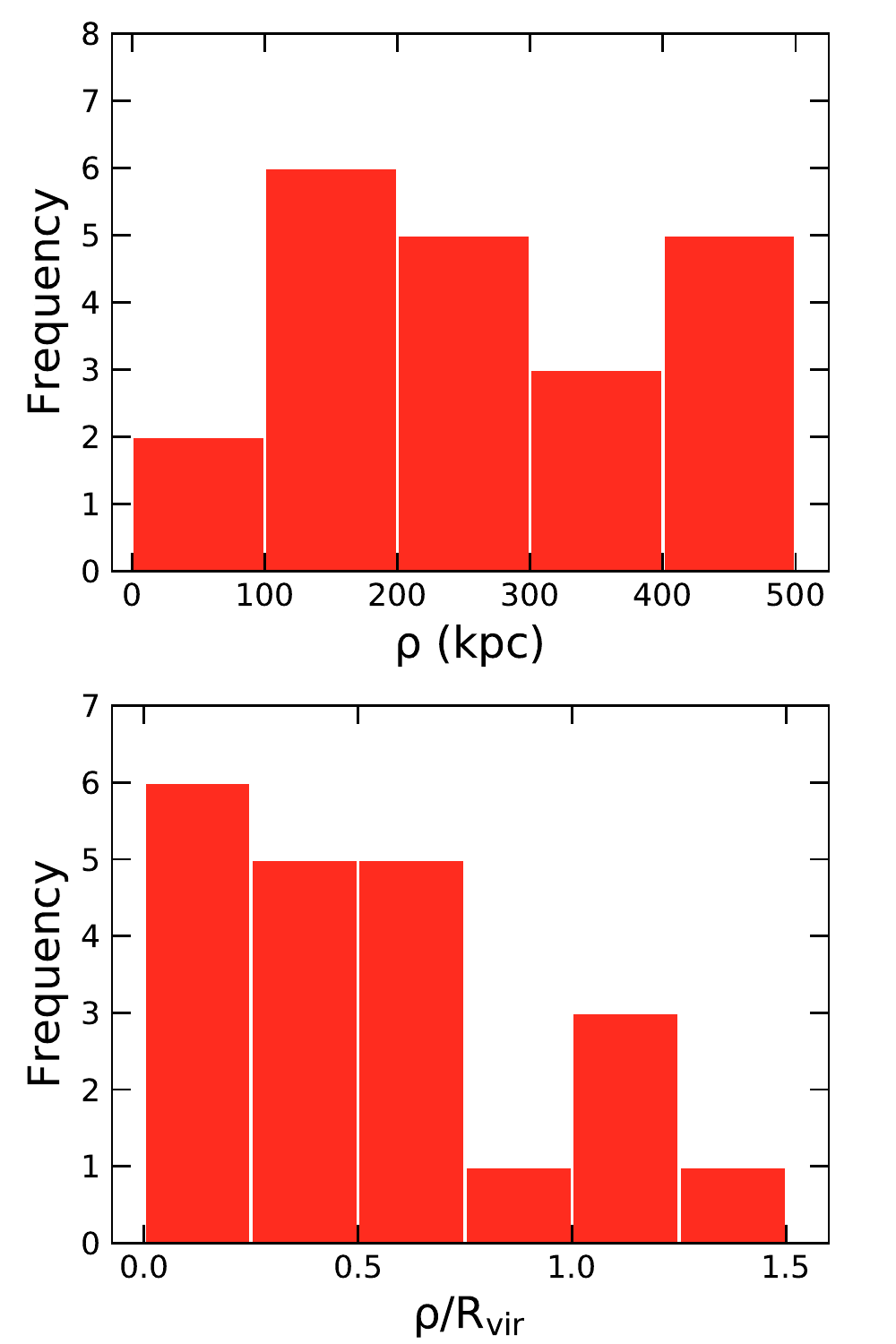}
\caption{Distributions of impact parameter, $\rho$ (top), and normalized impact
parameters, $\rho / \Rvir$ (bottom), for the QSO-LRG pairs in the RDR sample.}
\label{fig:rho_distributions}
\end{figure}

Table~\ref{tab:lrgs} summarizes the properties of the LRGs in the RDR sample. While
there are galaxies in our final sample potentially hosting star formation ($\sim$20\% show detectable
\OII\ emission), the defining characteristics of the galaxies are their redshift
range and large masses. The redshifts in Table~\ref{tab:lrgs} are adopted from
the SDSS pipeline for each LRG. The median redshift of our sample is $\zlrg = 0.46$, while the median impact
parameter is $\rho = 261$ kpc. The distributions of redshifts and several other
properties of our sample (described below) are shown in
Figures~\ref{fig:LRGdistributions} and \ref{fig:rho_distributions}.

For each target galaxy we estimate absolute \sdssr -band
magnitudes and stellar masses using the \kcorrect\ code (v4\_3)\footnote{Available through
\url{http://kcorrect.org} or \url{http://github.com/blanton144/kcorrect}.} of
\citet{blanton2007}. These SED fits make use of spectral templates from
\citet{bruzual2003} and assume the initial mass function of
\citet{chabrier2003}. We used Galactic extinction-corrected \citep{schlegel1998}
{\tt MODEL} magnitudes with the pipeline redshifts as inputs to \kcorrect\ to assess the $k$-corrections
and mass-to-light ratios of the galaxies. The results are given in
Table~\ref{tab:lrgs}. The output masses from the code are given in $\msun
h^{-1}$, which we convert to masses with units \msun\ in Table~\ref{tab:lrgs}
and all that follows. These masses likely have an uncertainty of at least
$\pm0.3$ dex, if not somewhat larger \citep{conroy2013}. Although we have not specifically fit the
full spectra for all of these galaxies, the redshifts are fixed and the galaxy
templates are well determined for these systems. We show the distribution of
stellar masses for our sample in Figure~\ref{fig:LRGdistributions} (bottom right panel). The galaxies
in our sample have a median stellar mass $\log M_\star = 11.4$ (in \msun) with a
standard deviation of 0.2 dex, both consistent with the sample studied by
\citet{huang2016}.

We estimate the halo mass, $M_{\rm h}$, of each LRG using the stellar
mass--halo mass (SMHM) relationship of \citet{rodriguez-puebla2017}, who combine
a large number of observational studies of the SMHM scalings. In particular,
they provide a prescription (their Equation 66) for assessing the mean halo mass
at a given stellar mass, $\langle \log M_{\rm h} (M_\star) \rangle$ \citep[which is
different than the inverse of the stellar mass-to-halo mass ratios given the
asymmetries in the scatter about the mean relationship;][]{behroozi2010}.
Recently \citet{tinker2017} have assessed the SMHM relationship in a sample of
CMASS galaxies drawn in much the same way as our sample, focusing on $\log
M_\star \ga 11.4$, though they extend their fits to somewhat lower masses. Their
predictions for $\langle \log M_{\rm h} (M_\star) \rangle$ yield halo masses lower by
roughly 0.3 dex (i.e., a factor of $\sim$2) at a given stellar mass than those of
\citet{rodriguez-puebla2017}. There are a number of other recent studies on the
SMHM relation at high stellar masses \citep[e.g.,][]{velander2014,shan2017}. We
adopt the \citet{rodriguez-puebla2017} results both because they account
for the asymmetric scatter in the $M_\star-M_{\rm h}$ relationship in order to
properly calculate $\langle \log M_{\rm h} (M_\star) \rangle$ and because their
analysis allows us to calculate the masses of the RDR LRGs and lower-mass
galaxies against which we will compare the LRGs with a consistent treatment. Our
final halo mass estimates are given in Table~\ref{tab:lrgs}; the median halo
mass derived for our sample is $\log M_{\rm h} = 13.4$ (in \msun).

For comparison with other galaxy absorber studies, it can be useful to consider
the impact parameters relative to the virial radii of each galaxy. We define the ``virial radius'' as $\Rvir = R_{200}$, the radius enclosing the halo mass for a mean density $200\times$ that of the critical density at the
redshift of each LRG. That is, $\Rvir = R_{200} \equiv (3 M_{\rm h} / 4 \pi \, \Delta \,
\rho_c)^{1/3}$, where $\Delta = 200$ and $\rho_c$ is the critical density. We
refer to this scale simply as \Rvir\ throughout this paper. Our estimates of \Rvir\
are given in Table~\ref{tab:lrgs}. Given the $M_{\rm h}^{1/3}$ dependence of these
values, they are not strongly affected by our choice of SMHM relation. Our
calculations give a median virial radius $\Rvir = 516$ kpc; the median
normalized impact parameter of our sightlines is $\rho / \Rvir =
0.44$.

To limit the star formation rates (SFRs) of the RDR sample, we use \OII\ 3727+3729 emission as an indicator of star formation in these galaxies. These results should be considered upper limits given the potential for non-star formation contributions to these lines \citep{huang2016}. We adopt the calibration of the \OII -SFR relationship from \citet{moustakas2006} for the highest-luminosity galaxies (which are appropriate
given the range in absolute magnitudes of our sample; see Table~\ref{tab:lrgs}).
Thus we assume $\log {\rm SFR \, (\msunyr) } = -40.24 + \log L_{\rm [O\, II]}\,
({\rm ergs\ s}^{-1})$. We adopt the SDSS pipeline fits to \OII\ emission from
the LRGs, correcting for Milky Way foreground extinction assuming
\citet{schlegel1998} extinction with $R_V = 3.1$ and an extinction curve
following \citet{cardelli1989}. We do not correct for internal extinction, as neither do \citet{moustakas2006} in their calibration, which partly causes the luminosity dependence in their \OII -SFR relationship. Where galaxies do not have detectable \OII\ emission, we adopt the
$2\sigma$ upper limits on the flux.

In two cases the SDSS fits gave uncertainties well above ($\ga 10\times$ higher)
those typical of the sample. In one case, it was not clear why the fits yielded such
high uncertainties; in the other, the \OII\ doublet was coincident with a
poorly-subtracted sky emission line. In these two cases, we used the {\tt pPXF}
spectral fitting code of \citet{cappellari2017} to refit the SDSS spectra. Our
fit for LRG SDSSJ125859.98+413128.3 yielded tighter limits on the \OII\ emission.
For LRG SDSSJ075217.92+273835.6, where the \OII\ line is contaminated, we had
good measurements for the fluxes of H$\beta$ and H$\alpha$. The limit on H$\beta$ is
stronger (as H$\alpha$ is in the forest of sky lines in the far red). For only
this galaxy we use the H$\beta$ line as an indicator of the galaxy's SFR,
adopting the H$\beta$-SFR calibration of $\log {\rm SFR \, (\msunyr) } = -40.11
+ \log L_{\rm H\beta}\, ({\rm ergs\ s}^{-1})$ from \citet{moustakas2006}.

The specific star formation rates, ${\rm sSFR} \equiv {\rm SFR}/M_\star$, for the RDR galaxies are
given in Table~\ref{tab:lrgs}. Because the SDSS fibers used to limit the SFR do
not encompass all of the stellar light from the galaxy, we compare our derived
star formation rates to the stellar masses contained within the fiber. We use
the ratio of the SDSS {\tt FIBER2FLUX} to the {\tt MODELFLUX} in the $r$-band to scale down
our derived stellar masses for this comparison. In most cases,
non-detections of \OII\ do not set stringent constraints on the sSFR. In
Figure~\ref{fig:LRGdistributions} we show the sSFRs calculated using the SFR
derived from a luminosity-weighted median stack of the LRG spectra showing no
detectable \OII\ emission, ${\rm SFR} = 0.49\pm0.23$ \msunyr. Thus even the
stacked spectrum of these galaxies gives only a marginal detection of \OII\ emission.

Even amongst the galaxies with detectable \OII\ emission, it is not necessarily connected to star formation. The \OII\ and \OIII\ emission of LRGs generally resembles
that of low-ionization nuclear emission-line region (LINER)-like galaxies
\citep{huang2016}, where the emission is suspected to come from either activity
from active galactic nuclei (AGN) or ionization from post-asymptotic giant
branch stars (e.g., \citealt{ho2008,yan2012}). \citet{belfiore2016} have shown more generally that LINER emission does not necessarily originate from the nucleus, but from spatially extended regions in both star-forming and passive galaxies. Furthermore, stacking of FIRST radio images suggests nearly
all LRGs house an AGN \citep{hodge2008,hodge2009}. In light of these considerations, all sSFRs should be considered upper limits.

For the most part we do not have extensive data on the environments of these specific galaxies. We expect all will be surrounded by an extensive suite of lower-mass satellites. At lower redshift ($0.16 < z < 0.33$), $\sim$70\% of LRGs are central galaxies in clusters with richness parameter $\lambda > 20$ in the redMaPPer cluster catalog (\citealt{rykoff2014,rykoff2016,hoshino2015}). We cross-matched our RDR sample of galaxies with the redMaPPer cluster catalog version 6.3\footnote{Available through \url{http://risa.stanford.edu/redmapper/}.} and found five matches. The LRGs SDSSJ124307.36+353926.3, SDSSJ141307.39+091956.7, SDSSJ171651.46+302649.0, SDSSJ075217.92+273835.6, and SDSSJ132457.98+271742.6 are all within $< 0.2"$ of the cluster center, indicating that they are either the central galaxy or the brightest cluster galaxy. LRGs SDSSJ124307.36+353926.3, SDSSJ141307.39+091956.7,
and SDSSJ075217.92+273835.6 are found in rich clusters ($\lambda > 20$), while
the other two are found in clusters with a smaller number of member galaxies ($\lambda \sim 10$).
The fact that the other galaxies are not specifically identified with overdensities like those in the redMaPPer catalog does not mean they lack satellites. \added{A detailed search and characterization of the LRG environment is beyond the scope of this paper. However, \citet{chen2019} characterized the environment around one of the \citetalias{chen2018} LRGs. Using integral field spectroscopy, they determined the LRG is located within a group of galaxies with a range of $10^7 \le M_\star \le 10^{11}$ M$_\odot$. We expect this is the typical environment for our sample as well.}

\added{It has also been shown that brightest cluster galaxies have extended stellar halos \citep{huang2018}. If the RDR LRGs exhibit such extended stellar halos, our derived stellar masses -- as well as halo masses and virial radii -- will be underestimated due to light being missed by the apertures. This does not effect our goal to observe the CGM about high-mass galaxies, but it affects the context of our sightlines, as we discuss their normalized impact parameter, $\rho / \Rvir$, throughout this paper. \citet{huang2018} determined the stellar mass could be underestimated by at most 0.2 dex when using the SDSS {\tt CMODEL} magnitudes to fit galaxy SEDs. In our analysis we utilized the {\tt MODEL} magnitudes, and the stellar mass estimates we derived using \kcorrect\ carry a $\pm$0.3 dex uncertainty. The error in the mass estimate due to missed stellar light in the aperture is within the uncertainty of our mass estimates. If our virial radii are underestimated, the pLLSs/LLSs we detect would be located at even lower normalized impact parameters. These issues neither change fundamentally our results nor our discussion of the origin and fate of the gas we detect.}

\section{Observations and Data Analysis}\label{sec:methods}

Our RDR sample comprises 21 QSOs with \hst\ UV spectroscopy that pierce the CGM of 21 LRGs within $\rho \la 500$ kpc. The use of UV spectroscopy allows us to identify the prevalence of \HI\ and associated metal ions in the CGM of these galaxies. We supplement these UV data with optical spectroscopy to measure \MgII\ associated with the LRGs. Below we describe the datasets we have used and our approach to measuring absorption from the LRG CGM.

\subsection{Ultraviolet Spectroscopy from HST}

A summary of the QSO observations is provided in Table~\ref{tab:qsoobs}. Seventeen of the QSOs were observed with the medium resolution mode ($R\sim 18,000$) of COS using the G130M and/or G160M gratings. The other five observations were obtained with low resolution ($R\simeq 1,300 - 2,000$) UV spectra from COS and FOS. All of the COS spectra were retrieved from the \hst\ Spectroscopic Legacy Archive (HSLA, \citealt{peeples2017}) except for SDSSJ112756.76+115427.1, SDSSJ124307.57+353907.1, and SDSSJ150527.60+294718.3, which were not available in the April 2017 data release. For these spectra, we use the routines in COS Tools \citep{danforth2010,danforth2016} to coadd the data. Differences exist between the two data reduction processes, but they do not impact our column density measurements \citep{lehner2018}. For the objects observed with FOS, we use the reduced spectra from \citet{ribaudo2011}.\footnote{Available through \href{http://vizier.u-strasbg.fr/viz-bin/VizieR?-source=J/ApJ/736/42}{Vizier}.} One object, SDSSJ125224.99+291321.1, was observed with both COS and FOS, with a gap of $\sim$140 \AA\ between the two spectral regions. The spectrum and analysis of SDSSJ171654.20+302701.4 is described in \citet{wotta2016}. We make use of their data and column densities for this sightline.

\subsection{Supporting Ground-based Observations}

Previous surveys have used \MgII\ to understand the presence of metal-enriched gas in LRG halos (e.g., \citealt{bowen2011,zhu2014,huang2016}). We also measure \MgII\ for several of our sightlines through LRG halos because \MgII\ often provides the best metallicity constraints for pLLSs and LLSs \citep{wotta2016, wotta2019}. High resolution spectroscopy yields the best constraints on the column density; two QSOs have spectroscopy from the High Resolution Echelle Spectrometer (HIRES) on Keck I in the Keck Observatory Archive. The spectra are reduced following \citet{lehner2016}, and one of these sightlines is reported in \citet{wotta2016}. \deleted{A second sightline was observed with the Multi-Object Double Spectrograph (MODS) on the Large Binocular Telescope (LBT). \citet{wotta2016} details the reduction procedure for this instrument.} We also analyze the SDSS spectra of the QSOs when available (four sightlines do not have \MgII\ coverage, \citealt{abolfathi2018}). Due to the limited sensitivity and resolution of SDSS (and the low \HI\ column densities along most of the sightlines), these columns are almost all upper limits.

\subsection{Absorption Line Search Methodology}

To find absorbing gas associated with the RDR LRGs, we conduct a search for hydrogen and metal-ion absorption in the QSO spectrum at the redshift of the LRG using a search window of $\pm1000$ \kms. Our choice of this large velocity search window follows from the escape velocity of the median LRG halo from our survey ($\sim$930 \kms\ assuming a Navarro-Frank-White potential, \citealt{nfw1996,nfw1997}). \added{In the absence of an AGN, winds from a galaxy with $M_h \ge 10^{13}$ M$_\odot$ will not be able to escape the potential well \citep{sharma2013}. We also consider} the \MgII\ survey in the CGM of LRGs by \citet[][see their Figure 4]{huang2016}. They found only 12\% of their absorbers lie at a velocity offset greater than $\pm 500$ \kms\ when considering impact parameters $<500$ kpc. However, they also found a few systems at $\pm 1000$ \kms, so we adopt the same window to ensure we consider all the gas that could be associated with each LRG.

When \HI\ absorption is detected within $\pm 1000$ \kms\ of the redshift of the LRG, we set the redshift of the absorbing gas using the Lyman series transitions. If a break in the QSO spectrum is observed, the redshift of the pLLS/LLS is determined from the centroids of the higher Lyman series lines. If a break is not present (due to spectral coverage or lower \HI\ column), we use Ly$\alpha$ or Ly$\beta$. In cases where no absorption is detected, we use the redshift of the LRG to set the center of the velocity integration window to estimate upper limits on the column densities.

\subsection{Column Density Measurements}\label{sec:coldens}

Our measurements of \HI\ and ion column densities are made with different techniques or assumptions depending on the column density of the system and the quality of the data. Due to the heterogeneous mix of resolution and signal-to-noise (S/N), our limits on \HI\ or metal absorption are not uniform across the sample. However, our focus in this work is on understanding the amount of high \HI\ column density gas present about LRGs. For this purpose, the mixture of data properties is not an issue because the Lyman break is covered in almost every sightline. Two of the three sightlines that do not have Lyman break coverage do not exhibit high \HI\ column absorbers; the final sightline can only be given a lower limit for the \HI\ column density.

Four of the LRGs show strong \HI\ systems ($\logNHI \ge 16.0$). Of these four systems, three have observable breaks in the QSO spectrum at the Lyman limit. The \HI\ column density can be determined from the break using $\NHI = \tau_{LL}/\sigma_{LL}$, where $\tau_{LL}$ is the optical depth at the Lyman limit, and $\sigma_{LL}$ is the absorption cross section of the hydrogen atom at the Lyman limit \citep{spitzer1978}. A composite QSO model is scaled to fit the continuum of the spectrum \citep{telfer2002}\footnote{\added{While the \citet{telfer2002} model does not match all the specific features in a given QSO spectrum, the features in the EUV (the regime in which we are most interested) are not very prominent. This means the selection of the template does not have a large influence in the resultant fit. We also mitigate the differences in spectral index between templates by allowing our model to have a tip/tilt that accounts for these variances and possible calibration/dust effects.}}; a break is then added to match the level of the absorbed continuum giving a measure of the optical depth and hence \NHI. To estimate the errors on \NHI\, we follow the methodology described in \citet{wotta2016}. In short, the errors are estimated by offsetting the model to a point in which the model and data are inconsistent. This column density value is taken to be the $2\sigma$ value. \added{We have included an example error fit to illustrate this inconsistency in Appendix~\ref{appa}.}

We determine the \HI\ column density for the absorbers associated with LRGs SDSSJ111132.33+554712.8 and SDSSJ171651.46+302649.0 (reported in \citealt{wotta2016}) from the optical depth at the break, but we can only estimate a lower limit for the absorber associated with LRG SDSSJ111508.24+023752.7 from this method due to the low level of flux recovery blueward of the Lyman limit. \replaced{To further constrain the column density of the latter absorber, we perform a Voigt profile fit on the Lyman series lines using an updated version of the profile fitting code from \citet{fs1997} (see \citealt{lehner2018} for more detail). This technique models the absorption line (with the ability of adding more components) and determines the best-fit model through $\chi^2$ minimization. Using an initial guess for the column density, velocity dispersion (Doppler $b$ parameter, and velocity centroid, it returns the best-fit values of the model profile for these three parameters and can be iterated on. We fit the lines \HI\ $\lambda \lambda$1025, 972, 937, 926, 923, 920, 919, 918, 917, 916 in this manner (see Appendix~\ref{appa}). We use two components to model the \HI\ profiles since the three stronger transitions show an additional component with $\logNHI =15.18 \pm 0.04$ and $b = 31.8 \pm 3.6$ \kms, shifted by $-110$ \kms\ from the redshift of the LLS. The LLS is reasonably well fit with $\logNHI =18.12 \pm 0.24$ and $b = 27.4 \pm 1.5$ \kms, which is consistent with the lower limit derived just using the break at the Lyman limit, $\logNHI >17.9$ (see also \citealt{lehner2018}).}{\citet{lehner2018} modeled the \HI\ profiles of this absorber with a Voigt profile fit using two components. However, revisiting this fit using the information from the velocity profiles of the metal lines (and in particular newly acquired Keck/HIRES spectra of \MgII\ and \FeII), it is apparent that at least three components are needed at $-113$, $-19$ (strongest component), and $+28$ \kms, the latter being the additional component observed in the metal lines. With now a lack of information on both the positive and negative wings of Ly$\beta$ in the strongest component at $-19$ \kms, there is a degeneracy in the solutions owing to poorly constrained individual $b$ and $N$ values. Thus, the strongest component cannot be constrained to better than $\logNHI = [18.00,19.60]$. The lowest value is from the break at the Lyman limit. The highest value is from a Voigt profile fit with three components to the \HI\ Lyman series using an updated version of the profile fitting code from \citet{fs1997} and the tabulated model of the line-spread COS function.\footnote{Available at \url{http://www.stsci.edu/hst/cos/performance/spectral_resolution/}.} Any value above $\logNHI \simeq 19.6$ would produce too strong damping wings in Ly$\beta$ and too strong absorption in other weaker transitions. Though the allowed range of \logNHI\ overlaps with the canonical super-LLSs regime (SLLSs, $19.0 \le \logNHI < 20.3$), we will refer to this absorber as an LLS throughout the rest of the paper for simplicity.}

No strong break in the QSO spectrum is observed for the other high \HI\ column density system. The pLLS associated with LRG SDSSJ141540.01+163336.4 is on the low end of the pLLS column density range, exhibiting a small partial break in the QSO spectrum. We measure the \HI\ column density from the higher Lyman series lines using the apparent optical depth method (AODM, \citealt{SS1991}). The AODM converts an absorption line to apparent column density per unit velocity interval. The total apparent column density is then determined by direct integration of the apparent column density per unit velocity interval: $N_{a} = \int^{v_{2}}_{v_{1}} N_{a}(v)dv$. The column density $N$ is assumed to be $N_{a}$ as long as there is no evidence of saturation (see below for our treatment of saturation). We also measure \NHI\ from the break in the QSO spectrum at the Lyman limit for this absorber and find the column densities from the two methods are consistent. The \NHI\ value measured from the break exhibits larger errors, so we adopt the value measured from the Lyman series lines.

An absorber is detected in the spectrum of QSO SDSSJ125224.99+291321.1, which has been observed with COS and FOS. Unfortunately, these data do not cover the Lyman limit and higher Lyman series lines associated with this absorber. With only Ly$\alpha$ and Ly$\beta$, a curve-of-growth analysis did not provide any robust results. Using the AODM on Ly$\beta$ observed with FOS, we could place only a lower limit on \NHI\ ($\logNHI >15.2$).

For the remaining absorbers, the absorption is much weaker, and we use only the AODM to estimate the column densities or limits on the column densities of the \HI\ and metal lines. The metal ion absorption line integration limits are set by the width of the associated hydrogen lines unless the metal ion profile is broader, contaminated, or shifted. If a line is contaminated, we shorten the integration limits. When no absorption is detected (for hydrogen or metals) we measure $2\sigma$ upper limits using the AODM by integrating over the section of the spectrum where the line would be located. As stated above, the redshift of the LRG is used as the center of our integration window in these cases, and the velocity integration limits are set by the smallest line width ($\Delta$v) that can still be detected in the respective grating. For the low-resolution observations, this equates to nine pixels for FOS/G190H and 12 pixels for COS/G140L.

To check for saturation in lines that do not reach zero flux but have significant peak optical depth, we employ two procedures. If the transition is part of a doublet, we compare the column density values of the stronger line to the weaker line. When the weaker line has a higher column density, the \citet{SS1991} saturation correction is applied. For other transitions where there are no other lines in the same ionization state of a species to use for comparison, other species in the same state are considered (e.g., \CIII\ and \ensuremath{\mbox{\ion{Si}{3}}}). If one of these lines is saturated, we also mark the other transitions as saturated if the peak optical depth is similar or larger. \added{For COS/G130M,G160M observations, transitions that reach a normalized residual flux of 0.25 and lower are definitely saturated. We assume transitions in the COS/G140L and FOS/G190H observations are saturated unless there are other transitions that can be used to check this assumption.}

We include in Appendix~\ref{appa} plots of the \HI\ and ionic absorption profiles for each sightline to show the velocity range over which the transitions are integrated. Table~\ref{tab:iontable} details the apparent column densities measured for each sightline. If an absorber shows more than one component in the profiles, we determine the column densities for each component, in an effort to compare the component metallicities. Since we will show there is no significant difference between the component metallicities in these cases (see Appendix~\ref{appc}), we quote total column densities throughout the paper. The three spectra in which we observe a break at the Lyman limit of the pLLS/LLS are also provided in Appendix~\ref{appa}\deleted{, along with the Voigt profile fits to the Lyman series lines for the absorber that could only be given a lower limit for \logNHI\ from the break in the QSO spectrum}.

\section{Characteristics of Strong \HI\ Absorption in the CGM About LRGs}\label{sec:general}

\subsection{Hydrogen and Metal-Line Absorption}\label{sec:abs}

\begin{figure*}
    \epsscale{1.}
    \plotone{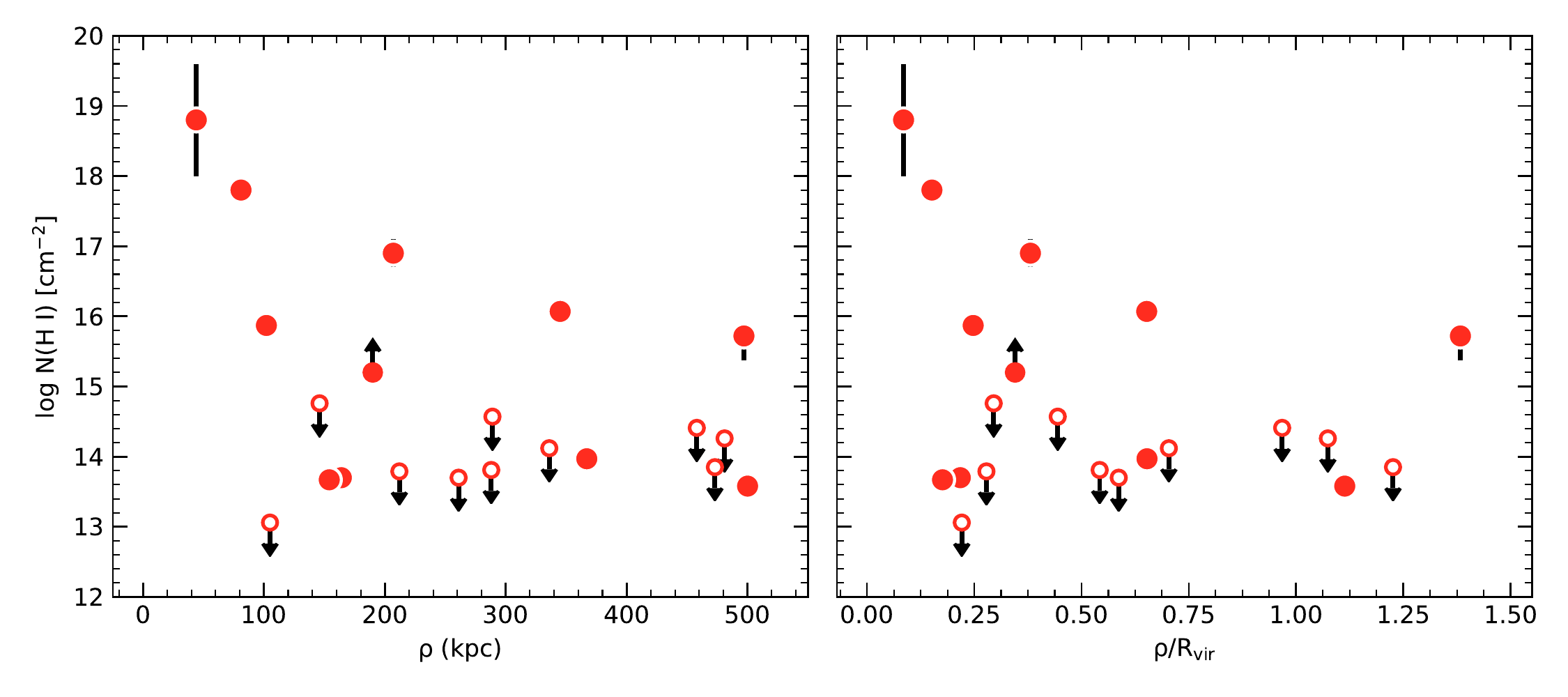}
    \caption{Distribution of \logNHI\ versus impact parameter (left) and normalized impact parameter (right) for the LRGs in the RDR sample. The strong \HI\ detections ($\logNHI \ga 16.0$) are preferentially found at low impact parameters, notably within $\rho \la \Rvir$. The detections of absorption with $\logNHI \le 14.5$ are consistent with interloping Ly$\alpha$ forest absorbers. We expected $\sim5$ such interlopers and observe five (only four data points are seen here, as one sightline has two such interlopers, and we have plotted the combined columns).}
    \label{fig:NHIrho}
\end{figure*}

\begin{figure}
    \epsscale{1.25}
    \plotone{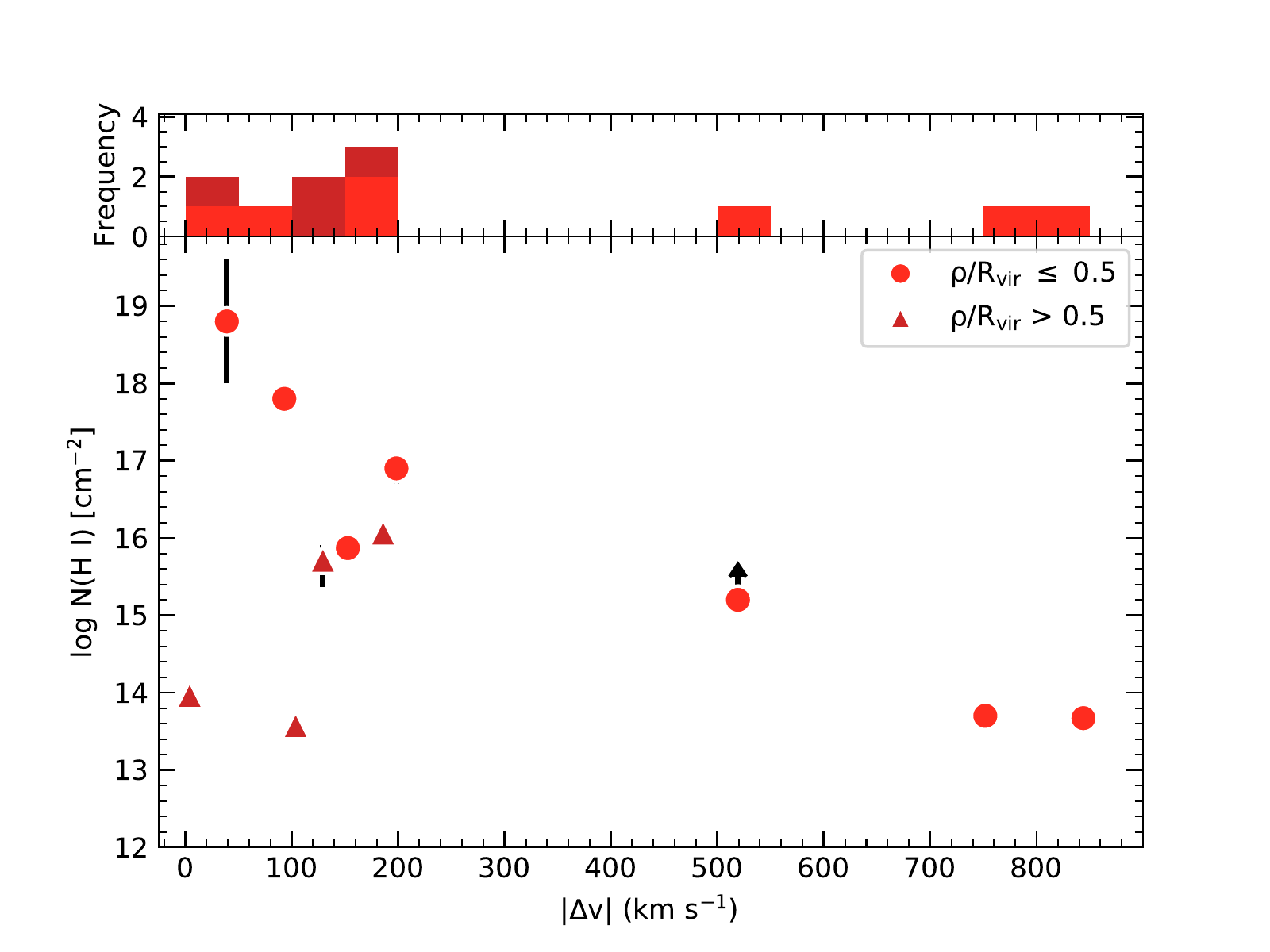}
    \caption{Distribution of \logNHI\ versus the absolute value of the absorber velocity offset relative to the LRG. We split the sample by normalized impact parameter and include a histogram of the absorber velocity offsets. Only four low \logNHI\ data points are seen here, as one sightline has two such absorbers, and we have plotted the combined columns.}
    \label{fig:NHIdeltav}
\end{figure}

Out of 21 sightlines, we find 11 detections of absorbing gas within $\pm1000$ \kms\ of the LRG central redshift. The hydrogen column densities of the detected absorption lines span a broad range from $\logNHI < 13.0$ to $> 18.0$, but the distribution is not continuous. The absorber properties are summarized in Figures~\ref{fig:NHIrho} and \ref{fig:NHIdeltav}. We show in Figure~\ref{fig:NHIrho} the \HI\ column density as a function of impact parameter for the RDR sample. We find four strong \HI\ absorbers (two pLLSs and two LLSs) in our sample with $\logNHI \ge 16.0$; four of the other \replaced{six}{seven} \HI\ detections are mostly at $\logNHI \le 14.0$ (likely Ly$\alpha$ forest interlopers; see below). It seems the LRGs exhibit either high \HI\ column density gas or not much at all. Most of the strong absorbers lie at small impact parameters, $\rho \la$ 0.5 \Rvir, consistent with results for other samples of galaxies (e.g., \citealt{tumlinson2013, heckman2017, keeney2017}; \citetalias{prochaska2017}).

The low \HI\ column density absorbers in Figure~\ref{fig:NHIrho} are likely dominated by Ly$\alpha$ forest contamination. As shown in Figure~\ref{fig:NHIdeltav}, a higher proportion of the weak absorbers exhibit large velocity offsets from the redshift of the LRGs ($|\Delta v| = |v_{\rm Ly\alpha} - v_{\rm LRG}|$) compared with the strong absorbers\added{. The expected velocity dispersion for virialized gas in the halos of LRGs is $\sigma \approx 260$ \kms\ \citep{zahedy2019}}, suggesting the \added{low \HI\ column density absorbers found at high velocity offsets} are not directly associated with the LRGs. We estimate the expected number of Ly$\alpha$ forest interlopers in our sample, $\mathcal{N}$, by integrating the differential column density distribution, $f(N,z)$, for each sightline $i$ that has sensitivity to $\logNHI \le 14.5$ absorption \added{(this includes sightlines with non-detections)}. We use the fit to $f(N,z)$ from \citet[][their equation 6]{lehner2007}, extrapolating below their sensitivity limit of $\logNHI \sim 13.2$ and using the corrected parameter values from their erratum for the column density range $\logNHI = [13.2, 14.4]$ with $b\le 40$ \kms. We perform the integration for each sightline from $\NHI_{\rm min}$ to $\NHI_{\rm max} = 10^{14.5}$ \column, where $\NHI_{\rm min}$ is the sensitivity to \HI\ column for each spectrum. Once $\mathcal{N}_{i}$ has been calculated for each sightline, we total the values to determine $\mathcal{N}$ for our survey. We expect $\mathcal{N}$ $\sim$ 5 Ly$\alpha$ forest interlopers in the RDR survey, equal to the number of $\logNHI \le 14.5$ systems we detect (Figures~\ref{fig:NHIrho} and \ref{fig:NHIdeltav} show only four as SDSSJ125901.67+413055.8 exhibits two systems that we sum for display in the figures; see Appendix~\ref{appa}). Thus, it is very likely that all of the low \HI\ column density systems are Ly$\alpha$ forest interlopers unrelated to the RDR galaxies themselves. Following a similar procedure, we expect to find zero random interlopers for $\logNHI > 14.5$ assuming the parameterization from \citet{shull2017}. Thus, the high column density systems are likely associated with the LRGs.

\added{For the absorbers with $\logNHI > 14.5$, the mean velocity offset from the LRG is $-70$ \kms. The \citetalias{chen2018} LRG sample exhibits a mean of +17 \kms\ and a dispersion of 147 \kms\ (kinematics for this sample are discussed in \citealt{zahedy2019}). Our low number of absorbers prevents us from fitting the velocity distribution to determine the dispersion and how it compares to the expected value for virialized motion in the LRG halo. \added{The integration range (i.e., the full width at zero absorption) of the metal ions} we observe are 
$\sim$\added{200} \kms\ in the COS/G130M,G160M gratings, with larger \added{spans} seen in the lower resolution observations. These small velocity spans indicate the gas is cool and there is an absence of winds. In no cases do we (or \citetalias{chen2018}) find metal-line absorption having the strength or velocity breadth of those seen around the two LRGs studied by \citet{smailagic2018}. It is unclear why they have detected such velocity spans \added{($>$750 \kms)} at $\rho=29$ kpc and 343 kpc. While we do not have a sightline at such a low impact parameter, we do have several sightlines at large impact parameters. So far, these two sightlines are the only instances with absorbers this large in the CGM of LRGs. A larger sample of sightlines through the halos of LRGs will be able to inform us how often these broad absorbers occur.}

\added{The RDR} LRGs exhibiting high \HI\ column densities ($\logNHI > 16.0$) \deleted{in our sample} have associated metal-line absorption. We do not have uniform spectral coverage for each sightline, so the ions we probe vary (see Appendix~\ref{appa}). For several of our sightlines (16/21) we cover \OVI\ and can estimate its column density, which is the focus of \citetalias{howk2019}. For pLLSs/LLSs, \CIII\ is often one of the most prominent ions. We find absorption limits or detections of \CIII\ spanning the range $12.58 \le \logNCIII \le 14.27$, with detections in four RDR LRGs. We calculate the covering factor (see below) for \CIII\ to be $f_c(\rho \le 0.5\, \Rvir) = 0.50 \pm 0.18$ and $f_c(\rho \le \Rvir) = 0.41^{+0.15}_{-0.14}$ (68\% confidence interval). We find no inconsistencies between our \MgII\ equivalent width distribution and that presented by \citet{huang2016}, although we have a limited sample. \deleted{In no cases do we find metal-line absorption having the strength or velocity breadth of those seen by \citet{smailagic2018}.}

\subsection{Covering Factor of Strong \HI\ About LRGs}\label{sec:fc}

Covering factors can be used to characterize the CGM about different classes of galaxies allowing for quantitative comparisons. In addition, simulations often report covering factors of different ions, and their treatment of the CGM can be assessed by comparing to observational results. We split the sightlines into bins of impact parameter and determine the ratio of absorbers detected, for a limiting \logNHI, to the total number of sightlines in each bin. An absorber that meets this \logNHI\ criterion is referred to as a ``hit," while those below this limit are ``misses." We calculate the covering factor, $f_c$, of strong \HI\ assuming a binomial distribution for the covering factor where the posterior distribution is described by a beta function. This follows from the discussion in \citet{cameron2011} who describes the posterior distribution of values in a Bayesian treatment of binomial-distributed data. The beta function also provides estimates of the Bayesian confidence intervals, which we utilize throughout this paper.

\begin{figure*}
    \epsscale{1.0}
    \plotone{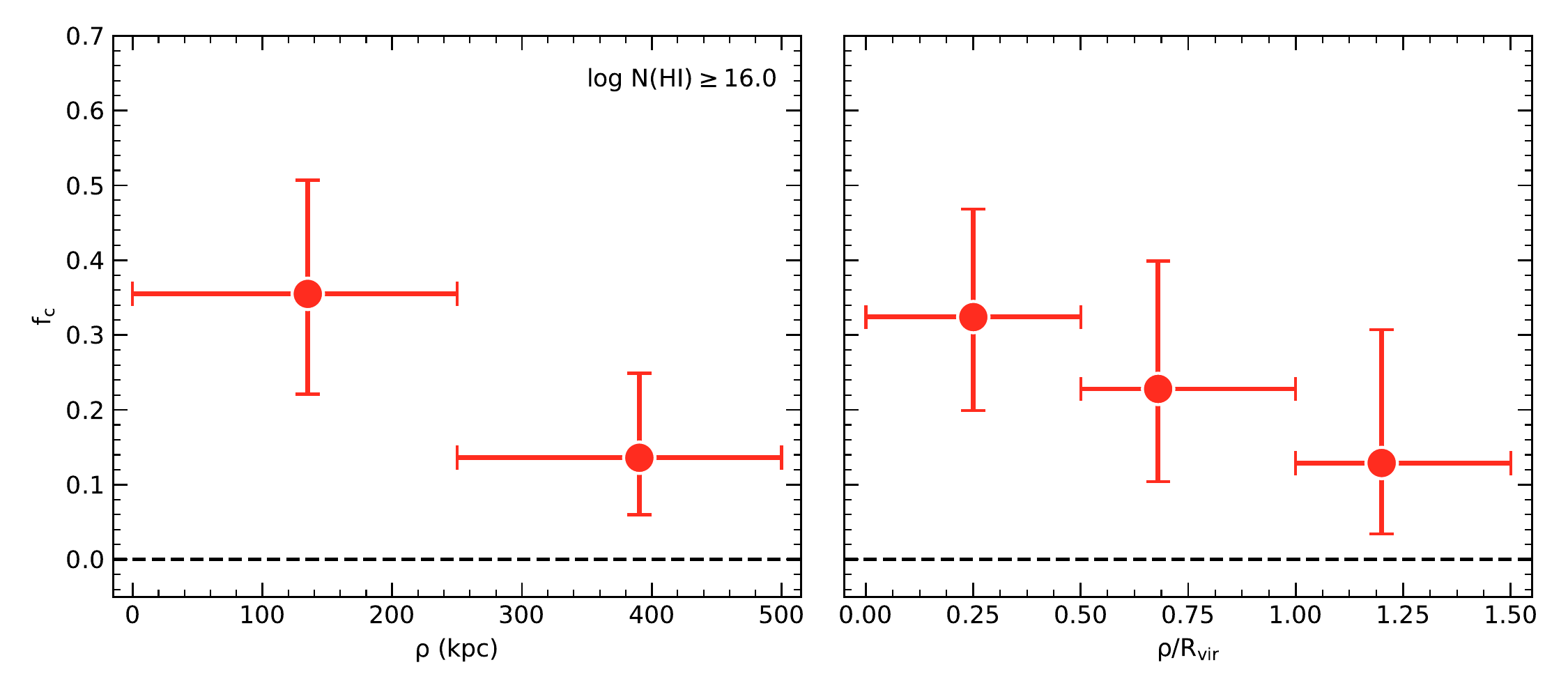}
    \caption{\HI\ covering factor about the RDR sample of LRGs for gas with $\logNHI \ge 16.0$ as a function of impact parameter (left) and normalized impact parameter (right). The vertical error bars show the 68\% confidence interval in $f_c$. The points are located at the mean impact parameter for each bin, while the horizontal error bars show the total extent of the bin. Note that unity covering factor is not shown.}
    \label{fig:coveringfactor}
\end{figure*}

We summarize our covering factor estimates in Table~\ref{tab:coveringfactor}, giving results for three limiting column densities: $\logNHI \ge 15.0$, 16.0, and 17.2. When imposing these limits we consider only the central value of our \HI\ measurements. Except for the covering factor calculation with $\logNHI \ge 15.0$, we do not include the absorber for which we could only derive a lower limit for \logNHI\ in all other covering factor calculations throughout this paper. Table~\ref{tab:coveringfactor} lists the impact parameter range for each bin, the mean impact parameter of sightlines within the bin, the covering factor with 68\% and 95\% confidence intervals, the number of sightlines probed, and the number of detected absorbers for all three \HI\ column density limits. We plot the covering factors as a function of impact parameter and normalized impact parameter for $\logNHI \ge 16.0$ in Figure~\ref{fig:coveringfactor}. The horizontal error bars show the width of the bins, while the vertical error bars show the 68\% confidence interval for the distribution. The covering factor within the inner 250 kpc of our sample of RDR LRGs is substantial at $f_c(\rho < 250\, \rm kpc) = 0.35^{+0.15}_{-0.13}$. The covering factor for $\logNHI \ge 16.0$ within the virial radius for our sample is $f_c(\rho \le \Rvir) = 0.27^{+0.11}_{-0.10}$. For LLS absorption with $\logNHI \ge 17.2$, we find $f_c(\rho \le \Rvir) = 0.15^{+0.10}_{-0.07}$.

The covering factor of high column density gas around LRGs is non-negligible, with 35\% of LRGs showing strong \HI\ at these columns within $\sim$0.5 \Rvir\ (the median \Rvir\ = 516 kpc for our sample). A similar result is found around the quiescent COS-Halos galaxies \citep[][see also \citetalias{prochaska2017}]{thom2012}. We calculate a covering factor within 160 kpc ($\sim$0.75 \Rvir) for the quiescent COS-Halos galaxies of $f_c(\rho \le 160\, \rm kpc) = 0.50 \pm 0.12$ at the same \logNHI\ cutoff. \citetalias{chen2018} recently reported a covering factor $f_c(\rho \le 165\, \rm kpc) = 0.44^{+0.12}_{-0.11}$ within 165 kpc ($\sim$0.33 \Rvir) of their sample of massive galaxies for $\logNHI \ge 17.2$, consistent with our results in Figure~\ref{fig:coveringfactor}.
While no direct simulation analogs of LRGs exist in the literature, we can extrapolate the results from \citet{rahmati2015} to lower \z. They use the EAGLE simulation suite to characterize the \HI\ distribution around galaxies from $z=4$--1. Though they do not extend the analysis to $z\sim0.5$, they simulate galaxies up to $\log M_{200} = 13.7$. In their Figure 5 they show a covering factor for $\logNHI > 17.2$ of $f_c(\rho < \Rvir) \sim 0.1$ for the LRG halo mass range. Our covering factor $f_c(\rho \le \Rvir) = 0.15^{+0.10}_{-0.07}$ at the same \HI\ cutoff is consistent with their result. \citet{rahmati2015} report a differential $f_c \sim 0.3$ centered at $\rho/\Rvir = 0.3$ for all galaxies with $\log M_{200}> 12$ (their Figure 7); over a similar normalized impact parameter ($0 \le \rho/\Rvir \le 0.25$) we find $f_c(\rho) = 0.36^{+0.18}_{-0.16}$ for the LRGs. For optically-thick gas, the LRG observations match those from the simulations at $z=1$. Future simulation work should consider the pLLS column density range at lower redshifts for covering factor calculations to aid in comparing with observations.

On the whole, LRGs do not make a strong contribution to the total population of pLLSs/LLSs. To assess their contribution to $d\mathcal{N}/dz$ we start with the absorber number per path for uniformly-distributed galaxies of cross section $\sigma = \pi\Rvir^2$ and number density $n_0$:

\begin{equation}
d\mathcal{N}/dX = \frac{c}{H_0}f_c(\rho \le \Rvir){\pi}R^2_{\rm vir}n_0,
\end{equation} where we assume covering factors from our measurements. Connecting absorber distance to redshift path,

\begin{equation}
dX = \frac{H_0}{H(z)}(1+z)^2 dz,
\end{equation} we find

\begin{eqnarray}
d\mathcal{N}/dz & = & \frac{c}{H_0}f_c(\rho \le \Rvir){\pi}R^2_{\rm vir} \nonumber \\
& & \times n_0(1+z)^2[\Omega_{\Lambda}+(1+z)^3\Omega_m]^{-1/2}
\end{eqnarray}
\citep{pad2002,prochaska2010,ribaudo2011}. In these equations $c$ is the speed of light, $f_c(\rho \le \Rvir)$ is the covering factor, \Rvir\ is the median virial radius for our sample, $n_0$ is the number density of LRGs, $H(z)$ is the Hubble parameter, and \z\ is the median redshift of our sample. All other variables are defined in \S~\ref{sec:intro}. The physical number density of LRGs at $z=0.5$ is listed in Table 1 of \citet{almeida2008}, and we adopt the fiducial value $n_0 \approx 10^{-4} h^3$ Mpc$^{-3}$. The contribution of LRGs to the combined pLLS and LLS population ($\logNHI \ge 16.0$) is $d\mathcal{N}/dz = 0.05 \pm 0.02$. The distribution calculated in \citet{shull2017} that includes pLLSs has a value of $d\mathcal{N}/dz$ = 2.24 at $z\sim0.5$. For LLS absorption ($\logNHI \ge 17.2$), we find $d\mathcal{N}/dz = 0.03^{+0.02}_{-0.01}$ for the LRG contribution. The line density of LLSs calculated by \citet{ribaudo2011} has a value of $d\mathcal{N}/dz = 0.5$ at $z\sim0.5$. Thus, LRGs contribute only a few percent of the population of strong \HI\ absorbers.

\subsection{An Extended Sample of Galaxy-Selected CGM Measurements}\label{sec:fcssfr}

\begin{figure*}
    \epsscale{1.15}
    \plotone{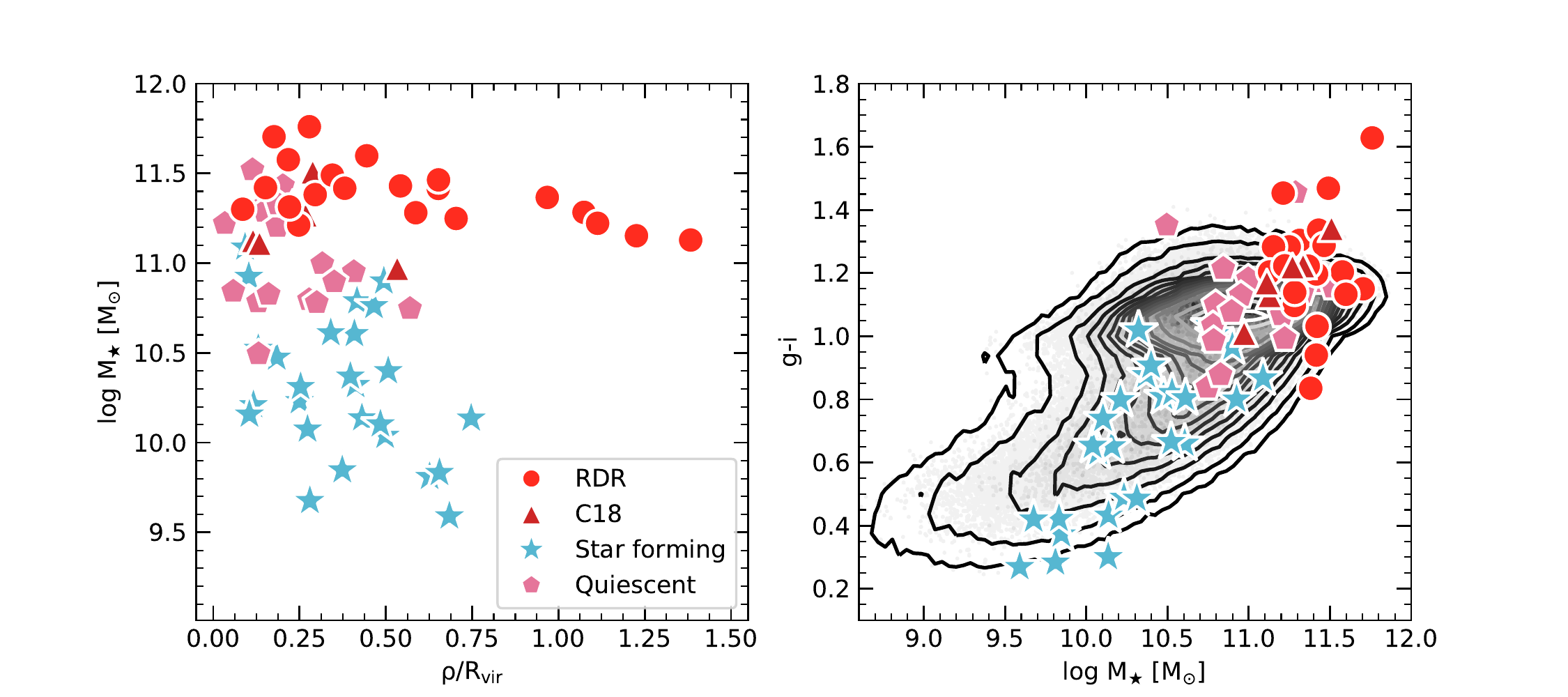}
    \caption{Distributions of stellar mass versus normalized impact parameter \added{(left) and color versus stellar mass (right)} for the combined sample: the RDR LRGs studied here, the COS-Halos sample, and the \citetalias{chen2018} sample of massive galaxies. The COS-Halos sample is split by sSFR (those with $\log$\,sSFR\,$> -11$ are marked as ``star forming''). We use our results and the COS-Halos results where they overlap the \citetalias{chen2018} sample. \added{The contours are made using the NYU-VAGC \citep{blanton2005}.}}
    \label{fig:stmassrho}
\end{figure*}

To understand the nature of strong CGM absorption systems, we compare and combine the RDR sample with the galaxy-selected COS-Halos survey (\citealt{tumlinson2013}; \citetalias{prochaska2017}) and the recent \citetalias{chen2018} LRG study. We show in Figure~\ref{fig:stmassrho} the distributions of stellar mass as a function of normalized impact parameter at which each of the galaxies is probed in this combined sample\added{, and the galaxy color versus stellar mass}. \added{We make use of the New York University Value Added Galaxy Catalog (NYU-VAGC) for the contours \citep{blanton2005}. The NYU-VAGC is a catalog of the photometric and spectroscopic properties (and additional derived quantities) of the galaxies in SDSS DR7.} We have recalculated the virial radii and normalized impact parameters for the COS-Halos and \citetalias{chen2018} samples using the same assumptions as for our sample, starting with the stellar masses given in those works. \added{For these calculations the COS-Halos galaxies are assumed to be centrals. As such, we estimate the satellite fraction for the survey to be 0.24 using the satellite fraction distribution from \citet[][their Figure 4]{wt2018}. This value does not include scatter between the stellar mass and halo properties.}

The COS-Halos sample comprises 44 star-forming and quiescent galaxies, with $\log M_\star = 9.5$--11.5, and a median galaxy redshift of $z\sim0.2$ \citep{werk2012}. As this sample is galaxy-selected, it has well-constrained host galaxy properties, and the sightlines extend to an impact parameter of 160 kpc (0.75 \Rvir\ for their mean galaxy). The COS-Halos galaxies were selected to be relatively isolated, but this did not always turn out to be the case \citep{werk2012}. The COS-Halos sample does not lie in the same redshift range as the RDR LRGs, so temporal evolution may affect our comparison. \added{We also note that the stellar population of the RDR LRGs is intrinsically different than that of the COS-Halos sample, as are their evolutionary histories \citep{eisenstein2001}.} The \citetalias{chen2018} sample is another galaxy-selected survey comprised of 16 galaxies with $\logMstar \ge 11$, a median redshift $z\sim0.4$, and impact parameters $\rho \le 165$ kpc. Our sample includes five galaxies in common with the \citetalias{chen2018} sample, and the COS-Halos sample has five galaxies in common with \citetalias{chen2018}. This leaves a total of six unique galaxies from \citetalias{chen2018} to include in the extended sample. In this section we focus on the COS-Halos sample, and in \S~\ref{sec:fcmass} we include the \citetalias{chen2018} results.

The \NHI\ measurements for much of the COS-Halos sample are well constrained; however, several sightlines have \HI\ absorption that can only be constrained by upper and lower bounds \citepalias{prochaska2017}. In what follows we remove three absorbers in the COS-Halos sample that have upper limits of $\logNHI \le 16.24$, 16.65, and 17.01. These upper limits are not constraining for our covering factor calculations for absorbers having $\logNHI \ge 16.0$. \added{(Removing these sightlines gives a marginal difference in the covering factor calculations. The values decrease slightly, but they are still consistent with those reported here.)} Following \citealt{tumlinson2011}, we cut the COS-Halos galaxies into two subsamples, defining quiescent and star forming galaxies as those with $\log {\rm sSFR } \le -11$ and $> -11$, respectively.

The way in which we assess covering factors for the COS-Halos sample is a bit different than the approach in \S~\ref{sec:fc}. For sightlines with only \HI\ bounds, the probability distribution function (PDF) of \logNHI\ is described by a top-hat distribution between the two limits. This occurs for sightlines where the break at the Lyman limit is saturated (taking the QSO spectrum to zero flux and giving only a lower limit) and there are no detectable damping wings on Ly$\alpha$ (giving an upper limit). There is also one sightline where the lower limit on \NHI\ comes from saturated Ly$\alpha$ and Ly$\beta$; in this case, we impose an upper bound of $\logNHI < 18.0$ due to the lack of damping wings (J.X. Prochaska 2018, priv. comm.). When determining if these absorbers are hits or misses, we must consider the entire range of allowed values for these bounded systems. If the \logNHI\ value at the low end of the interval is above 16.0, we count this absorber as a hit. However, the column density interval for several of these systems includes $\logNHI = 16.0$. In these cases, if less than 50\% of the interval is above $\logNHI = 16.0$, we count the absorber as a miss; if it is $\ge50\%$, we count it as a hit. All other absorbers are treated as discussed in \S~\ref{sec:fc}. We note that \citet{howk2017} performed a similar covering factor analysis of the COS-Halos results using a Monte Carlo approach; our approach does not produce significantly different results from theirs.

\begin{figure}
    \epsscale{1.25}
    \plotone{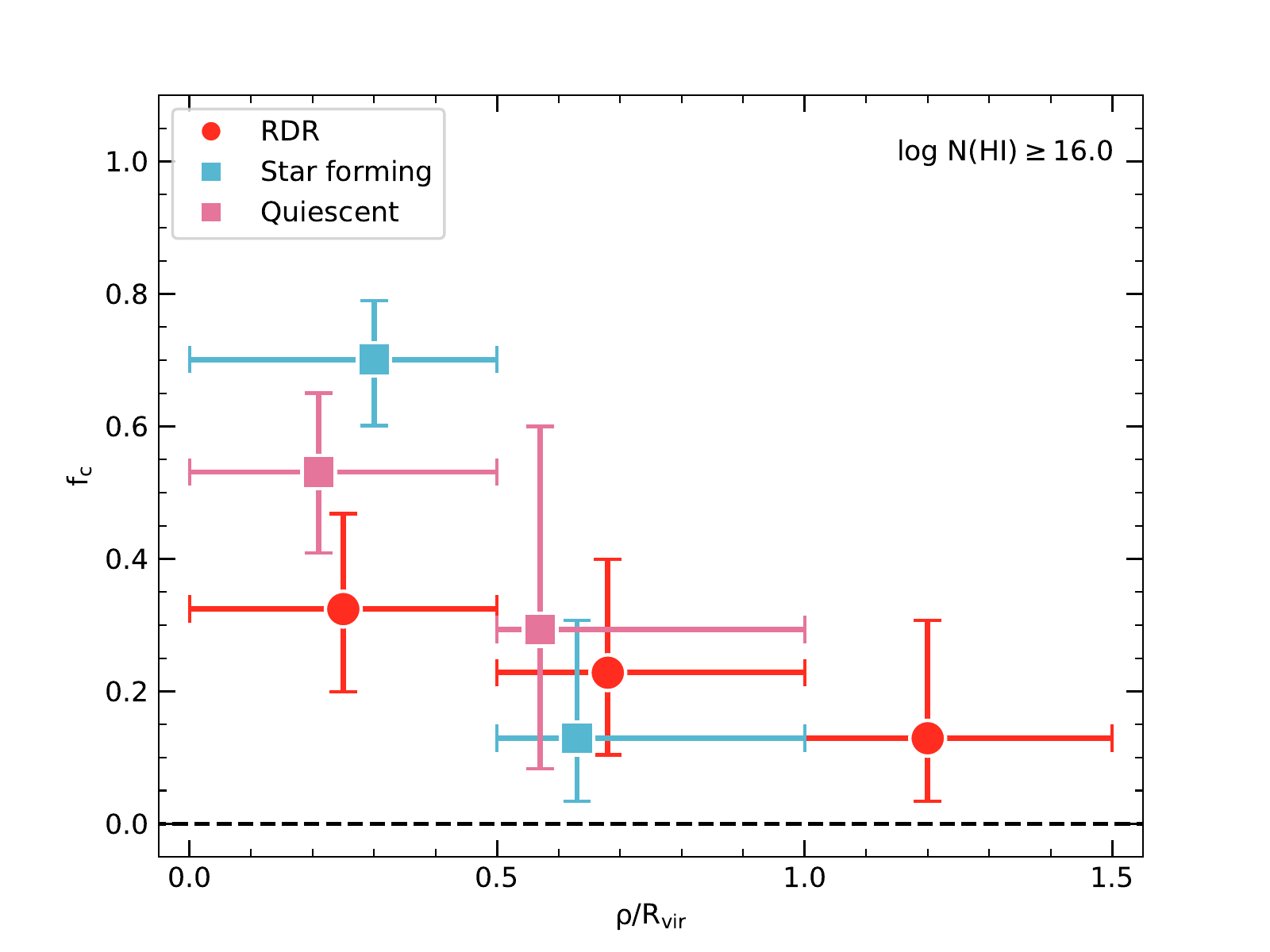}
    \caption{\HI\ covering factor distributed by sample for gas with $\logNHI \ge 16.0$ as a function of normalized impact parameter. The vertical error bars show the 68\% confidence interval for the covering factor. The horizontal error bars show the extent of each bin, while the location of the data points represents the mean normalized impact parameter. Galaxies from the COS-Halos sample make up the star-forming and quiescent samples \citepalias{prochaska2017}. We adopt their characterization of galaxies with $\log$ sSFR $> -11$ as ``star forming."}
    \label{fig:coveringfactor_ssfr}
\end{figure}

We compare the covering factors of our RDR sample with the quiescent and star-forming COS-Halos samples in Figure~\ref{fig:coveringfactor_ssfr}, and Table~\ref{tab:coveringfactorssfr} summarizes our derivation of covering factors for the COS-Halos galaxies as in Table~\ref{tab:coveringfactor} (we include a complementary figure and table in Appendix~\ref{appb} for results using optically-thick absorption). We use normalized impact parameter, $\rho/\Rvir$, in all comparisons. We note that the sample of star-forming galaxies is almost twice as large as the RDR sample for the inner bin. Within 0.5 \Rvir, the star-forming galaxies have a significantly higher covering factor of high \HI\ column gas than the LRGs with $f_c^{\rm SF} \simeq 2f_c^{\rm RDR}$. However, both the COS-Halos quiescent galaxies and RDR LRGs have non-negligible covering factors. All three samples are consistent at larger impact parameters, $0.5 < \rho/\Rvir \le 1.0$.

\begin{figure}
    \epsscale{1.25}
    \plotone{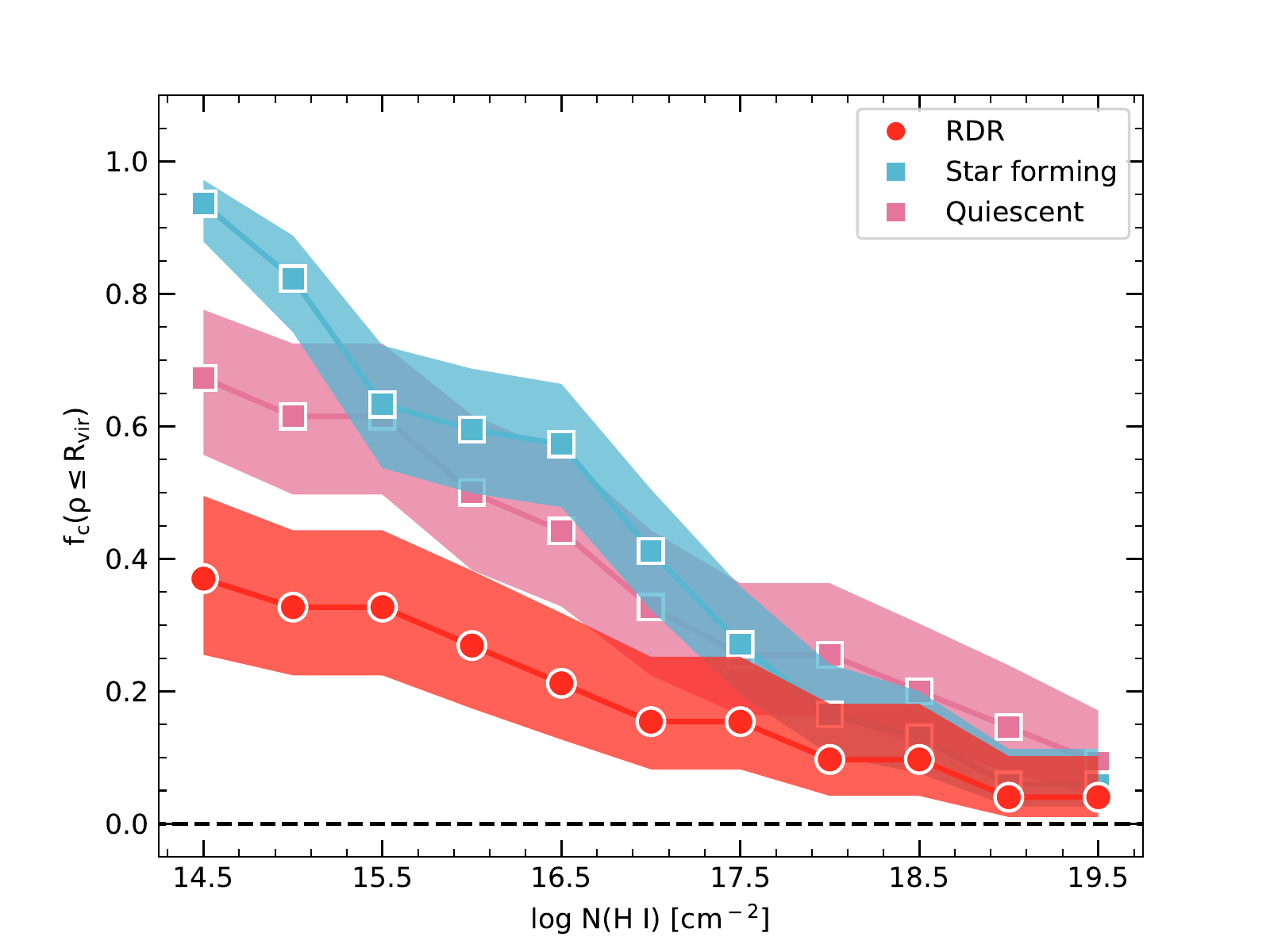}
    \caption{\HI\ covering factor for $\rho \le \Rvir$ as a function of limiting \logNHI\ for the RDR LRGs and COS-Halos galaxies. The shaded region shows the 68\% confidence interval for the distribution. The COS-Halos sample is split by sSFR ($\log$ sSFR $> -11$ is ``star forming"). The RDR LRGs show a relative paucity of gas with $14.5 \le \logNHI \le 16.0$.}
    \label{fig:coveringfactor_compare}
\end{figure}

Figure~\ref{fig:coveringfactor_compare} shows $f_c(\rho \le \Rvir)$ as a function of limiting \HI\ column for our RDR sample compared with the COS-Halos samples. We calculate the covering factors as discussed above, using the limiting \HI\ column density given along the horizontal axis. The RDR LRGs do not have many detections of gas within $\logNHI \le 16.0$, and the covering factor distribution is relatively flat for lower limiting columns. The COS-Halos quiescent galaxies have several detections at lower columns and a few detections of LLSs with column densities $\ge 18.5$\deleted{, while the highest column in the RDR sample is $\logNHI \simeq 18.1$}. Among these three samples, the LRGs are unique in the lack of gas with $\logNHI \la 16.0$ (a result that can also be seen in the \citetalias{chen2018} sample), while the passive galaxies in COS-Halos do show intermediate \HI\ column density gas ($14.5 \le \logNHI \le 16.0$). This difference may be in part due to the differing mass distributions between the two samples or the difference in redshift. It could also be the environment of these galaxies plays a role: most of the COS-Halos galaxies are selected (imperfectly) to be isolated, while the RDR LRGs are either in clusters or dense groups (see \S~ \replaced{\ref{sec:samselec}}{\ref{sec:galprop}}).

\subsection{The Mass Dependence of Strong \HI\ Absorption}\label{sec:fcmass}

\begin{figure}
    \epsscale{1.25}
    \plotone{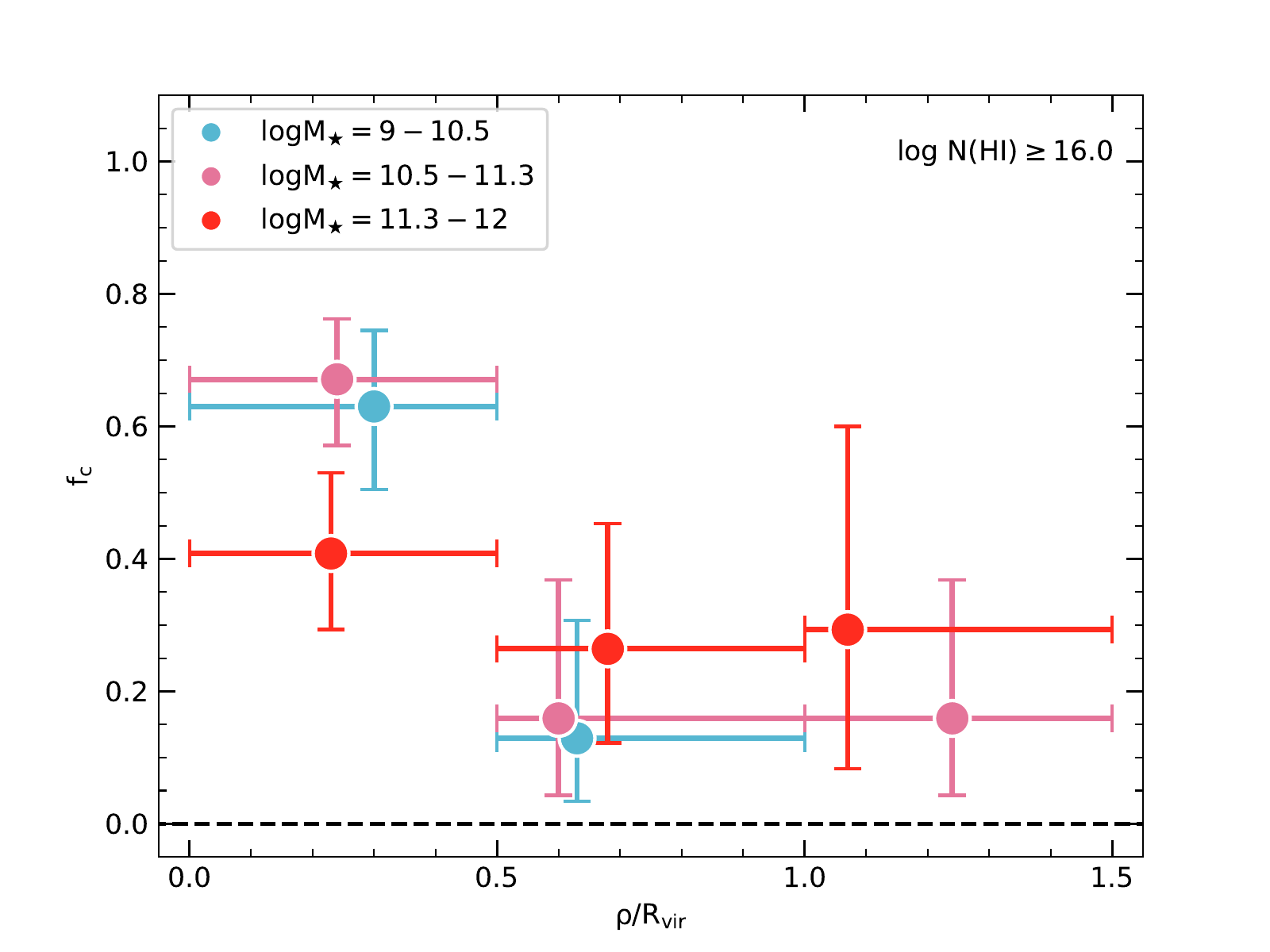}
    \caption{Stellar mass dependence of the \HI\ covering factor ($\logNHI \ge 16.0$) as a function of normalized impact parameter in our combined sample of galaxies incorporating our results with those from \citetalias{prochaska2017} and \citetalias{chen2018}. The vertical error bars show the 68\% confidence interval in $f_c$. The horizontal error bars show the extent of each bin, while the location of the data points represents the mean normalized impact parameter. We do not find a statistically significant difference in covering factors between host galaxy masses for these column densities.}
    \label{fig:coveringfactor_stellar}
\end{figure}

\begin{figure}
    \epsscale{1.25}
    \plotone{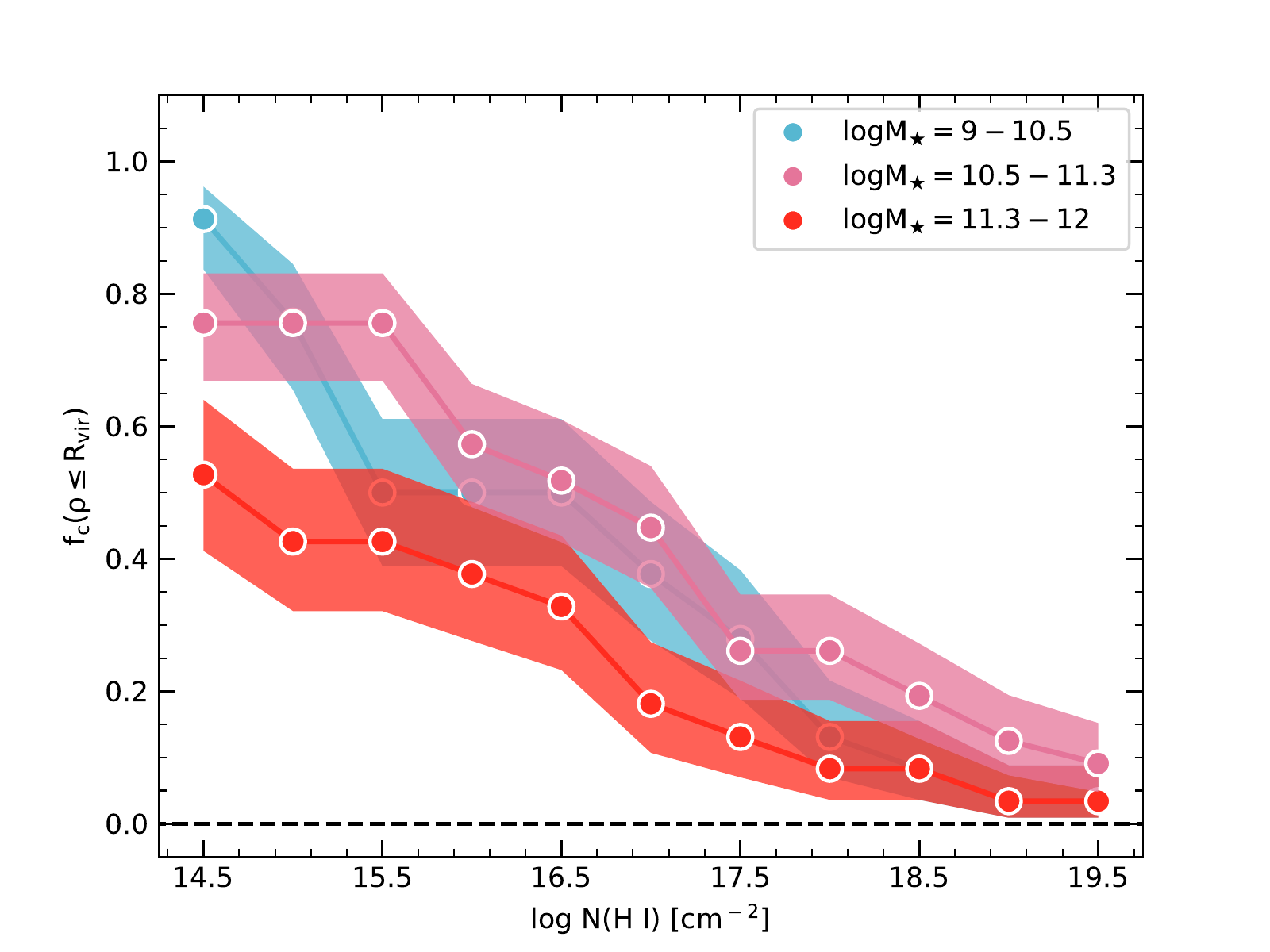}
    \caption{\HI\ covering factor for $\rho \le \Rvir$ as a function of limiting \logNHI\ for our combined sample of galaxies, including our results with data from \citetalias{prochaska2017} and \citetalias{chen2018}. The shaded region shows the 68\% confidence interval for the distribution. The highest-mass galaxies (dominated by the RDR LRGs) show a relative paucity of gas with $14.5 \le \logNHI \le 16.0$. There is no significant difference in the distributions between the low- and intermediate-mass ranges in this plot. If there is a critical mass above which cold gas becomes less common, it is significantly higher than the canonical value of $\logMstar \approx 10.5$.}
    \label{fig:coveringfactor_stellarcompare}
\end{figure}

In this section we combine the RDR sample with the COS-Halos and \citetalias{chen2018} samples to assess the CGM properties across this combined sample of 71 galaxies. All of the sightlines probed were selected based on galaxy properties. We break the entire sample into bins of stellar mass and halo mass to investigate whether the properties of CGM \HI\ absorption depend on galaxy mass. We first consider the covering factors for the combined sample in three bins of stellar mass: $\log M_\star = [$9.0--10.5), [10.5--11.3), [11.3--12.0]. We choose one of the boundaries at $\log M_\star = 10.5$ because this is the stellar mass at which early theoretical arguments suggested galaxies transition from cold-mode accretion to hot-mode accretion (\citealt{cattaneo2006,db2006}, but see \citealt{nelson2015}). We split the sample at $\log M_\star = 11.3$ in an effort to have a similar number of galaxies in the two higher mass bins. Figure~\ref{fig:coveringfactor_stellar} displays the covering factors for these mass bins, and Table~\ref{tab:coveringfactorstellar} summarizes these results (we include a complementary figure and table in Appendix~\ref{appb} for optically-thick absorption). There is no clear difference in the covering factors across the predicted critical value of stellar mass at $\Rvir \le 0.5$ (even comparing all masses above $\log M_\star = 10.5$ with those below). Figure~\ref{fig:coveringfactor_stellarcompare} shows the covering factors $f_c(\rho \le \Rvir)$ as a function of limiting \logNHI\ for each mass bin, as we showed in Figure~\ref{fig:coveringfactor_compare}. This figure shows that the lack of lower \HI\ column absorption is limited to the highest-mass galaxies. The good agreement in the curves for the two lower-mass bins suggests that any transition in CGM \HI\ properties occurs at masses above $\log M_\star \ge 11$. Indeed, galaxies with $10.5 \le \log M_\star \le 11.3$ show higher covering factors at $\logNHI \ge 18$ than the lower-mass galaxies.

\begin{figure}
    \epsscale{1.25}
    \plotone{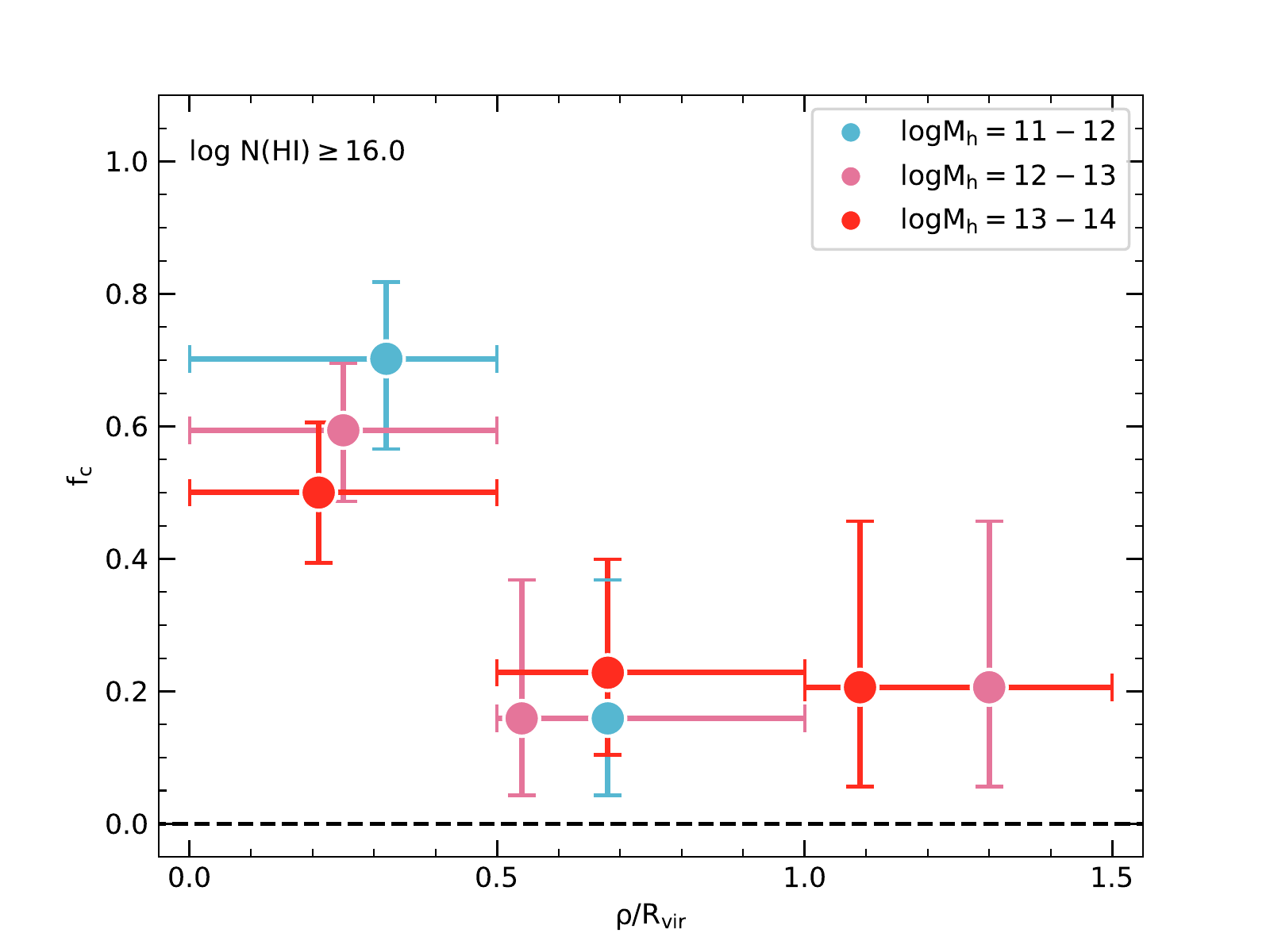}
    \caption{Halo mass dependence of the \HI\ covering factor ($\logNHI \ge 16.0$) as a function of normalized impact parameter in our combined sample of galaxies incorporating our results with those from \citetalias{prochaska2017} and \citetalias{chen2018}. The vertical error bars show the 68\% confidence interval in $f_c$. The horizontal error bars show the extent of each bin, while the location of the data points represents the mean normalized impact parameter. There is no statistically significant difference in the covering factors of galaxies with halo mass.}
    \label{fig:coveringfactor_halos}
\end{figure}

Figure~\ref{fig:coveringfactor_halos} shows the covering factors in bins of halo mass (using the scalings from stellar mass in \citealt{rodriguez-puebla2017}). \added{Figures~\ref{fig:coveringfactor_stellar} and \ref{fig:coveringfactor_halos} display the same information (with different bin boundaries), but we include both for the readers that are more familiar with galaxy properties in terms of halo mass.} We bin the combined RDR, COS-Halos, and \citetalias{chen2018} sample by $\log M_{\rm h} = [$11.0--12.0), [12.0--13.0), [13.0--14.0]. Table~\ref{tab:coveringfactorhalo} gives a summary of the information from the figure (we include a complementary figure and table in Appendix~\ref{appb} for optically-thick absorption). All three halo mass subsamples show consistent covering factors for $\logNHI \ge 16.0$. The lowest halo mass sample almost exclusively comprises star-forming galaxies, so this high covering factor result is consistent with Figure~\ref{fig:coveringfactor_ssfr}. There are several galaxies in the [12.0--13.0) halo mass range that are still forming a modest amount of stars, while those in the [13.0--14.0] halo mass range are all considered quiescent. These covering factors show there is a significant amount of cool gas in galaxy halos of all masses probed in this combined sample.

We can extend to even higher halo masses using galaxy clusters, the largest (often) virialized structures in the universe. These structures are roughly a factor of 10 higher in mass than the LRGs and extend to Mpc scales for \Rvir. \replaced{Clusters are dominated by hot X-ray emitting gas in}{The baryon budget of clusters is expected to be dominated by hot X-ray emitting gas called} the intra-cluster medium (ICM, the same phase that should dominate in the LRG CGM), and many satellite galaxies are undergoing gas stripping in these clusters \added{\citep{gonzalez2007,gonzalez2013}}. \citet{yoon2012} and \citet{yp2017} studied the distribution of Ly$\alpha$ absorbers in the Virgo and Coma clusters. They found no detections of strong or optically-thick \HI\ within \Rvir\ when considering absorption within $\pm1\sigma$ of the cluster velocity; most of their detections lie below $\logNHI \le 14.5$. Their detections at $\rho > \Rvir$ are more numerous, yet still below what we consider strong \HI\ gas.\footnote{There are two detections of strong \HI\ gas outside of the Virgo cluster virial radius and within $\pm2\sigma$ of the cluster velocity. Both detections are in areas of un-virialized substructure that may represent a recent merger event \citep{yp2017}.} The covering factors within \Rvir\ for the Virgo and Coma clusters at our adopted limit of $\logNHI \ge 16.0$ are $f_c^{\rm Virgo}(\rho \le \Rvir) = 0.05^{+0.08}_{-0.04}$ and $f_c^{\rm Coma}(\rho \le \Rvir) = 0.13^{+0.18}_{-0.10}$ (68\% confidence interval). \citet{burchett2018} also conducted a survey of absorption from cool gas \replaced{in}{within} the ICM of five X-ray detected clusters. As with the two other cluster surveys, no strong \HI\ was detected. We calculate the covering factor of $\logNHI \ge 16.0$ within \Rvir\ for the X-ray clusters to be $f_c^{\rm X-ray}(\rho \le \Rvir) = 0.16^{+0.21}_{-0.12}$. These results, coupled with that of Figure~\ref{fig:coveringfactor_halos}, suggest that the increase in mass between LRGs and clusters creates conditions that disfavor strong \HI. A similar result is found in the COS-Clusters survey (Tejos et al. in prep.).

In these low-redshift clusters, sightlines passing through the ICM do not exhibit strong \HI\ absorption (but see \citealt{muzahid2017}). It may be that the physical properties of the ICM are preventing the formation of cool, dense gas that would give rise to pLLSs and LLSs (e.g., as discussed in \citealt{yp2013} and \citealt{burchett2018}), perhaps in an analogous manner to the suppression of $\logNHI \la 16$ gas in LRGs. Having said that, we note that one of the pLLSs in our sample (associated with LRG SDSSJ171651.46+302649.0) is in a redMaPPer cluster with $\lambda < 20$ (see \S~\ref{sec:galprop}), so the suppression of such gas is not complete.

\section{Metallicities of Strong \HI\ Absorbers about LRGs}\label{sec:metallicity}

\subsection{Metallicity Methodology}

In the previous section, we demonstrated that the massive RDR galaxies have substantial \HI\ gas in their CGM. By determining the metal-enrichment levels of this gas, we can start to differentiate between its plausible origins. The gas probed by pLLSs and LLSs is predominantly ionized, and large ionization corrections are required to derive the metallicity since we compare metal-ions with \HI\ \citep[e.g.,][]{lehner2013,fumagalli2016,lehner2018,wotta2019}. To derive the metallicity, we follow the methodology of the COS CGM Compendium (CCC) where \citet{wotta2019} derived the metallicities of 82 pLLSs and 29 LLSs using Bayesian techniques and Markov-chain Monte Carlo (MCMC) sampling of a grid of photoionization models \citep[see][]{fumagalli2016}. \added{Using the MCMC sampler, we determine the posterior distribution function (PDF) of the metallicity of an absorber based on the likelihoods of the models in the grid to match the observed column densities. The error bars for these values are the Bayesian confidence intervals.} The grid of photoionization models \added{that we sample} is made using Cloudy \citep[version C13.02, see][]{ferland2013}, assuming a uniform slab geometry in thermal and ionization equilibrium. The slab is illuminated with a Haardt--Madau extragalactic ultravoilet background (EUVB) radiation field from quasars and galaxies \added{with no other local sources of ionization} (we adopt \citetalias{hm1996} as implemented in Cloudy, see \citealt{hm1996}). The grid parameters are summarized in Table~\ref{tab:grid}. \added{We also undertook models using the \citetalias{hm2012} EUVB and found the metallicities systematically increase by 0.1--0.5 dex, similar to the findings of \citealt{wotta2019}. This is most likely due to the harder slope of the QSO spectrum in this model.}

Most of the diagnostics available for the pLLSs and LLSs at $z\la 1$ probe the $\alpha$ elements (O, Mg, Si) as well as carbon. The $\alpha$ elements are produced by similar nucleosynthesis processes and, since dust depletion is negligible for the densities of the pLLSs and LLSs, their relative abundances are expected to be approximately solar \citep{lehner2013,lehner2018}. To describe the metallicity of the gas, we use the standard notation ${\rm [X/H]} = \log(N({\rm X})/N({\rm H})) - \log(N({\rm X})/N({\rm H}))_\odot$, where X here represents an $\alpha$-process element (e.g., O, Mg, Si). For the pLLSs and LLSs, [C/$\alpha$] is on average solar \citep{lehner2018,wotta2019}, but this ratio can depart from a solar value with $\langle {\rm [C/\alpha]} \rangle = -0.05\pm 0.35$ \citep{wotta2019}. In our photoionization models, we therefore use a flat prior on [C/$\alpha$] following \citet{wotta2019}. In certain cases, the metallicity and density of an absorber cannot be constrained from the observations without additional assumptions due to either not enough available metal ion measurements or the dominance of upper limits. \citet[][and see also \citealt{lehner2013,wotta2016,wotta2019}]{lehner2019} showed that the ionization parameter \added{($U = n_{\gamma}/\nH$ = ionizing photon density/hydrogen number density)} has a normal distribution in log-space for low-redshift \added{$z \lesssim 1$} pLLSs and LLSs, which can be used as a prior for the absorbers that do not have sufficient constraints; this method provides what we describe as ``low-resolution" metallicities \citep{wotta2016}. \added{These types of absorbers are associated with galaxies of a broad range of masses/luminosities, from sub-L* to super-L* galaxies \citep{lehner2013}. Due to the large span in host galaxy mass, this method is viable for the absorbers associated with LRGs.}

We run the MCMC analysis for the \replaced{four}{three} strong \HI\ absorbers associated with the LRGs \deleted{SDSSJ111508.24+023752.7,} SDSSJ111132.33+554712.8, SDSSJ171651.46+302649.0, and SDSSJ141540.01+163336.4. \added{We use the low-resolution method for the absorber associated with LRG SDSSJ171651.46+302649.0.} \added{Due to the range of allowed \logNHI\ for the LLS associated with LRG SDSSJ111508.24+023752.7, a satisfactory solution using the MCMC analysis could not be reached. Instead we estimate the metallicity using [O/H].} Metallicity estimates could not be determined for the weak \HI\ absorbers ($\logNHI < 16.0$) because they have only upper limits for their metal column densities or metal ion detections that will not give reliable metallicities. \deleted{This result is unsurprising, as COS is not sensitive to metal lines in absorbers with $\logNHI < 15.0$ \citep{lehner2018,lehner2019}.}

\subsection{Metallicities of LRG CGM Absorbers}\label{sec:lrgmet}

The metallicities for the four pLLSs/LLSs detected about RDR LRGs are summarized in Table~\ref{tab:met}. For each system the metallicity from our Cloudy/MCMC modeling is listed as the median value of the PDF with uncertainties reflecting the 68\% confidence interval, along with the bounds of the 95\% confidence interval. There are \replaced{three}{two} high-metallicity absorbers with [X/H] $\ge-1$ among the four RDR systems\added{, and one with unconstrained metallicity}. The last system, associated with LRG SDSSJ171651.46+302649.0, is a low-metallicity absorber with [X/H] $= -1.87^{+0.24}_{-0.25}$ (``low-resolution" method applied, see also \citealt{wotta2016}). ``Corner" plots summarizing the constraints on the properties of the four pLLSs/LLSs associated with the RDR LRGs and the modeling details for each absorber are included in Appendix~\ref{appc}.

\begin{figure}
    \epsscale{1.25}
    \plotone{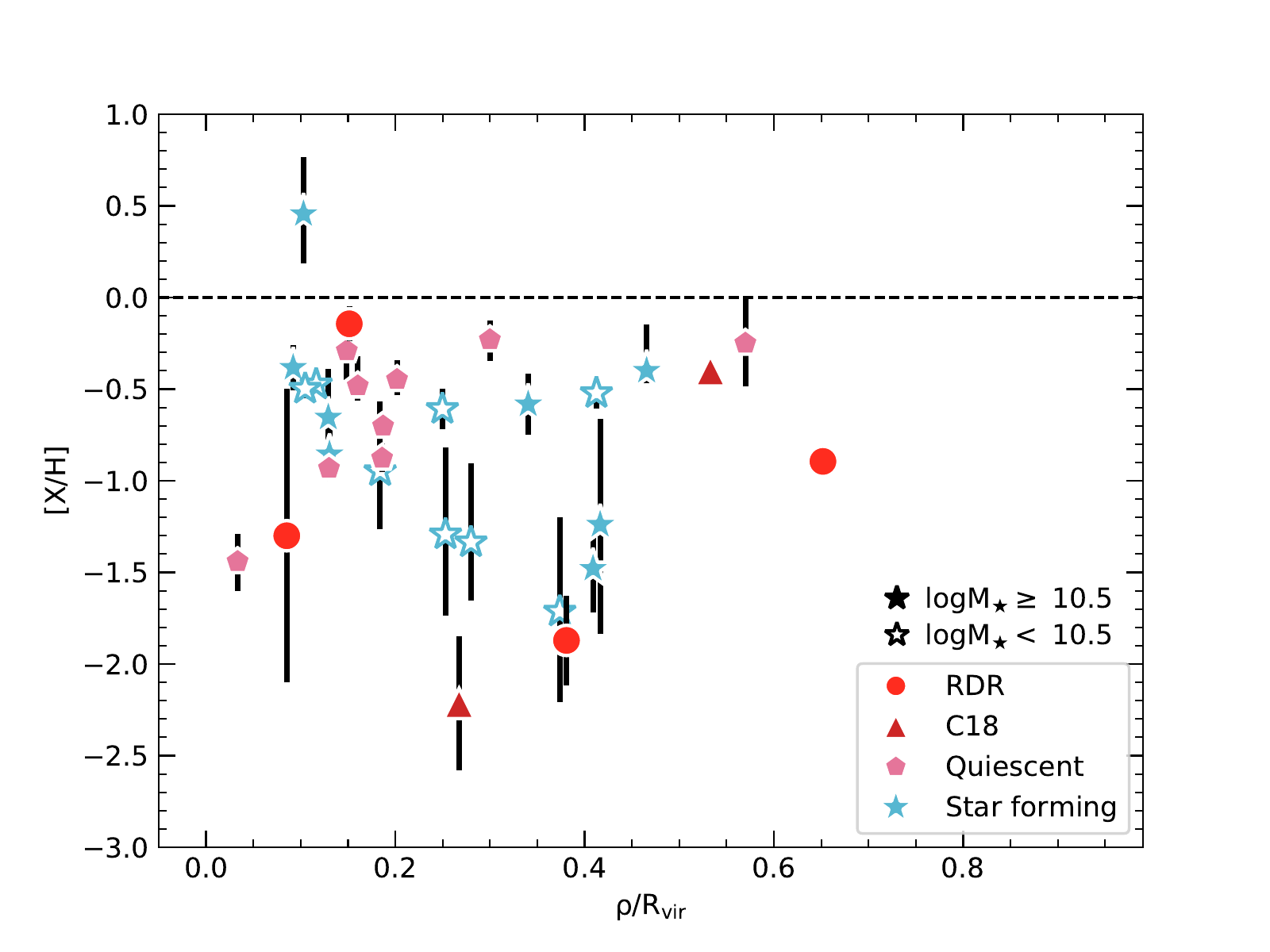}
    \caption{Metallicities of our combined sample of galaxy-selected pLLSs/LLSs versus normalized impact parameter.
    COS-Halos galaxies are split between star-forming and quiescent subsets at $\log {\rm sSFR } = -11$. Open symbols represent galaxies with $\logMstar < 10.5$. We include only absorbers with $\logNHI \ge 15.5$ with sufficient constraints to robustly calculate the metallicities. The metallicities for the COS-Halos and\added{one} \citetalias{chen2018} absorber are drawn from the CCC \citep{lehner2018,wotta2019}; thus, they are computed in a manner consistent with the results presented here. \added{See Appendix~\ref{appd} for information on the low-metallicity absorber in \citetalias{chen2018}.}}
    \label{fig:metrho}
\end{figure}

We plot the metallicity of the RDR absorbers versus normalized impact parameter in Figure~\ref{fig:metrho}; we also show two absorbers from the \citetalias{chen2018} sample with metallicities from the CCC \added{(one with new observations, see below)} and the metallicity measurements for absorbers associated with the COS-Halos galaxies (\citetalias{prochaska2017}). Out of 17 quiescent galaxies in the COS-Halos sample (defined by $\log {\rm sSFR } \le -11$), only nine have $\logNHI \ge 15.5$ and sufficient data to calculate an absorber metallicity. Out of 27 star-forming galaxies in the COS-Halos sample, only 16 have sufficient data and meet this \logNHI\ criterion. The COS-Halos metallicities in \citetalias{prochaska2017} were originally estimated using the \citetalias{hm2012} EUVB.\footnote{\deleted{J1009+0713\_170\_9 was originally reported with one value of \NHI, but we report \NHI\ and [X/H] for each of the two components. }The metallicity solution for J1241+5721\_199\_6 is not consistent with all of the observed column densities. For the SLLSs J0925+4004\_196\_22, J0928+6025\_110\_35, and J1435+3604\_68\_12 we use the column densities reported in \citet{battisti2012} from their Voigt profile fits. We could not reach a satisfactory solution for J0925+4004\_196\_22 and instead adopt the reported [O/H] metallicity from \citet{battisti2012} for this absorber.} To remove the systematic offset introduced when using different ionizing radiation fields to derive metallicities, all the metallicities for the COS-Halos absorbers have been recalculated with the \citetalias{hm1996} EUVB and are reported in \citet{wotta2019} and \citet{lehner2019}.

Figure~\ref{fig:metrho} shows both passive and star-forming galaxies have a similar broad range of metallicities. The COS-Halos sample includes a low-metallicity absorber with [X/H] $= -1.4 \pm 0.2$ in the CGM of a massive quiescent galaxy. Like the [X/H] $\approx -1.8$ system associated with our RDR LRG, this absorber is projected deep within its halo. These systems suggest low-metallicity absorbers might not be rare around quiescent galaxies. This result is also bolstered by a low-metallicity LLS (${\rm [X/H]} < -1.5$) reported around an LRG in the \citetalias{chen2018} sample. \replaced{In fact, using additional constraints from Keck/HIRES detections of \MgII, \citet{wotta2019} report the metallicity of this absorber to be extremely low, ${\rm [X/H]} = -2.45 \pm 0.24$.}{We have obtained Keck/HIRES observations that constrain the metallicity to the range $[-2.58,-1.85]$.} This absorber is not pristine, despite the lack of metal-line absorption in the COS spectrum analyzed by \citetalias{chen2018} (see \replaced{\citealt{lehner2018}}{Appendix~\ref{appd} for our treatment of this absorber}). \added{The line of sight velocities of these three absorbers are all within $\pm250$ \kms\ of their galaxy, signifying a high likelihood of being associated.}

\begin{figure}
    \epsscale{1.15}
    \plotone{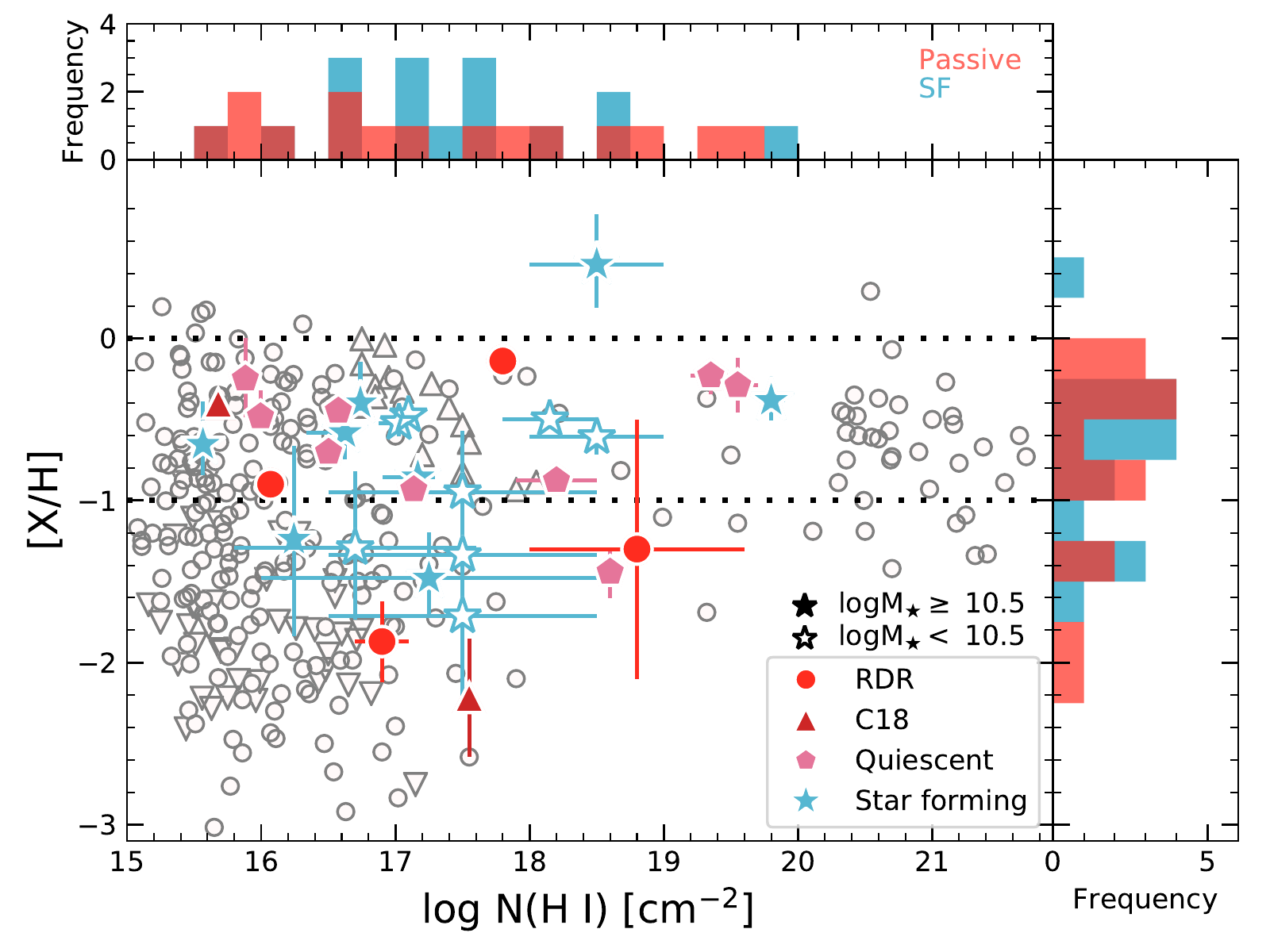}
    \caption{Metallicities versus \logNHI\ for absorption-selected systems from the CCC (grey symbols with downward- and upward-facing triangles being upper and lower limits, \citealt{lehner2019,wotta2019}) and our combined sample of galaxy-selected systems (colored, consisting of the RDR LRGs, \citetalias{chen2018}, and COS-Halos samples). All absorbers are at $z\la 1$. The dotted lines mark solar and 10\% solar metallicity. The vertical error bars for the galaxy-selected absorbers represent 68\% confidence intervals \added{(except for the two absorbers previously noted to be confined to a range of values)}. The error bars for the absorption-selected systems are not plotted, but they are similar to those from the galaxy-selected systems. The metallicities for the COS-Halos and \added{one of the} \citetalias{chen2018} absorbers are drawn from the CCC \citep{lehner2019,wotta2019}. The top sub-panel shows a histogram of the \logNHI\ distributions for the galaxy-selected absorbers, and the right sub-panel shows a histogram of the metallicity distributions for the galaxy-selected absorbers (central values are used for the histograms). The red histograms are a combination of all three passive galaxy samples (RDR LRGs, \citetalias{chen2018}, and the COS-Halos quiescent galaxies).}
    \label{fig:metnhi}
\end{figure}

Figure~\ref{fig:metnhi} shows the metallicity versus \logNHI\ for our sample of galaxy-selected clouds compared with \HI\ absorption-selected systems at $z < 1$. The blind \HI\ absorption-selected systems are shown in grey and are from the CCC \citep{lehner2019,wotta2019}, while the galaxy-selected absorbers (RDR LRGs, \citetalias{chen2018}, and COS-Halos) are shown in color. We also plot the \logNHI\ and metallicity histograms of the galaxy-selected absorbers of star-forming galaxies (blue) and passive galaxies (red, which includes the RDR LRGs, \citetalias{chen2018}, and COS-Halos quiescent galaxies) in the top and right sub-panels respectively.

Overall, the LRG absorbers do not seem to be fundamentally different in metallicity from the other galaxy-selected pLLSs/LLSs in spite of the different characteristics of their host galaxies. There is also no appreciable difference between the metallicity distributions of the star-forming and passive samples; most of the metallicity values lie between solar and a tenth solar, which can be explained by enrichment from stellar outflows or tidal material \citep{lehner2013}, and both samples have a similar fraction ($\sim$30\%) of low-metallicity absorbers ([X/H] $< -1$). There is also no appreciable difference between the distributions of star-forming galaxies split by stellar mass. Using a two-sample Kolmogorov-Smirnov test and K-sample Anderson-Darling test, we cannot reject the null hypothesis that the star-forming and passive metallicity distributions are drawn from the same population at better than 0.5 significance. We cannot reject the null hypothesis that the \logNHI\ values are drawn from the same distribution either. Thus, there is no evidence the metallicity of the cool CGM\textemdash at least for absorbers in the range $15.5 \la \logNHI \la 20$\textemdash is correlated with the galaxies' star-forming properties (although the samples are small). The LRG absorbers also have metallicities and \HI\ column densities consistent with the distributions in pure \HI\ absorption-selected systems. We find no significant difference between the absorption-selected and galaxy-selected absorbers.

\subsection{Relative Abundances of \FeII\ and \MgII}\label{sec:relabs}

For four of the LRG-associated absorbers, we have measurements of \logNFeII\ and \logNMgII\ from Keck/HIRES. We find a large spread in the values of [\FeII/\MgII] $\equiv \logNFeII - \logNMgII - \log ({\rm Fe/Mg})_\odot$ in these absorbers. This ratio does not include ionization corrections; \citet{zahedy2017a} have argued ionization corrections should be small given the similarity of the ionization potentials between \FeII\ and \MgII\ (although see Appendix \ref{appd}). Two of these absorbers exhibit super-solar values: LRG SDSSJ111508.24+023752.7 and LRG SDSSJ135726.27+043541.4 (a galaxy from \citetalias{chen2018}; cf., Appendix \ref{appd}) have [\FeII/\MgII$] = +0.37 \pm 0.03$ and [\FeII/\MgII$] = +0.50 \pm 0.08$, respectively. The other two sightlines can only be given upper limits: LRG SDSSJ111132.33+554712.8 and LRG SDSSJ171651.46+302649.0 show [\FeII/\MgII$] < -0.59$ (saturated \MgII) and [\FeII/\MgII$] < +0.03$ (\FeII\ upper limit), respectively. The impact parameters for these absorbers are $\rho$ = 44, 130, 81, and 207 kpc.

The observed [\FeII/\MgII] ratios are not correlated with metallicity.  The low-metallicity absorber associated with LRG SDSSJ135726.27+043541.4 displays a highly super-solar [\FeII/\MgII] ratio, whereas LRG SDSSJ171651.46+302649.0 exhibits a near solar value of [\FeII/\MgII], despite having a metallicity that is 1.5\% solar. The absorber associated with LRG SDSSJ111132.33+554712.8 has the highest metallicity ([X/H] $\approx -0.14$), yet it has the lowest [\FeII/\MgII] of all four sightlines, a highly sub-solar value. (A broad range of metallicities are allowed for the absorber associated with LRG SDSSJ111508.24+023752.7.)

\citet{zahedy2017a} have discussed enhanced \FeII\ columns compared to $\alpha$ element columns around massive red galaxies, arguing that some enhancement can be explained due to the enhanced contribution from Type Ia supernova ejecta from these galaxies' evolved stellar populations. They found individual components with significant Fe enhancement, though their average value was [\FeII/\MgII$] \approx +0.02$ for sightlines within $\sim$60 kpc of their galaxies. Our two detected absorbers are at significantly higher [\FeII/\MgII] than their mean, and only one sightline is located at a low impact parameter. The absorbers in our sample do not seem to follow a radial trend in enhancement.

Though we do not know the specifics of the satellite galaxies around the RDR LRGs, \citet{chen2019} investigated the environment around LRG SDSSJ135726.27+043541.4. This LRG is housed within a group, but it is the only massive galaxy. Since this absorber exhibits the highest [\FeII/\MgII] ratio, it is unclear how such an enhancement occurred at such large distances from the central, massive galaxy. Future studies of LRGs and other massive galaxies will need to measure this ratio and the absorber metallicity in order to increase our understanding of the Fe enrichment of their CGM.

\section{Discussion}\label{sec:discussion}

Using a sample of 21 LRGs with UV observations of background QSOs with
$\rho \la 500$ kpc, we have mapped the distribution of cool, neutral
hydrogen absorption in the CGM of these massive galaxies. We have measured the covering factor and metallicity of gas
with $\logNHI \ge 16.0$ in the CGM of LRGs. We find substantial covering
factors, especially in the inner regions of the CGM. The covering
factor within \Rvir\ for our sample is $f_c(\rho \le \Rvir) =
0.27^{+0.11}_{-0.10}$. The metallicities of the absorbers are mixed,
with both metal-rich systems at [X/H]$\, \ga -1$ and metal poor with
metallicities as low as [X/H] $\approx -1.8$. The covering factors
observed about the RDR LRGs are about 1/2 of those seen in lower-mass star-forming galaxies; the metallicity distributions of strong \HI\ absorbers about LRGs are consistent with those observed about lower-mass galaxies.

\subsection{Mass of Cool Gas in the LRG CGM}\label{sec:coolmass}

To assess the relative importance of this cool \HI\ to the overall mass budget
of the LRGs, we calculate the mass of cool gas associated with the measured \HI\ as:
\begin{equation}
M^{\rm cool}_{\rm CGM}(\HI) \simeq
    \int^{\rho_{max}}_{\rho_{min}} m_p N(\mbox{\ion{H}{1}})\relax \, 2{\pi}{\rho}f_c(\rho)\, d{\rho},
\end{equation}
where \NHI\ is a function of the impact parameter $\rho$. We do not have good constraints on the functional
distribution of \NHI\ or the covering factor for $\logNHI \ge 16.0$, so we
instead break this integral up into two discrete bins $[\rho_{min}$,
$\rho_{max}]$ = [0, 0.5 \Rvir] and (0.5 \Rvir, \Rvir], assuming constant
$\langle \NHI \rangle$ and $f_c(\rho)$ in each bin. \added{For this calculation and those that follow we use the median \logNHI\ value for the LLS associated with LRG SDSSJ111508.24+023752.7.} The sum of these two bins
gives the typical total \HI\ mass per LRG, $M_{\rm CGM}^{\rm cool}(\HI) \approx
1\times10^9$ \msun. We estimate the total cool gas mass as $M_{\rm CGM}^{\rm cool} =
{\mu}M_{\rm CGM}^{\rm cool}(\HI)/x(\HI)$, where the average ionization fraction of \HI\
for the pLLSs/LLSs comes from the best-fit Cloudy models, and $\mu = 1.4$ to account for
helium. The typical LRG has a total cool CGM mass of $M_{\rm CGM}^{\rm cool} \approx
1\times10^{11}$ \msun\ in its halo, assuming the average value $x(\HI) = 0.01$ from our Cloudy calculations (our absorbers exhibit a range $0.002 \le x(\HI) \le 0.02$ consistent with \citealt{lehner2019}). \added{This mass is 2.5 times larger than the estimate in \citet{zahedy2019}.} Compared to the expected
baryonic mass for these galaxies,\added{$M_b = M_h(\Omega_b/\Omega_m) \approx 4\times 10^{12}$ \msun}, only
$\sim$3\% of the baryons is in the cool CGM, while $\sim$6\% is in stars. If the
hot gas in the LRG halo dominates the remaining baryon mass, it could represent
$> 10^{12}$ \msun.

The total cool CGM mass has also been calculated around $L$* galaxies. The COS-Halos sample, treated as one halo with many sightlines, has $M_{\rm CGM}^{\rm cool} \approx 9\times10^{10}$ \msun\ within $\rho = 160$ kpc (\citetalias{prochaska2017}, but see \citealt{bregman2018}). \citet{keeney2017} supplemented the COS-Halos sample with measurements extending to \Rvir; they adopted different assumptions for the structure and ionization of the gas to determine $M_{\rm CGM}^{\rm cool} \approx 3\times10^{10}$ \msun. The cool CGM mass of M31 is estimated to be $M_{\rm CGM}^{\rm cool} \gtrsim 10^9$ \msun\ within \Rvir\ assuming a solar metallicity for the gas \citep{lehner2015}.\footnote{The erratum for \citet{lehner2015} lists this updated cool CGM mass estimate.} However, while the cool CGM mass in LRGs is comparable to that for lower-mass galaxies, it is a much smaller fraction of the total baryons given their larger halo mass. For the RDR sample, the cool CGM is $\sim$40\% of the stellar mass of the galaxy. M31 shows a 10\% fraction, while the COS-Halos galaxies (and the extended \citealt{keeney2017} results) have almost twice as much mass in the cool CGM as the stellar disk. That said, the cool CGM gas in LRGs is not unimportant as we show below.

\subsection{Origins and Fate of Cool Gas in LRG Halos}
\label{sec:origins}

Cool gas is clearly found in the halos of LRGs and other massive, quiescent
galaxies (e.g., \citealt{thom2012}; \citetalias{prochaska2017,chen2018}; this work).
Several models predict cool gas can form through
cooling instabilities in the hot corona and fall onto the central galaxy in
high-mass halos (e.g., \citealt{mb2004,voit2017,correa2018}). Metal-rich gas can
more efficiently cool and condense out of the hot halo due to the increased
number of cooling channels. Coupled with the shorter cooling time associated
with denser gas, we expect condensation is most prevalent at smaller radii
\citep[e.g., see the definitions of the cooling or precipitation
radii in][]{voit2015, correa2018}. For metal-rich gas with $[{\rm X/H}]\approx0$, the ratio of the cooling to free-fall time is $t_{\rm cool} / t_{\rm ff} \la 10$\textemdash a characteristic criterion for the
formation of thermal instabilities in galaxy halos \citep{sharma2012}\textemdash at $r
\la$ 0.5 \Rvir\ (assuming a gas profile following \citealt{mathews2017}, cooling
losses from \citealt{gnat2007}, and a \citealt{nfw1996} mass profile). For metallicities
below $[{\rm X/H}] \la -1$, gas within $r \la$ 0.25 \Rvir\ typically meets these
criteria. Certainly the absorbers we find around LRGs
have higher column densities on average at smaller impact parameters, and the
highest metallicity gas is seen within $\rho \la$ 0.25 \Rvir\ in Figure~\ref{fig:metrho}. Thus, the high metallicity absorbers could plausibly represent gas precipitated from the corona.

However, the lowest metallicity gas in our combined sample of massive galaxies
is also seen in the inner CGM within $\pm$250 \kms\ of the galaxy. The [X/H] $\approx -1.8$ absorber reported here
is found at $\rho \approx$ 0.4 \Rvir, while the [X/H] = $[-2.58,-1.85]$ system drawn from the
\citetalias{chen2018} sample is at an impact parameter $\rho \approx$ 0.3 \Rvir. It is difficult to imagine such low-metallicity gas could have condensed out of the halo, as the typical corona about galaxies with masses similar to our RDR sample are expected to be significantly more metal rich (e.g., the ICM of clusters is often quite metal rich, \citealt{mernier2018}, while the Milky Way corona is thought to have $[{\rm X/H}] \ga -0.7$, \citealt{miller2013}).
As such, this low-metallicity material could be material accreted from the IGM or stripped from an infalling satellite galaxy.
As we discuss below, cold gas like this would be susceptible to relatively rapid evaporation
if embedded within a diffuse, hot corona, which makes the direct accretion
scenario problematic. On the other hand, extremely-metal-poor dwarf galaxies with $[{\rm
X/H}] \la -1.5$ are rare enough \citep{izotov2018} that it is also difficult to
explain the prevalence of very-metal-poor pLLSs/LLSs seen at low redshift
through satellite stripping \citep{lehner2013,wotta2019}. Given these difficulties, the origins of the gas are not yet clear. While it is possible to explain the more metal-rich gas as material
directly condensed from a hot corona (similar to hot-mode accretion or
precipitation models) or through satellite stripping, this cannot universally
explain the cold, dense gas we see in the CGM of LRGs.

\added{Another potential source for the absorbers we observe is recycled winds. Stellar winds do not seem to be a likely source for this gas, as the LRGs have had no star formation for the past Gyr \citep{gauthier2011}. We also do not see any signatures of stellar winds in the kinematics of the gas (e.g., extended velocity profiles) like those observed around starburst galaxies \citep{heckman2017}. AGN winds are also unlikely since we do not see a strong population of AGN in LRGs, indicating the duty cycle is long. In addition, the timescale for recycling an AGN wind for the LRG mass scale is $10^8-10^9$ yr \citep{oppenheimer2010}. The most recent AGN outburst would need to be $\sim$Gyr ago for us to be detecting the gas now. This AGN activity would be less cyclic and more sporadic (see also \citealt{afruni2019}). Even though we can rule out recent winds as a potential source, it is possible the absorbers are material recycling from an ancient outflow that was blown out of the galaxy when the AGN was active or star formation was occurring. This could be the origin of the super-solar [\FeII/\MgII] abundances observed in two of the absorbers.}

The ultimate fate of the cool gas we detect is clearly not surviving the fall into the center of the galaxy and fueling star formation. If all of the mass of the cool CGM were to collapse to the central galaxy within a free-fall time (see below) from \Rvir,
$t_{\rm ff}(\Rvir)$, then the galaxy would have $\rm sSFR \sim 10^{-10} \ {\rm
yr^{-1}}$, assuming maximum efficiency. We do not see sSFRs this high for the LRGs frequently (and the \OII\ emission seen in these galaxies may not trace star formation). It appears that much of the cool CGM gas does not actually make it to the
galaxy cold, or it is somehow kept from forming stars if it does reach the central galaxy.

It is not clear that we expect the gas to survive such a fall through the corona of the LRGs. Here we consider the evaporative destruction timescales for dense, cool gas in a hot LRG corona, comparing it with typical free-fall times. We calculate the expected evaporation timescale for a spherical cloud in the halo of an LRG following \citet{mb2004} and the discussion in \citet{gc2011}. Specifically, we evaluate equation 35 from \citet{mb2004} as
\begin{eqnarray}
\tau_{\rm evap} & \simeq & (16\, {\rm Gyr}) \left(\frac{n_Hm_p(4/3){\pi}r^3_{\rm cl}}{10^6\, \msun}\right)^{2/3}  \nonumber \\
  & & \times \left(\frac{T}{10^6\, {\rm K}}\right)^{-3/2}  \left(\Lambda_z\frac{t_f}{8\, {\rm Gyr}}\right)^{-1/3}.
\end{eqnarray}
We calculate the virial temperature of the median LRG halo \citep[e.g., equation
1 of][]{oppenheimer2016} $T_{\rm vir} = 6\times10^6$ K. We adopt $\Lambda_z =
1.0$ for the metallicity scaling of the cooling function (Appendix A in \citealt{mb2004}), which is appropriate for gas with 0.1
Z$_{\odot}$, close to the mean metallicity of the LRG absorbers in our
study (0.11 Z$_{\odot}$). We assume $t_f$, the halo formation time, is the age of the universe at $z \sim
0.5$. We adopt the average density for the absorbers calculated from the
best-fit Cloudy models, $\nH = 10^{-2.4}$ \percc; for the radius of the cloud we
use half of the average length scale of the absorbers, $r_{\rm cl} = 0.5 \times l
= 0.5 \times N_{\rm HI}/x(\HI)\nH \approx 0.8$ kpc. These values are comparable to the median
values seen in the \citet{lehner2019} sample of pLLSs/LLSs. These assumptions
give an evaporation timescale of $\tau_{\rm evap} \sim440$ Myr.

For these clouds to survive to accrete into the center of the galaxy, the
evaporation timescale should be greater than the free-fall timescale, $t_{\rm ff}$.
We assess $t_{\rm ff}$ at \Rvir\ and 0.5 \Rvir\ for an LRG halo, evaluating the
expression
\begin{equation}
    t_{\rm ff} =
        \left(\frac{3\pi}{32G\bar{\rho}}\right)^{1/2},
\end{equation}
where the average density $\bar{\rho} = 3 M_{\rm encl}/4{\pi}R^3$. For the
calculation at $R = \Rvir$ we adopt the median $M_{\rm h}$ of our sample for
$M_{\rm encl}$. This yields $t_{\rm ff}(\Rvir) \sim 1.3$ Gyr. For $R = 0.5\, \Rvir$ we
integrate the Navarro-Frank-White density profile \citep{nfw1996,nfw1997} for a
dark matter halo to determine the mass enclosed within a radius
$r$. Using a concentration parameter, \added{$c = R_{200}/R_s = 3.7$}, from \citet[][their Equation 8 and the Table 2 parameterizations for $0.4 \le z \le 0.6$]{shan2017} and our values for $\Rvir \equiv R_{200}$, we find $t_{\rm ff}(0.5\, \Rvir) \sim 600$ Myr.

These considerations suggest that if a cloud falls into the halo from the IGM it
is unlikely to survive to the center of the LRG before evaporating. As discussed
above, however, there is not another obvious way to produce cool gas with small velocity offsets and
metallicities as low as we observe deep in the halos of quiescent, massive
galaxies (but see \citealt{peeples2019}). More metal-rich gas that condenses
from the corona within $\rho \lesssim 0.5$ \Rvir\ may roughly remain intact
to the centers of the LRGs.
However, even if the cool gas from the inner CGM makes it to the center of the
galaxy, it evidently does not fuel significant star formation.

AGN feedback could plausibly play a role in this. While LRGs do not house many active AGN
\citep{sadler2007,hodge2008,hodge2009}, the duty cycle of such AGN is not
well-known. Most LRG AGNs could be in a quiescent state awaiting the build-up
of a cool gas reservoir to fuel its ignition\added{, as in the precipitation model proposed by \citet{voit2017}}. \added{However, due to the short feedback cycle in this model, it seems unlikely AGN would be the cause of the quenched star formation we currently see.} We note that these calculations have, of
course, been done in the absence of magnetic fields, which could significantly
affect the conductive evaporation timescales (e.g., \citealt{lm2019}). However, even if the timescale is increased and the gas is able to survive into the halo core, we still do not see the gas being transformed into stars due to the low SFRs.

There may be evidence in the distribution of \HI\ column
densities in the most massive halos that evaporation or some other destructive
mechanism is at work. The paucity of \HI\ absorption systems below $\logNHI \la 16$
about the most massive galaxies may be due to the destruction of the clouds that would lead to such absorption. There is no lack of such absorption in lower-mass galaxies (e.g.,
\citealt{keeney2017}, \citetalias{prochaska2017}). This difference can be seen in Figure~\ref{fig:coveringfactor_stellarcompare}, where the covering factor continues to
rise towards lower column densities in the two lower bins of galaxy mass
($\logMstar \la 11.3$), while it is consistent with a flat profile for the
highest-mass galaxies. If the lower column density gas is on average lower
density, it may be more susceptible to evaporative destruction in the dense, hot
coronae about the most massive galaxies. Radiative and other feedback effects,
especially from AGN, could also play a role in suppressing the amount of low
column density gas in the halos of massive galaxies, although this effect should
also be present in lower-mass systems.

\section{Summary and Concluding Remarks} \label{sec:conclusion}

This paper presents the first results from the RDR survey. We analyze the CGM of 21 LRGs at $z\sim0.5$ using UV spectroscopy of background QSOs projected within $\rho \le 500$ kpc. We measure the covering factor of \HI\ about the RDR LRGs and determine the metallicity of the strong \HI\ absorbers. Our main results are as follows.

\begin{enumerate}

    \item We detect \HI\ absorption in 11/21 sightlines in the CGM of LRGs. We
    detect four strong \HI\ absorbers: two pLLSs and two LLSs. There is a dearth of
    absorption between $14.5 \le \logNHI \le 16.0$, and we show that any absorbers with $\logNHI \le 14.5$ most likely arise in the IGM.
    The covering factor of strong \HI\ absorption about the RDR LRGs at small impact parameters is $\sim$1/2 that seen about
    lower-mass star-forming galaxies. The covering factor for
    $\logNHI \ge 16.0$ about LRGs is $f_c(\rho \le \Rvir) =
    0.27^{+0.11}_{-0.10}$.

    \item Combining our data with previous galaxy-selected samples
    \citepalias{prochaska2017,chen2018}, we estimate the covering factors of \HI\
    as a function of galaxy mass. The covering factor of gas with $\logNHI \ga
    16.0$ in the highest-mass galaxies ($\log M_\star \ga 11.3$) is marginally
    smaller than that of lower-mass galaxies (differences at only the
    $\sim$1$\sigma$ level). If there is a transition between hot- and cold-mode
    accretion with mass, it shows little signature in strong \HI\ systems. Lower-mass galaxies have significantly higher covering factors for gas with $14.5
    \le \logNHI \le 16.0$, which is generally not seen in the highest-mass
    galaxies.

    \item We estimate the metallicity of the four strong \HI\ absorbers about the RDR
    LRGs. \replaced{Three}{Two} of these are metal-rich with [X/H] $\ge -1$, \added{one is unconstrained, and} the fourth
    absorber is metal-poor at [X/H] $\approx -1.8$.
    Other identifications of metal-poor absorbers in the inner halos of
    massive galaxies have been reported in the literature. Although the sample is still
    small, these metal-poor absorbers do not appear to be rare in the CGM of LRGs. The frequency of low-metallicity systems is also similar between
    star-forming and quiescent galaxies. There is no statistically
    significant difference in the metallicity distributions of cool CGM gas as a function of
    stellar mass.

\end{enumerate}

Our observations represent tests of our understanding of the gaseous
environments of massive galaxies, including our prevailing theories of how
galaxies accrete gas. We have found plentiful dense, cold gas surviving deep
in the halos of very massive galaxies. The mass of this gas is similar to that
found in lower-mass star-forming galaxies, although it represents a lower
fraction of the total baryonic mass in the high-mass halos.

The metal-rich cool gas we observe could plausibly have condensed from the dense
halos in these galaxies. If the coronae of these galaxies contain a majority
of their halo's baryons, the inner regions may have conditions in which sustained formation of cool clouds via thermal instabilities is possible. The high column density (high density)
gas we observe is concentrated toward the inner regions (at least in
projection), as expected if the gas is condensing from the highest-density (and
perhaps highest-metallicity) regions of the corona, where the cooling times are
shortest. The relatively small velocity offsets between the cold gas and the
central galaxies is also consistent with this origin, as has been noted previously for more
massive, quiescent galaxies \citep[e.g.,][]{thom2012,tumlinson2013}.

At the same time, cooling from the corona cannot be the entire story. The
metal-poor absorbers seen in our RDR sample and others
\citepalias{prochaska2017,chen2018}, with metallicities below $[{\rm X/H}] \la
-1.8$, cannot be explained in this way, as the coronae of these massive galaxies
should have metallicities in excess of $[{\rm X/H}] \ga -1$, if not
substantially higher \citep{miller2013,mernier2018}. While in principle such gas could arise as matter being stripped from very-low-mass dwarf satellites, we argue that this is unlikely given their frequency and the small impact parameters at which we see such gas (and it would require the galaxy to maintain its gas to small impact parameters
against ram-pressure stripping and probably the dwarf to be on its first pass
close to the central halo). Alternatively, the low-metallicity cold gas may
have arisen in the IGM and remained cold through a putative accretion
shock, surviving against evaporation as it fell. This is not entirely
satisfying either, unless we have misunderstood the importance of these
phenomena.

If there is a critical mass above which massive cold clouds are unable to exist for a
significant amount of time, it is not clear from our measurements. We find similar covering factors with mass and find metal-poor absorbers in high-mass galaxies with roughly the same frequency as seen in
lower-mass systems. The only clear distinction between clouds in the highest-mass halos and those about
lower-mass galaxies is at column densities $\logNHI \la 16.0$, which are
essentially absent in the highest-mass galaxies ($\logMstar \ga 11.3$). The lack
of \OVI\ in these halos (see \citetalias{howk2019}) also indicates there is
little cool gas at very low densities, $\log n_{\rm H} \la -4$
\citep[e.g.,][]{stern2018,rocafabrega2019}, which would give rise to photoionized \OVI.
This may imply gas below some critical density is readily evaporated, heated, and
ionized to a point where it is not detected in our \HI\ selection, while the
higher density gas can survive. However, the survival of cold gas at small
impact parameters may be telling us that the evaporation within a diffuse hot
corona must consider other mitigating factors, such as magnetic fields.

Even though the LRGs have a supply of cool gas at low impact parameters,
some process must be at work \added{either} keeping it from reaching the central galaxy \added{or from cooling further}
and forming stars. The presence of these
clouds coupled with the lack of significant star formation in the central
regions implies there is a mechanism actively quenching the formation of stars
from this material \added{on the timescale of $<$ 1 Gyr}.

In the second paper in this series \citepalias{howk2019}, we focus on the high-ionization phase of the CGM traced by \OVI\ for the RDR sample of LRGs \citep[see also][]{zahedy2019}. We are also conducting a survey to observe the CGM of 50 LRGs within $\sim$0.3 \Rvir\ with \hst/COS. We will employ these data to further delineate the metallicity distribution function of CGM absorbers about LRGs.

\acknowledgments

Support for this research was provided by NASA through grants HST-AR-12854 and HST-GO-15075 from the Space Telescope Science Institute, which is operated by the Association of Universities for Research in Astronomy, Incorporated, under NASA contract NAS5-26555. Support for the KODIAQ project that enabled some of this work was provided by the NSF through grant AST-1517353 and by NASA SMD through grant NNX16AF52G.
Funding for the Sloan Digital Sky Survey IV has been provided by the Alfred P. Sloan Foundation, the U.S. Department of Energy Office of Science, and the Participating Institutions. SDSS-IV acknowledges
support and resources from the Center for High-Performance Computing at
the University of Utah. The SDSS web site is www.sdss.org. SDSS-IV is managed by the Astrophysical Research Consortium for the
Participating Institutions of the SDSS Collaboration including the
Brazilian Participation Group, the Carnegie Institution for Science, Carnegie Mellon University, the Chilean Participation Group, the French Participation Group, Harvard-Smithsonian Center for Astrophysics,
Instituto de Astrof\'isica de Canarias, The Johns Hopkins University,
Kavli Institute for the Physics and Mathematics of the Universe (IPMU) /
University of Tokyo, the Korean Participation Group, Lawrence Berkeley National Laboratory,
Leibniz Institut f\"ur Astrophysik Potsdam (AIP),
Max-Planck-Institut f\"ur Astronomie (MPIA Heidelberg),
Max-Planck-Institut f\"ur Astrophysik (MPA Garching),
Max-Planck-Institut f\"ur Extraterrestrische Physik (MPE),
National Astronomical Observatories of China, New Mexico State University,
New York University, University of Notre Dame,
Observat\'ario Nacional / MCTI, The Ohio State University,
Pennsylvania State University, Shanghai Astronomical Observatory,
United Kingdom Participation Group,
Universidad Nacional Aut\'onoma de M\'exico, University of Arizona,
University of Colorado Boulder, University of Oxford, University of Portsmouth,
University of Utah, University of Virginia, University of Washington, University of Wisconsin,
Vanderbilt University, and Yale University.
Some of the data presented herein were obtained at the W. M. Keck Observatory, which is operated as a scientific partnership among the California Institute of Technology, the University of California and the National Aeronautics and Space Administration. The Observatory was made possible by the generous financial support of the W. M. Keck Foundation. The authors wish to recognize and acknowledge the very significant cultural role and reverence that the summit of Maunakea has always had within the indigenous Hawaiian community. We are most fortunate to have the opportunity to conduct observations from this mountain.
This research was supported in part by the Notre Dame Center for Research Computing through the Grid Engine software and, together with the Notre Dame Cooperative Computing Lab, through the HTCondor software.


\facilities{HST (COS, FOS), Keck:I (HIRES), Sloan}

\software{astropy \citep{robitaille2013,astropy2018},
          Cloudy \citep{ferland2013}, Matplotlib \citep{hunter2007}}



\bibliography{lrgrefs}


\clearpage

\input{Table_ObjectLocation}
\clearpage

\input{Table_LRGproperties}
\input{Table_QSOproperties}
\input{LRGcovering_columns}
\input{coveringfactor_ssfr}
\input{coveringfactor_stellar}
\input{coveringfactor_halos}
\input{Table_GridParameters}
\input{Table_metallicities}
\clearpage

\clearpage
\appendix

\makeatletter
\renewcommand{\thefigure}{A\@arabic\c@figure}
\renewcommand{\thetable}{A\@arabic\c@table}
\makeatother
\setcounter{table}{0}
\setcounter{figure}{0}

In these appendices, we provide ion information for each sightline and the ion profiles from the QSO spectra, the results of covering factor calculations at the optically-thick absorption limit ($\logNHI \ge 17.2$), detailed metallicity results for each absorber \added{, and our metallicity estimate for the low-metallicity absorber reported in \citetalias{chen2018}}.

\section{Sightline Information}\label{appa}

\begin{figure*}
    \epsscale{1.0}
     \plotone{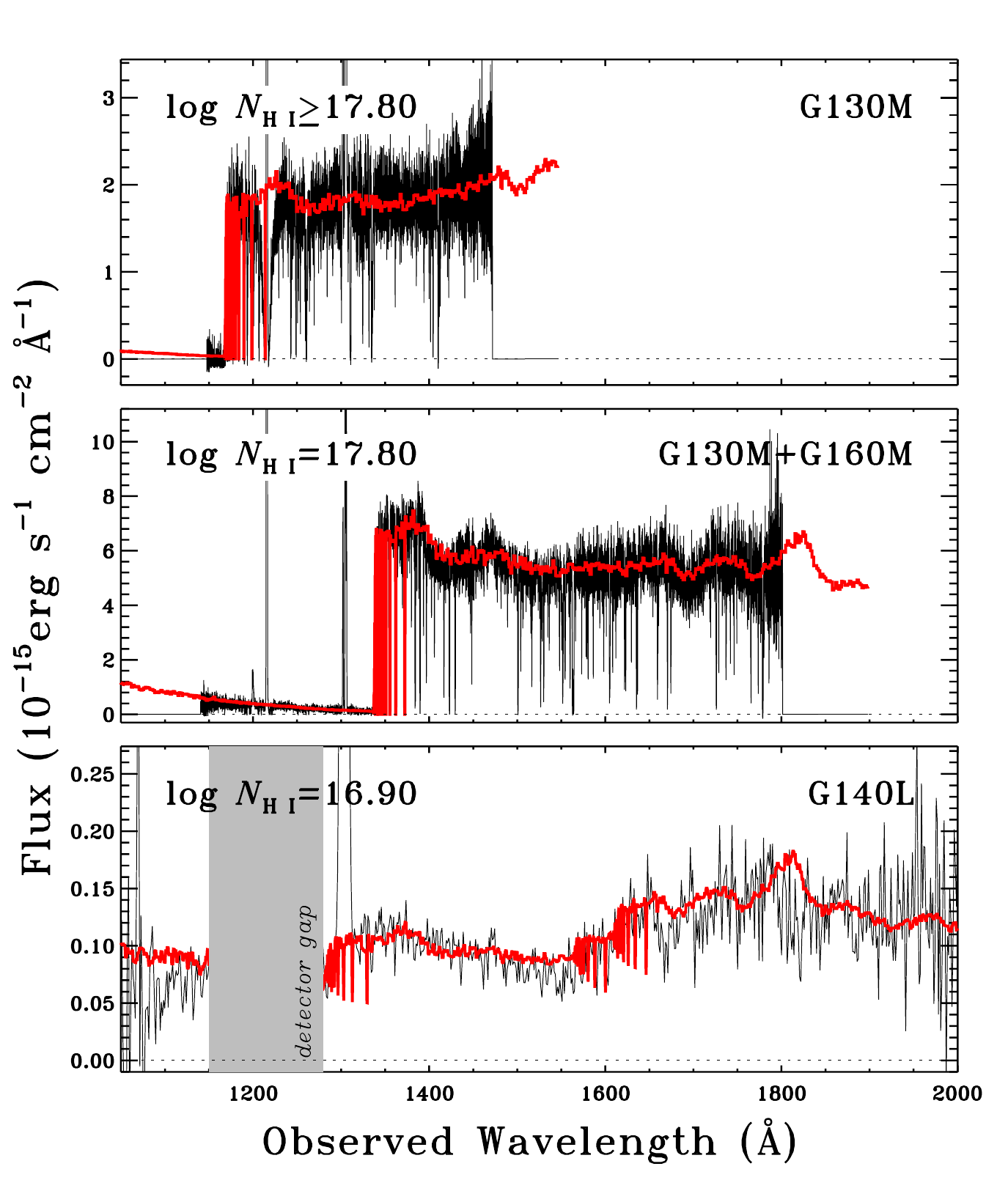}
    \caption{Spectrum of SDSSJ111507.65+023757.5 (top), SDSSJ111132.18+554726.1 (middle), and SDSSJ171654.20+302701.4 (bottom) with \added{\citet{telfer2002}} model overlaid. We calculate a lower limit on the hydrogen column density for the absorber along the SDSSJ111507.65+023757.5 sightline from this method because there is not enough flux recovery blueward of the Lyman limit to get a more accurate measurement.\deleted{A Voigt profile fit of the Lyman series lines gives a constrained column density for this absorber.} For the SDSSJ171654.20+302701.4 sightline, we show the composite model from three pLLSs in this spectrum as reported in \citet{wotta2016}. The break in the QSO spectrum at the Lyman limit for the absorber associated with an RDR LRG is located at $\lambda \sim 1300$ \AA.}
    \label{fig:breaks}
\end{figure*}

\begin{figure*}
    \epsscale{1.0}
     \plotone{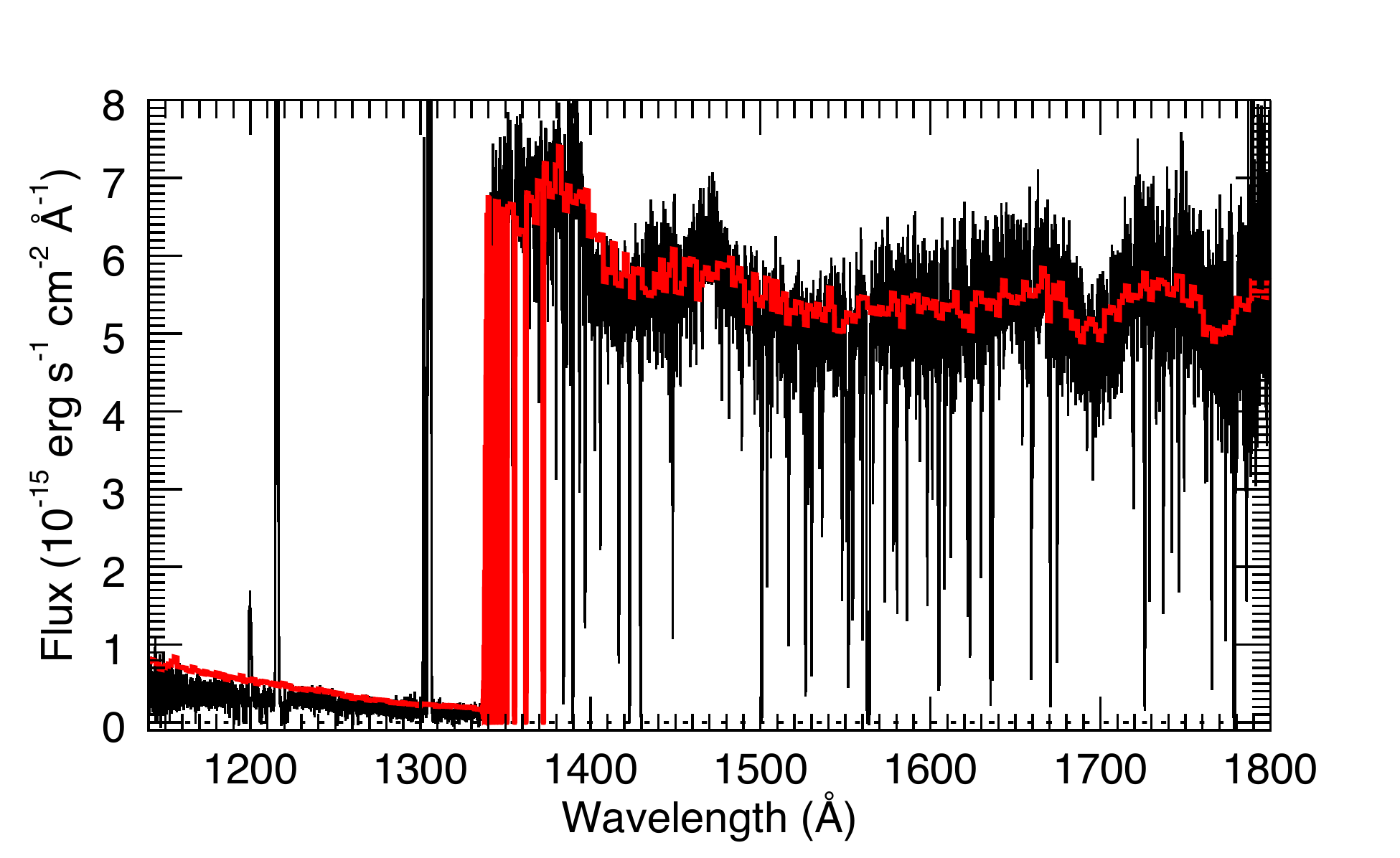}
    \caption{Spectrum of SDSSJ111132.18+554726.1 with \citet{telfer2002} model overlaid. In the flux recovery section of the QSO spectrum, we have decreased the value of \logNHI\ for the absorber to the point where the model now rests on top of the spectrum. This \logNHI\ value is taken to be the 2$\sigma$ value.}
    \label{fig:breakerror}
\end{figure*}

\begin{figure*}
    \epsscale{0.5}
     \plotone{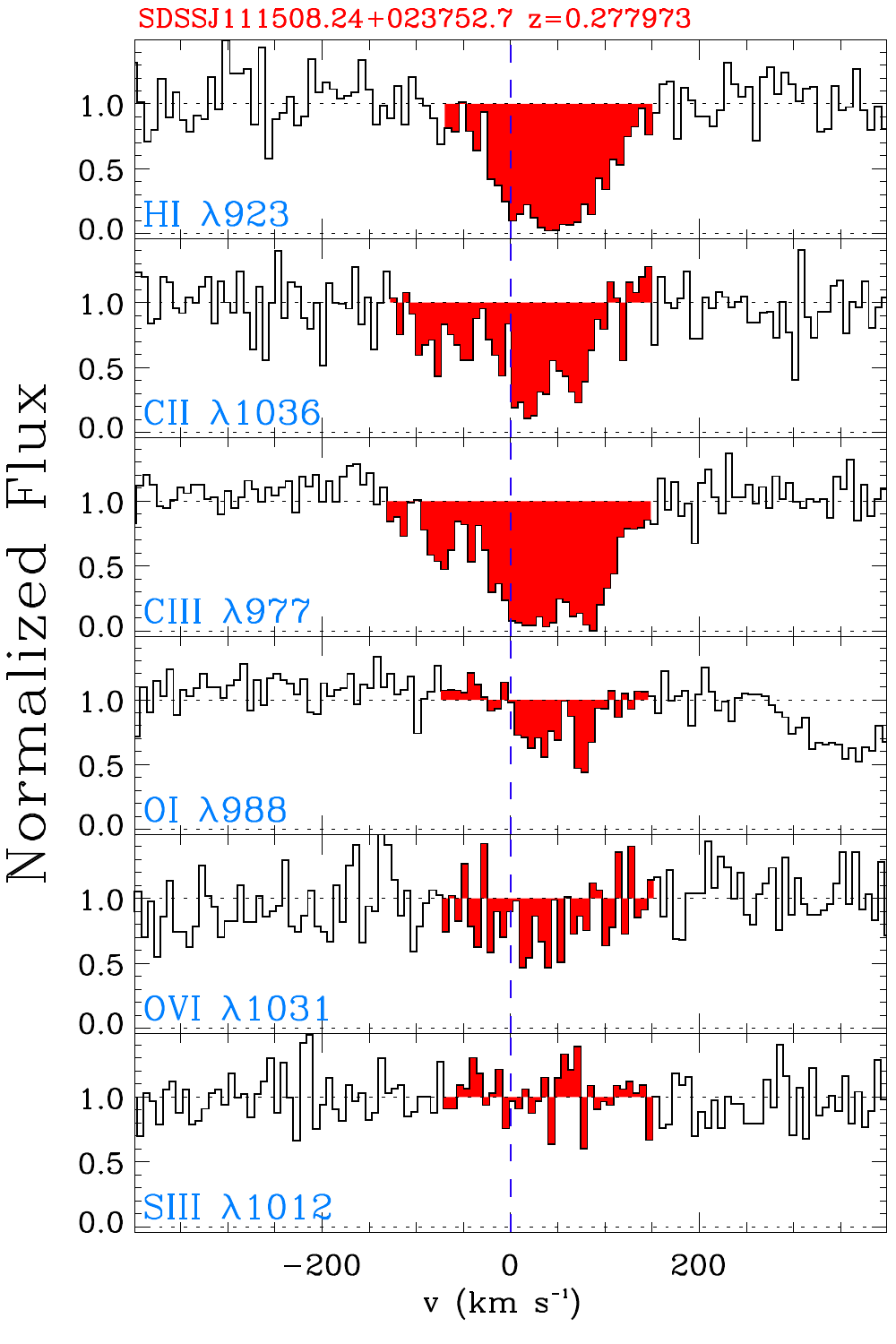}
    \caption{Ion profiles for absorption associated with LRG SDSSJ111508.24+023752.7 at an impact parameter of $\rho = 44$ kpc. The profiles are shown in the reference frame of the LRG, and the red shading shows the section of the spectrum integrated to determine the ion column density.}
    \label{fig:start}
\end{figure*}

\begin{figure*}
    \epsscale{1}
     \plotone{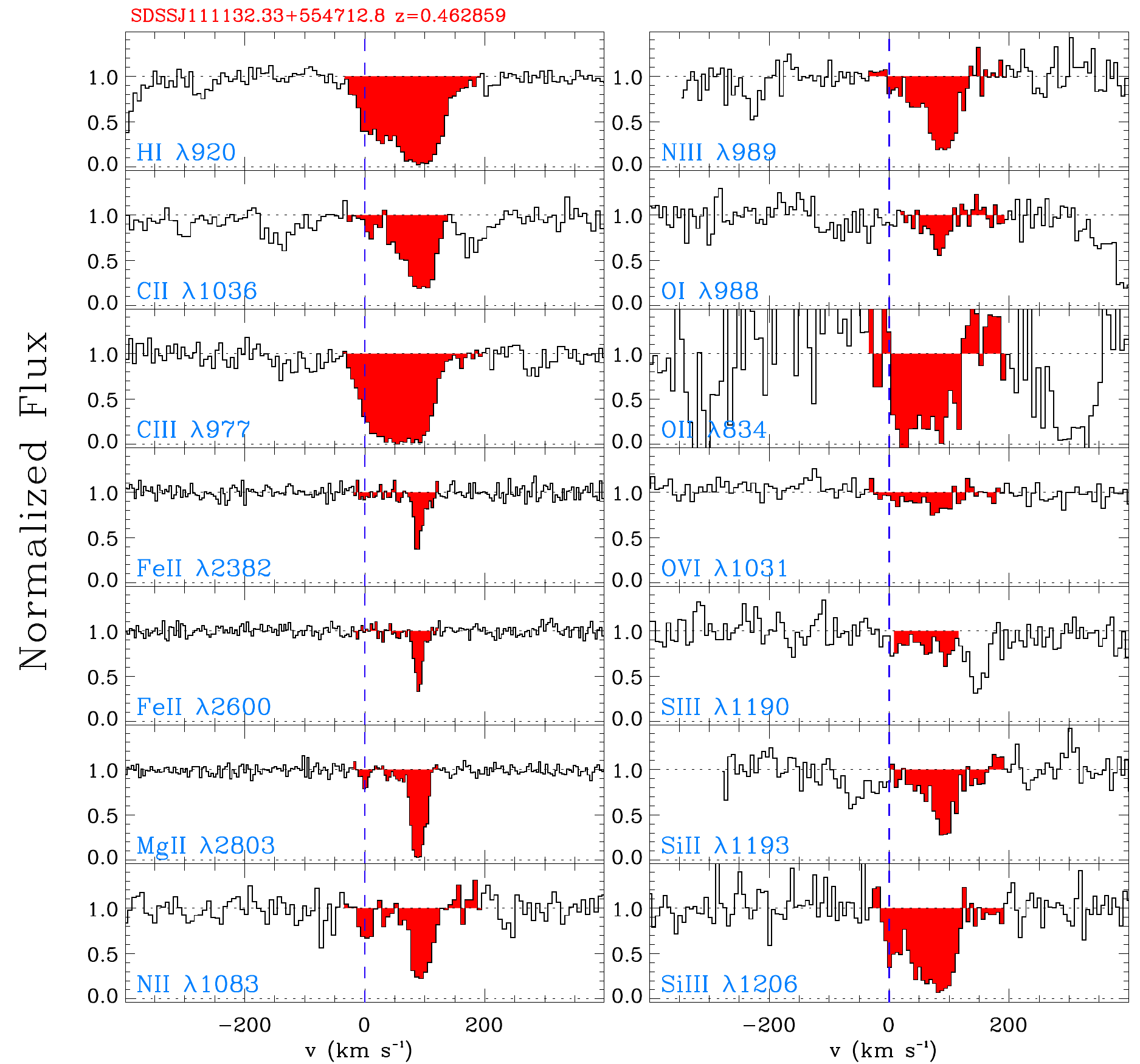}
    \caption{Same as Figure~\ref{fig:start}, but for absorption associated with LRG SDSSJ111132.33+554712.8 at an impact parameter of $\rho = 81$ kpc. The weak \HI\ component of this absorber is integrated from -31 to 28 \kms.}
    \label{fig:appc2}
\end{figure*}

Here we provide the ion information for each sightline. The three spectra for which we fit a break in the QSO spectrum at the Lyman limit to determine \logNHI\ are presented in Figure~\ref{fig:breaks}\added{, and an example fit of how we determine the error for \logNHI\ is presented in Figure~\ref{fig:breakerror}}. \deleted{The Voigt profile fits of the Lyman series lines for the absorber that could only be given a lower limit for \logNHI\ from the break in the QSO spectrum are presented in Figure (see \S\ref{sec:coldens} for the description of the profile fitting).} The ion profiles for each sightline follow in Figures~\ref{fig:start}-\ref{fig:stop}. Table~\ref{tab:iontable} details the integration limits, \added{average} velocity, and column density for the ions measured for each sightline.

\begin{figure*}
    \epsscale{0.5}
     \plotone{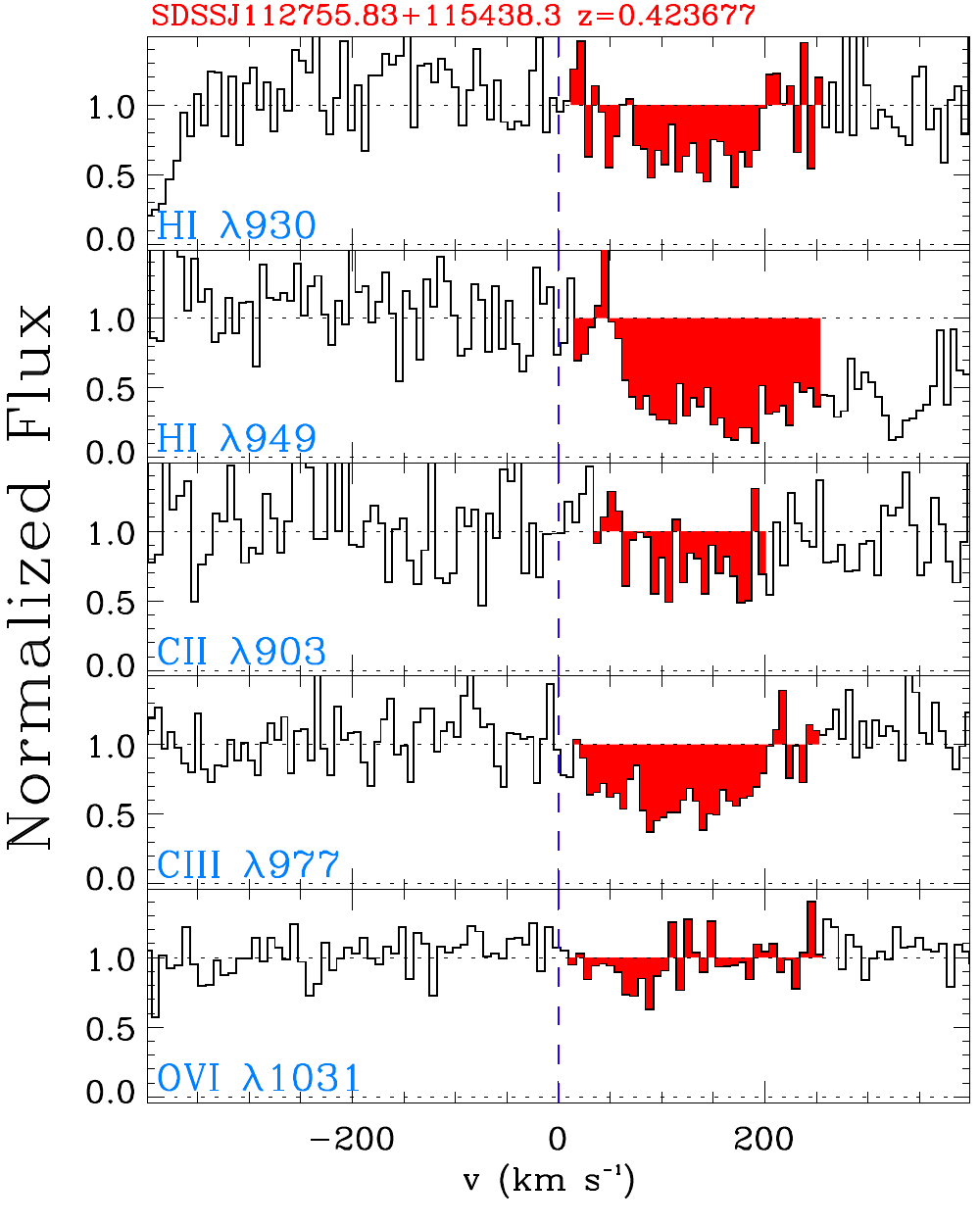}
    \caption{Same as Figure~\ref{fig:start}, but for absorption associated with LRG SDSSJ112755.83+115438.3 at an impact parameter of $\rho = 102$ kpc.}
\end{figure*}

\begin{figure}
    \epsscale{0.5}
     \plotone{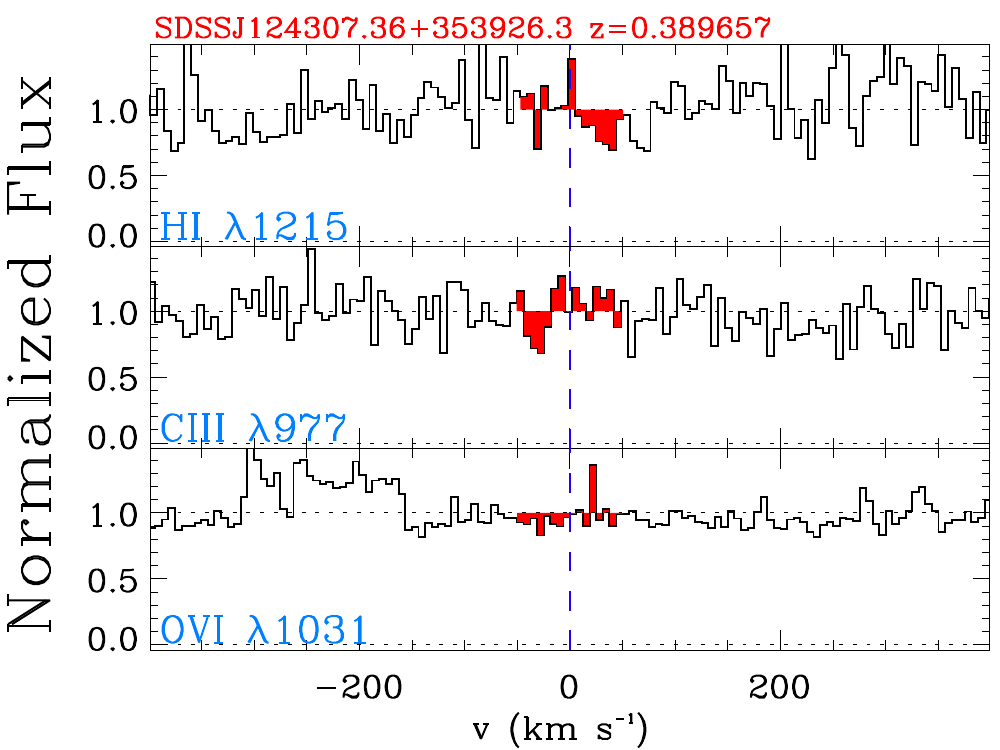}
    \caption{Same as Figure~\ref{fig:start}, but for absorption associated with LRG SDSSJ124307.36+353926.3 at an impact parameter of $\rho = 105$ kpc.}
\end{figure}

\begin{figure}
    \epsscale{0.5}
     \plotone{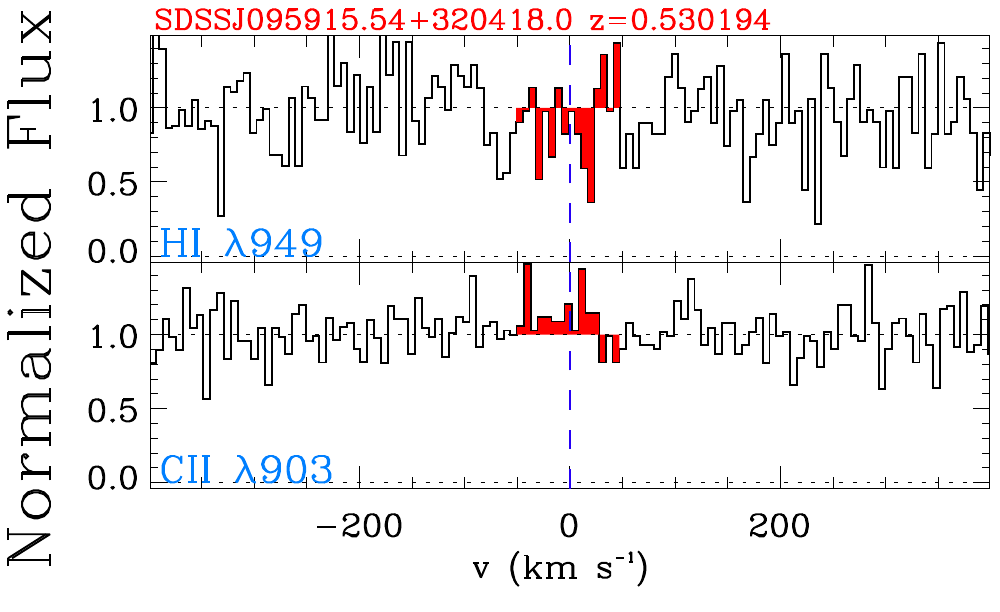}
    \caption{Same as Figure~\ref{fig:start}, but for absorption associated with LRG SDSSJ095915.54+320418.0 at an impact parameter of $\rho = 146$ kpc.}
\end{figure}

\begin{figure}
    \epsscale{0.5}
     \plotone{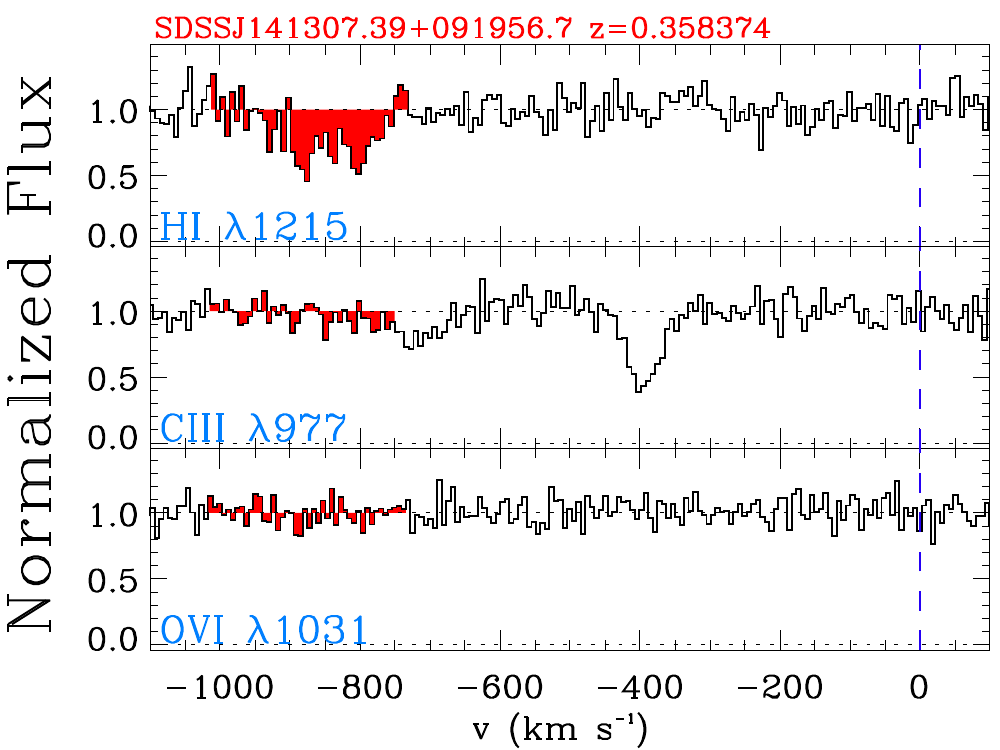}
    \caption{Same as Figure~\ref{fig:start}, but for absorption associated with LRG SDSSJ141307.39+091956.7 at an impact parameter of $\rho = 154$ kpc.}
\end{figure}

\begin{figure}
    \epsscale{0.5}
     \plotone{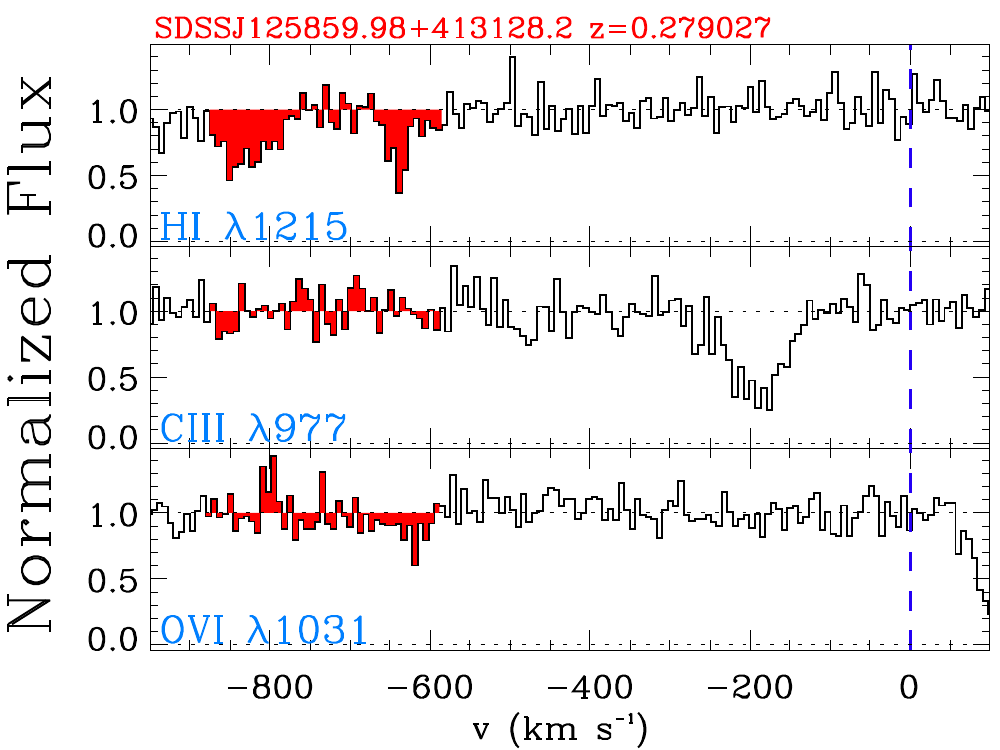}
    \caption{Same as Figure~\ref{fig:start}, but for absorption associated with LRG SDSSJ125859.98+413128.2 at an impact parameter of $\rho = 164$ kpc. We report the sum of the two \HI\ components in this paper.}
\end{figure}

\begin{figure}
    \epsscale{0.5}
     \plotone{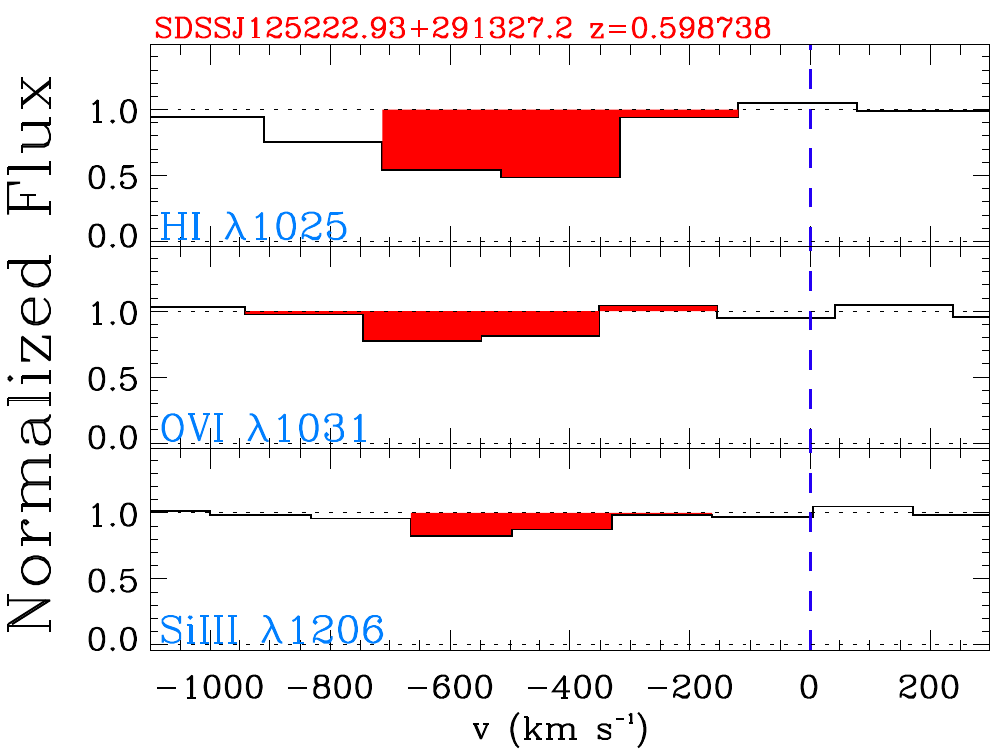}
    \caption{Same as Figure~\ref{fig:start}, but for absorption associated with LRG SDSSJ125222.93+291327.2 at an impact parameter of $\rho = 190$ kpc.}
\end{figure}

\begin{figure}
    \epsscale{0.5}
     \plotone{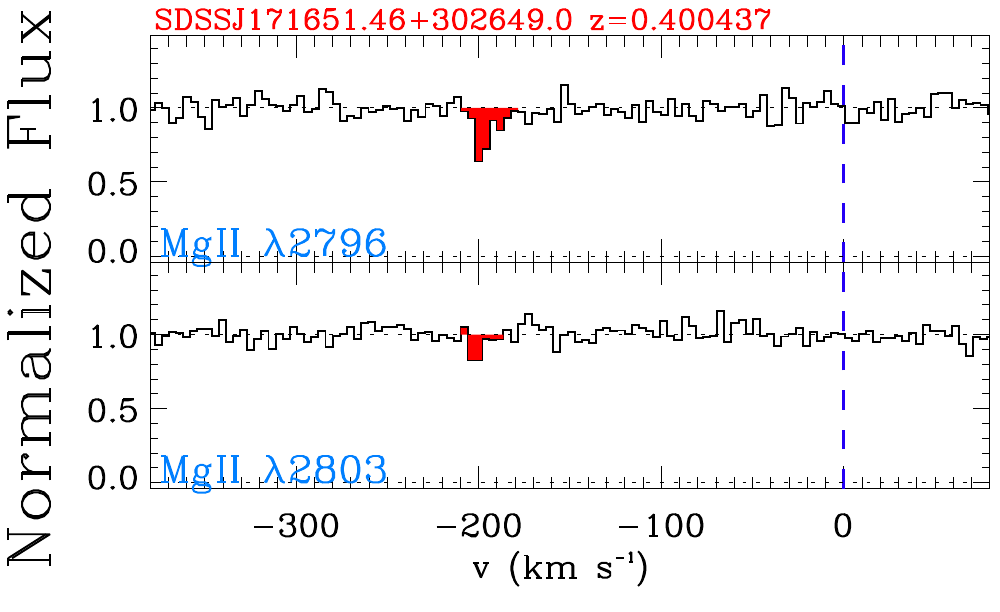}
    \caption{Same as Figure~\ref{fig:start}, but for absorption associated with LRG SDSSJ171651.46+302649.0 at an impact parameter of $\rho = 207$ kpc.}
    \label{fig:appc3}
\end{figure}

\clearpage

\begin{figure}
    \epsscale{0.5}
     \plotone{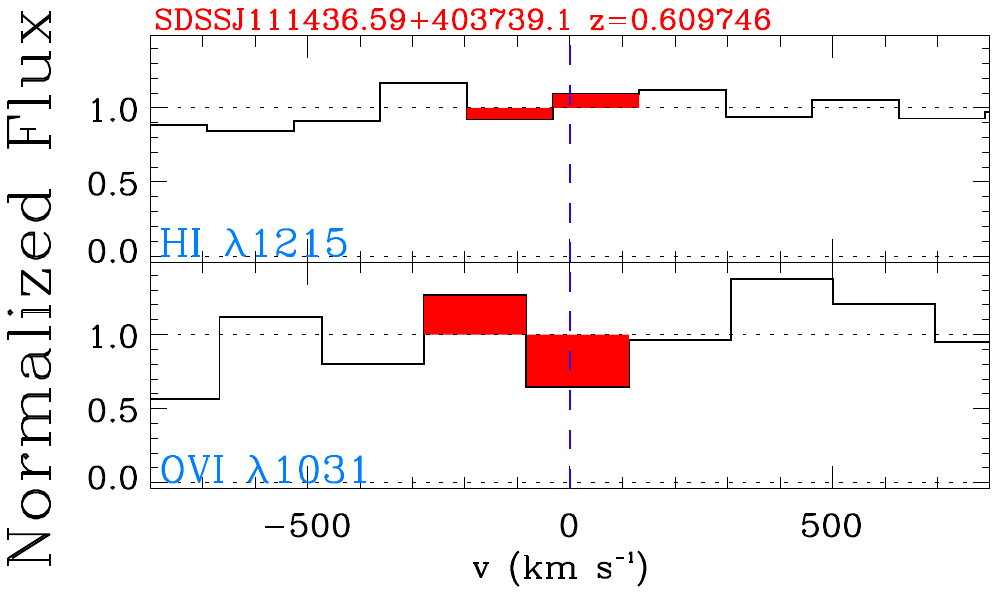}
    \caption{Same as Figure~\ref{fig:start}, but for absorption associated with LRG SDSSJ111436.59+403739.1 at an impact parameter of $\rho = 212$ kpc.}
\end{figure}

\begin{figure}
    \epsscale{0.5}
     \plotone{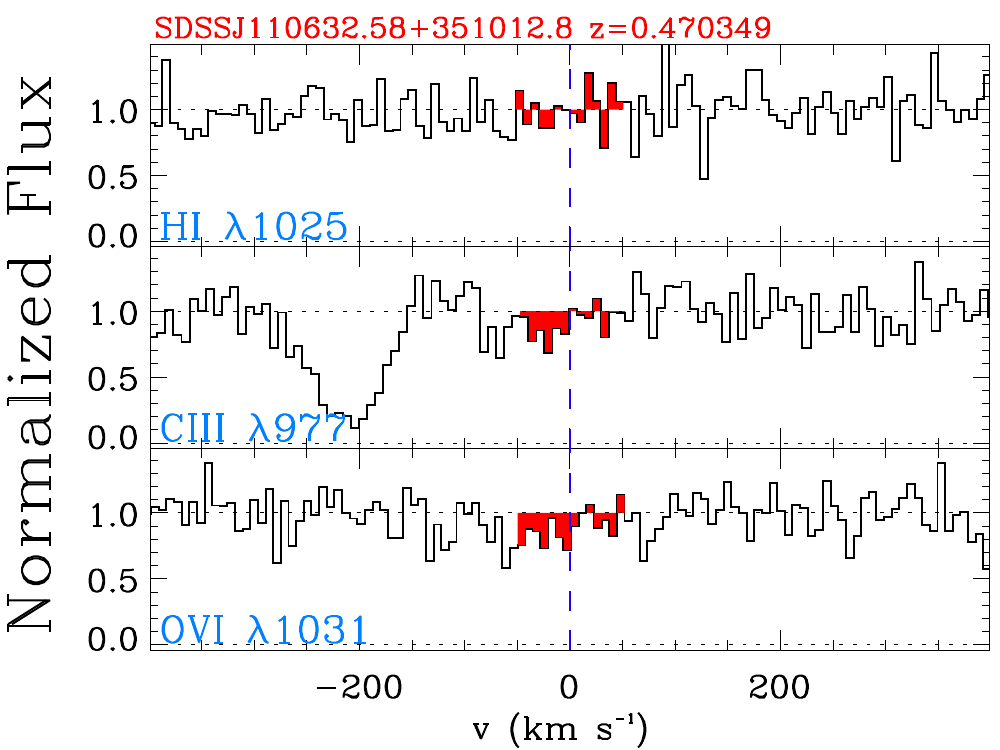}
    \caption{Same as Figure~\ref{fig:start}, but for absorption associated with LRG SDSSJ110632.58+351012.8 at an impact parameter of $\rho = 261$ kpc.}
\end{figure}

\begin{figure}
    \epsscale{0.5}
     \plotone{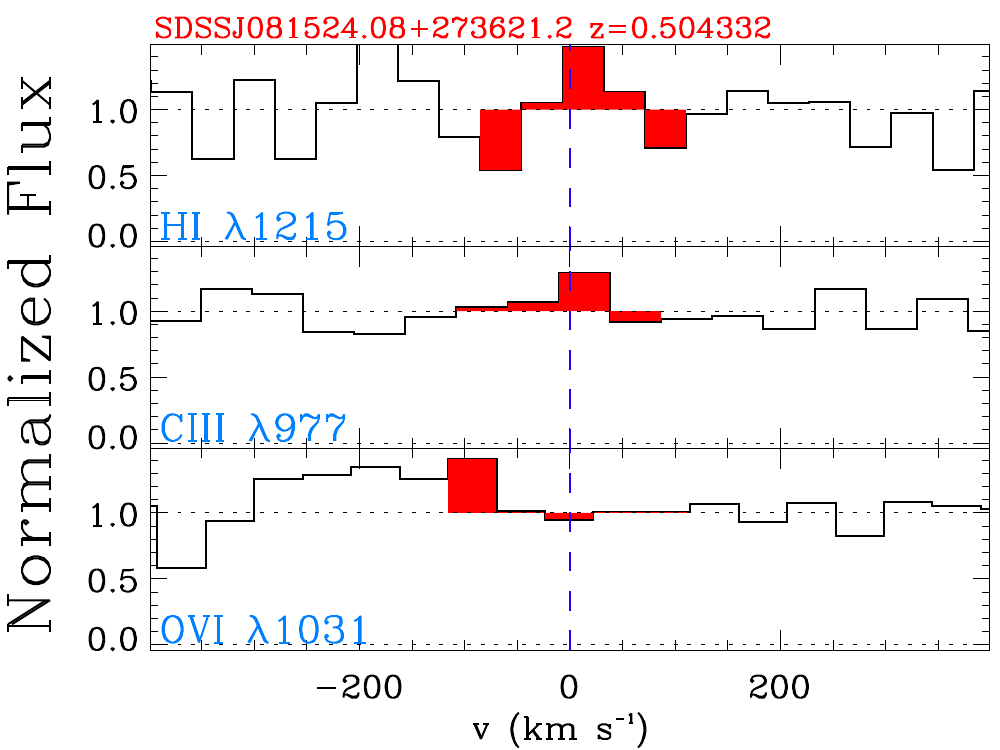}
    \caption{Same as Figure~\ref{fig:start}, but for absorption associated with LRG SDSSJ081524.08+273621.2 at an impact parameter of $\rho = 288$ kpc.}
\end{figure}
\clearpage

\begin{figure}
    \epsscale{0.5}
     \plotone{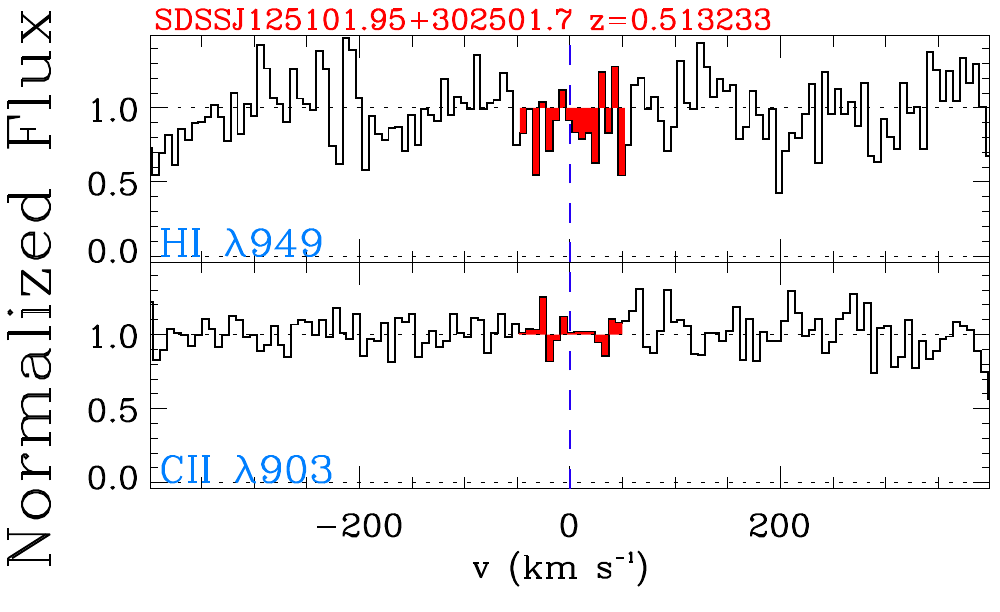}
    \caption{Same as Figure~\ref{fig:start}, but for absorption associated with LRG SDSSJ125101.95+302501.7 at an impact parameter of $\rho = 289$ kpc.}
\end{figure}

\begin{figure}
    \epsscale{0.5}
     \plotone{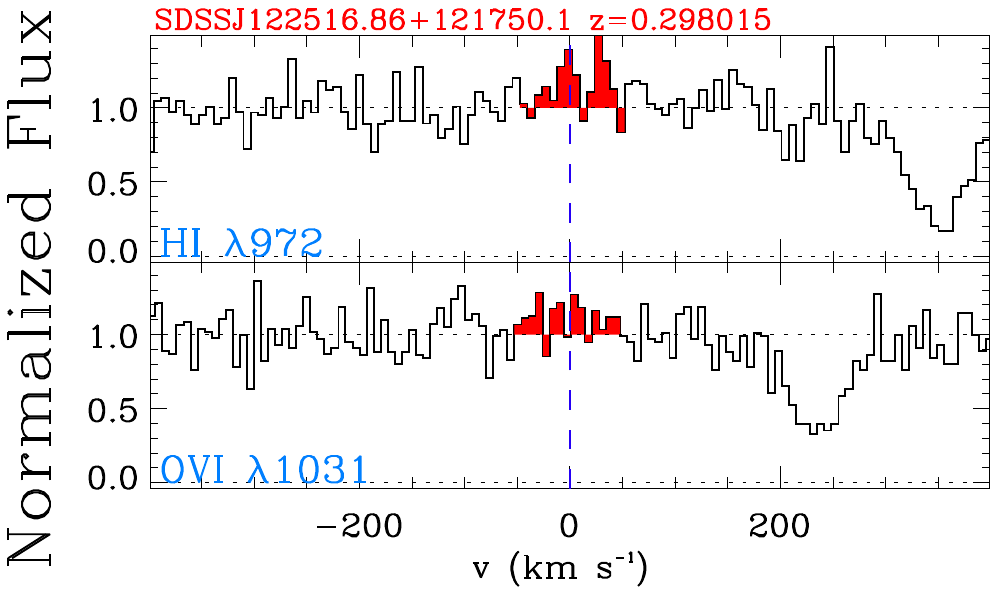}
    \caption{Same as Figure~\ref{fig:start}, but for absorption associated with LRG SDSSJ122516.86+121750.1 at an impact parameter of $\rho = 336$ kpc.}
\end{figure}

\begin{figure*}
    \epsscale{1}
     \plotone{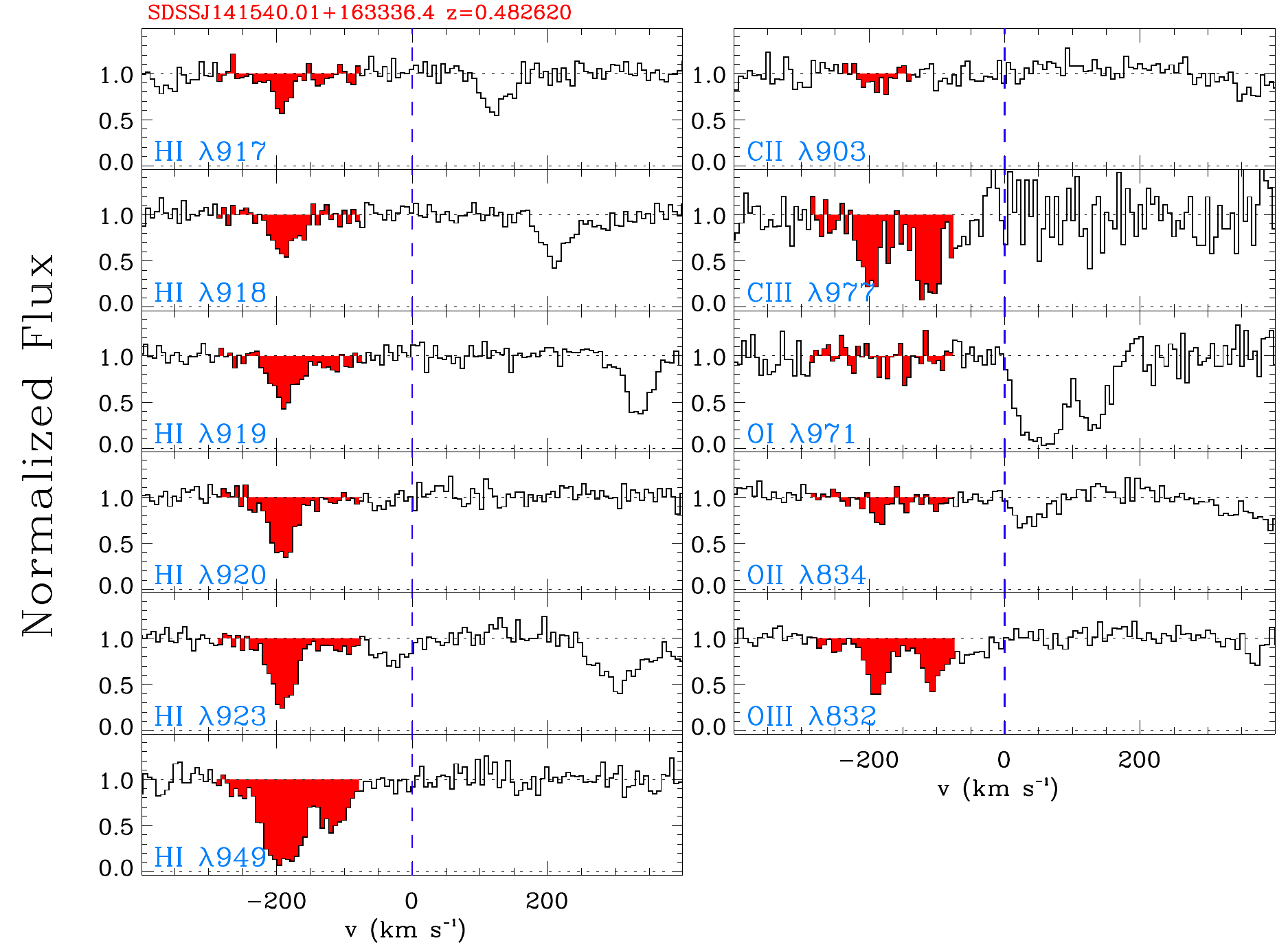}
    \caption{Same as Figure~\ref{fig:start}, but for absorption associated with LRG SDSSJ141540.01+163336.4 at an impact parameter of $\rho = 345$ kpc. The weak \HI\ component is integrated from -146 to -75 \kms.}
    \label{fig:appc4}
\end{figure*}

\begin{figure}
    \epsscale{0.5}
     \plotone{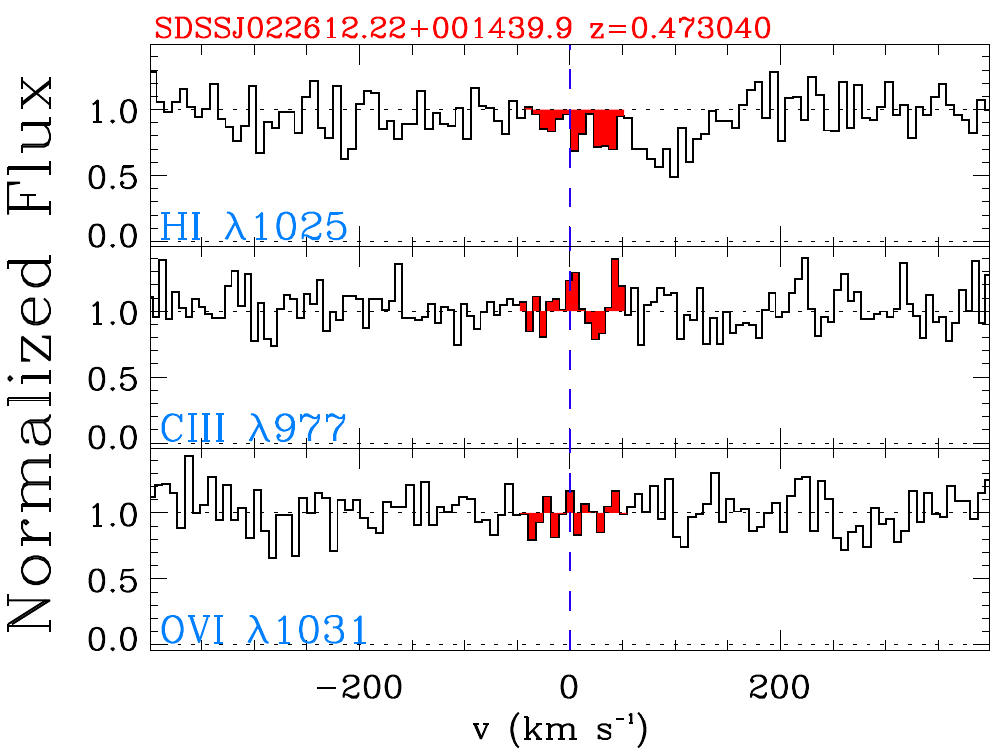}
    \caption{Same as Figure~\ref{fig:start}, but for absorption associated with LRG SDSSJ022612.22+001439.9 at an impact parameter of $\rho = 367$ kpc.}
\end{figure}

\clearpage

\begin{figure}
    \epsscale{0.5}
     \plotone{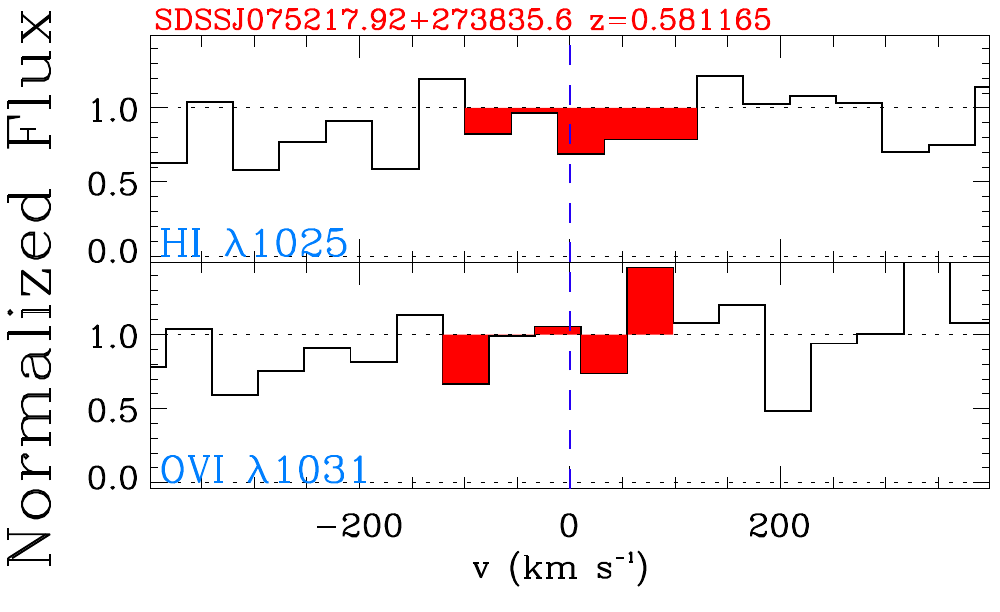}
    \caption{Same as Figure~\ref{fig:start}, but for absorption associated with LRG SDSSJ075217.92+273835.6 at an impact parameter of $\rho = 458$ kpc.}
\end{figure}

\begin{figure}
    \epsscale{0.5}
     \plotone{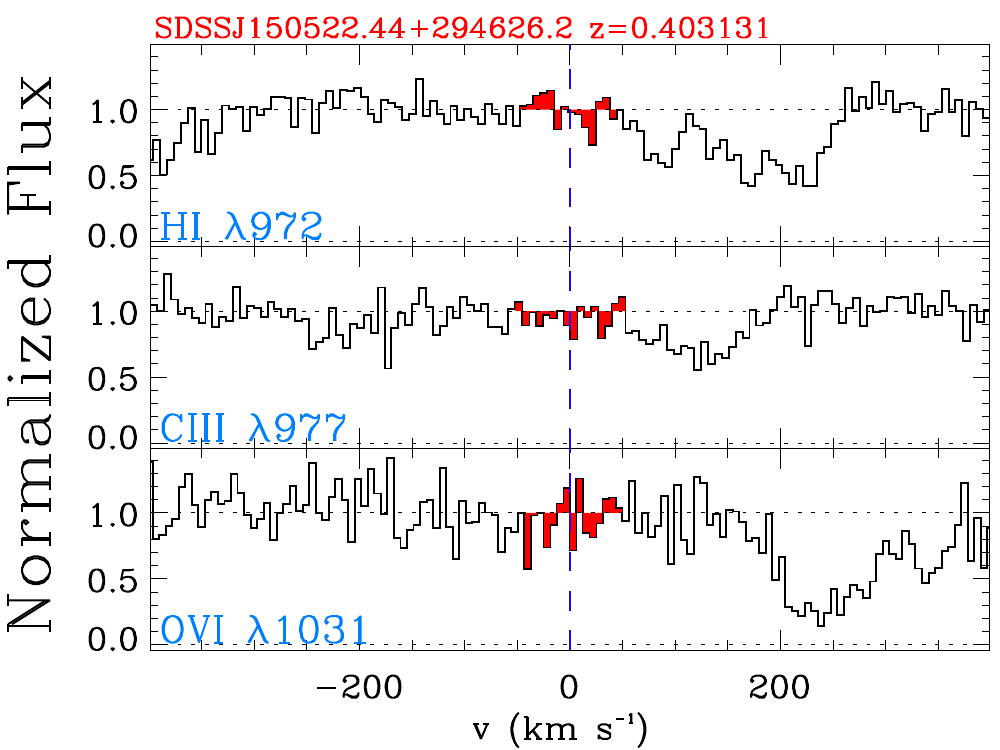}
    \caption{Same as Figure~\ref{fig:start}, but for absorption associated with LRG SDSSJ150522.44+294626.2 at an impact parameter of $\rho = 473$ kpc.}
\end{figure}

\begin{figure}
    \epsscale{0.5}
     \plotone{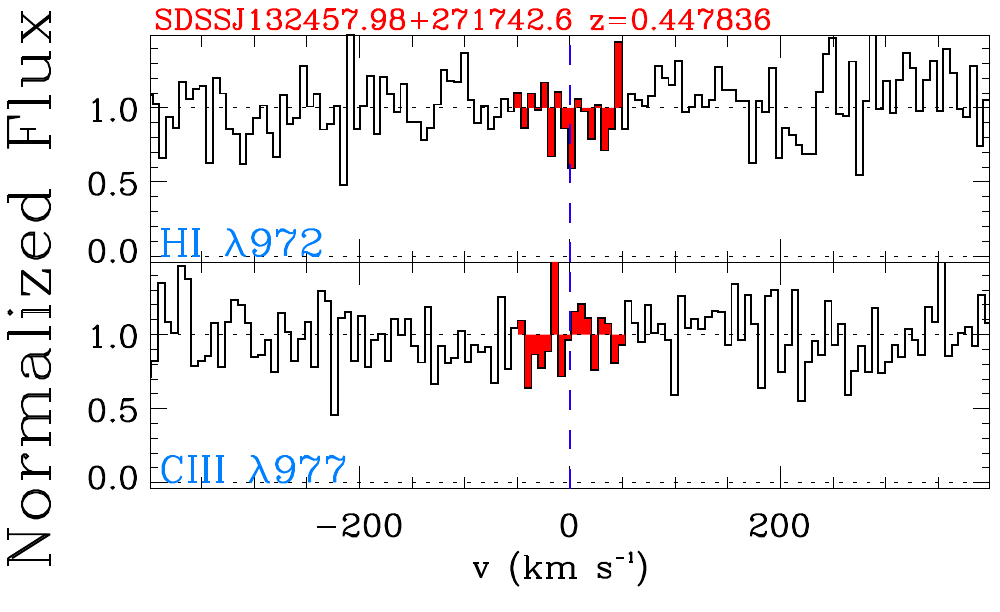}
    \caption{Same as Figure~\ref{fig:start}, but for absorption associated with LRG SDSSJ132457.98+271742.6 at an impact parameter of $\rho = 481$ kpc.}
\end{figure}

\begin{figure}
    \epsscale{0.5}
     \plotone{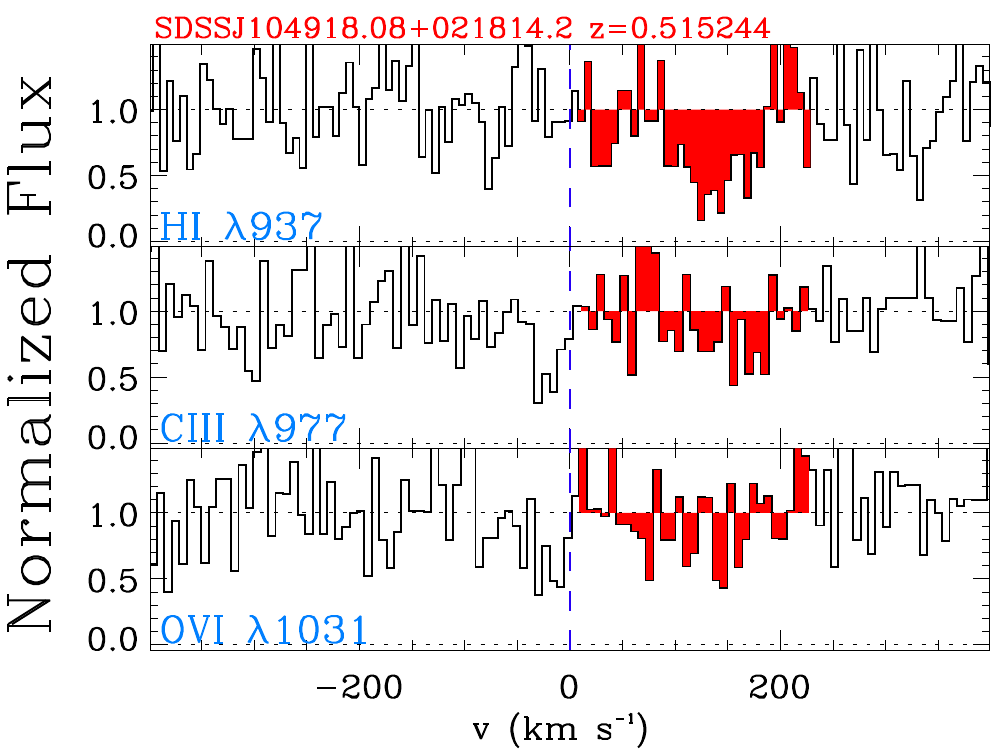}
    \caption{Same as Figure~\ref{fig:start}, but for absorption associated with LRG SDSSJ104918.08+021814.2 at an impact parameter of $\rho = 497$ kpc.}
\end{figure}

\begin{figure}
    \epsscale{0.5}
     \plotone{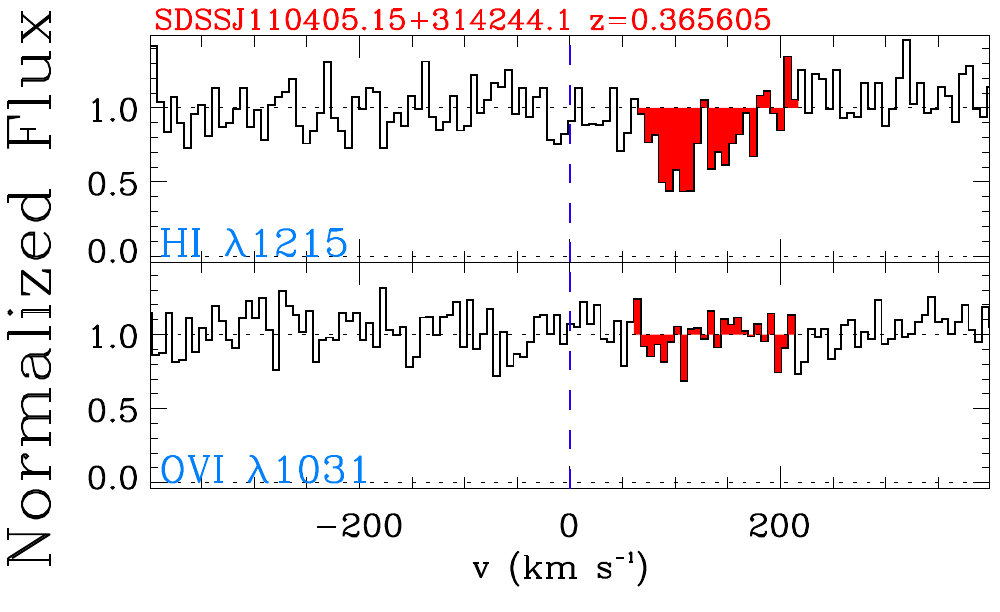}
    \caption{Same as Figure~\ref{fig:start}, but for absorption associated with LRG SDSSJ110405.15+314244.1 at an impact parameter of $\rho = 500$ kpc.}
    \label{fig:stop}
\end{figure}

\input{Table_ion_column}

\clearpage

\section{Covering Factors for Optically-Thick Absorption}\label{appb}

\begin{figure}
    \epsscale{1.0}
    \plotone{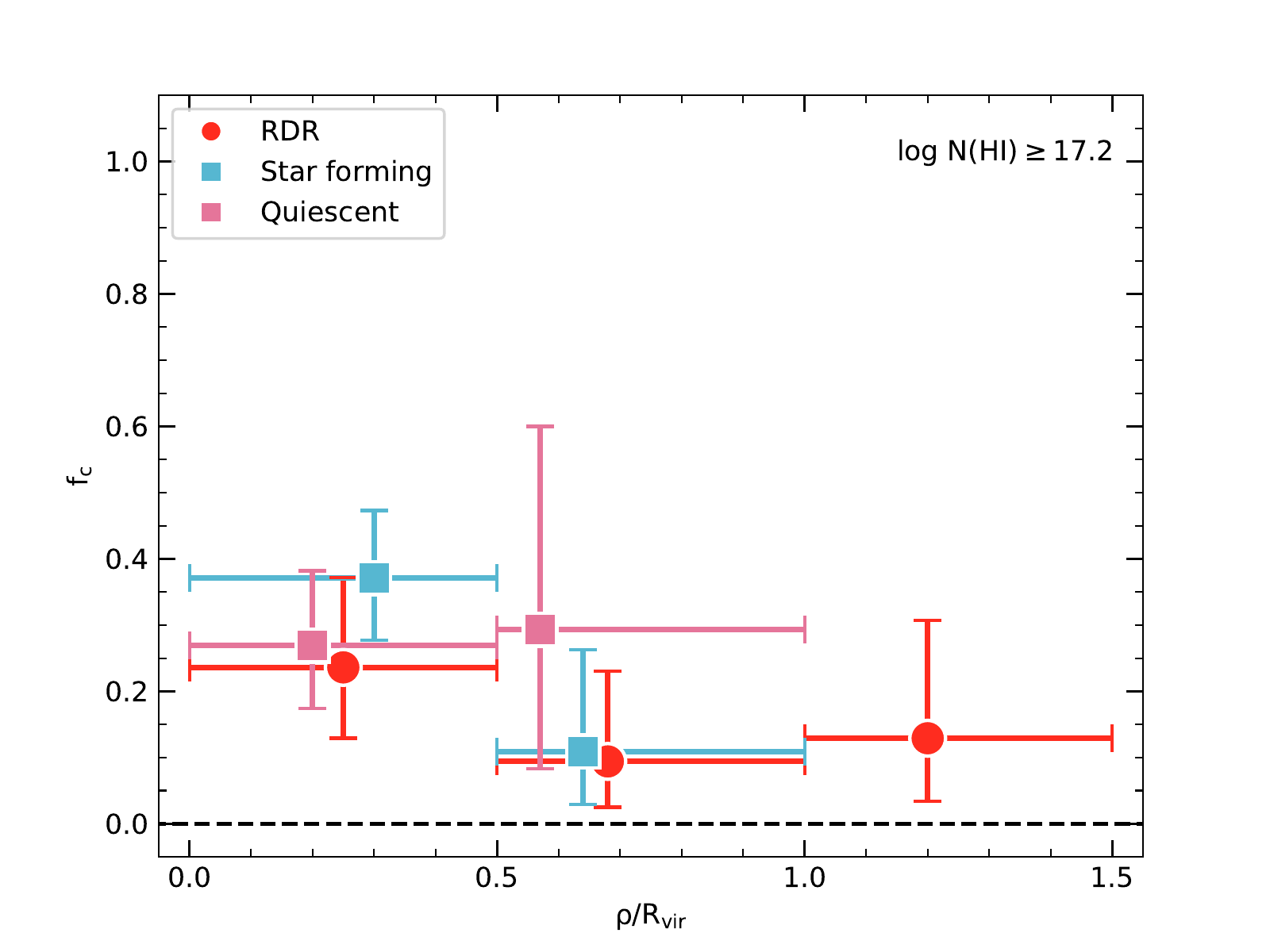}
    \caption{\HI\ covering factor distributed by sample for gas with $\logNHI \ge 17.2$ as a function of normalized impact parameter. The vertical error bars show the 68\% confidence interval for the covering factor. The horizontal error bars show the extent of each bin, while the location of the data points represents the mean normalized impact parameter. Galaxies from the COS-Halos sample make up the star-forming and quiescent samples \citepalias{prochaska2017}. We adopt their characterization of galaxies with $\log$\,sSFR\,$> -11$ as ``star forming."}
    \label{fig:coveringfactor_ssfr_lls}
\end{figure}

\begin{figure}
    \epsscale{1.0}
    \plotone{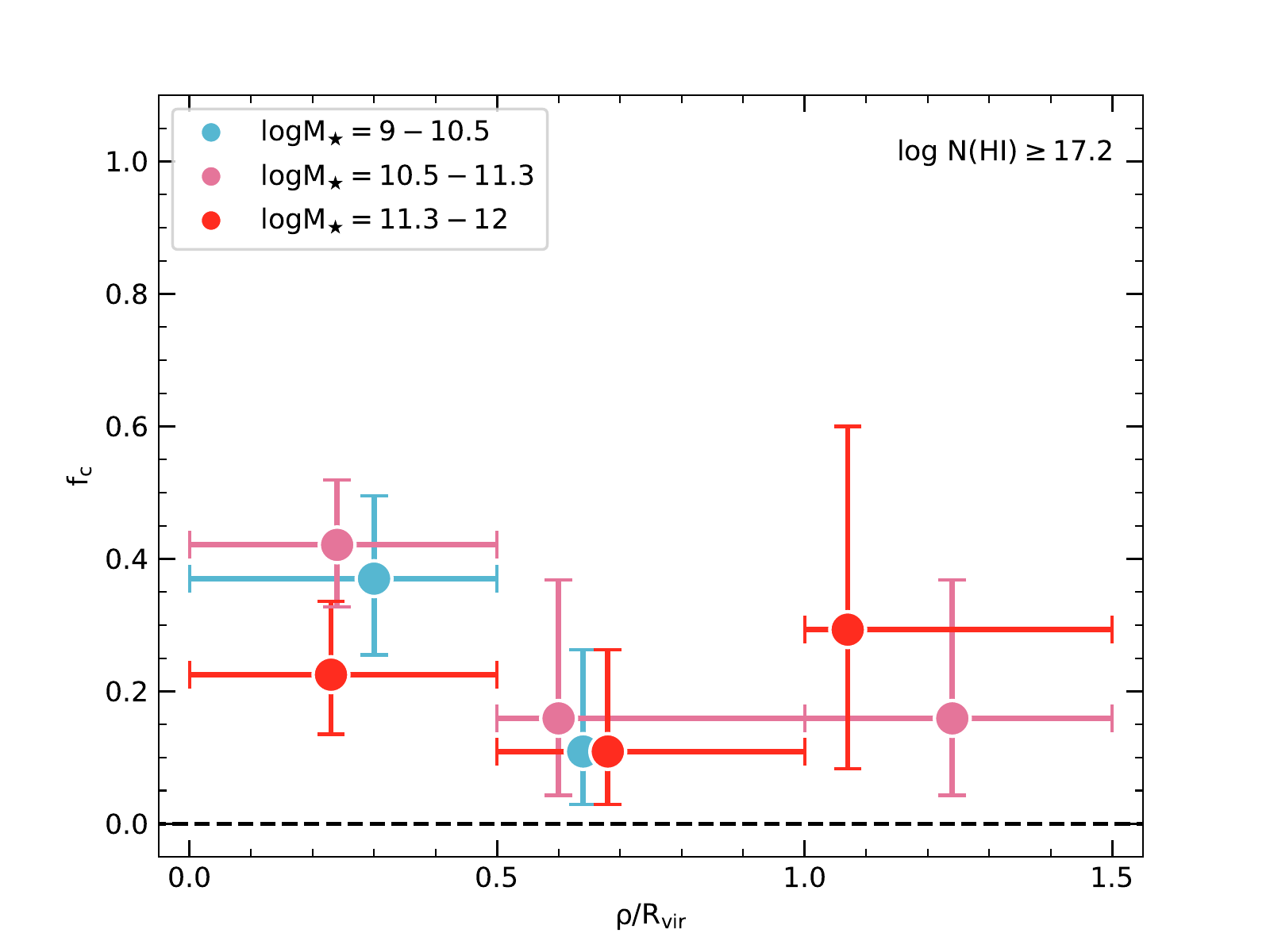}
    \caption{Stellar mass dependence of the \HI\ covering factor ($\logNHI \ge 17.2$) as a function of normalized impact parameter in our combined sample of galaxies incorporating our results with those from \citetalias{prochaska2017} and \citetalias{chen2018}. The vertical error bars show the 68\% confidence interval in $f_c$. The horizontal error bars show the extent of each bin, while the location of the data points represents the mean normalized impact parameter. We find no statistically significant difference in covering factors between mass bins for these column densities.}
    \label{fig:coveringfactor_stellar_lls}
\end{figure}

\begin{figure}
    \epsscale{1.0}
    \plotone{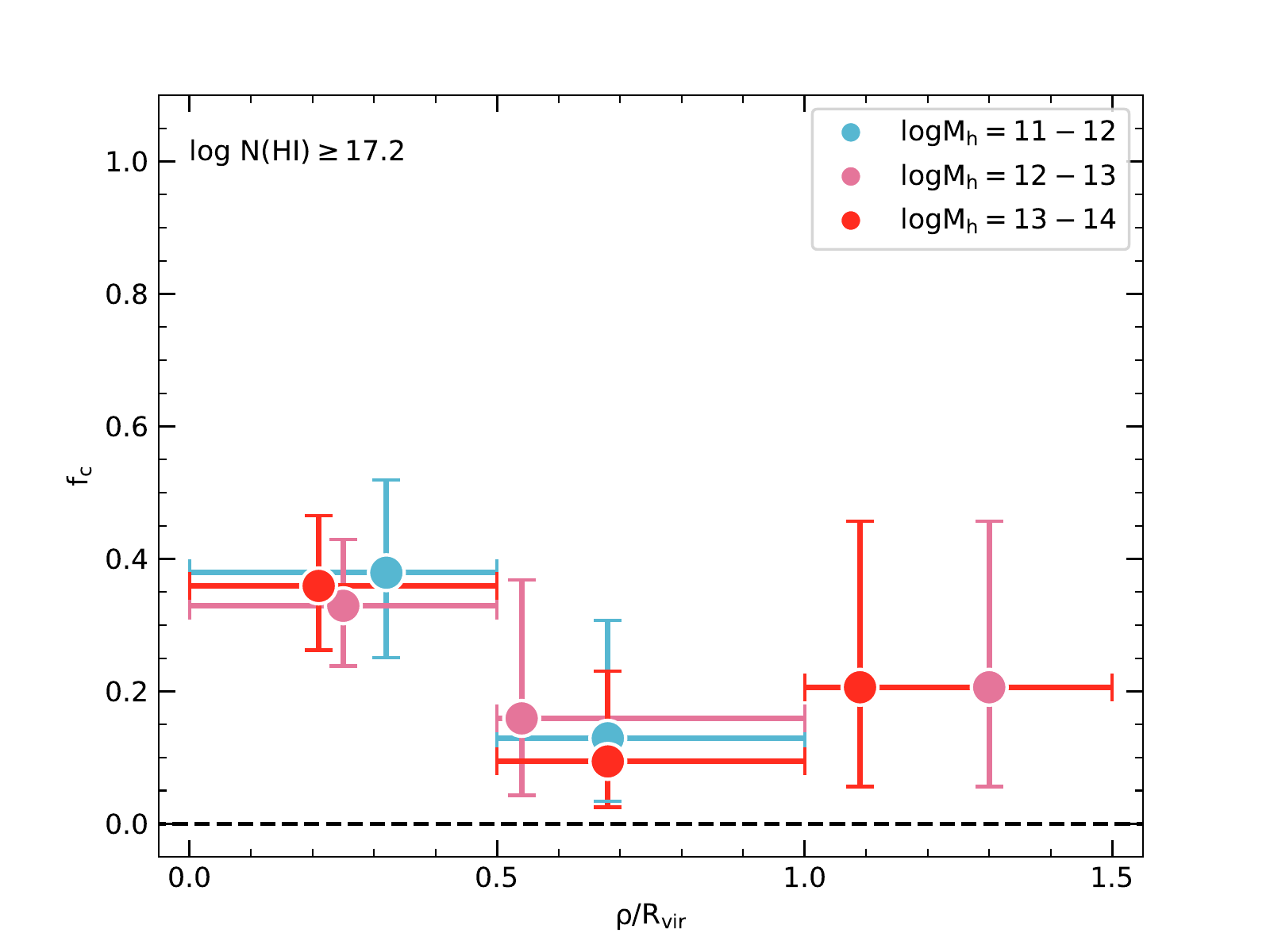}
    \caption{Halo mass dependence of the \HI\ covering factor ($\logNHI \ge 17.2$) as a function of normalized impact parameter in our combined sample of galaxies incorporating our results with those from \citetalias{prochaska2017} and \citetalias{chen2018}. The vertical error bars show the 68\% confidence interval in $f_c$. The horizontal error bars show the extent of each bin, while the location of the data points represents the mean normalized impact parameter. We find no statistically significant difference in the covering factors of galaxies between bins of halo mass.}
    \label{fig:coveringfactor_halos_lls}
\end{figure}

In this appendix we present the results of the covering factor calculations from \S~\ref{sec:general} at the optically-thick absorption limit: $\logNHI \ge 17.2$. Figures~\ref{fig:coveringfactor_ssfr_lls}-\ref{fig:coveringfactor_halos_lls} display the covering factors by sample, stellar mass, and halo mass. Tables~\ref{tab:coveringfactorssfr_lls}-\ref{tab:coveringfactorhalo_lls} detail the information about the calculations.

\input{coveringfactor_ssfr_LLS}

\input{coveringfactor_stellar_LLS}

\input{coveringfactor_halos_LLS}

\clearpage

\section{Metallicity Results for Individual Absorbers}\label{appc}

We present the details of our metallicity analysis for each of the four strong \HI\ absorbers along with the corner plots of the Cloudy model grid parameters from our MCMC analysis.

{\it LRG SDSSJ111508.24+023752.7} \replaced{We constrain the metallicity of the LLS associated with this galaxy using \ensuremath{\mbox{\ion{C}{2}}}, \ensuremath{\mbox{\ion{C}{3}}}, \ensuremath{\mbox{\ion{O}{1}}}, and \ensuremath{\mbox{\ion{S}{3}}} (see Figure~\ref{fig:start}). We do not allow the [C/$\alpha$] ratio to vary for this absorber because this variable is unconstrained when the carbon input ions are lower limits. A prior on the log $U$ distribution of this absorber is unnecessary. The corner plot of the MCMC PDFs for this absorber is presented in Figure.}{For this absorber, we estimate $\logNOI = 14.22^{+0.09}_{-0.12}$ in the strongest component that dominates the \HI\ absorption. \ion{O}{1} is the best metal proxy for \HI\ since its ionization potential and charge exchange reactions with hydrogen ensure the ionization of \HI\ and \ion{O}{1} are strongly coupled. With $\logNHI = [18.00,19.60]$, this yields a metallicity range for this absorber of ${\rm [X/H]} = [-2.10,-0.50]$. It is poorly constrained owing to the poorly constrained \NHI, but it is unlikely to be solar. We note that we also ran an MCMC analysis with a flat prior on the \HI\ column density and used \ion{C}{2}, \ion{C}{3}, \ion{O}{1}, and \ion{S}{3} to constrain the metallicity (see Figure~\ref{fig:start}). The analysis systematically converged to the low metallicity value, i.e., favored a high value of \NHI. We feel that in this case the use of \ion{O}{1} to derive the metallicity provides a more conservative range of metallicity.}

\begin{figure}
    \epsscale{1.0}
    \plotone{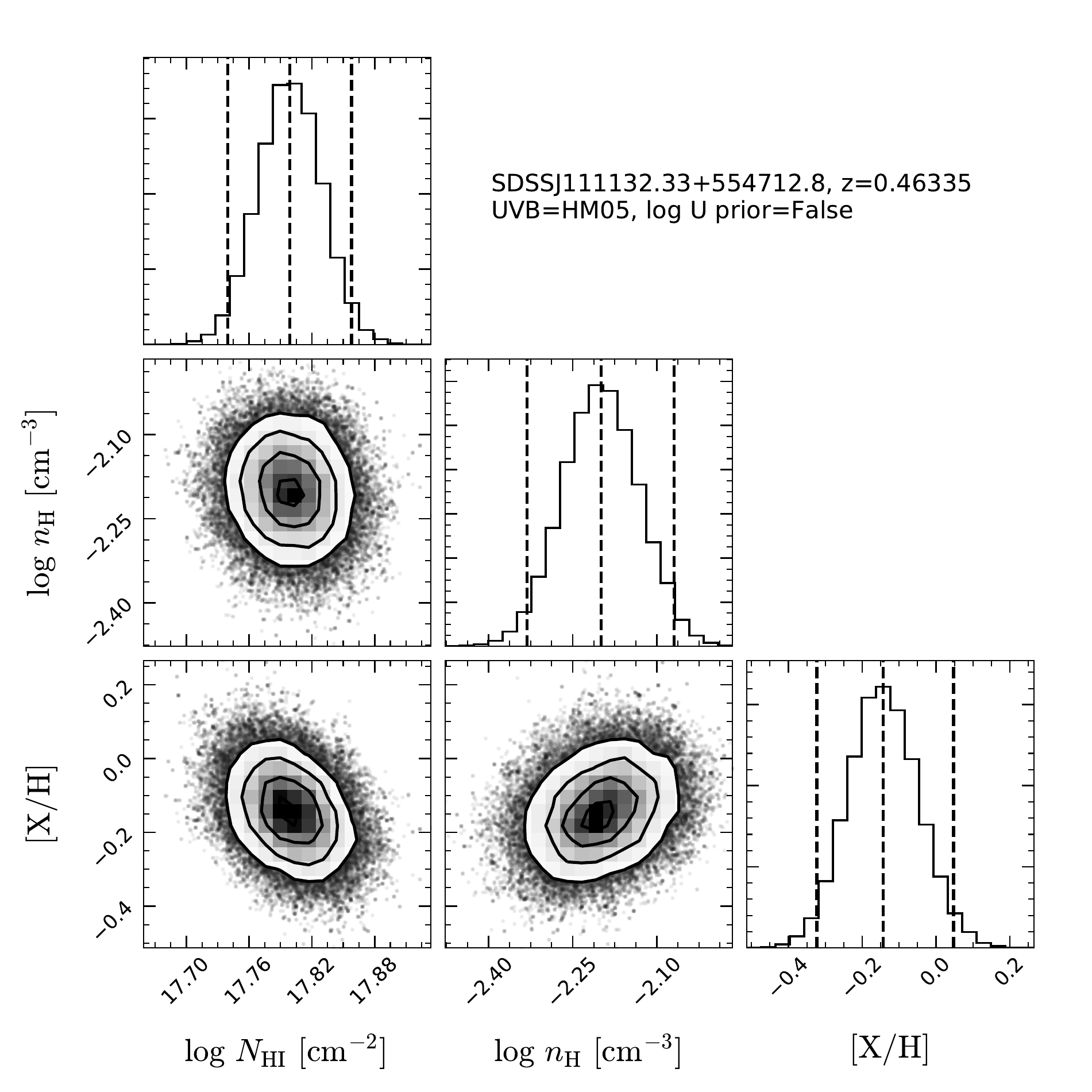}
    \caption{Corner plot of the MCMC PDFs for the LLS associated with LRG SDSSJ111132.33+554712.8. The PDFs for each variable are plotted along the diagonal, and the other panels are the 2D distributions of the final MCMC walkers chains. The vertical dashed lines show the median value of the distribution and the bounds of the 95\% confidence interval.}
    \label{fig:corner2}
\end{figure}

{\it LRG SDSSJ111132.33+554712.8} The LLS associated with this galaxy exhibits two absorbing components (see Figure~\ref{fig:appc2}). We calculate the metallicity of each of the components separately as well as the metallicity implied by the combined components. For the weak \HI\ component, we constrain the metallicity using the ions \ensuremath{\mbox{\ion{C}{2}}}, \ensuremath{\mbox{\ion{N}{2}}}, \ensuremath{\mbox{\ion{N}{3}}}, \ensuremath{\mbox{\ion{O}{1}}}, \ensuremath{\mbox{\ion{Si}{3}}}, and \ensuremath{\mbox{\ion{S}{3}}}. We allow the [C/$\alpha$] ratio to vary. There is enough information on this component that applying our Gaussian prior on the log $U$ distribution is unnecessary. The weak \HI\ component of the LLS has a metallicity [X/H] $= -0.15^{+0.14}_{-0.21}$. For the strong \HI\ component, we use the ions \ensuremath{\mbox{\ion{C}{2}}}, \ensuremath{\mbox{\ion{C}{3}}}, \ensuremath{\mbox{\ion{N}{1}}}, \ensuremath{\mbox{\ion{N}{2}}}, \ensuremath{\mbox{\ion{N}{3}}}, \ensuremath{\mbox{\ion{O}{1}}}, \ensuremath{\mbox{\ion{O}{2}}}, \ensuremath{\mbox{\ion{Si}{2}}}, \ensuremath{\mbox{\ion{Si}{3}}}, and \ensuremath{\mbox{\ion{S}{3}}}. We do not allow the [C/$\alpha$] ratio to vary for this component because this variable is unconstrained when the carbon input ions are lower limits. A Gaussian prior on the log $U$ distribution of this component is unnecessary. The strong \HI\ component has a metallicity [X/H] $= -0.20 \pm 0.08$. The similarity in metallicity suggests these components may have a common origin. For the full absorber, we constrain the metallicity using the ions \ensuremath{\mbox{\ion{C}{2}}}, \ensuremath{\mbox{\ion{C}{3}}}, \MgII, \ensuremath{\mbox{\ion{N}{2}}}, \ensuremath{\mbox{\ion{N}{3}}}, \ensuremath{\mbox{\ion{O}{1}}}, \ensuremath{\mbox{\ion{O}{2}}}, \ensuremath{\mbox{\ion{Si}{2}}}, \ensuremath{\mbox{\ion{Si}{3}}}, and \ensuremath{\mbox{\ion{S}{3}}}. We do not allow the [C/$\alpha$] ratio to vary for this absorber, nor is a Gaussian prior on the log $U$ distribution necessary. The corner plot of the MCMC PDFs for this absorber is presented in Figure~\ref{fig:corner2}. \added{The implied metallicity of the combined components is [X/H] = $-0.14^{+0.09}_{-0.10}$.}

\begin{figure}
    \epsscale{1.0}
    \plotone{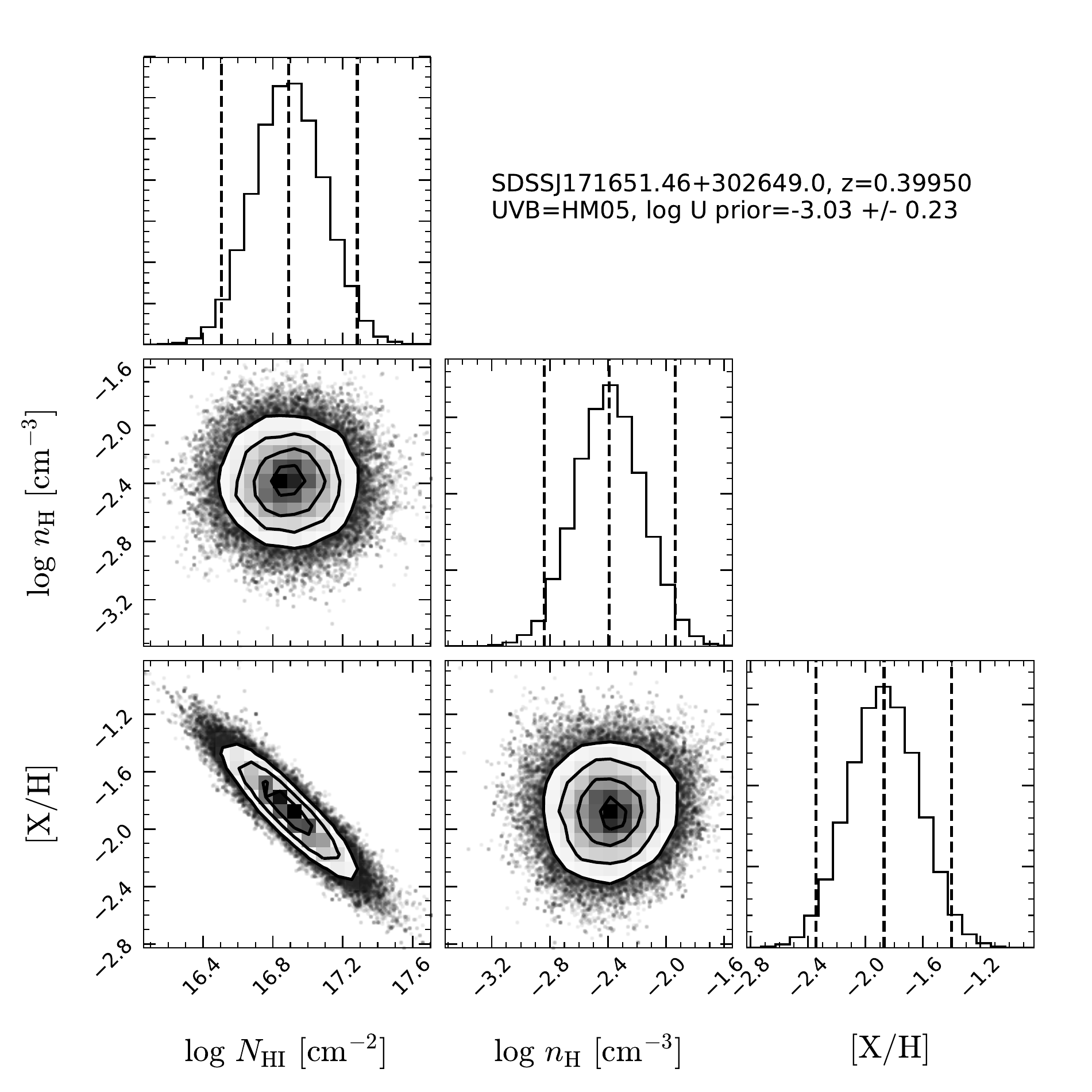}
    \caption{Same as in Fig.~\ref{fig:corner2}, but for the pLLS associated with LRG SDSSJ171651.46+302649.0. As this absorber only has an \HI\ and \MgII\ measurement, we apply a Gaussian $\log U$ constraint.}
    \label{fig:corner3}
\end{figure}

{\it LRG SDSSJ171651.46+302649.0} The pLLS associated with this galaxy has only \HI\ and \MgII\ column density measurements (see Figure~\ref{fig:appc3}). A Gaussian prior on the log $U$ distribution of this absorber is necessary to constrain the metallicity (i.e., we use the low-resolution method of \citealt{wotta2016, wotta2019}). The corner plot of the MCMC PDFs for this absorber is presented in Figure~\ref{fig:corner3}. \added{The metallicity of this absorber is [X/H] $= -1.87^{+0.24}_{-0.25}$.}

\begin{figure}
    \epsscale{1.0}
    \plotone{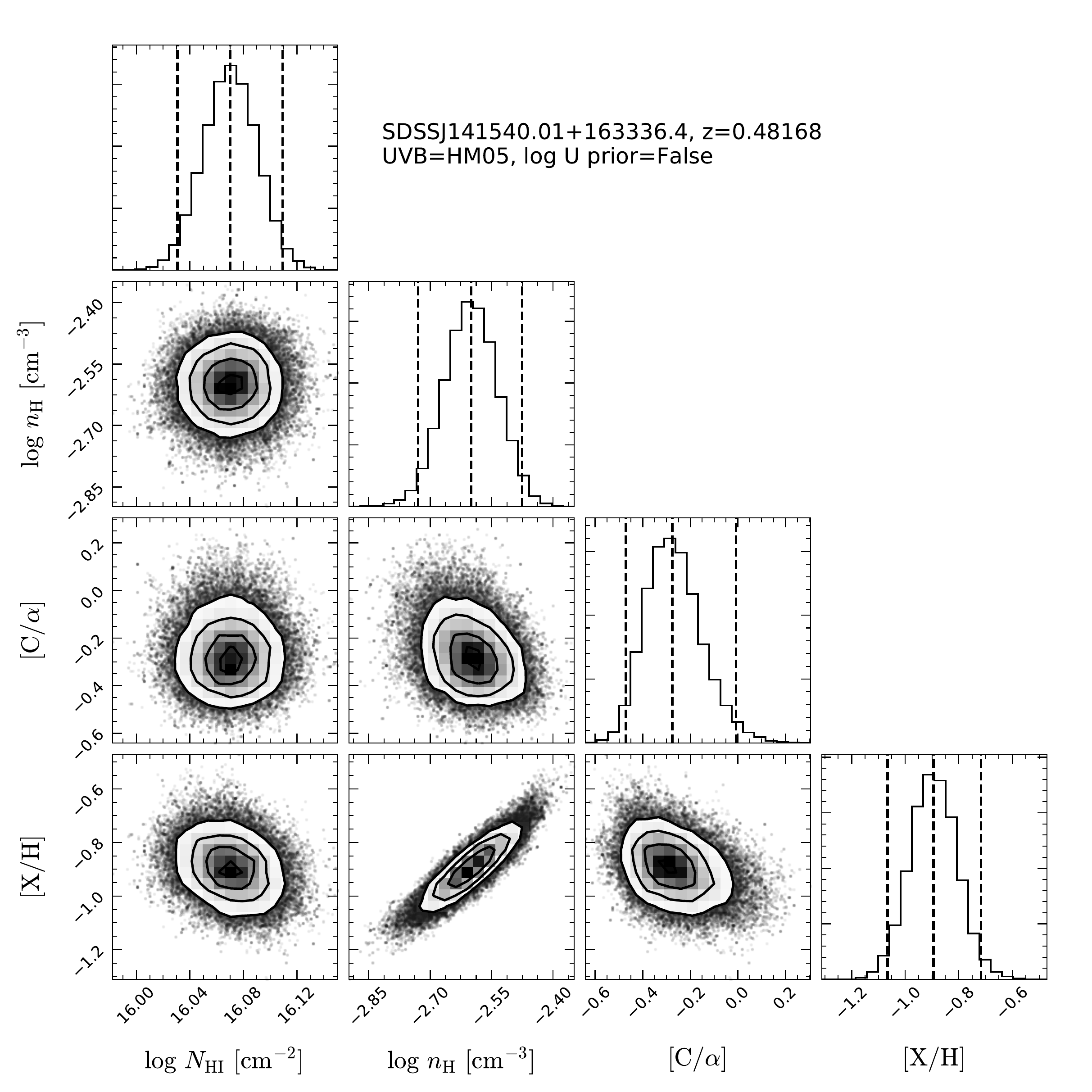}
    \caption{Same as in Fig.~\ref{fig:corner2}, but for the pLLS associated with LRG SDSSJ141540.01+163336.4. We allow the [C/$\alpha$] ratio to vary for this absorber.}
    \label{fig:corner4}
\end{figure}

{\it LRG SDSSJ141540.01+163336.4} The pLLS associated with this galaxy exhibits two absorbing components (see Figure~\ref{fig:appc4}). We calculate the metallicity of each of the components separately as well as the implied metallicity of the total absorber. For the weak \HI\ component, we constrain the metallicity using the ions \ensuremath{\mbox{\ion{C}{2}}}, \ensuremath{\mbox{\ion{C}{3}}}, \ensuremath{\mbox{\ion{O}{1}}}, \ensuremath{\mbox{\ion{O}{2}}}, and \ensuremath{\mbox{\ion{O}{3}}}. We allow the [C/$\alpha$] ratio to vary; a Gaussian prior on the log $U$ distribution of this component is unnecessary. We find a metallicity of [X/H] $= -0.50^{+0.26}_{-0.24}$ for the weak \HI\ component. For the strong \HI\ component, we use the ions \ensuremath{\mbox{\ion{C}{2}}}, \ensuremath{\mbox{\ion{C}{3}}}, \ensuremath{\mbox{\ion{O}{1}}}, \ensuremath{\mbox{\ion{O}{2}}}, and \ensuremath{\mbox{\ion{O}{3}}}. We allow the [C/$\alpha$] ratio to vary; a Gaussian prior on the log $U$ distribution of this component is unnecessary. The strong \HI\ component has a metallicity [X/H] $= -1.02^{+0.11}_{-0.10}$. The weak \HI\ component contributes more to the total \CIII\ column density of this absorber than the strong \HI\ component, causing the marginal difference in metallicity. The velocity offset between the two components is modest ($\sim$75 \kms). It is not clear if these components should be considered independently or if they are tracing a common structure within the CGM. For the full absorber (combining the components), we constrain the metallicity using the ions \ensuremath{\mbox{\ion{C}{2}}}, \ensuremath{\mbox{\ion{C}{3}}}, \MgII, \ensuremath{\mbox{\ion{O}{1}}}, \ensuremath{\mbox{\ion{O}{2}}}, and \ensuremath{\mbox{\ion{O}{3}}}. We allow the [C/$\alpha$] ratio to vary; a Gaussian prior on the log $U$ distribution of this absorber is unnecessary. The corner plot of the MCMC PDFs for this absorber is presented in Figure~\ref{fig:corner4}. \added{The implied metallicity of the combined components is [X/H] $= -0.89^{+0.08}_{-0.09}$.}

\section{Metallicity of the LLS associated with LRG SDSSJ135726.27+043541.4}\label{appd}

The absorber associated with LRG SDSSJ135726.27+043541.4 has been
discussed at length in \citetalias{chen2018} and is in the CCC sample of
\citet{wotta2019}. Both works identify this absorber as a low-metallicity
system, and it shows no metal absorption in the available COS data (\citetalias{chen2018}; \citealt{lehner2018}). However, it shows absorption in \MgII\ and \FeII\ in the Keck/HIRES observations \citep{lehner2018}.

\begin{figure}
    \epsscale{1.0}
    \plotone{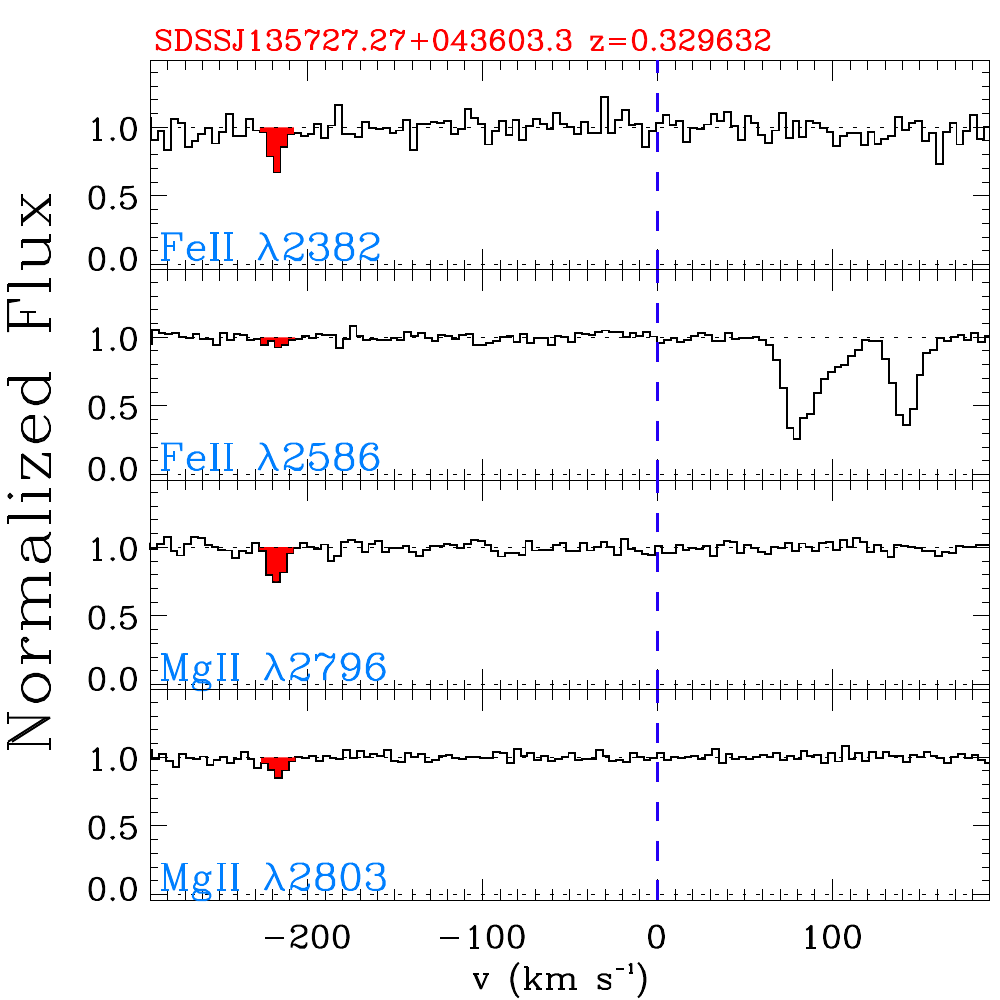}
    \caption{Ion profiles for absorption associated with LRG SDSSJ135727.27+043603.3 from the \citetalias{chen2018} sample at an impact parameter of $\rho = 130$ kpc. The profiles are shown in the reference frame of the LRG, and the red shading shows the section of the spectrum integrated to determine the ion column density.}
    \label{fig:plotstackJ1357}
\end{figure}

The Magellan/MIKE observations reported by \citet{chen2019} were inconclusive regarding the detection of \MgII\ or \FeII. In order to improve the \MgII\ and especially \FeII\ reported in \citet{lehner2018}, we obtained new Keck/HIRES observations of QSO SDSSJ135726.27+043541.4. Figure~\ref{fig:plotstackJ1357} displays the ion profiles from our new HIRES observations for this LLS, showing unambiguous detections of
aligned \MgII\ and \FeII\ in both the weak and strong transitions of each ion. Thus this absorber is not ``pristine,'' even though there are no metal-line detection in the COS spectrum. We measure the column densities using the AODM for each transition (see \S~\ref{sec:coldens}), giving a consistent result between the weak and strong transitions. Averaging out the results, we find \logNMgII\ = $11.84 \pm 0.04$ and \logNFeII\ = $12.26 \pm 0.07$.

\begin{figure}
    \epsscale{1.0}
    \plotone{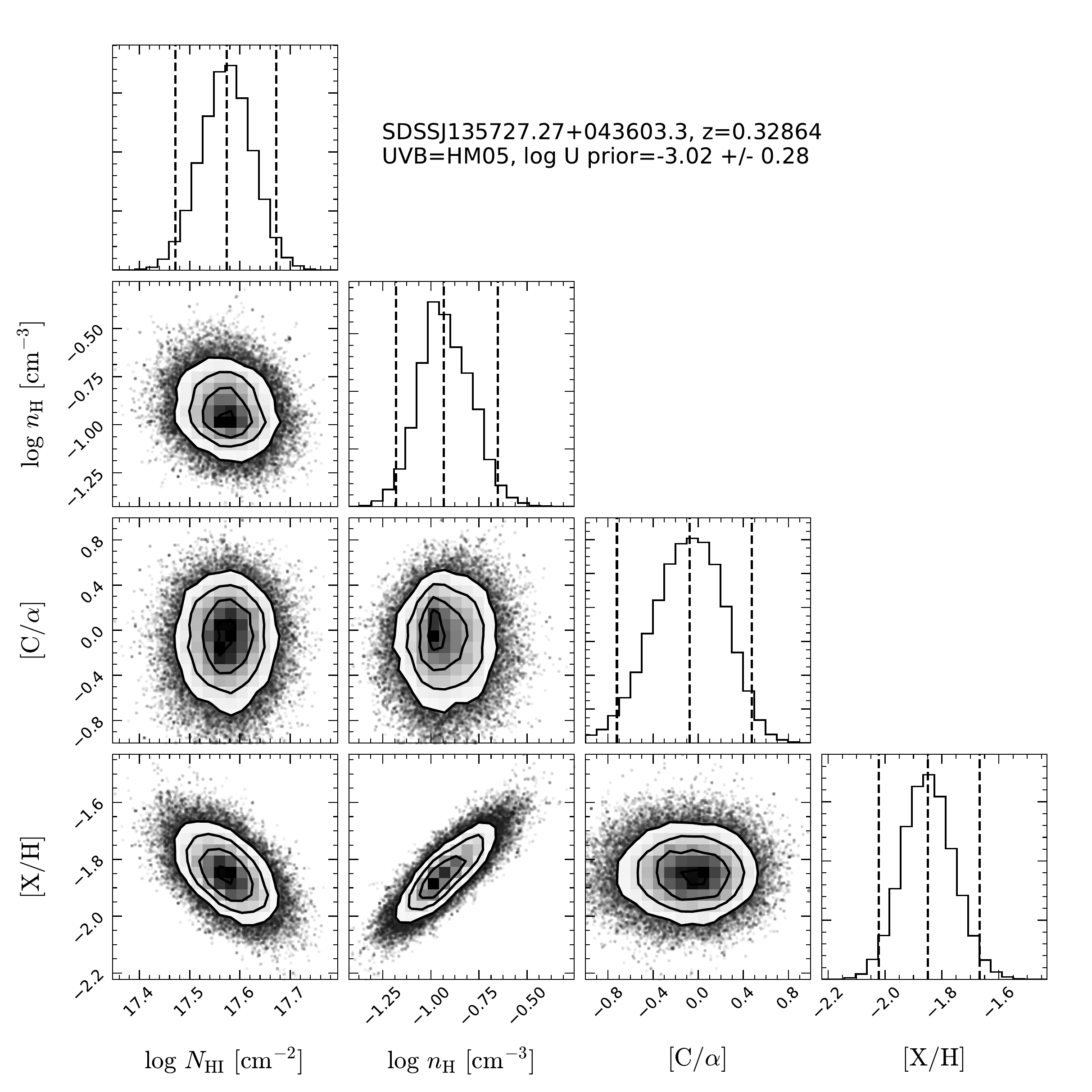}
    \caption{Corner plot of the MCMC PDFs for the LLS associated with LRG SDSSJ135727.27+043603.3. The PDFs for each variable are plotted along the diagonal, and the other panels are the 2D distributions of the final MCMC walkers chains. The vertical dashed lines show the median value of the distribution and the bounds of the 95\% confidence interval. \FeII\ is included in this run.}
    \label{fig:corner_fe}
\end{figure}

\begin{figure}
    \epsscale{1.0}
    \plotone{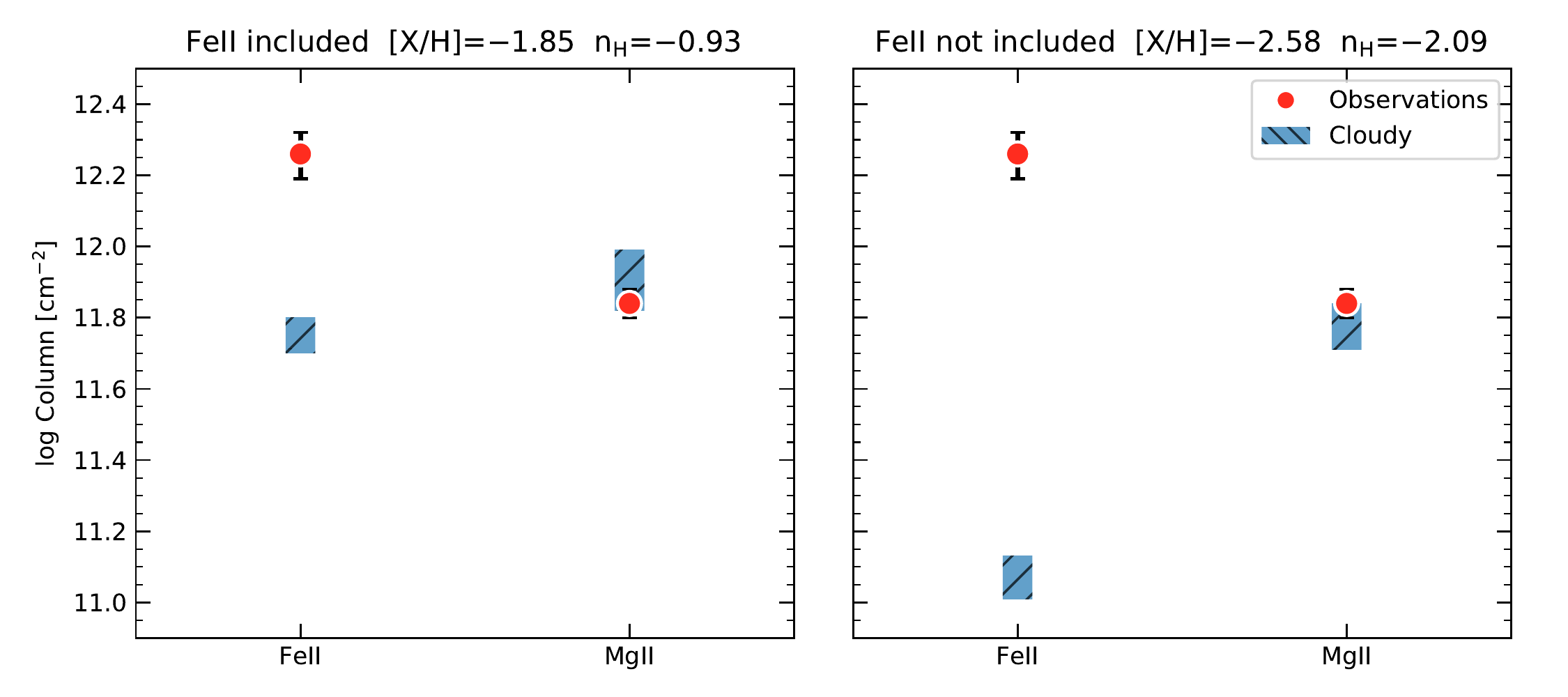}
    \caption{Plot of observed \FeII\ and \MgII\ column densities compared to the column density ranges expected from Cloudy models. The median metallicity and 68\% confidence interval of the log $U$ distribution from the MCMC runs were used to create these models.}
    \label{fig:residal}
\end{figure}

\begin{figure}
    \epsscale{1.0}
    \plotone{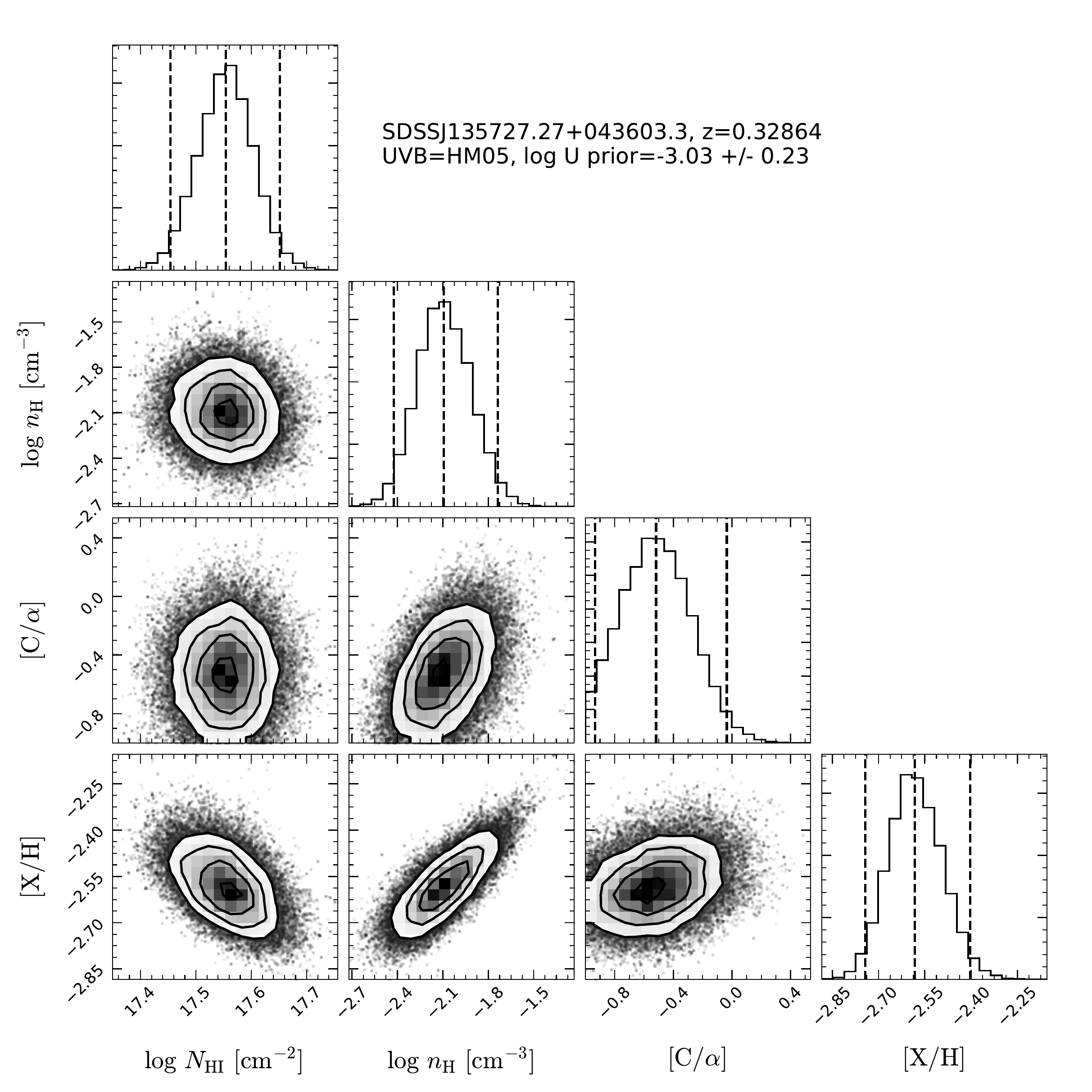}
    \caption{Same as in Fig.~\ref{fig:corner_fe}, but \FeII\ is not included in this run.}
    \label{fig:corner_nofe}
\end{figure}

We have reanalyzed the metallicity of this system using the new information from our HIRES observations with the CCC results \citep[upper limits on metal-ion column densities and a measurement of \logNHI,][]{wotta2019}. With only \MgII, \FeII, and hydrogen detected, we need to apply a Gaussian prior on the $\log U$ distribution. We allow, however, the [C/$\alpha$] ratio to vary. We first consider the modeling with \FeII. The corner plot is shown in Figure~\ref{fig:corner_fe}, and the comparison between the observed and modeled detected metal column densities is shown in the left panel of Figure~\ref{fig:residal}. The model results show the range of column densities encompassed by models representing the 16\% and 84\% confidence values in $\log U$. \FeII\ is underproduced, which is not surprising since without any ionization modeling, the column densities of \FeII\ and \MgII\ imply a super-solar value [\FeII/\MgII$] = +0.50 \pm 0.08$. In this case the median metallicity value is [X/H$]= -1.85$. We also ran a model without \FeII, and in this case the median metallicity value is [X/H$]= -2.58$ (see Figure~\ref{fig:corner_nofe}). In the right panel of Figure~\ref{fig:residal}, we show the predicted value of \FeII\ from this model, and the discrepancy is even larger.

Given the large Fe enhancement relative to $\alpha$ elements in this system, it is unclear how to treat the metallicity of this absorber. Our clear detection of iron makes it unlikely this absorber has a metallicity $<1\%$ of the solar metallicity. However, we are wary about assuming the ionization balance of Fe is well-described: given the extensive network of energy levels for Fe$^+$ and Fe$^{+2}$ below their ionization energies, the unaccounted-for autoionizing levels may play an important role in its behavior \citep{ferland2017}. Because no other metal ions are detected, the models using \MgII\ or \MgII +\FeII\ are the best constraints. They give different results, and neither reproduces the observed \FeII\ column densities. Thus, we report the metallicity of this system using only the range bracketed by these two results, [X/H] = $[-2.58,-1.85]$. This absorber has a low metallicity (i.e., [X/H] $< -1$). However, given the detection of \MgII\ and \FeII, it is not ``pristine.''

\end{document}

%% file: Table_ObjectLocation.tex
\begin{deluxetable*}{lcccc}
\tablecaption{RDR Object Information \label{tab:objinfo}}
\tablehead{\colhead{LRG Name} & \colhead{$z_{\rm LRG}$} & \colhead{QSO Target}  & \colhead{$z_{\rm em}$} & \colhead{$\rho$ (kpc)}}
\startdata
SDSSJ111508.24+023752.7 & 0.27797 & SDSSJ111507.65+023757.5 & 0.567 & 44 \\
SDSSJ111132.33+554712.8 & 0.46286 & SDSSJ111132.18+554726.1 & 0.766 & 81 \\
SDSSJ112755.83+115438.3 & 0.42368 & SDSSJ112756.76+115427.1 & 0.509 & 102 \\
SDSSJ124307.36+353926.3 & 0.38966 & SDSSJ124307.57+353907.1 & 0.547 & 105 \\
SDSSJ095915.54+320418.0 & 0.53019 & SDSSJ095914.84+320357.2 & 0.564 & 146 \\
SDSSJ141307.39+091956.7 & 0.35837 & SDSSJ141309.14+092011.2 & 0.460 & 154 \\
SDSSJ125859.98+413128.2 & 0.27903 & SDSSJ125901.67+413055.8 & 0.745 & 164 \\
SDSSJ125222.93+291327.2 & 0.59874 & SDSSJ125224.99+291321.1 & 0.823 & 190 \\
SDSSJ171651.46+302649.0 & 0.40044 & SDSSJ171654.20+302701.4 & 0.752 & 207 \\
SDSSJ111436.59+403739.1 & 0.60975 & SDSSJ111438.71+403720.3 & 0.736 & 212 \\
SDSSJ110632.58+351012.8 & 0.47035 & SDSSJ110631.05+351051.3 & 0.485 & 261 \\
SDSSJ081524.08+273621.2 & 0.50433 & SDSSJ081520.66+273616.9 & 0.907 & 288 \\
SDSSJ125101.95+302501.7 & 0.51323 & SDSSJ125100.31+302541.8 & 0.652 & 289 \\
SDSSJ122516.86+121750.1 & 0.29801 & SDSSJ122512.93+121835.6 & 0.411 & 336 \\
SDSSJ141540.01+163336.4 & 0.48262 & SDSSJ141542.90+163413.7 & 0.743 & 345 \\
SDSSJ022612.22+001439.9 & 0.47304 & SDSSJ022614.46+001529.7 & 0.615 & 367 \\
SDSSJ075217.92+273835.6 & 0.58117 & SDSSJ075222.91+273823.1 & 1.056 & 458 \\
SDSSJ150522.44+294626.2 & 0.40313 & SDSSJ150527.60+294718.3 & 0.526 & 473 \\
SDSSJ132457.98+271742.6 & 0.44784 & SDSSJ132503.79+271718.7 & 0.522 & 481 \\
SDSSJ104918.08+021814.2 & 0.51524 & SDSSJ104923.24+021806.0 & 0.749 & 497 \\
SDSSJ110405.15+314244.1 & 0.36560 & SDSSJ110406.94+314111.4 & 0.434 & 500 \\
\enddata
\end{deluxetable*}

%% file: Table_LRGproperties.tex
\begin{deluxetable*}{lcccccccc}
\tablecaption{LRG and Absorber Properties \label{tab:lrgs}}
\tablehead{\colhead{LRG Name} & \colhead{$z_{\rm LRG}$} & \colhead{$\rho$} & \colhead{$M_r$} & \colhead{$\log M_{\star}$} & \colhead{$\log M_{\rm h}$} & \colhead{$R_{\rm vir}$} & \colhead{$\log {\rm sSFR}$} & \colhead{\logNHI}}
\startdata
SDSSJ111508.24+023752.7 & 0.27797 & 44 & $-$22.9 & 11.3 & 13.3 & 516 & $< -10.98$ & $[18.00,19.60]$ \\
SDSSJ111132.33+554712.8 & 0.46286 & 81 & $-$23.2 & 11.4 & 13.4 & 535 & $-10.11 \pm 0.13$ & $17.80 \pm 0.03$ \\
SDSSJ112755.83+115438.3 & 0.42368 & 102 & $-$22.5 & 11.2 & 13.1 & 412 & $< -10.69$ & $15.87^{+0.05}_{-0.06}$ \\
SDSSJ124307.36+353926.3 & 0.38966 & 105 & $-$22.9 & 11.3 & 13.2 & 475 & $< -9.91$ & $< 13.06$ \\
SDSSJ095915.54+320418.0 & 0.53019 & 146 & $-$23.2 & 11.4 & 13.4 & 494 & $< -9.98$ & $< 14.76$ \\
SDSSJ141307.39+091956.7 & 0.35837 & 154 & $-$23.9 & 11.7 & 14.0 & 875 & $< -10.16$ & $13.67^{+0.05}_{-0.06}$ \\
SDSSJ125859.98+413128.2 & 0.27903 & 164 & $-$23.5 & 11.6 & 13.8 & 754 & $< -10.96$ & $13.70 \pm 0.05$ \\
SDSSJ125222.93+291327.2 & 0.59874 & 190 & $-$23.4 & 11.5 & 13.5 & 550 & $< -10.67$ & $> 15.20$ \\
SDSSJ171651.46+302649.0 & 0.40044 & 207 & $-$23.2 & 11.4 & 13.4 & 544 & $< -9.77$ & $16.90 \pm 0.20$ \\
SDSSJ111436.59+403739.1 & 0.60975 & 212 & $-$24.1 & 11.8 & 14.0 & 762 & $< -10.30$ & $< 13.79$ \\
SDSSJ110632.58+351012.8 & 0.47035 & 261 & $-$22.8 & 11.3 & 13.2 & 445 & $< -10.32$ & $< 13.70$ \\
SDSSJ081524.08+273621.2 & 0.50433 & 288 & $-$23.3 & 11.4 & 13.4 & 531 & $-9.69 \pm 0.17$ & $< 13.81$ \\
SDSSJ125101.95+302501.7 & 0.51323 & 289 & $-$23.6 & 11.6 & 13.7 & 650 & $< -10.95$ & $< 14.57$ \\
SDSSJ122516.86+121750.1 & 0.29801 & 336 & $-$22.7 & 11.2 & 13.2 & 478 & $< -10.31$ & $< 14.12$ \\
SDSSJ141540.01+163336.4 & 0.48262 & 345 & $-$23.3 & 11.4 & 13.4 & 529 & $-10.24 \pm 0.13$ & $16.07 \pm 0.02$ \\
SDSSJ022612.22+001439.9 & 0.47304 & 367 & $-$23.3 & 11.5 & 13.5 & 562 & $-10.16 \pm 0.10$ & $13.97^{+0.11}_{-0.15}$ \\
SDSSJ075217.92+273835.6 & 0.58117 & 458 & $-$23.0 & 11.4 & 13.3 & 473 & $< -10.23$ & $< 14.41$ \\
SDSSJ150522.44+294626.2 & 0.40313 & 473 & $-$22.4 & 11.2 & 13.0 & 386 & $-10.28 \pm 0.17$ & $< 13.85$ \\
SDSSJ132457.98+271742.6 & 0.44784 & 481 & $-$22.8 & 11.3 & 13.2 & 448 & $< -9.92$ & $< 14.26$ \\
SDSSJ104918.08+021814.2 & 0.51524 & 497 & $-$22.4 & 11.1 & 12.9 & 359 & $< -10.13$ & $15.72^{+0.19}_{-0.35}$ \\
SDSSJ110405.15+314244.1 & 0.36560 & 500 & $-$22.7 & 11.2 & 13.2 & 449 & $< -9.87$ & $13.58^{+0.08}_{-0.10}$ \\
\enddata
\end{deluxetable*}

%% file: Table_QSOproperties.tex
\begin{deluxetable}{lc}
\tablecaption{QSO UV Spectroscopy \label{tab:qsoobs}}
\tablehead{\colhead{QSO Target} & \colhead{Instrument/Grating}}
\startdata
SDSSJ111507.65+023757.5 & COS/G130M \\
SDSSJ111132.18+554726.1 & COS/G130M,G160M \\
SDSSJ112756.76+115427.1 & COS/G130M,G160M \\
SDSSJ124307.57+353907.1 & COS/G130M,G160M \\
SDSSJ095914.84+320357.2 & COS/G130M \\
SDSSJ141309.14+092011.2 & COS/G130M,G160M \\
SDSSJ125901.67+413055.8 & COS/G130M,G160M \\
SDSSJ125224.99+291321.1 & COS/G130M;FOS/G190H \\
SDSSJ171654.20+302701.4 & COS/G140L \\
SDSSJ111438.71+403720.3 & FOS/G190H \\
SDSSJ110631.05+351051.3 & COS/G160M \\
SDSSJ081520.66+273616.9 & COS/G140L \\
SDSSJ125100.31+302541.8 & COS/G130M \\
SDSSJ122512.93+121835.6 & COS/G130M \\
SDSSJ141542.90+163413.7 & COS/G130M \\
SDSSJ022614.46+001529.7 & COS/G130M,G160M \\
SDSSJ075222.91+273823.1 & COS/G140L \\
SDSSJ150527.60+294718.3 & COS/G130M \\
SDSSJ132503.79+271718.7 & COS/G130M \\
SDSSJ104923.24+021806.0 & COS/G130M,G160M \\
SDSSJ110406.94+314111.4 & COS/G130M,G160M \\
\enddata
\tablecomments{The grating resolutions are as follows: G130M/G160M: $R=18,000$; G190H: $R=1,300$; G140L: $R=2,300$.}
\end{deluxetable}

%% file: LRGcovering_columns.tex
\begin{deluxetable*}{lccccc}
\tablecaption{\HI\ Covering Factors for RDR LRGs\label{tab:coveringfactor}}
\tablehead{\colhead{$\rho$ (kpc)} & \colhead{$\langle\rho\rangle$ (kpc)} & \colhead{$f_c$\tablenotemark{a}} & \colhead{95\% C.I.} & \colhead{Sightlines\tablenotemark{b}} & \colhead{Detections\tablenotemark{c}}}
\startdata
\hline
\multicolumn{6}{c}{\logNHI\ $\ge$ 15.0} \\
\hline
$[  0,250]$ & 140 & $0.41^{+0.15}_{-0.14}$ & $[0.17,0.69]$ & 10 & 4 \\
$(250,500]$ & 390 & $0.22^{+0.13}_{-0.10}$ & $[0.06,0.48]$ & 11 & 2 \\
\hline
\multicolumn{6}{c}{\logNHI\ $\ge$ 16.0} \\
\hline
$[  0,250]$ & 135 & $0.35^{+0.15}_{-0.13}$ & $[0.12,0.65]$ & 9 & 3 \\
$(250,500]$ & 390 & $0.14^{+0.11}_{-0.08}$ & $[0.02,0.39]$ & 11 & 1 \\
\hline
\multicolumn{6}{c}{\logNHI\ $\ge$ 17.2} \\
\hline
$[  0,250]$ & 135 & $0.26^{+0.15}_{-0.12}$ & $[0.07,0.56]$ & 9 & 2 \\
$(250,500]$ & 390 & $0.06^{+0.09}_{-0.04}$ & $[0.00,0.27]$ & 11 & 0 \\
\enddata
\tablenotetext{a}{The error bars on the covering factor represent the 68\% confidence interval.}
\tablenotetext{b}{Number of sightlines considered in covering factor calculation.}
\tablenotetext{c}{Number of absorbers detected above the limiting \HI\ value.}
\end{deluxetable*}

%% file: coveringfactor_ssfr.tex
\begin{deluxetable*}{lccccc}
\tablecaption{\HI\ Covering Factors by Sample\label{tab:coveringfactorssfr}}
\tablehead{\colhead{$\rho/\Rvir$} & \colhead{$\langle\rho/\Rvir\rangle$} & \colhead{$f_c$\tablenotemark{a}} & \colhead{95\% C.I.} & \colhead{Sightlines\tablenotemark{b}} & \colhead{Detections\tablenotemark{c}}}
\startdata
\hline
\multicolumn{6}{c}{RDR} \\
\hline
$[0.0,0.5]$ & 0.25 & $0.32^{+0.14}_{-0.12}$ & $[0.11,0.61]$ & 10 & 3 \\
$(0.5,1.0]$ & 0.68 & $0.23^{+0.17}_{-0.12}$ & $[0.04,0.58]$ & 6 & 1 \\
$(1.0,1.5]$ & 1.20 & $0.13^{+0.18}_{-0.10}$ & $[0.01,0.52]$ & 4 & 0 \\
\hline
\multicolumn{6}{c}{Star-Forming Galaxies} \\
\hline
$[0.0,0.5]$ & 0.30 & $0.70^{+0.09}_{-0.10}$ & $[0.50,0.86]$ & 21 & 15 \\
$(0.5,1.0]$ & 0.63 & $0.13^{+0.18}_{-0.10}$ & $[0.01,0.52]$ & 4 & 0 \\
\hline
\multicolumn{6}{c}{Quiescent Galaxies} \\
\hline
$[0.0,0.5]$ & 0.21 & $0.53 \pm 0.12$ & $[0.30,0.75]$ & 15 & 8 \\
$(0.5,1.0]$ & 0.57 & $0.29^{+0.31}_{-0.21}$ & $[0.01,0.84]$ & 1 & 0 \\
\enddata
\tablecomments{The limiting \HI\ value for the covering factor calculation is \logNHI\ $\ge$ 16 [\column]. The star-forming and quiescent galaxy samples are from the \citetalias{prochaska2017} sample.}
\tablenotetext{a}{The error bars on the covering factor represent the 68\% confidence interval.}
\tablenotetext{b}{Number of sightlines considered in covering factor calculation.}
\tablenotetext{c}{Number of absorbers detected above the limiting \HI\ value.}
\end{deluxetable*}

%% file: coveringfactor_stellar.tex
\begin{deluxetable*}{lccccc}
\tablecaption{\HI\ Covering Factors by Stellar Mass\label{tab:coveringfactorstellar}}
\tablehead{\colhead{$\rho/\Rvir$} & \colhead{$\langle\rho/\Rvir\rangle$} & \colhead{$f_c$\tablenotemark{a}} & \colhead{95\% C.I.} & \colhead{Sightlines\tablenotemark{b}} & \colhead{Detections\tablenotemark{c}}}
\startdata
\hline
\multicolumn{6}{c}{$9.0 \le \log M_{\star} < 10.5$} \\
\hline
$[0.0,0.5]$ & 0.30 & $0.63 \pm 0.12$ & $[0.38,0.84]$ & 14 & 9 \\
$(0.5,1.0]$ & 0.63 & $0.13^{+0.18}_{-0.10}$ & $[0.01,0.52]$ & 4 & 0 \\
\hline
\multicolumn{6}{c}{$10.5 \le \log M_{\star} < 11.3$} \\
\hline
$[0.0,0.5]$ & 0.24 & $0.67^{+0.09}_{-0.10}$ & $[0.47,0.84]$ & 22 & 15 \\
$(0.5,1.0]$ & 0.60 & $0.16^{+0.21}_{-0.12}$ & $[0.01,0.60]$ & 3 & 0 \\
$(1.0,1.5]$ & 1.24 & $0.16^{+0.21}_{-0.12}$ & $[0.01,0.60]$ & 3 & 0 \\
\hline
\multicolumn{6}{c}{$11.3 \le \log M_{\star} \le 12.0$} \\
\hline
$[0.0,0.5]$ & 0.23 & $0.41^{+0.12}_{-0.11}$ & $[0.20,0.65]$ & 15 & 6 \\
$(0.5,1.0]$ & 0.68 & $0.26^{+0.19}_{-0.14}$ & $[0.04,0.64]$ & 5 & 1 \\
$(1.0,1.5]$ & 1.07 & $0.29^{+0.31}_{-0.21}$ & $[0.01,0.84]$ & 1 & 0 \\
\enddata
\tablecomments{The limiting \HI\ value for the covering factor calculation is \logNHI\ $\ge$ 16 [\column]. A combination of the RDR LRGs, the galaxies in the \citetalias{prochaska2017} sample, and the galaxies in the \citetalias{chen2018} sample are used for this covering factor calculation. We do not include sightlines in common between the \citetalias{chen2018} sample and the RDR LRGs or the \citetalias{prochaska2017} sample.}
\tablenotetext{a}{The error bars on the covering factor represent the 68\% confidence interval.}
\tablenotetext{b}{Number of sightlines considered in covering factor calculation.}
\tablenotetext{c}{Number of absorbers detected above the limiting \HI\ value.}
\end{deluxetable*}

%% file: coveringfactor_halos.tex
\begin{deluxetable*}{lccccc}
\tablecaption{\HI\ Covering Factors by Halo Mass\label{tab:coveringfactorhalo}}
\tablehead{\colhead{$\rho/\Rvir$} & \colhead{$\langle\rho/\Rvir\rangle$} & \colhead{$f_c$\tablenotemark{a}} & \colhead{95\% C.I.} & \colhead{Sightlines\tablenotemark{b}} & \colhead{Detections\tablenotemark{c}}}
\startdata
\hline
\multicolumn{6}{c}{$11.0 \le \log M_{\rm h} < 12.0$} \\
\hline
$[0.0,0.5]$ & 0.32 & $0.70^{+0.12}_{-0.14}$ & $[0.43,0.90]$ & 11 & 8 \\
$(0.5,1.0]$ & 0.68 & $0.16^{+0.21}_{-0.12}$ & $[0.01,0.60]$ & 3 & 0 \\
\hline
\multicolumn{6}{c}{$12.0 \le \log M_{\rm h} < 13.0$} \\
\hline
$[0.0,0.5]$ & 0.25 & $0.59^{+0.10}_{-0.11}$ & $[0.38,0.78]$ & 20 & 12 \\
$(0.5,1.0]$ & 0.54 & $0.16^{+0.21}_{-0.12}$ & $[0.01,0.60]$ & 3 & 0 \\
$(1.0,1.5]$ & 1.30 & $0.21^{+0.25}_{-0.15}$ & $[0.01,0.71]$ & 2 & 0 \\
\hline
\multicolumn{6}{c}{$13.0 \le \log M_{\rm h} \le 14.0$} \\
\hline
$[0.0,0.5]$ & 0.21 & $0.50 \pm 0.11$ & $[0.30,0.70]$ & 20 & 10 \\
$(0.5,1.0]$ & 0.68 & $0.23^{+0.17}_{-0.12}$ & $[0.04,0.58]$ & 6 & 1 \\
$(1.0,1.5]$ & 1.09 & $0.21^{+0.25}_{-0.15}$ & $[0.01,0.71]$ & 2 & 0 \\
\enddata
\tablecomments{The limiting \HI\ value for the covering factor calculation is \logNHI\ $\ge$ 16 [\column]. A combination of the RDR LRGs, the galaxies in the \citetalias{prochaska2017} sample, and the galaxies in the \citetalias{chen2018} sample are used for this covering factor calculation. We do not include sightlines in common between the \citetalias{chen2018} sample and the RDR LRGs or the \citetalias{prochaska2017} sample.}
\tablenotetext{a}{The error bars on the covering factor represent the 68\% confidence interval.}
\tablenotetext{b}{Number of sightlines considered in covering factor calculation.}
\tablenotetext{c}{Number of absorbers detected above the limiting \HI\ value.}
\end{deluxetable*}

%% file: Table_GridParameters.tex
\begin{deluxetable}{lccc}
\tablecaption{Cloudy Grid Parameters \label{tab:grid}}
\tablehead{\colhead{Parameter} & \colhead{Minimum} & \colhead{Maximum} & \colhead{Step Size\tablenotemark{a}}}
\startdata
\logNHI\ [\column] & 12.0 & 20.0 & 0.25 \\
\z\ & 0.0 & 5.0 & 0.25 \\
${\rm [X/H]}$ & $-5.0$ & $+2.5$ & 0.25 \\
$\log$ \nH\ [\percc] & $-4.5$ & 0.0 & 0.25 \\
${\rm [C/\alpha]}$ & $-1.0$ & 1.0 & 0.20 \\
\enddata
\tablenotetext{a}{Even though the grid is run using these step sizes, the MCMC sampler interpolates between the grid points.}
\end{deluxetable}

%% file: Table_metallicities.tex
\begin{deluxetable*}{lccccccc}
\tablecaption{Metallicity Constraints on High \HI\ Column Absorbers \label{tab:met}}
\tablehead{\colhead{LRG Name} & \colhead{$z_{\rm LRG}$} & \colhead{$\rho$} & \colhead{\logNHI} & \colhead{Metallicity\tablenotemark{a}} & \colhead{95\% C.I.} & \colhead{Metallicity\tablenotemark{a}} & \colhead{95\% C.I.} \\ %
\colhead{} &\colhead{} & \colhead{(kpc)} &\colhead{}  &\colhead{(HM05)} & \colhead{(HM05)} &\colhead{(HM12)} &\colhead{(HM12)} %
}
\startdata
SDSSJ111508.24+023752.7 & 0.27797 & 44 & $[18.00,19.60]$ & $[-2.10,-0.50]\tablenotemark{b}$ & \nodata & \nodata & \nodata \\
SDSSJ111132.33+554712.8 & 0.46286 & 81 & $17.80 \pm 0.03$ & $-0.14^{+0.09}_{-0.10}$ & $[-0.32,+0.05]$ & $-0.06 \pm 0.09$ & $[-0.23,+0.12]$ \\
SDSSJ171651.46+302649.0 & 0.40044 & 207 & $16.90 \pm 0.20$ & $-1.87^{+0.24}_{-0.25}$ & $[-2.34,-1.40]$ & $-1.36^{+0.33}_{-0.28}$ & $[-1.90,-0.44]$ \\
SDSSJ141540.01+163336.4 & 0.48262 & 345 & $16.07 \pm 0.02$ & $-0.89^{+0.08}_{-0.09}$ & $[-1.07,-0.72]$ & $-0.59 \pm 0.07$ & $[-0.72,-0.45]$ \\
\enddata
\tablenotetext{a}{The error bars on the median metallicity represent the 68\% confidence interval.}
\tablenotetext{b}{The metallicity for this absorber is determined from [O/H] using \logNOI\ from the strongest component and the range of allowed \logNHI\ values.}
\end{deluxetable*}

%% file: Table_ion_column.tex
\startlongtable
\begin{deluxetable}{lccc}
\tablecaption{Average velocities and column densities of the hydrogen and metal ions in the absorbers associated with the RDR LRGs \label{tab:iontable}}
\tablehead{
\colhead{Species} & \colhead{$[v_1,v_2]$} & \colhead{$\langle v \rangle$} & \colhead{$\log N$}\\
\colhead{} & \colhead{(\kms)} & \colhead{(\kms)}& \colhead{[cm$^{-2} $]}
}
\startdata
\hline
\multicolumn{4}{c}{SDSSJ111508.24+023752.7, $z= 0.277973$ } \\
\hline
\ensuremath{\mbox{\ion{C}{2}}} $\lambda$1036    &  [$ -131, 148 $]     &  $ 12.8 \pm 34.1 $   &                   $ > 14.75 $ \\
\ensuremath{\mbox{\ion{C}{3}}} $\lambda$977     &  [$ -131, 148 $]     &  $ 33.7 \pm 34.1 $   &                   $ > 14.27 $ \\
\ensuremath{\mbox{\ion{Fe}{2}}} $\lambda$1144   &  [$ -71, 148 $]      &  \nodata             &                   $ < 13.97 $ \\
\ensuremath{\mbox{\ion{Fe}{3}}} $\lambda$1122   &  [$ -71, 148 $]      &  \nodata             &                   $ < 13.97 $ \\
\ensuremath{\mbox{\ion{H}{1}}} Break+Voigt            & \nodata              & \nodata              &               $[18.00,19.60]$ \\
\ensuremath{\mbox{\ion{N}{2}}} $\lambda$1083    &  [$ -71, 148 $]      &  $ 35.6 \pm 34.0 $   &             $14.24 \pm 0.08 $ \\
\ensuremath{\mbox{\ion{N}{3}}} $\lambda$989     &  [$ -71, 148 $]      &  $ 50.3 \pm 33.8 $   &             $14.24 \pm 0.06 $ \\
\ensuremath{\mbox{\ion{O}{1}}} $\lambda$971     &  [$ -71, 118 $]      &  \nodata             &                   $ < 14.60 $ \\
\ensuremath{\mbox{\ion{O}{1}}} $\lambda$988     &  [$ -71, 148 $]      &  $ 59.5 \pm 35.4 $   &             $14.45 \pm 0.11 $ \\
\ensuremath{\mbox{\ion{O}{1}}}                  &                      &  $ 59.5 \pm 35.4 $   &             $14.45 \pm 0.11 $ \\
\ensuremath{\mbox{\ion{O}{6}}} $\lambda$1037    &  [$ -71, 148 $]      &  \nodata             &                   $ < 14.07 $ \\
\ensuremath{\mbox{\ion{O}{6}}} $\lambda$1031    &  [$ -71, 148 $]      &  $ 29.3 \pm 35.8 $   &             $14.21 \pm 0.15 $ \\
\ensuremath{\mbox{\ion{O}{6}}}                  &                      &  $ 29.3 \pm 35.8 $   &             $14.21 \pm 0.15 $ \\
\ensuremath{\mbox{\ion{S}{3}}} $\lambda$1012    &  [$ -71, 148 $]      &  \nodata             &                   $ < 14.20 $ \\
\ensuremath{\mbox{\ion{S}{4}}} $\lambda$1062    &  [$ -71, 148 $]      &  \nodata             &                   $ < 13.99 $ \\
\hline
\multicolumn{4}{c}{SDSSJ111132.33+554712.8, $z= 0.462859$ } \\
\hline
\ensuremath{\mbox{\ion{C}{2}}} $\lambda$1036    &  [$ -31, 138 $]      &  $ 83.4 \pm 29.6 $   &                   $ > 14.48 $ \\
\ensuremath{\mbox{\ion{C}{3}}} $\lambda$977     &  [$ -34, 193 $]      &  $ 57.4 \pm 27.7 $   &                   $ > 14.27 $ \\
\ensuremath{\mbox{\ion{Fe}{2}}} $\lambda$2382 &  [$   -16,  123 $]     &  $ 83.0  \pm 27.9 $  &          $12.98  \pm 0.04 $ \\
\ensuremath{\mbox{\ion{Fe}{2}}} $\lambda$2586 &  [$   -16,  123 $]     &  $ 91.7  \pm 30.0 $  &          $13.06  \pm 0.11 $ \\
\ensuremath{\mbox{\ion{Fe}{2}}} $\lambda$2600 &  [$   -16,  123 $]     &  $ 87.2  \pm 27.8 $  &          $13.00  \pm 0.03 $ \\
\ensuremath{\mbox{\ion{Fe}{2}}}               &                      &  $ 87.1  \pm 16.5 $   &                              $13.00  \pm 0.02 $ \\
\ensuremath{\mbox{\ion{H}{1}}} Break            & \nodata              & \nodata              &               $17.80  \pm 0.03 $ \\ 
\ensuremath{\mbox{\ion{Mg}{2}}} $\lambda$2796 &  [$   -16,  123 $]    &  $ 84.3  \pm 27.6 $  &          $13.34  \pm 0.05 $ \\
\ensuremath{\mbox{\ion{Mg}{2}}} $\lambda$2803 &  [$   -16,  123 $]    &  $ 84.5  \pm 27.6 $  &          $13.52  \pm 0.04 $ \\
\ensuremath{\mbox{\ion{Mg}{2}}}               &                     &  $ 84.5  \pm 27.6 $   &                                $ > 13.67 $ \\
\ensuremath{\mbox{\ion{N}{1}}} $\lambda$1199    &  [$ -31, 193 $]      &  \nodata             &                   $ < 13.47 $ \\
\ensuremath{\mbox{\ion{N}{2}}} $\lambda$1083    &  [$ -31, 193 $]      &  $ 73.2 \pm 28.6 $   &                   $ > 14.31 $ \\
\ensuremath{\mbox{\ion{N}{3}}} $\lambda$989     &  [$ -31, 193 $]      &  $ 80.3 \pm 28.0 $   &                   $ > 14.48 $ \\
\ensuremath{\mbox{\ion{O}{1}}} $\lambda$971     &  [$ -31, 173 $]      &  \nodata             &                   $ < 14.31 $ \\
\ensuremath{\mbox{\ion{O}{1}}} $\lambda$988     &  [$ 18, 193 $]       &  $ 79.5 \pm 30.6 $   &             $14.23 \pm 0.13 $ \\
\ensuremath{\mbox{\ion{O}{1}}}                  &                      &  $ 79.5 \pm 30.6 $   &             $14.23 \pm 0.13 $ \\
\ensuremath{\mbox{\ion{O}{2}}} $\lambda$834     &  [$ -31, 193 $]      &  $ 52.2 \pm 37.7 $   &                   $ > 14.61 $ \\
\ensuremath{\mbox{\ion{O}{4}}} $\lambda$787     &  [$ -31, 193 $]      &  \nodata             &                   $ < 14.52 $ \\
\ensuremath{\mbox{\ion{O}{6}}} $\lambda$1031    &  [$ -31, 193 $]      &  $ 67.6 \pm 30.4 $   &             $13.70 \pm 0.10 $ \\
\ensuremath{\mbox{\ion{O}{6}}} $\lambda$1037    &  [$ -31, 193 $]      &  $ 69.2 \pm 36.0 $   &             $13.74 \pm 0.20 $ \\
\ensuremath{\mbox{\ion{O}{6}}}                  &                      &  $ 68.2 \pm 23.2 $   &             $13.71 \pm 0.09 $ \\
\ensuremath{\mbox{\ion{S}{3}}} $\lambda$1012    &  [$ -31, 193 $]      &  $ 85.5 \pm 28.8 $   &             $14.34 \pm 0.06 $ \\
\ensuremath{\mbox{\ion{S}{4}}} $\lambda$809     &  [$ -31, 193 $]      &  \nodata             &                   $ < 14.28 $ \\
\ensuremath{\mbox{\ion{S}{6}}} $\lambda$933     &  [$ -31, 193 $]      &  \nodata             &                   $ < 12.70 $ \\
\ensuremath{\mbox{\ion{S}{3}}} $\lambda$1190    &  [$ 11, 113 $]       &  $ 67.4 \pm 28.2 $   &             $14.48 \pm 0.09 $ \\
\ensuremath{\mbox{\ion{Si}{2}}} $\lambda$1193   &  [$ 3, 193 $]        &  $ 83.1 \pm 28.1 $   &                   $ > 13.51 $ \\
\ensuremath{\mbox{\ion{Si}{3}}} $\lambda$1206   &  [$ -31, 193 $]      &  $ 70.1 \pm 27.9 $   &                   $ > 13.54 $ \\
\hline
\multicolumn{4}{c}{SDSSJ112755.83+115438.3, $z= 0.423677$ } \\
\hline
\ensuremath{\mbox{\ion{C}{2}}} $\lambda$903     &  [$ 32, 202 $]       &  $ 141.6 \pm 31.7 $  &             $14.11 \pm 0.15 $ \\
\ensuremath{\mbox{\ion{C}{3}}} $\lambda$977     &  [$ 12, 252 $]       &  $ 115.0 \pm 29.6 $  &             $13.73 \pm 0.05 $ \\
\ensuremath{\mbox{\ion{Fe}{2}}} $\lambda$1144   &  [$ 12, 252 $]       &  \nodata             &                   $ < 13.94 $ \\
\ensuremath{\mbox{\ion{Fe}{3}}} $\lambda$1122   &  [$ 12, 252 $]       &  \nodata             &                   $ < 14.18 $ \\
\ensuremath{\mbox{\ion{H}{1}}} $\lambda$930     &  [$ 12, 252 $]       &  $ 130.8 \pm 31.7 $  &             $15.88 \pm 0.11 $ \\
\ensuremath{\mbox{\ion{H}{1}}} $\lambda$949     &  [$ 12, 252 $]       &  $ 156.3 \pm 29.9 $  &             $15.87 \pm 0.08 $ \\
\ensuremath{\mbox{\ion{H}{1}}}                  &                      &  $ 144.3 \pm 21.7 $  &             $15.87 \pm 0.06 $ \\
\ensuremath{\mbox{\ion{N}{2}}} $\lambda$1083    &  [$ 12, 252 $]       &  \nodata             &                   $ < 13.58 $ \\
\ensuremath{\mbox{\ion{O}{1}}} $\lambda$1039    &  [$ 12, 252 $]       &  \nodata             &                   $ < 14.67 $ \\
\ensuremath{\mbox{\ion{O}{2}}} $\lambda$834     &  [$ 12, 252 $]       &  \nodata             &                   $ < 14.08 $ \\
\ensuremath{\mbox{\ion{O}{6}}} $\lambda$1031    &  [$ 12, 252 $]       &  \nodata             &                   $ < 13.49 $ \\
\ensuremath{\mbox{\ion{S}{3}}} $\lambda$1012    &  [$ 12, 252 $]       &  \nodata             &                   $ < 13.94 $ \\
\ensuremath{\mbox{\ion{S}{4}}} $\lambda$1062    &  [$ 12, 252 $]       &  \nodata             &                   $ < 13.97 $ \\
\ensuremath{\mbox{\ion{S}{6}}} $\lambda$944     &  [$ 12, 252 $]       &  \nodata             &                   $ < 13.59 $ \\
\ensuremath{\mbox{\ion{S}{3}}} $\lambda$1190    &  [$ 12, 252 $]       &  \nodata             &                   $ < 14.60 $ \\
\ensuremath{\mbox{\ion{Si}{2}}} $\lambda$1193   &  [$ 12, 252 $]       &  \nodata             &                   $ < 13.24 $ \\
\ensuremath{\mbox{\ion{Si}{3}}} $\lambda$1206   &  [$ 12, 252 $]       &  \nodata             &                   $ < 12.78 $ \\
\hline
\multicolumn{4}{c}{SDSSJ124307.36+353926.3, $z= 0.389657$ } \\
\hline
\ensuremath{\mbox{\ion{C}{2}}} $\lambda$903     &  [$ -50, 50 $]       &  \nodata             &                   $ < 12.94 $ \\
\ensuremath{\mbox{\ion{C}{3}}} $\lambda$977     &  [$ -50, 50 $]       &  \nodata             &                   $ < 12.65 $ \\
\ensuremath{\mbox{\ion{Fe}{2}}} $\lambda$1144   &  [$ -50, 50 $]       &  \nodata             &                   $ < 13.45 $ \\
\ensuremath{\mbox{\ion{Fe}{3}}} $\lambda$1122   &  [$ -50, 50 $]       &  \nodata             &                   $ < 13.92 $ \\
\ensuremath{\mbox{\ion{H}{1}}} $\lambda$1215    &  [$ -50, 50 $]       &  \nodata             &                   $ < 13.06 $ \\
\ensuremath{\mbox{\ion{Mg}{2}}} $\lambda$1239   &  [$ -50, 50 $]       &  \nodata             &                   $ < 15.91 $ \\
\ensuremath{\mbox{\ion{N}{2}}} $\lambda$1083    &  [$ -50, 50 $]       &  \nodata             &                   $ < 13.41 $ \\
\ensuremath{\mbox{\ion{N}{3}}} $\lambda$989     &  [$ -50, 50 $]       &  \nodata             &                   $ < 13.72 $ \\
\ensuremath{\mbox{\ion{N}{5}}} $\lambda$1238    &  [$ -50, 50 $]       &  \nodata             &                   $ < 13.56 $ \\
\ensuremath{\mbox{\ion{O}{1}}} $\lambda$988     &  [$ -50, 50 $]       &  \nodata             &                   $ < 13.86 $ \\
\ensuremath{\mbox{\ion{O}{2}}} $\lambda$834     &  [$ -50, 50 $]       &  \nodata             &                   $ < 13.73 $ \\
\ensuremath{\mbox{\ion{O}{6}}} $\lambda$1031    &  [$ -50, 50 $]       &  \nodata             &                   $ < 13.24 $ \\
\ensuremath{\mbox{\ion{S}{2}}} $\lambda$1259    &  [$ -50, 50 $]       &  \nodata             &                   $ < 14.69 $ \\
\ensuremath{\mbox{\ion{S}{3}}} $\lambda$1012    &  [$ -50, 50 $]       &  \nodata             &                   $ < 13.61 $ \\
\ensuremath{\mbox{\ion{S}{4}}} $\lambda$1062    &  [$ -50, 50 $]       &  \nodata             &                   $ < 13.79 $ \\
\ensuremath{\mbox{\ion{S}{6}}} $\lambda$933     &  [$ -50, 50 $]       &  \nodata             &                   $ < 12.85 $ \\
\ensuremath{\mbox{\ion{S}{3}}} $\lambda$1190    &  [$ -50, 50 $]       &  \nodata             &                   $ < 14.23 $ \\
\ensuremath{\mbox{\ion{Si}{2}}} $\lambda$1260   &  [$ -50, 50 $]       &  \nodata             &                   $ < 12.88 $ \\
\ensuremath{\mbox{\ion{Si}{3}}} $\lambda$1206   &  [$ -50, 50 $]       &  \nodata             &                   $ < 12.47 $ \\
\hline
\multicolumn{4}{c}{SDSSJ095915.54+320418.0, $z= 0.530194$ } \\
\hline
\ensuremath{\mbox{\ion{C}{2}}} $\lambda$903     &  [$ -50, 50 $]       &  \nodata             &                   $ < 13.19 $ \\
\ensuremath{\mbox{\ion{H}{1}}} $\lambda$949     &  [$ -50, 50 $]       &  \nodata             &                   $ < 14.76 $ \\
\ensuremath{\mbox{\ion{N}{3}}} $\lambda$763     &  [$ -50, 50 $]       &  \nodata             &                   $ < 14.11 $ \\
\ensuremath{\mbox{\ion{N}{4}}} $\lambda$765     &  [$ -50, 50 $]       &  \nodata             &                   $ < 13.21 $ \\
\ensuremath{\mbox{\ion{Ne}{8}}} $\lambda$770    &  [$ -50, 50 $]       &  \nodata             &                   $ < 13.78 $ \\
\ensuremath{\mbox{\ion{O}{1}}} $\lambda$948     &  [$ -50, 50 $]       &  \nodata             &                   $ < 15.11 $ \\
\ensuremath{\mbox{\ion{O}{2}}} $\lambda$834     &  [$ -50, 50 $]       &  \nodata             &                   $ < 13.53 $ \\
\ensuremath{\mbox{\ion{O}{3}}} $\lambda$832     &  [$ -50, 50 $]       &  \nodata             &                   $ < 13.61 $ \\
\ensuremath{\mbox{\ion{O}{4}}} $\lambda$787     &  [$ -50, 50 $]       &  \nodata             &                   $ < 13.66 $ \\
\ensuremath{\mbox{\ion{S}{4}}} $\lambda$809     &  [$ -50, 50 $]       &  \nodata             &                   $ < 13.63 $ \\
\ensuremath{\mbox{\ion{S}{5}}} $\lambda$786     &  [$ -50, 50 $]       &  \nodata             &                   $ < 12.55 $ \\
\ensuremath{\mbox{\ion{S}{6}}} $\lambda$933     &  [$ -50, 50 $]       &  \nodata             &                   $ < 13.09 $ \\
\hline
\multicolumn{4}{c}{SDSSJ141307.39+091956.7, $z= 0.358374$ } \\
\hline
\ensuremath{\mbox{\ion{C}{2}}} $\lambda$903     &  [$ -893, -733 $]    &  \nodata             &                   $ < 12.87 $ \\
\ensuremath{\mbox{\ion{C}{3}}} $\lambda$977     &  [$ -1013, -753 $]   &  \nodata             &                   $ < 12.61 $ \\
\ensuremath{\mbox{\ion{Fe}{2}}} $\lambda$1133   &  [$ -1013, -733 $]   &  \nodata             &                   $ < 14.77 $ \\
\ensuremath{\mbox{\ion{H}{1}}} $\lambda$1025    &  [$ -1013, -733 $]   &  \nodata             &                   $ < 13.61 $ \\
\ensuremath{\mbox{\ion{H}{1}}} $\lambda$1215    &  [$ -1013, -733 $]   &  $ -847.1 \pm 21.5 $ &             $13.67 \pm 0.06 $ \\
\ensuremath{\mbox{\ion{H}{1}}}                  &                      &  $ -847.1 \pm 21.5 $ &             $13.67 \pm 0.06 $ \\
\ensuremath{\mbox{\ion{N}{1}}} $\lambda$1199    &  [$ -933, -773 $]    &  \nodata             &                   $ < 13.31 $ \\
\ensuremath{\mbox{\ion{N}{5}}} $\lambda$1238    &  [$ -1013, -733 $]   &  \nodata             &                   $ < 13.47 $ \\
\ensuremath{\mbox{\ion{O}{1}}} $\lambda$988     &  [$ -1013, -733 $]   &  \nodata             &                   $ < 13.83 $ \\
\ensuremath{\mbox{\ion{O}{6}}} $\lambda$1031    &  [$ -1013, -733 $]   &  \nodata             &                   $ < 13.41 $ \\
\ensuremath{\mbox{\ion{S}{2}}} $\lambda$1259    &  [$ -943, -733 $]    &  \nodata             &                   $ < 14.35 $ \\
\ensuremath{\mbox{\ion{S}{3}}} $\lambda$1012    &  [$ -1013, -733 $]   &  \nodata             &                   $ < 13.84 $ \\
\ensuremath{\mbox{\ion{S}{6}}} $\lambda$933     &  [$ -1013, -733 $]   &  \nodata             &                   $ < 12.88 $ \\
\ensuremath{\mbox{\ion{S}{3}}} $\lambda$1190    &  [$ -1013, -733 $]   &  \nodata             &                   $ < 14.21 $ \\
\ensuremath{\mbox{\ion{Si}{2}}} $\lambda$1260   &  [$ -1013, -733 $]   &  \nodata             &                   $ < 12.57 $ \\
\ensuremath{\mbox{\ion{Si}{3}}} $\lambda$1206   &  [$ -1013, -733 $]   &  \nodata             &                   $ < 12.35 $ \\
\hline
\multicolumn{4}{c}{SDSSJ125859.98+413128.2, $z= 0.279027$ } \\
\hline
\ensuremath{\mbox{\ion{C}{2}}} $\lambda$1036    &  [$ -878, -587 $]    &  \nodata             &                   $ < 13.59 $ \\
\ensuremath{\mbox{\ion{C}{3}}} $\lambda$977     &  [$ -878, -587 $]    &  \nodata             &                   $ < 12.83 $ \\
\ensuremath{\mbox{\ion{H}{1}}} $\lambda$1025    &  [$ -878, -587 $]    &  \nodata             &                   $ < 13.80 $ \\
\ensuremath{\mbox{\ion{H}{1}}} $\lambda$1215    &  [$ -878, -587 $]    &  $ -751.8 \pm 9.7 $  &             $13.70 \pm 0.05 $ \\
\ensuremath{\mbox{\ion{H}{1}}}                  &                      &  $ -751.8 \pm 9.7 $  &             $13.70 \pm 0.05 $ \\
\ensuremath{\mbox{\ion{N}{2}}} $\lambda$1083    &  [$ -878, -587 $]    &  \nodata             &                   $ < 13.62 $ \\
\ensuremath{\mbox{\ion{N}{3}}} $\lambda$989     &  [$ -878, -587 $]    &  \nodata             &                   $ < 13.62 $ \\
\ensuremath{\mbox{\ion{N}{5}}} $\lambda$1238    &  [$ -878, -587 $]    &  \nodata             &                   $ < 13.69 $ \\
\ensuremath{\mbox{\ion{O}{1}}} $\lambda$1302    &  [$ -878, -587 $]    &  \nodata             &                   $ < 14.03 $ \\
\ensuremath{\mbox{\ion{O}{6}}} $\lambda$1031    &  [$ -878, -587 $]    &  \nodata             &                   $ < 13.53 $ \\
\ensuremath{\mbox{\ion{S}{2}}} $\lambda$1250    &  [$ -878, -587 $]    &  \nodata             &                   $ < 15.13 $ \\
\ensuremath{\mbox{\ion{S}{6}}} $\lambda$944     &  [$ -878, -587 $]    &  \nodata             &                   $ < 13.47 $ \\
\ensuremath{\mbox{\ion{S}{3}}} $\lambda$1190    &  [$ -878, -587 $]    &  \nodata             &                   $ < 14.13 $ \\
\ensuremath{\mbox{\ion{Si}{2}}} $\lambda$1193   &  [$ -878, -587 $]    &  \nodata             &                   $ < 12.75 $ \\
\hline
\multicolumn{4}{c}{SDSSJ125222.93+291327.2, $z= 0.598738$ } \\
\hline
\ensuremath{\mbox{\ion{C}{2}}} $\lambda$903   &  [$  -569, -469 $]    &  \nodata             &                    $ < 13.38 $ \\
\ensuremath{\mbox{\ion{Fe}{2}}} $\lambda$1144   &  [$ -739, -189 $]    &  \nodata             &                   $ < 13.96 $ \\
\ensuremath{\mbox{\ion{Fe}{3}}} $\lambda$1122   &  [$ -739, -189 $]    &  \nodata             &                   $ < 14.09 $ \\
\ensuremath{\mbox{\ion{H}{1}}} $\lambda$1025    &  [$ -739, -189 $]    &  $ -498.4 \pm 50.2 $ &                   $ > 15.20 $ \\
\ensuremath{\mbox{\ion{N}{2}}} $\lambda$1083    &  [$ -739, -189 $]    &  \nodata             &                   $ < 13.92 $ \\
\ensuremath{\mbox{\ion{N}{5}}} $\lambda$1242    &  [$ -739, -189 $]    &  \nodata             &                   $ < 13.87 $ \\
\ensuremath{\mbox{\ion{O}{1}}} $\lambda$1355    &  [$ -739, -189 $]    &  \nodata             &                   $ < 18.49 $ \\
\ensuremath{\mbox{\ion{O}{2}}} $\lambda$834   &  [$  -569, -469 $]    &  \nodata             &                   $ < 13.62 $ \\
\ensuremath{\mbox{\ion{O}{3}}} $\lambda$832   &  [$  -569, -469 $]    &  \nodata             &                   $ < 13.73 $ \\
\ensuremath{\mbox{\ion{O}{6}}} $\lambda$1037    &  [$ -919, -189 $]    &  \nodata             &                   $ < 14.63 $ \\
\ensuremath{\mbox{\ion{O}{6}}} $\lambda$1031    &  [$ -919, -189 $]    &  $ -569.9 \pm 87.3 $ &   $14.41 \,^{+0.16}_{-0.25} $ \\
\ensuremath{\mbox{\ion{O}{6}}}                  &                      &  $ -569.9 \pm 87.3 $ &   $14.41 \,^{+0.16}_{-0.25} $ \\
\ensuremath{\mbox{\ion{S}{2}}} $\lambda$1259    &  [$ -739, -189 $]    &  \nodata             &                   $ < 14.55 $ \\
\ensuremath{\mbox{\ion{S}{3}}} $\lambda$1012    &  [$ -739, -189 $]    &  \nodata             &                   $ < 15.08 $ \\
\ensuremath{\mbox{\ion{S}{4}}} $\lambda$1062    &  [$ -739, -189 $]    &  \nodata             &                   $ < 14.29 $ \\
\ensuremath{\mbox{\ion{Si}{2}}} $\lambda$1260   &  [$ -739, -189 $]    &  \nodata             &                   $ < 12.67 $ \\
\ensuremath{\mbox{\ion{Si}{3}}} $\lambda$1206   &  [$ -739, -189 $]    &  $ -515.3 \pm 53.1 $ &             $(>)13.09: $ \\
\ensuremath{\mbox{\ion{Si}{4}}} $\lambda$1393   &  [$ -739, -189 $]    &  \nodata             &                   $ < 13.11 $ \\
\hline
\multicolumn{4}{c}{SDSSJ171651.46+302649.0, $z= 0.400437$ } \\
\hline
\ensuremath{\mbox{\ion{Fe}{2}}} $\lambda$2382    &  [$ -210, -177 $]    &  \nodata            &                   $ < 11.91 $ \\
\ensuremath{\mbox{\ion{H}{1}}} Break            & \nodata              & \nodata              &               $16.90  \pm 0.20 $ \\
\ensuremath{\mbox{\ion{Mg}{2}}} $\lambda$2796   &  [$ -210, -177 $]    &  $ -196.1 \pm 18.7 $ &             $12.01 \pm 0.08 $ \\
\ensuremath{\mbox{\ion{Mg}{2}}} $\lambda$2803   &  [$ -209, -185 $]    &  $ -200.3 \pm 18.9 $ &             $11.85 \pm 0.19 $ \\
\ensuremath{\mbox{\ion{Mg}{2}}}                 &                      &  $ -198.2 \pm 13.3 $ &             $11.94 \pm 0.09 $ \\
\hline
\multicolumn{4}{c}{SDSSJ111436.59+403739.1, $z= 0.609746$ } \\
\hline
\ensuremath{\mbox{\ion{C}{2}}} $\lambda$1036    &  [$ -200, 200 $]     &  \nodata             &                   $ < 14.61 $ \\
\ensuremath{\mbox{\ion{Fe}{2}}} $\lambda$1608   &  [$ -200, 200 $]     &  \nodata             &                   $ < 14.12 $ \\
\ensuremath{\mbox{\ion{Fe}{3}}} $\lambda$1122   &  [$ -200, 200 $]     &  \nodata             &                   $ < 14.58 $ \\
\ensuremath{\mbox{\ion{H}{1}}} $\lambda$1215    &  [$ -200, 200 $]     &  \nodata             &                   $ < 13.79 $ \\
\ensuremath{\mbox{\ion{N}{2}}} $\lambda$1083    &  [$ -200, 200 $]     &  \nodata             &                   $ < 14.36 $ \\
\ensuremath{\mbox{\ion{N}{5}}} $\lambda$1242    &  [$ -200, 200 $]     &  \nodata             &                   $ < 14.32 $ \\
\ensuremath{\mbox{\ion{O}{1}}} $\lambda$1302    &  [$ -200, 200 $]     &  \nodata             &                   $ < 14.19 $ \\
\ensuremath{\mbox{\ion{O}{6}}} $\lambda$1031    &  [$ -200, 200 $]     &  \nodata             &                   $ < 14.63 $ \\
\ensuremath{\mbox{\ion{S}{2}}} $\lambda$1259    &  [$ -200, 200 $]     &  \nodata             &                   $ < 14.99 $ \\
\ensuremath{\mbox{\ion{S}{3}}} $\lambda$1012    &  [$ -200, 200 $]     &  \nodata             &                   $ < 15.51 $ \\
\ensuremath{\mbox{\ion{S}{4}}} $\lambda$1062    &  [$ -200, 200 $]     &  \nodata             &                   $ < 14.81 $ \\
\ensuremath{\mbox{\ion{S}{3}}} $\lambda$1190    &  [$ -200, 200 $]     &  \nodata             &                   $ < 14.94 $ \\
\ensuremath{\mbox{\ion{Si}{2}}} $\lambda$1260   &  [$ -200, 200 $]     &  \nodata             &                   $ < 13.09 $ \\
\ensuremath{\mbox{\ion{Si}{3}}} $\lambda$1206   &  [$ -200, 200 $]     &  \nodata             &                   $ < 13.07 $ \\
\ensuremath{\mbox{\ion{Si}{4}}} $\lambda$1393   &  [$ -200, 200 $]     &  \nodata             &                   $ < 13.44 $ \\
\hline
\multicolumn{4}{c}{SDSSJ110632.58+351012.8, $z= 0.470349$ } \\
\hline
\ensuremath{\mbox{\ion{C}{2}}} $\lambda$1036    &  [$ -50, 50 $]       &  \nodata             &                   $ < 13.46 $ \\
\ensuremath{\mbox{\ion{C}{3}}} $\lambda$977     &  [$ -50, 50 $]       &  \nodata             &                   $ < 12.72 $ \\
\ensuremath{\mbox{\ion{Fe}{2}}} $\lambda$1144   &  [$ -50, 50 $]       &  \nodata             &                   $ < 13.97 $ \\
\ensuremath{\mbox{\ion{Fe}{3}}} $\lambda$1122   &  [$ -50, 50 $]       &  \nodata             &                   $ < 14.13 $ \\
\ensuremath{\mbox{\ion{H}{1}}} $\lambda$1025    &  [$ -50, 50 $]       &  \nodata             &                   $ < 13.70 $ \\
\ensuremath{\mbox{\ion{N}{1}}} $\lambda$1134    &  [$ -20, 20 $]       &  \nodata             &                   $ < 14.01 $ \\
\ensuremath{\mbox{\ion{N}{2}}} $\lambda$1083    &  [$ -50, 50 $]       &  \nodata             &                   $ < 13.68 $ \\
\ensuremath{\mbox{\ion{O}{1}}} $\lambda$971     &  [$ -50, 50 $]       &  \nodata             &                   $ < 14.54 $ \\
\ensuremath{\mbox{\ion{O}{6}}} $\lambda$1031    &  [$ -50, 50 $]       &  \nodata             &                   $ < 13.65 $ \\
\ensuremath{\mbox{\ion{S}{3}}} $\lambda$1012    &  [$ -50, 50 $]       &  \nodata             &                   $ < 13.97 $ \\
\ensuremath{\mbox{\ion{S}{6}}} $\lambda$944     &  [$ -50, 50 $]       &  \nodata             &                   $ < 13.44 $ \\
\ensuremath{\mbox{\ion{S}{3}}} $\lambda$1190    &  [$ -50, 50 $]       &  \nodata             &                   $ < 14.73 $ \\
\ensuremath{\mbox{\ion{Si}{2}}} $\lambda$1190   &  [$ -50, 50 $]       &  \nodata             &                   $ < 13.73 $ \\
\hline
\multicolumn{4}{c}{SDSSJ081524.08+273621.2, $z= 0.504332$ } \\
\hline
\ensuremath{\mbox{\ion{C}{2}}} $\lambda$903     &  [$ -50, 50 $]       &  \nodata             &                   $ < 13.37 $ \\
\ensuremath{\mbox{\ion{C}{3}}} $\lambda$977     &  [$ -100, 100 $]     &  \nodata             &                   $ < 13.25 $ \\
\ensuremath{\mbox{\ion{C}{4}}} $\lambda$1548    &  [$ -100, 100 $]     &  \nodata             &                   $ < 14.56 $ \\
\ensuremath{\mbox{\ion{Fe}{2}}} $\lambda$1144   &  [$ -100, 100 $]     &  \nodata             &                   $ < 14.42 $ \\
\ensuremath{\mbox{\ion{Fe}{3}}} $\lambda$1122   &  [$ -100, 100 $]     &  \nodata             &                   $ < 14.58 $ \\
\ensuremath{\mbox{\ion{H}{1}}} $\lambda$1215    &  [$ -100, 100 $]     &  \nodata             &                   $ < 13.81 $ \\
\ensuremath{\mbox{\ion{N}{1}}} $\lambda$1199    &  [$ -60, 60 $]       &  \nodata             &                   $ < 14.16 $ \\
\ensuremath{\mbox{\ion{N}{2}}} $\lambda$1083    &  [$ -100, 100 $]     &  \nodata             &                   $ < 14.26 $ \\
\ensuremath{\mbox{\ion{N}{3}}} $\lambda$989     &  [$ -100, 100 $]     &  \nodata             &                   $ < 14.01 $ \\
\ensuremath{\mbox{\ion{N}{5}}} $\lambda$1238    &  [$ -100, 100 $]     &  \nodata             &                   $ < 14.25 $ \\
\ensuremath{\mbox{\ion{O}{1}}} $\lambda$988     &  [$ -100, 100 $]     &  \nodata             &                   $ < 14.43 $ \\
\ensuremath{\mbox{\ion{O}{3}}} $\lambda$702     &  [$ -100, 100 $]     &  \nodata             &                   $ < 15.12 $ \\
\ensuremath{\mbox{\ion{O}{6}}} $\lambda$1031    &  [$ -100, 100 $]     &  \nodata             &                   $ < 14.08 $ \\
\ensuremath{\mbox{\ion{S}{2}}} $\lambda$1259    &  [$ -100, 100 $]     &  \nodata             &                   $ < 15.26 $ \\
\ensuremath{\mbox{\ion{S}{3}}} $\lambda$677     &  [$ -100, 100 $]     &  \nodata             &                   $ < 14.12 $ \\
\ensuremath{\mbox{\ion{S}{4}}} $\lambda$744     &  [$ -100, 100 $]     &  \nodata             &                   $ < 14.04 $ \\
\ensuremath{\mbox{\ion{S}{6}}} $\lambda$933     &  [$ -100, 100 $]     &  \nodata             &                   $ < 13.46 $ \\
\ensuremath{\mbox{\ion{S}{3}}} $\lambda$1190    &  [$ -100, 100 $]     &  \nodata             &                   $ < 14.97 $ \\
\ensuremath{\mbox{\ion{Si}{2}}} $\lambda$1260   &  [$ -100, 100 $]     &  \nodata             &                   $ < 13.40 $ \\
\ensuremath{\mbox{\ion{Si}{3}}} $\lambda$1206   &  [$ -100, 100 $]     &  \nodata             &                   $ < 13.22 $ \\
\ensuremath{\mbox{\ion{Si}{4}}} $\lambda$1393   &  [$ -100, 100 $]     &  \nodata             &                   $ < 14.90 $ \\
\hline
\multicolumn{4}{c}{SDSSJ125101.95+302501.7, $z= 0.513233$ } \\
\hline
\ensuremath{\mbox{\ion{C}{2}}} $\lambda$903     &  [$ -50, 50 $]       &  \nodata             &                   $ < 12.98 $ \\
\ensuremath{\mbox{\ion{H}{1}}} $\lambda$949     &  [$ -50, 50 $]       &  \nodata             &                   $ < 14.57 $ \\
\ensuremath{\mbox{\ion{N}{4}}} $\lambda$765     &  [$ -50, 50 $]       &  \nodata             &                   $ < 12.83 $ \\
\ensuremath{\mbox{\ion{O}{1}}} $\lambda$936     &  [$ -50, 50 $]       &  \nodata             &                   $ < 14.98 $ \\
\ensuremath{\mbox{\ion{O}{2}}} $\lambda$834     &  [$ -50, 50 $]       &  \nodata             &                   $ < 13.28 $ \\
\ensuremath{\mbox{\ion{O}{4}}} $\lambda$787     &  [$ -50, 50 $]       &  \nodata             &                   $ < 13.53 $ \\
\hline
\multicolumn{4}{c}{SDSSJ122516.86+121750.1, $z= 0.298015$ } \\
\hline
\ensuremath{\mbox{\ion{C}{2}}} $\lambda$903     &  [$ -50, 50 $]       &  \nodata             &                   $ < 13.20 $ \\
\ensuremath{\mbox{\ion{H}{1}}} $\lambda$972     &  [$ -50, 50 $]       &  \nodata             &                   $ < 14.12 $ \\
\ensuremath{\mbox{\ion{N}{2}}} $\lambda$1083    &  [$ -50, 50 $]       &  \nodata             &                   $ < 13.50 $ \\
\ensuremath{\mbox{\ion{O}{1}}} $\lambda$988     &  [$ -50, 50 $]       &  \nodata             &                   $ < 14.05 $ \\
\ensuremath{\mbox{\ion{O}{6}}} $\lambda$1031    &  [$ -50, 50 $]       &  \nodata             &                   $ < 13.46 $ \\
\ensuremath{\mbox{\ion{S}{3}}} $\lambda$1012    &  [$ -50, 50 $]       &  \nodata             &                   $ < 13.93 $ \\
\ensuremath{\mbox{\ion{S}{6}}} $\lambda$933     &  [$ -50, 50 $]       &  \nodata             &                   $ < 13.18 $ \\
\ensuremath{\mbox{\ion{Si}{2}}} $\lambda$1020   &  [$ -50, 50 $]       &  \nodata             &                   $ < 14.34 $ \\
\hline
\multicolumn{4}{c}{SDSSJ141540.01+163336.4, $z= 0.482620$ } \\
\hline
\ensuremath{\mbox{\ion{C}{2}}} $\lambda$903     &  [$ -235, -135 $]    &  \nodata             &                   $ < 13.03 $ \\
\ensuremath{\mbox{\ion{C}{2}}} $\lambda$903     &  [$ -235, -135 $]    &  $ -183.0 \pm 26.2 $ &   $12.89 \,^{+0.13}_{-0.19} $ \\
\ensuremath{\mbox{\ion{C}{2}}}                  &                      &  $ -183.0 \pm 26.2 $ &   $12.89 \,^{+0.13}_{-0.19} $ \\
\ensuremath{\mbox{\ion{C}{3}}} $\lambda$977     &  [$ -285, -75 $]     &  $ -149.3 \pm 25.4 $ &                   $ > 13.85 $ \\
\ensuremath{\mbox{\ion{H}{1}}} $\lambda$917     &  [$ -285, -75 $]     &  $ -182.8 \pm 26.0 $ &             $16.06 \pm 0.08 $ \\
\ensuremath{\mbox{\ion{H}{1}}} $\lambda$918     &  [$ -285, -75 $]     &  $ -181.9 \pm 25.3 $ &             $16.06 \pm 0.06 $ \\
\ensuremath{\mbox{\ion{H}{1}}} $\lambda$919     &  [$ -285, -75 $]     &  $ -173.8 \pm 24.7 $ &             $16.13 \pm 0.04 $ \\
\ensuremath{\mbox{\ion{H}{1}}} $\lambda$920     &  [$ -285, -75 $]     &  $ -183.0 \pm 24.6 $ &             $16.05 \pm 0.04 $ \\
\ensuremath{\mbox{\ion{H}{1}}} $\lambda$923     &  [$ -285, -75 $]     &  $ -182.6 \pm 24.4 $ &             $16.05 \pm 0.03 $ \\
\ensuremath{\mbox{\ion{H}{1}}}                  &                      &  $ -180.8 \pm 11.2 $ &             $16.07 \pm 0.02 $ \\
\ensuremath{\mbox{\ion{O}{1}}} $\lambda$971     &  [$ -285, -75 $]     &  \nodata             &                   $ < 14.56 $ \\
\ensuremath{\mbox{\ion{O}{2}}} $\lambda$834     &  [$ -285, -75 $]     &  $ -159.4 \pm 28.7 $ &             $13.68 \pm 0.13 $ \\
\ensuremath{\mbox{\ion{O}{3}}} $\lambda$832     &  [$ -285, -75 $]     &  $ -156.7 \pm 24.5 $ &             $14.44 \pm 0.03 $ \\
\ensuremath{\mbox{\ion{S}{5}}} $\lambda$786     &  [$ -285, -75 $]     &  \nodata             &                   $ < 12.44 $ \\
\ensuremath{\mbox{\ion{S}{6}}} $\lambda$933     &  [$ -285, -75 $]     &  \nodata             &                   $ < 12.80 $ \\
\hline
\multicolumn{4}{c}{SDSSJ022612.22+001439.9, $z= 0.473040$ } \\
\hline
\ensuremath{\mbox{\ion{C}{2}}} $\lambda$903     &  [$ -45, 54 $]       &  \nodata             &                   $ < 12.97 $ \\
\ensuremath{\mbox{\ion{C}{3}}} $\lambda$977     &  [$ -45, 54 $]       &  \nodata             &                   $ < 12.67 $ \\
\ensuremath{\mbox{\ion{H}{1}}} $\lambda$972     &  [$ -45, 54 $]       &  \nodata             &                   $ < 14.08 $ \\
\ensuremath{\mbox{\ion{H}{1}}} $\lambda$1025    &  [$ -45, 54 $]       &  $ 9.8 \pm 17.8 $    &             $13.97 \pm 0.15 $ \\
\ensuremath{\mbox{\ion{H}{1}}}                  &                      &  $ 9.8 \pm 17.8 $    &             $13.97 \pm 0.15 $ \\
\ensuremath{\mbox{\ion{N}{2}}} $\lambda$1083    &  [$ -45, 54 $]       &  \nodata             &                   $ < 13.75 $ \\
\ensuremath{\mbox{\ion{N}{3}}} $\lambda$989     &  [$ -45, 54 $]       &  \nodata             &                   $ < 13.44 $ \\
\ensuremath{\mbox{\ion{O}{1}}} $\lambda$988     &  [$ -45, 54 $]       &  \nodata             &                   $ < 13.86 $ \\
\ensuremath{\mbox{\ion{O}{2}}} $\lambda$834     &  [$ -45, 54 $]       &  \nodata             &                   $ < 13.33 $ \\
\ensuremath{\mbox{\ion{O}{4}}} $\lambda$787     &  [$ -45, 54 $]       &  \nodata             &                   $ < 13.88 $ \\
\ensuremath{\mbox{\ion{O}{6}}} $\lambda$1031    &  [$ -45, 54 $]       &  \nodata             &                   $ < 13.42 $ \\
\ensuremath{\mbox{\ion{S}{3}}} $\lambda$1012    &  [$ -45, 54 $]       &  \nodata             &                   $ < 13.94 $ \\
\ensuremath{\mbox{\ion{S}{4}}} $\lambda$809     &  [$ -45, 54 $]       &  \nodata             &                   $ < 13.58 $ \\
\ensuremath{\mbox{\ion{S}{5}}} $\lambda$786     &  [$ -45, 54 $]       &  \nodata             &                   $ < 12.79 $ \\
\ensuremath{\mbox{\ion{S}{6}}} $\lambda$933     &  [$ -45, 54 $]       &  \nodata             &                   $ < 12.86 $ \\
\ensuremath{\mbox{\ion{Si}{2}}} $\lambda$1193   &  [$ -45, 54 $]       &  \nodata             &                   $ < 13.04 $ \\
\ensuremath{\mbox{\ion{Si}{3}}} $\lambda$1206   &  [$ -45, 54 $]       &  \nodata             &                   $ < 12.82 $ \\
\hline
\multicolumn{4}{c}{SDSSJ075217.92+273835.6, $z= 0.581165$ } \\
\hline
\ensuremath{\mbox{\ion{C}{2}}} $\lambda$903     &  [$ -50, 50 $]       &  \nodata             &                   $ < 13.46 $ \\
\ensuremath{\mbox{\ion{H}{1}}} $\lambda$1025    &  [$ -100, 100 $]     &  \nodata             &                   $ < 14.41 $ \\
\ensuremath{\mbox{\ion{N}{2}}} $\lambda$1083    &  [$ -100, 100 $]     &  \nodata             &                   $ < 14.42 $ \\
\ensuremath{\mbox{\ion{N}{3}}} $\lambda$989     &  [$ -100, 100 $]     &  \nodata             &                   $ < 14.26 $ \\
\ensuremath{\mbox{\ion{O}{1}}} $\lambda$988     &  [$ -100, 100 $]     &  \nodata             &                   $ < 14.64 $ \\
\ensuremath{\mbox{\ion{O}{2}}} $\lambda$834     &  [$ -100, 100 $]     &  \nodata             &                   $ < 14.09 $ \\
\ensuremath{\mbox{\ion{O}{3}}} $\lambda$832     &  [$ -100, 100 $]     &  \nodata             &                   $ < 14.16 $ \\
\ensuremath{\mbox{\ion{O}{6}}} $\lambda$1031    &  [$ -100, 100 $]     &  \nodata             &                   $ < 14.26 $ \\
\ensuremath{\mbox{\ion{S}{3}}} $\lambda$724     &  [$ -100, 100 $]     &  \nodata             &                   $ < 13.85 $ \\
\ensuremath{\mbox{\ion{S}{4}}} $\lambda$1062    &  [$ -100, 100 $]     &  \nodata             &                   $ < 14.73 $ \\
\ensuremath{\mbox{\ion{S}{6}}} $\lambda$933     &  [$ -100, 100 $]     &  \nodata             &                   $ < 13.65 $ \\
\ensuremath{\mbox{\ion{Si}{2}}} $\lambda$1020   &  [$ -100, 100 $]     &  \nodata             &                   $ < 15.10 $ \\
\hline
\multicolumn{4}{c}{SDSSJ150522.44+294626.2, $z= 0.403131$ } \\
\hline
\ensuremath{\mbox{\ion{C}{2}}} $\lambda$903     &  [$ -50, 50 $]       &  \nodata             &                   $ < 12.69 $ \\
\ensuremath{\mbox{\ion{C}{3}}} $\lambda$977     &  [$ -50, 50 $]       &  \nodata             &                   $ < 12.58 $ \\
\ensuremath{\mbox{\ion{H}{1}}} $\lambda$972     &  [$ -50, 50 $]       &  \nodata             &                   $ < 13.85 $ \\
\ensuremath{\mbox{\ion{N}{3}}} $\lambda$989     &  [$ -50, 50 $]       &  \nodata             &                   $ < 13.24 $ \\
\ensuremath{\mbox{\ion{O}{1}}} $\lambda$988     &  [$ -50, 50 $]       &  \nodata             &                   $ < 13.65 $ \\
\ensuremath{\mbox{\ion{O}{2}}} $\lambda$834     &  [$ -50, 50 $]       &  \nodata             &                   $ < 13.30 $ \\
\ensuremath{\mbox{\ion{O}{3}}} $\lambda$832     &  [$ -50, 50 $]       &  \nodata             &                   $ < 13.41 $ \\
\ensuremath{\mbox{\ion{O}{6}}} $\lambda$1031    &  [$ -50, 50 $]       &  \nodata             &                   $ < 13.38 $ \\
\ensuremath{\mbox{\ion{S}{6}}} $\lambda$933     &  [$ -50, 50 $]       &  \nodata             &                   $ < 12.74 $ \\
\ensuremath{\mbox{\ion{Si}{2}}} $\lambda$1020   &  [$ -50, 50 $]       &  \nodata             &                   $ < 14.13 $ \\
\hline
\multicolumn{4}{c}{SDSSJ132457.98+271742.6, $z= 0.447836$ } \\
\hline
\ensuremath{\mbox{\ion{C}{3}}} $\lambda$977     &  [$ -50, 50 $]       &  \nodata             &                   $ < 12.83 $ \\
\ensuremath{\mbox{\ion{H}{1}}} $\lambda$972     &  [$ -50, 50 $]       &  \nodata             &                   $ < 14.26 $ \\
\ensuremath{\mbox{\ion{N}{3}}} $\lambda$989     &  [$ -50, 50 $]       &  \nodata             &                   $ < 13.73 $ \\
\ensuremath{\mbox{\ion{O}{1}}} $\lambda$988     &  [$ -50, 50 $]       &  \nodata             &                   $ < 14.09 $ \\
\ensuremath{\mbox{\ion{O}{2}}} $\lambda$834     &  [$ -50, 50 $]       &  \nodata             &                   $ < 13.84 $ \\
\ensuremath{\mbox{\ion{O}{4}}} $\lambda$787     &  [$ -50, 50 $]       &  \nodata             &                   $ < 14.47 $ \\
\ensuremath{\mbox{\ion{S}{3}}} $\lambda$1012    &  [$ -50, 50 $]       &  \nodata             &                   $ < 14.30 $ \\
\ensuremath{\mbox{\ion{S}{4}}} $\lambda$809     &  [$ -50, 50 $]       &  \nodata             &                   $ < 14.13 $ \\
\ensuremath{\mbox{\ion{S}{5}}} $\lambda$786     &  [$ -50, 50 $]       &  \nodata             &                   $ < 13.31 $ \\
\ensuremath{\mbox{\ion{S}{6}}} $\lambda$933     &  [$ -50, 50 $]       &  \nodata             &                   $ < 13.09 $ \\
\hline
\multicolumn{4}{c}{SDSSJ104918.08+021814.2, $z= 0.515244$ } \\
\hline
\ensuremath{\mbox{\ion{C}{2}}} $\lambda$903     &  [$ 79, 179 $]       &  \nodata             &                   $ < 13.46 $ \\
\ensuremath{\mbox{\ion{C}{3}}} $\lambda$977     &  [$ 9, 229 $]        &  \nodata             &                   $ < 13.26 $ \\
\ensuremath{\mbox{\ion{H}{1}}} $\lambda$926     &  [$ 9, 229 $]        &  \nodata             &                   $ < 15.73 $ \\
\ensuremath{\mbox{\ion{H}{1}}} $\lambda$937     &  [$ 9, 229 $]        &  $ 120.8 \pm 51.6 $  &   $15.72 \,^{+0.19}_{-0.35} $ \\
\ensuremath{\mbox{\ion{H}{1}}}                  &                      &  $ 120.8 \pm 51.6 $  &   $15.72 \,^{+0.19}_{-0.35} $ \\
\ensuremath{\mbox{\ion{N}{2}}} $\lambda$1083    &  [$ 9, 229 $]        &  \nodata             &                   $ < 14.24 $ \\
\ensuremath{\mbox{\ion{N}{3}}} $\lambda$989     &  [$ 9, 229 $]        &  \nodata             &                   $ < 14.11 $ \\
\ensuremath{\mbox{\ion{N}{4}}} $\lambda$765     &  [$ 9, 229 $]        &  \nodata             &                   $ < 13.62 $ \\
\ensuremath{\mbox{\ion{O}{1}}} $\lambda$971     &  [$ 9, 229 $]        &  \nodata             &                   $ < 15.09 $ \\
\ensuremath{\mbox{\ion{O}{2}}} $\lambda$834     &  [$ 9, 229 $]        &  \nodata             &                   $ < 14.11 $ \\
\ensuremath{\mbox{\ion{O}{3}}} $\lambda$832     &  [$ 9, 229 $]        &  \nodata             &                   $ < 14.22 $ \\
\ensuremath{\mbox{\ion{O}{6}}} $\lambda$1031    &  [$ 9, 229 $]        &  \nodata             &                   $ < 14.08 $ \\
\ensuremath{\mbox{\ion{S}{3}}} $\lambda$1012    &  [$ 9, 229 $]        &  \nodata             &                   $ < 14.51 $ \\
\ensuremath{\mbox{\ion{S}{4}}} $\lambda$809     &  [$ 9, 229 $]        &  \nodata             &                   $ < 14.21 $ \\
\ensuremath{\mbox{\ion{S}{5}}} $\lambda$786     &  [$ 9, 229 $]        &  \nodata             &                   $ < 13.26 $ \\
\ensuremath{\mbox{\ion{S}{6}}} $\lambda$933     &  [$ 9, 229 $]        &  \nodata             &                   $ < 13.58 $ \\
\ensuremath{\mbox{\ion{Si}{2}}} $\lambda$1020   &  [$ 9, 229 $]        &  \nodata             &                   $ < 14.97 $ \\
\hline
\multicolumn{4}{c}{SDSSJ110405.15+314244.1, $z= 0.365605$ } \\
\hline
\ensuremath{\mbox{\ion{C}{2}}} $\lambda$903     &  [$ 63, 173 $]       &  \nodata             &                   $ < 13.31 $ \\
\ensuremath{\mbox{\ion{H}{1}}} $\lambda$1025    &  [$ 63, 213 $]       &  \nodata             &                   $ < 13.65 $ \\
\ensuremath{\mbox{\ion{H}{1}}} $\lambda$1215    &  [$ 63, 213 $]       &  $ 114.1 \pm 21.3 $  &             $13.58 \pm 0.10 $ \\
\ensuremath{\mbox{\ion{H}{1}}}                  &                      &  $ 114.1 \pm 21.3 $  &             $13.58 \pm 0.10 $ \\
\ensuremath{\mbox{\ion{N}{1}}} $\lambda$1199    &  [$ 63, 183 $]       &  \nodata             &                   $ < 13.48 $ \\
\ensuremath{\mbox{\ion{N}{3}}} $\lambda$989     &  [$ 63, 213 $]       &  \nodata             &                   $ < 13.47 $ \\
\ensuremath{\mbox{\ion{N}{5}}} $\lambda$1238    &  [$ 63, 213 $]       &  \nodata             &                   $ < 13.45 $ \\
\ensuremath{\mbox{\ion{O}{1}}} $\lambda$988     &  [$ 63, 213 $]       &  \nodata             &                   $ < 13.88 $ \\
\ensuremath{\mbox{\ion{O}{6}}} $\lambda$1031    &  [$ 63, 213 $]       &  \nodata             &                   $ < 13.45 $ \\
\ensuremath{\mbox{\ion{S}{2}}} $\lambda$1259    &  [$ 63, 213 $]       &  \nodata             &                   $ < 14.33 $ \\
\ensuremath{\mbox{\ion{S}{3}}} $\lambda$1012    &  [$ 63, 213 $]       &  \nodata             &                   $ < 13.94 $ \\
\ensuremath{\mbox{\ion{S}{4}}} $\lambda$1062    &  [$ 63, 213 $]       &  \nodata             &                   $ < 13.66 $ \\
\ensuremath{\mbox{\ion{S}{3}}} $\lambda$1190    &  [$ 63, 213 $]       &  \nodata             &                   $ < 14.27 $ \\
\ensuremath{\mbox{\ion{Si}{2}}} $\lambda$1193   &  [$ 63, 213 $]       &  \nodata             &                   $ < 12.87 $ \\
\hline
\enddata
\tablecomments{$\langle v \rangle$ is the average velocity of the absorption line calculated as the first moment with respect to optical depth. Upper limits ($<$) are non-detections quoted at the 2$\sigma$ level. Column densities preceded by $>$ are lower limits owing to saturation in the absorption. If a column density is followed by a : this value is uncertain. For a given atom or ion with more than one transition, we list in the row with no wavelength information the adopted weighted average column densities and velocities. All ions are measured using the AODM unless otherwise labelled. \MgII\ is not detected in the SDSS spectra for the absorbers with no other ground-based observations.}
\end{deluxetable}

%% file: coveringfactor_ssfr_LLS.tex
\begin{deluxetable*}{lccccc}
\tablecaption{\HI\ Covering Factors by Sample at LLS Limit\label{tab:coveringfactorssfr_lls}}
\tablehead{\colhead{$\rho/\Rvir$} & \colhead{$\langle\rho/\Rvir\rangle$} & \colhead{$f_c$\tablenotemark{a}} & \colhead{95\% C.I.} & \colhead{Sightlines\tablenotemark{b}} & \colhead{Detections\tablenotemark{c}}}
\startdata
\hline
\multicolumn{6}{c}{RDR} \\
\hline
$[0.0,0.5]$ & 0.25 & $0.24^{+0.14}_{-0.11}$ & $[0.06,0.52]$ & 10 & 2 \\
$(0.5,1.0]$ & 0.68 & $0.09^{+0.14}_{-0.07}$ & $[0.00,0.41]$ & 6 & 0 \\
$(1.0,1.5]$ & 1.20 & $0.13^{+0.18}_{-0.10}$ & $[0.01,0.52]$ & 4 & 0 \\
\hline
\multicolumn{6}{c}{Star-Forming Galaxies} \\
\hline
$[0.0,0.5]$ & 0.30 & $0.37^{+0.10}_{-0.09}$ & $[0.20,0.57]$ & 22 & 8 \\
$(0.5,1.0]$ & 0.64 & $0.11^{+0.15}_{-0.08}$ & $[0.00,0.46]$ & 5 & 0 \\
\hline
\multicolumn{6}{c}{Quiescent Galaxies} \\
\hline
$[0.0,0.5]$ & 0.20 & $0.27^{+0.11}_{-0.10}$ & $[0.10,0.50]$ & 16 & 4 \\
$(0.5,1.0]$ & 0.57 & $0.29^{+0.31}_{-0.21}$ & $[0.01,0.84]$ & 1 & 0 \\
\enddata
\tablecomments{The limiting \HI\ value for the covering factor calculation is \logNHI\ $\ge$ 17.2 [\column]. The star-forming and quiescent galaxy samples are from the \citetalias{prochaska2017} sample.}
\tablenotetext{a}{The error bars on the covering factor represent the 68\% confidence interval.}
\tablenotetext{b}{Number of sightlines considered in covering factor calculation.}
\tablenotetext{c}{Number of absorbers detected above the limiting \HI\ value.}
\end{deluxetable*}

%% file: coveringfactor_stellar_LLS.tex
\begin{deluxetable*}{lccccc}
\tablecaption{\HI\ Covering Factors by Stellar Mass at LLS Limit\label{tab:coveringfactorstellar_lls}}
\tablehead{\colhead{$\rho/\Rvir$} & \colhead{$\langle\rho/\Rvir\rangle$} & \colhead{$f_c$\tablenotemark{a}} & \colhead{95\% C.I.} & \colhead{Sightlines\tablenotemark{b}} & \colhead{Detections\tablenotemark{c}}}
\startdata
\hline
\multicolumn{6}{c}{$9.0 \le \log M_{\star} < 10.5$} \\
\hline
$[0.0,0.5]$ & 0.30 & $0.37 \pm 0.12$ & $[0.16,0.62]$ & 14 & 5 \\
$(0.5,1.0]$ & 0.64 & $0.11^{+0.15}_{-0.08}$ & $[0.00,0.46]$ & 5 & 0 \\
\hline
\multicolumn{6}{c}{$10.5 \le \log M_{\star} < 11.3$} \\
\hline
$[0.0,0.5]$ & 0.24 & $0.42^{+0.10}_{-0.09}$ & $[0.24,0.61]$ & 24 & 10 \\
$(0.5,1.0]$ & 0.60 & $0.16^{+0.21}_{-0.12}$ & $[0.01,0.60]$ & 3 & 0 \\
$(1.0,1.5]$ & 1.24 & $0.16^{+0.21}_{-0.12}$ & $[0.01,0.60]$ & 3 & 0 \\
\hline
\multicolumn{6}{c}{$11.3 \le \log M_{\star} \le 12.0$} \\
\hline
$[0.0,0.5]$ & 0.23 & $0.23^{+0.11}_{-0.09}$ & $[0.07,0.46]$ & 15 & 3 \\
$(0.5,1.0]$ & 0.68 & $0.11^{+0.15}_{-0.08}$ & $[0.00,0.46]$ & 5 & 0 \\
$(1.0,1.5]$ & 1.07 & $0.29^{+0.31}_{-0.21}$ & $[0.01,0.84]$ & 1 & 0 \\
\enddata
\tablecomments{The limiting \HI\ value for the covering factor calculation is \logNHI\ $\ge$ 17.2 [\column]. A combination of the RDR LRGs, the galaxies in the \citetalias{prochaska2017} sample, and the galaxies in the \citetalias{chen2018} sample are used for this covering factor calculation. We do not include sightlines in common between the \citetalias{chen2018} sample and the RDR LRGs or the \citetalias{prochaska2017} sample.}
\tablenotetext{a}{The error bars on the covering factor represent the 68\% confidence interval.}
\tablenotetext{b}{Number of sightlines considered in covering factor calculation.}
\tablenotetext{c}{Number of absorbers detected above the limiting \HI\ value.}
\end{deluxetable*}

%% file: coveringfactor_halos_LLS.tex
\begin{deluxetable*}{lccccc}
\tablecaption{\HI\ Covering Factors by Halo Mass at LLS Limit\label{tab:coveringfactorhalo_lls}}
\tablehead{\colhead{$\rho/\Rvir$} & \colhead{$\langle\rho/\Rvir\rangle$} & \colhead{$f_c$\tablenotemark{a}} & \colhead{95\% C.I.} & \colhead{Sightlines\tablenotemark{b}} & \colhead{Detections\tablenotemark{c}}}
\startdata
\hline
\multicolumn{6}{c}{$11.0 \le \log M_{\rm h} < 12.0$} \\
\hline
$[0.0,0.5]$ & 0.32 & $0.38^{+0.14}_{-0.13}$ & $[0.15,0.65]$ & 11 & 4 \\
$(0.5,1.0]$ & 0.68 & $0.13^{+0.18}_{-0.10}$ & $[0.01,0.52]$ & 4 & 0 \\
\hline
\multicolumn{6}{c}{$12.0 \le \log M_{\rm h} < 13.0$} \\
\hline
$[0.0,0.5]$ & 0.25 & $0.33^{+0.10}_{-0.09}$ & $[0.16,0.53]$ & 22 & 7 \\
$(0.5,1.0]$ & 0.54 & $0.16^{+0.21}_{-0.12}$ & $[0.01,0.60]$ & 3 & 0 \\
$(1.0,1.5]$ & 1.30 & $0.21^{+0.25}_{-0.15}$ & $[0.01,0.71]$ & 2 & 0 \\
\hline
\multicolumn{6}{c}{$13.0 \le \log M_{\rm h} \le 14.0$} \\
\hline
$[0.0,0.5]$ & 0.21 & $0.36^{+0.11}_{-0.10}$ & $[0.18,0.57]$ & 20 & 7 \\
$(0.5,1.0]$ & 0.68 & $0.09^{+0.14}_{-0.07}$ & $[0.00,0.41]$ & 6 & 0 \\
$(1.0,1.5]$ & 1.09 & $0.21^{+0.25}_{-0.15}$ & $[0.01,0.71]$ & 2 & 0 \\
\enddata
\tablecomments{The limiting \HI\ value for the covering factor calculation is \logNHI\ $\ge$ 17.2 [\column]. A combination of the RDR LRGs, the galaxies in the \citetalias{prochaska2017} sample, and the galaxies in the \citetalias{chen2018} sample are used for this covering factor calculation. We do not include sightlines in common between the \citetalias{chen2018} sample and the RDR LRGs or the \citetalias{prochaska2017} sample.}
\tablenotetext{a}{The error bars on the covering factor represent the 68\% confidence interval.}
\tablenotetext{b}{Number of sightlines considered in covering factor calculation.}
\tablenotetext{c}{Number of absorbers detected above the limiting \HI\ value.}
\end{deluxetable*}

%% file: LRG.bbl
\begin{thebibliography}{}
\expandafter\ifx\csname natexlab\endcsname\relax\def\natexlab#1{#1}\fi
\providecommand{\url}[1]{\href{#1}{#1}}

\bibitem[{{Abolfathi} {et~al.}(2018){Abolfathi}, {Aguado}, {Aguilar}, {Allende
  Prieto}, {Almeida}, {Ananna}, {Anders}, {Anderson}, {Andrews}, {Anguiano}, \&
  et~al.}]{abolfathi2018}
{Abolfathi}, B., {Aguado}, D.~S., {Aguilar}, G., {et~al.} 2018, \apjs, 235, 42

\bibitem[{{Afruni} {et~al.}(2019){Afruni}, {Fraternali}, \&
  {Pezzulli}}]{afruni2019}
{Afruni}, A., {Fraternali}, F., \& {Pezzulli}, G. 2019, \aap, 625, A11

\bibitem[{{Albareti} {et~al.}(2017){Albareti}, {Allende Prieto}, {Almeida},
  {Anders}, {Anderson}, {Andrews}, {Arag{\'o}n-Salamanca},
  {Argudo-Fern{\'a}ndez}, {Armengaud}, {Aubourg}, \& et~al.}]{albareti2017}
{Albareti}, F.~D., {Allende Prieto}, C., {Almeida}, A., {et~al.} 2017, \apjs,
  233, 25

\bibitem[{{Almeida} {et~al.}(2008){Almeida}, {Baugh}, {Wake}, {Lacey},
  {Benson}, {Bower}, \& {Pimbblet}}]{almeida2008}
{Almeida}, C., {Baugh}, C.~M., {Wake}, D.~A., {et~al.} 2008, \mnras, 386, 2145

\bibitem[{{Anderson} {et~al.}(2013){Anderson}, {Bregman}, \&
  {Dai}}]{anderson2013}
{Anderson}, M.~E., {Bregman}, J.~N., \& {Dai}, X. 2013, \apj, 762, 106

\bibitem[{{Astropy Collaboration} {et~al.}(2013){Astropy Collaboration},
  {Robitaille}, {Tollerud}, {Greenfield}, {Droettboom}, {Bray}, {Aldcroft},
  {Davis}, {Ginsburg}, {Price-Whelan}, {Kerzendorf}, {Conley}, {Crighton},
  {Barbary}, {Muna}, {Ferguson}, {Grollier}, {Parikh}, {Nair}, {Unther},
  {Deil}, {Woillez}, {Conseil}, {Kramer}, {Turner}, {Singer}, {Fox}, {Weaver},
  {Zabalza}, {Edwards}, {Azalee Bostroem}, {Burke}, {Casey}, {Crawford},
  {Dencheva}, {Ely}, {Jenness}, {Labrie}, {Lim}, {Pierfederici}, {Pontzen},
  {Ptak}, {Refsdal}, {Servillat}, \& {Streicher}}]{robitaille2013}
{Astropy Collaboration}, {Robitaille}, T.~P., {Tollerud}, E.~J., {et~al.} 2013,
  \aap, 558, A33

\bibitem[{{Banerji} {et~al.}(2010){Banerji}, {Ferreras}, {Abdalla}, {Hewett},
  \& {Lahav}}]{banerji2010}
{Banerji}, M., {Ferreras}, I., {Abdalla}, F.~B., {Hewett}, P., \& {Lahav}, O.
  2010, \mnras, 402, 2264

\bibitem[{{Battisti} {et~al.}(2012){Battisti}, {Meiring}, {Tripp}, {Prochaska},
  {Werk}, {Jenkins}, {Lehner}, {Tumlinson}, \& {Thom}}]{battisti2012}
{Battisti}, A.~J., {Meiring}, J.~D., {Tripp}, T.~M., {et~al.} 2012, \apj, 744,
  93

\bibitem[{Behroozi {et~al.}(2010)Behroozi, Conroy, \& Wechsler}]{behroozi2010}
Behroozi, P.~S., Conroy, C., \& Wechsler, R.~H. 2010, The Astrophysical
  Journal, 717, 379

\bibitem[{{Belfiore} {et~al.}(2016){Belfiore}, {Maiolino}, {Maraston},
  {Emsellem}, {Bershady}, {Masters}, {Yan}, {Bizyaev}, {Boquien}, {Brownstein},
  {Bundy}, {Drory}, {Heckman}, {Law}, {Roman-Lopes}, {Pan}, {Stanghellini},
  {Thomas}, {Weijmans}, \& {Westfall}}]{belfiore2016}
{Belfiore}, F., {Maiolino}, R., {Maraston}, C., {et~al.} 2016, \mnras, 461,
  3111

\bibitem[{{Birnboim} \& {Dekel}(2003)}]{birnboim2003}
{Birnboim}, Y., \& {Dekel}, A. 2003, \mnras, 345, 349

\bibitem[{{Blanton} \& {Roweis}(2007)}]{blanton2007}
{Blanton}, M.~R., \& {Roweis}, S. 2007, \aj, 133, 734

\bibitem[{{Blanton} {et~al.}(2005){Blanton}, {Schlegel}, {Strauss},
  {Brinkmann}, {Finkbeiner}, {Fukugita}, {Gunn}, {Hogg}, {Ivezi{\'c}}, {Knapp},
  {Lupton}, {Munn}, {Schneider}, {Tegmark}, \& {Zehavi}}]{blanton2005}
{Blanton}, M.~R., {Schlegel}, D.~J., {Strauss}, M.~A., {et~al.} 2005, \aj, 129,
  2562

\bibitem[{{Bouch{\'e}} {et~al.}(2004){Bouch{\'e}}, {Murphy}, \&
  {P{\'e}roux}}]{bouche2004}
{Bouch{\'e}}, N., {Murphy}, M.~T., \& {P{\'e}roux}, C. 2004, \mnras, 354, L25

\bibitem[{{Bowen} \& {Chelouche}(2011)}]{bowen2011}
{Bowen}, D.~V., \& {Chelouche}, D. 2011, \apj, 727, 47

\bibitem[{{Bregman} {et~al.}(2018){Bregman}, {Anderson}, {Miller},
  {Hodges-Kluck}, {Dai}, {Li}, {Li}, \& {Qu}}]{bregman2018}
{Bregman}, J.~N., {Anderson}, M.~E., {Miller}, M.~J., {et~al.} 2018, \apj, 862,
  3

\bibitem[{{Bruzual} \& {Charlot}(2003)}]{bruzual2003}
{Bruzual}, G., \& {Charlot}, S. 2003, \mnras, 344, 1000

\bibitem[{{Burchett} {et~al.}(2018){Burchett}, {Tripp}, {Wang}, {Willmer},
  {Bowen}, \& {Jenkins}}]{burchett2018}
{Burchett}, J.~N., {Tripp}, T.~M., {Wang}, Q.~D., {et~al.} 2018, \mnras, 475,
  2067

\bibitem[{{Cameron}(2011)}]{cameron2011}
{Cameron}, E. 2011, \pasa, 28, 128

\bibitem[{{Cappellari}(2017)}]{cappellari2017}
{Cappellari}, M. 2017, \mnras, 466, 798

\bibitem[{{Cardelli} {et~al.}(1989){Cardelli}, {Clayton}, \&
  {Mathis}}]{cardelli1989}
{Cardelli}, J.~A., {Clayton}, G.~C., \& {Mathis}, J.~S. 1989, \apj, 345, 245

\bibitem[{{Cattaneo} {et~al.}(2006){Cattaneo}, {Dekel}, {Devriendt},
  {Guiderdoni}, \& {Blaizot}}]{cattaneo2006}
{Cattaneo}, A., {Dekel}, A., {Devriendt}, J., {Guiderdoni}, B., \& {Blaizot},
  J. 2006, \mnras, 370, 1651

\bibitem[{{Chabrier}(2003)}]{chabrier2003}
{Chabrier}, G. 2003, \pasp, 115, 763

\bibitem[{{Chen}(2017)}]{chen2017}
{Chen}, H.-W. 2017, in Astrophysics and Space Science Library, Vol. 430, Gas
  Accretion onto Galaxies, ed. A.~{Fox} \& R.~{Dav{\'e}}, 167

\bibitem[{{Chen} {et~al.}(2010{\natexlab{a}}){Chen}, {Helsby}, {Gauthier},
  {Shectman}, {Thompson}, \& {Tinker}}]{chen2010}
{Chen}, H.-W., {Helsby}, J.~E., {Gauthier}, J.-R., {et~al.} 2010{\natexlab{a}},
  \apj, 714, 1521

\bibitem[{{Chen} {et~al.}(2010{\natexlab{b}}){Chen}, {Wild}, {Tinker},
  {Gauthier}, {Helsby}, {Shectman}, \& {Thompson}}]{chen2010a}
{Chen}, H.-W., {Wild}, V., {Tinker}, J.~L., {et~al.} 2010{\natexlab{b}}, \apjl,
  724, L176

\bibitem[{{Chen} {et~al.}(2018){Chen}, {Zahedy}, {Johnson}, {Pierce}, {Huang},
  {Weiner}, \& {Gauthier}}]{chen2018}
{Chen}, H.-W., {Zahedy}, F.~S., {Johnson}, S.~D., {et~al.} 2018, \mnras, 479,
  2547

\bibitem[{{Chen} {et~al.}(2019){Chen}, {Johnson}, {Straka}, {Zahedy}, {Schaye},
  {Muzahid}, {Bouch{\'e}}, {Cantalupo}, {Marino}, \& {Wendt}}]{chen2019}
{Chen}, H.-W., {Johnson}, S.~D., {Straka}, L.~A., {et~al.} 2019, \mnras, 484,
  431

\bibitem[{{Conroy}(2013)}]{conroy2013}
{Conroy}, C. 2013, \araa, 51, 393

\bibitem[{{Correa} {et~al.}(2018){Correa}, {Schaye}, {van de Voort}, {Duffy},
  \& {Wyithe}}]{correa2018}
{Correa}, C.~A., {Schaye}, J., {van de Voort}, F., {Duffy}, A.~R., \& {Wyithe},
  J.~S.~B. 2018, \mnras, 478, 255

\bibitem[{{Crain} {et~al.}(2010){Crain}, {McCarthy}, {Frenk}, {Theuns}, \&
  {Schaye}}]{crain2010}
{Crain}, R.~A., {McCarthy}, I.~G., {Frenk}, C.~S., {Theuns}, T., \& {Schaye},
  J. 2010, \mnras, 407, 1403

\bibitem[{{Danforth} {et~al.}(2010){Danforth}, {Keeney}, {Stocke}, {Shull}, \&
  {Yao}}]{danforth2010}
{Danforth}, C.~W., {Keeney}, B.~A., {Stocke}, J.~T., {Shull}, J.~M., \& {Yao},
  Y. 2010, \apj, 720, 976

\bibitem[{{Danforth} {et~al.}(2016){Danforth}, {Keeney}, {Tilton}, {Shull},
  {Stocke}, {Stevans}, {Pieri}, {Savage}, {France}, {Syphers}, {Smith},
  {Green}, {Froning}, {Penton}, \& {Osterman}}]{danforth2016}
{Danforth}, C.~W., {Keeney}, B.~A., {Tilton}, E.~M., {et~al.} 2016, \apj, 817,
  111

\bibitem[{Dawson {et~al.}(2013)Dawson, Schlegel, Ahn, Anderson, Aubourg,
  Bailey, Barkhouser, Bautista, Beifiori, Berlind, \& et~al.}]{dawson2013}
Dawson, K.~S., Schlegel, D.~J., Ahn, C.~P., {et~al.} 2013, The Astronomical
  Journal, 145, 10

\bibitem[{Dawson {et~al.}(2016)Dawson, Kneib, Percival, Alam, Albareti,
  Anderson, Armengaud, Aubourg, Bailey, Bautista, \& et~al.}]{dawson2016}
Dawson, K.~S., Kneib, J.-P., Percival, W.~J., {et~al.} 2016, The Astronomical
  Journal, 151, 44

\bibitem[{{Dekel} \& {Birnboim}(2006)}]{db2006}
{Dekel}, A., \& {Birnboim}, Y. 2006, \mnras, 368, 2

\bibitem[{{Eisenstein} {et~al.}(2001){Eisenstein}, {Annis}, {Gunn}, {Szalay},
  {Connolly}, {Nichol}, {Bahcall}, {Bernardi}, {Burles}, {Castander},
  {Fukugita}, {Hogg}, {Ivezi{\'c}}, {Knapp}, {Lupton}, {Narayanan}, {Postman},
  {Reichart}, {Richmond}, {Schneider}, {Schlegel}, {Strauss}, {SubbaRao},
  {Tucker}, {Vanden Berk}, {Vogeley}, {Weinberg}, \& {Yanny}}]{eisenstein2001}
{Eisenstein}, D.~J., {Annis}, J., {Gunn}, J.~E., {et~al.} 2001, \aj, 122, 2267

\bibitem[{{Eisenstein} {et~al.}(2005){Eisenstein}, {Zehavi}, {Hogg},
  {Scoccimarro}, {Blanton}, {Nichol}, {Scranton}, {Seo}, {Tegmark}, {Zheng},
  {Anderson}, {Annis}, {Bahcall}, {Brinkmann}, {Burles}, {Castander},
  {Connolly}, {Csabai}, {Doi}, {Fukugita}, {Frieman}, {Glazebrook}, {Gunn},
  {Hendry}, {Hennessy}, {Ivezi{\'c}}, {Kent}, {Knapp}, {Lin}, {Loh}, {Lupton},
  {Margon}, {McKay}, {Meiksin}, {Munn}, {Pope}, {Richmond}, {Schlegel},
  {Schneider}, {Shimasaku}, {Stoughton}, {Strauss}, {SubbaRao}, {Szalay},
  {Szapudi}, {Tucker}, {Yanny}, \& {York}}]{eisenstein2005}
{Eisenstein}, D.~J., {Zehavi}, I., {Hogg}, D.~W., {et~al.} 2005, \apj, 633, 560

\bibitem[{{Fakhouri} {et~al.}(2010){Fakhouri}, {Ma}, \&
  {Boylan-Kolchin}}]{fakhouri2010}
{Fakhouri}, O., {Ma}, C.-P., \& {Boylan-Kolchin}, M. 2010, \mnras, 406, 2267

\bibitem[{{Ferland} {et~al.}(2013){Ferland}, {Porter}, {van Hoof}, {Williams},
  {Abel}, {Lykins}, {Shaw}, {Henney}, \& {Stancil}}]{ferland2013}
{Ferland}, G.~J., {Porter}, R.~L., {van Hoof}, P.~A.~M., {et~al.} 2013, \rmxaa,
  49, 137

\bibitem[{{Ferland} {et~al.}(2017){Ferland}, {Chatzikos}, {Guzm{\'a}n},
  {Lykins}, {van Hoof}, {Williams}, {Abel}, {Badnell}, {Keenan}, {Porter}, \&
  {Stancil}}]{ferland2017}
{Ferland}, G.~J., {Chatzikos}, M., {Guzm{\'a}n}, F., {et~al.} 2017, \rmxaa, 53,
  385

\bibitem[{{Fitzpatrick} \& {Spitzer}(1997)}]{fs1997}
{Fitzpatrick}, E.~L., \& {Spitzer}, Jr., L. 1997, \apj, 475, 623

\bibitem[{{Fox} \& {Dav{\'e}}(2017)}]{fox2017}
{Fox}, A., \& {Dav{\'e}}, R., eds. 2017, Astrophysics and Space Science
  Library, Vol. 430, {Gas Accretion onto Galaxies}

\bibitem[{{Fumagalli} {et~al.}(2016){Fumagalli}, {O'Meara}, \&
  {Prochaska}}]{fumagalli2016}
{Fumagalli}, M., {O'Meara}, J.~M., \& {Prochaska}, J.~X. 2016, \mnras, 455,
  4100

\bibitem[{{Gauthier} \& {Chen}(2011{\natexlab{a}})}]{gc2011}
{Gauthier}, J.-R., \& {Chen}, H.-W. 2011{\natexlab{a}}, \mnras, 418, 2730

\bibitem[{{Gauthier} \& {Chen}(2011{\natexlab{b}})}]{gauthier2011}
---. 2011{\natexlab{b}}, \mnras, 418, 2730

\bibitem[{Gauthier {et~al.}(2009)Gauthier, Chen, \& Tinker}]{gauthier2009}
Gauthier, J.-R., Chen, H.-W., \& Tinker, J.~L. 2009, The Astrophysical Journal,
  702, 50

\bibitem[{{Gnat} \& {Sternberg}(2007)}]{gnat2007}
{Gnat}, O., \& {Sternberg}, A. 2007, The Astrophysical Journal Supplement
  Series, 168, 213

\bibitem[{{Gonzalez} {et~al.}(2013){Gonzalez}, {Sivanandam}, {Zabludoff}, \&
  {Zaritsky}}]{gonzalez2013}
{Gonzalez}, A.~H., {Sivanandam}, S., {Zabludoff}, A.~I., \& {Zaritsky}, D.
  2013, \apj, 778, 14

\bibitem[{{Gonzalez} {et~al.}(2007){Gonzalez}, {Zaritsky}, \&
  {Zabludoff}}]{gonzalez2007}
{Gonzalez}, A.~H., {Zaritsky}, D., \& {Zabludoff}, A.~I. 2007, \apj, 666, 147

\bibitem[{{Haardt} \& {Madau}(1996)}]{hm1996}
{Haardt}, F., \& {Madau}, P. 1996, \apj, 461, 20

\bibitem[{{Haardt} \& {Madau}(2012)}]{hm2012}
---. 2012, \apj, 746, 125

\bibitem[{{Hausammann} {et~al.}(2019){Hausammann}, {Revaz}, \&
  {Jablonka}}]{hausammann2019}
{Hausammann}, L., {Revaz}, Y., \& {Jablonka}, P. 2019, \aap, 624, A11

\bibitem[{{Heckman} {et~al.}(2017){Heckman}, {Borthakur}, {Wild},
  {Schiminovich}, \& {Bordoloi}}]{heckman2017}
{Heckman}, T., {Borthakur}, S., {Wild}, V., {Schiminovich}, D., \& {Bordoloi},
  R. 2017, \apj, 846, 151

\bibitem[{{Ho}(2008)}]{ho2008}
{Ho}, L.~C. 2008, \araa, 46, 475

\bibitem[{{Hodge} {et~al.}(2008){Hodge}, {Becker}, {White}, \& {de
  Vries}}]{hodge2008}
{Hodge}, J.~A., {Becker}, R.~H., {White}, R.~L., \& {de Vries}, W.~H. 2008,
  \aj, 136, 1097

\bibitem[{{Hodge} {et~al.}(2009){Hodge}, {Zeimann}, {Becker}, \&
  {White}}]{hodge2009}
{Hodge}, J.~A., {Zeimann}, G.~R., {Becker}, R.~H., \& {White}, R.~L. 2009, \aj,
  138, 900

\bibitem[{{Hoshino} {et~al.}(2015){Hoshino}, {Leauthaud}, {Lackner}, {Hikage},
  {Rozo}, {Rykoff}, {Mandelbaum}, {More}, {More}, {Saito}, \&
  {Vulcani}}]{hoshino2015}
{Hoshino}, H., {Leauthaud}, A., {Lackner}, C., {et~al.} 2015, \mnras, 452, 998

\bibitem[{{Howk} {et~al.}(2017){Howk}, {Wotta}, {Berg}, {Lehner}, {Lockman},
  {Hafen}, {Pisano}, {Faucher-Gigu{\`e}re}, {Wakker}, {Prochaska}, {Wolfe},
  {Ribaudo}, {Barger}, {Corlies}, {Fox}, {Guhathakurta}, {Jenkins}, {Kalirai},
  {O'Meara}, {Peeples}, {Stewart}, \& {Strader}}]{howk2017}
{Howk}, J.~C., {Wotta}, C.~B., {Berg}, M.~A., {et~al.} 2017, \apj, 846, 141

\bibitem[{{Howk} {et~al.}(2019){Howk}, {Berg}, {Lehner}, {Wotta}, {O'Meara},
  {Bowen}, {Burchett}, {Peeples}, \& {Tejos}}]{howk2019}
{Howk}, J.~C., {Berg}, M.~A., {Lehner}, N., {et~al.} 2019, in preparation

\bibitem[{{Huang} {et~al.}(2018){Huang}, {Leauthaud}, {Greene}, {Bundy}, {Lin},
  {Tanaka}, {Miyazaki}, \& {Komiyama}}]{huang2018}
{Huang}, S., {Leauthaud}, A., {Greene}, J.~E., {et~al.} 2018, \mnras, 475, 3348

\bibitem[{{Huang} {et~al.}(2016){Huang}, {Chen}, {Johnson}, \&
  {Weiner}}]{huang2016}
{Huang}, Y.-H., {Chen}, H.-W., {Johnson}, S.~D., \& {Weiner}, B.~J. 2016,
  \mnras, 455, 1713

\bibitem[{Hunter(2007)}]{hunter2007}
Hunter, J.~D. 2007, Computing In Science \& Engineering, 9, 90

\bibitem[{{Izotov} {et~al.}(2018){Izotov}, {Thuan}, {Guseva}, \&
  {Liss}}]{izotov2018}
{Izotov}, Y.~I., {Thuan}, T.~X., {Guseva}, N.~G., \& {Liss}, S.~E. 2018,
  \mnras, 473, 1956

\bibitem[{{Keeney} {et~al.}(2017){Keeney}, {Stocke}, {Danforth}, {Shull},
  {Pratt}, {Froning}, {Green}, {Penton}, \& {Savage}}]{keeney2017}
{Keeney}, B.~A., {Stocke}, J.~T., {Danforth}, C.~W., {et~al.} 2017, \apjs, 230,
  6

\bibitem[{{Kere{\v s}} {et~al.}(2005){Kere{\v s}}, {Katz}, {Weinberg}, \&
  {Dav{\'e}}}]{keres2005}
{Kere{\v s}}, D., {Katz}, N., {Weinberg}, D.~H., \& {Dav{\'e}}, R. 2005,
  \mnras, 363, 2

\bibitem[{{Lan} \& {Mo}(2018)}]{lan2018}
{Lan}, T.-W., \& {Mo}, H. 2018, \apj, 866, 36

\bibitem[{{Lan} \& {Mo}(2019)}]{lm2019}
---. 2019, \mnras, 486, 608

\bibitem[{{Lehner} {et~al.}(2015){Lehner}, {Howk}, \& {Wakker}}]{lehner2015}
{Lehner}, N., {Howk}, J.~C., \& {Wakker}, B.~P. 2015, \apj, 804, 79 (LHW15)

\bibitem[{{Lehner} {et~al.}(2016){Lehner}, {O'Meara}, {Howk}, {Prochaska}, \&
  {Fumagalli}}]{lehner2016}
{Lehner}, N., {O'Meara}, J.~M., {Howk}, J.~C., {Prochaska}, J.~X., \&
  {Fumagalli}, M. 2016, \apj, 833, 283

\bibitem[{{Lehner} {et~al.}(2007){Lehner}, {Savage}, {Richter}, {Sembach},
  {Tripp}, \& {Wakker}}]{lehner2007}
{Lehner}, N., {Savage}, B.~D., {Richter}, P., {et~al.} 2007, \apj, 658, 680

\bibitem[{{Lehner} {et~al.}(2018){Lehner}, {Wotta}, {Howk}, {O'Meara},
  {Oppenheimer}, \& {Cooksey}}]{lehner2018}
{Lehner}, N., {Wotta}, C.~B., {Howk}, J.~C., {et~al.} 2018, \apj, 866, 33

\bibitem[{{Lehner} {et~al.}(2019){Lehner}, {Wotta}, {Howk}, {O'Meara},
  {Oppenheimer}, \& {Cooksey}}]{lehner2019}
---. 2019, arXiv e-prints, arXiv:1902.10147

\bibitem[{{Lehner} {et~al.}(2013){Lehner}, {Howk}, {Tripp}, {Tumlinson},
  {Prochaska}, {O'Meara}, {Thom}, {Werk}, {Fox}, \& {Ribaudo}}]{lehner2013}
{Lehner}, N., {Howk}, J.~C., {Tripp}, T.~M., {et~al.} 2013, \apj, 770, 138

\bibitem[{{Lundgren} {et~al.}(2009){Lundgren}, {Brunner}, {York}, {Ross},
  {Quashnock}, {Myers}, {Schneider}, {Al Sayyad}, \& {Bahcall}}]{lundgren2009}
{Lundgren}, B.~F., {Brunner}, R.~J., {York}, D.~G., {et~al.} 2009, \apj, 698,
  819

\bibitem[{{Maller} \& {Bullock}(2004)}]{mb2004}
{Maller}, A.~H., \& {Bullock}, J.~S. 2004, \mnras, 355, 694

\bibitem[{Maraston {et~al.}(2013)Maraston, Pforr, Henriques, Thomas, Wake,
  Brownstein, Capozzi, Tinker, Bundy, Skibba, \& et~al.}]{maraston2013}
Maraston, C., Pforr, J., Henriques, B.~M., {et~al.} 2013, Monthly Notices of
  the Royal Astronomical Society, 435, 2764

\bibitem[{{Martin} {et~al.}(2005){Martin}, {Fanson}, {Schiminovich},
  {Morrissey}, {Friedman}, {Barlow}, {Conrow}, {Grange}, {Jelinsky},
  {Milliard}, {Siegmund}, {Bianchi}, {Byun}, {Donas}, {Forster}, {Heckman},
  {Lee}, {Madore}, {Malina}, {Neff}, {Rich}, {Small}, {Surber}, {Szalay},
  {Welsh}, \& {Wyder}}]{martin2005}
{Martin}, D.~C., {Fanson}, J., {Schiminovich}, D., {et~al.} 2005, \apjl, 619,
  L1

\bibitem[{{Masters} {et~al.}(2011){Masters}, {Maraston}, {Nichol}, {Thomas},
  {Beifiori}, {Bundy}, {Edmondson}, {Higgs}, {Leauthaud}, {Mandelbaum},
  {Pforr}, {Ross}, {Ross}, {Schneider}, {Skibba}, {Tinker}, {Tojeiro}, {Wake},
  {Brinkmann}, \& {Weaver}}]{masters2011}
{Masters}, K.~L., {Maraston}, C., {Nichol}, R.~C., {et~al.} 2011, \mnras, 418,
  1055

\bibitem[{{Mathews} \& {Prochaska}(2017)}]{mathews2017}
{Mathews}, W.~G., \& {Prochaska}, J.~X. 2017, \apj, 846, L24

\bibitem[{{McCourt} {et~al.}(2012){McCourt}, {Sharma}, {Quataert}, \&
  {Parrish}}]{mccourt2012}
{McCourt}, M., {Sharma}, P., {Quataert}, E., \& {Parrish}, I.~J. 2012, \mnras,
  419, 3319

\bibitem[{{Mernier} {et~al.}(2018){Mernier}, {Biffi}, {Yamaguchi}, {Medvedev},
  {Simionescu}, {Ettori}, {Werner}, {Kaastra}, {de Plaa}, \&
  {Gu}}]{mernier2018}
{Mernier}, F., {Biffi}, V., {Yamaguchi}, H., {et~al.} 2018, \ssr, 214, 129

\bibitem[{{Miller} \& {Bregman}(2013)}]{miller2013}
{Miller}, M.~J., \& {Bregman}, J.~N. 2013, \apj, 770, 118

\bibitem[{{Moustakas} {et~al.}(2006){Moustakas}, {Kennicutt}, \&
  {Tremonti}}]{moustakas2006}
{Moustakas}, J., {Kennicutt}, Jr., R.~C., \& {Tremonti}, C.~A. 2006, \apj, 642,
  775

\bibitem[{{Muzahid} {et~al.}(2017){Muzahid}, {Charlton}, {Nagai}, {Schaye}, \&
  {Srianand}}]{muzahid2017}
{Muzahid}, S., {Charlton}, J., {Nagai}, D., {Schaye}, J., \& {Srianand}, R.
  2017, \apjl, 846, L8

\bibitem[{{Navarro} {et~al.}(1996){Navarro}, {Frenk}, \& {White}}]{nfw1996}
{Navarro}, J.~F., {Frenk}, C.~S., \& {White}, S.~D.~M. 1996, \apj, 462, 563

\bibitem[{{Navarro} {et~al.}(1997){Navarro}, {Frenk}, \& {White}}]{nfw1997}
---. 1997, \apj, 490, 493

\bibitem[{{Nelson} {et~al.}(2015){Nelson}, {Genel}, {Vogelsberger}, {Springel},
  {Sijacki}, {Torrey}, \& {Hernquist}}]{nelson2015}
{Nelson}, D., {Genel}, S., {Vogelsberger}, M., {et~al.} 2015, \mnras, 448, 59

\bibitem[{{Nelson} {et~al.}(2018){Nelson}, {Pillepich}, {Springel},
  {Weinberger}, {Hernquist}, {Pakmor}, {Genel}, {Torrey}, {Vogelsberger},
  {Kauffmann}, {Marinacci}, \& {Naiman}}]{nelson2018}
{Nelson}, D., {Pillepich}, A., {Springel}, V., {et~al.} 2018, \mnras, 475, 624

\bibitem[{{Ocvirk} {et~al.}(2008){Ocvirk}, {Pichon}, \&
  {Teyssier}}]{ocvirk2008}
{Ocvirk}, P., {Pichon}, C., \& {Teyssier}, R. 2008, \mnras, 390, 1326

\bibitem[{{Oppenheimer} {et~al.}(2010){Oppenheimer}, {Dav{\'e}}, {Kere{\v s}},
  {Fardal}, {Katz}, {Kollmeier}, \& {Weinberg}}]{oppenheimer2010}
{Oppenheimer}, B.~D., {Dav{\'e}}, R., {Kere{\v s}}, D., {et~al.} 2010, \mnras,
  406, 2325

\bibitem[{{Oppenheimer} \& {Schaye}(2013)}]{oppenheimer2013}
{Oppenheimer}, B.~D., \& {Schaye}, J. 2013, \mnras, 434, 1043

\bibitem[{{Oppenheimer} {et~al.}(2016){Oppenheimer}, {Crain}, {Schaye},
  {Rahmati}, {Richings}, {Trayford}, {Tumlinson}, {Bower}, {Schaller}, \&
  {Theuns}}]{oppenheimer2016}
{Oppenheimer}, B.~D., {Crain}, R.~A., {Schaye}, J., {et~al.} 2016, \mnras, 460,
  2157

\bibitem[{{Oser} {et~al.}(2010){Oser}, {Ostriker}, {Naab}, {Johansson}, \&
  {Burkert}}]{oser2010}
{Oser}, L., {Ostriker}, J.~P., {Naab}, T., {Johansson}, P.~H., \& {Burkert}, A.
  2010, \apj, 725, 2312

\bibitem[{{O'Sullivan} {et~al.}(2018){O'Sullivan}, {Combes}, {Salom{\'e}},
  {David}, {Babul}, {Vrtilek}, {Lim}, {Olivares}, {Raychaudhury}, \&
  {Schellenberger}}]{osullivan2018}
{O'Sullivan}, E., {Combes}, F., {Salom{\'e}}, P., {et~al.} 2018, \aap, 618,
  A126

\bibitem[{{Padmanabhan}(2002)}]{pad2002}
{Padmanabhan}, T. 2002, {Theoretical Astrophysics - Volume 3, Galaxies and
  Cosmology}, 638, doi:10.2277/0521562422

\bibitem[{{Peeples} {et~al.}(2017){Peeples}, {Tumlinson}, {Fox}, {Aloisi},
  {Fleming}, {Jedrzejewski}, {Oliveira}, {Ayres}, {Danforth}, {Keeney}, \&
  {Jenkins}}]{peeples2017}
{Peeples}, M., {Tumlinson}, J., {Fox}, A., {et~al.} 2017, {The Hubble
  Spectroscopic Legacy Archive}, Tech. rep.

\bibitem[{{Peeples} {et~al.}(2019){Peeples}, {Corlies}, {Tumlinson}, {O'Shea},
  {Lehner}, {O'Meara}, {Howk}, {Earl}, {Smith}, {Wise}, \&
  {Hummels}}]{peeples2019}
{Peeples}, M.~S., {Corlies}, L., {Tumlinson}, J., {et~al.} 2019, \apj, 873, 129

\bibitem[{{P{\'e}rez-R{\`a}fols} {et~al.}(2015){P{\'e}rez-R{\`a}fols},
  {Miralda-Escud{\'e}}, {Lundgren}, {Ge}, {Petitjean}, {Schneider}, {York}, \&
  {Weaver}}]{perez-rafols2015}
{P{\'e}rez-R{\`a}fols}, I., {Miralda-Escud{\'e}}, J., {Lundgren}, B., {et~al.}
  2015, \mnras, 447, 2784

\bibitem[{{Planck Collaboration} {et~al.}(2016){Planck Collaboration}, {Ade},
  {Aghanim}, {Arnaud}, {Ashdown}, {Aumont}, {Baccigalupi}, {Banday},
  {Barreiro}, {Bartlett}, \& et~al.}]{planck2016}
{Planck Collaboration}, {Ade}, P.~A.~R., {Aghanim}, N., {et~al.} 2016, \aap,
  594, A13

\bibitem[{Prakash {et~al.}(2016)Prakash, Licquia, Newman, Ross, Myers, Dawson,
  Kneib, Percival, Bautista, Comparat, \& et~al.}]{prakash2016}
Prakash, A., Licquia, T.~C., Newman, J.~A., {et~al.} 2016, The Astrophysical
  Journal Supplement Series, 224, 34

\bibitem[{{Prochaska} {et~al.}(2010){Prochaska}, {O'Meara}, \&
  {Worseck}}]{prochaska2010}
{Prochaska}, J.~X., {O'Meara}, J.~M., \& {Worseck}, G. 2010, \apj, 718, 392

\bibitem[{{Prochaska} {et~al.}(2017){Prochaska}, {Werk}, {Worseck}, {Tripp},
  {Tumlinson}, {Burchett}, {Fox}, {Fumagalli}, {Lehner}, {Peeples}, \&
  {Tejos}}]{prochaska2017}
{Prochaska}, J.~X., {Werk}, J.~K., {Worseck}, G., {et~al.} 2017, \apj, 837, 169

\bibitem[{{Rahmati} {et~al.}(2015){Rahmati}, {Schaye}, {Bower}, {Crain},
  {Furlong}, {Schaller}, \& {Theuns}}]{rahmati2015}
{Rahmati}, A., {Schaye}, J., {Bower}, R.~G., {et~al.} 2015, \mnras, 452, 2034

\bibitem[{{Ribaudo} {et~al.}(2011){Ribaudo}, {Lehner}, \& {Howk}}]{ribaudo2011}
{Ribaudo}, J., {Lehner}, N., \& {Howk}, J.~C. 2011, \apj, 736, 42

\bibitem[{{Roca-F{\`a}brega} {et~al.}(2019){Roca-F{\`a}brega}, {Dekel},
  {Faerman}, {Gnat}, {Strawn}, {Ceverino}, {Primack}, {Macci{\`o}}, {Dutton},
  {Prochaska}, \& {Stern}}]{rocafabrega2019}
{Roca-F{\`a}brega}, S., {Dekel}, A., {Faerman}, Y., {et~al.} 2019, \mnras, 484,
  3625

\bibitem[{{Rodr{\'{\i}}guez-Puebla} {et~al.}(2017){Rodr{\'{\i}}guez-Puebla},
  {Primack}, {Avila-Reese}, \& {Faber}}]{rodriguez-puebla2017}
{Rodr{\'{\i}}guez-Puebla}, A., {Primack}, J.~R., {Avila-Reese}, V., \& {Faber},
  S.~M. 2017, \mnras, 470, 651

\bibitem[{{Rykoff} {et~al.}(2014){Rykoff}, {Rozo}, {Busha}, {Cunha},
  {Finoguenov}, {Evrard}, {Hao}, {Koester}, {Leauthaud}, {Nord}, {Pierre},
  {Reddick}, {Sadibekova}, {Sheldon}, \& {Wechsler}}]{rykoff2014}
{Rykoff}, E.~S., {Rozo}, E., {Busha}, M.~T., {et~al.} 2014, \apj, 785, 104

\bibitem[{{Rykoff} {et~al.}(2016){Rykoff}, {Rozo}, {Hollowood},
  {Bermeo-Hernandez}, {Jeltema}, {Mayers}, {Romer}, {Rooney}, {Saro}, {Vergara
  Cervantes}, {Wechsler}, {Wilcox}, {Abbott}, {Abdalla}, {Allam}, {Annis},
  {Benoit-L{\'e}vy}, {Bernstein}, {Bertin}, {Brooks}, {Burke}, {Capozzi},
  {Carnero Rosell}, {Carrasco Kind}, {Castander}, {Childress}, {Collins},
  {Cunha}, {D'Andrea}, {da Costa}, {Davis}, {Desai}, {Diehl}, {Dietrich},
  {Doel}, {Evrard}, {Finley}, {Flaugher}, {Fosalba}, {Frieman}, {Glazebrook},
  {Goldstein}, {Gruen}, {Gruendl}, {Gutierrez}, {Hilton}, {Honscheid}, {Hoyle},
  {James}, {Kay}, {Kuehn}, {Kuropatkin}, {Lahav}, {Lewis}, {Lidman}, {Lima},
  {Maia}, {Mann}, {Marshall}, {Martini}, {Melchior}, {Miller}, {Miquel},
  {Mohr}, {Nichol}, {Nord}, {Ogando}, {Plazas}, {Reil}, {Sahl{\'e}n},
  {Sanchez}, {Santiago}, {Scarpine}, {Schubnell}, {Sevilla-Noarbe}, {Smith},
  {Soares-Santos}, {Sobreira}, {Stott}, {Suchyta}, {Swanson}, {Tarle},
  {Thomas}, {Tucker}, {Uddin}, {Viana}, {Vikram}, {Walker}, {Zhang}, \& {DES
  Collaboration}}]{rykoff2016}
{Rykoff}, E.~S., {Rozo}, E., {Hollowood}, D., {et~al.} 2016, \apjs, 224, 1

\bibitem[{{Sadler} {et~al.}(2007){Sadler}, {Cannon}, {Mauch}, {Hancock},
  {Wake}, {Ross}, {Croom}, {Drinkwater}, {Edge}, {Eisenstein}, {Hopkins},
  {Johnston}, {Nichol}, {Pimbblet}, {de Propris}, {Roseboom}, {Schneider}, \&
  {Shanks}}]{sadler2007}
{Sadler}, E.~M., {Cannon}, R.~D., {Mauch}, T., {et~al.} 2007, \mnras, 381, 211

\bibitem[{{Savage} \& {Sembach}(1991)}]{SS1991}
{Savage}, B.~D., \& {Sembach}, K.~R. 1991, \apj, 379, 245

\bibitem[{{Schlegel} {et~al.}(1998){Schlegel}, {Finkbeiner}, \&
  {Davis}}]{schlegel1998}
{Schlegel}, D.~J., {Finkbeiner}, D.~P., \& {Davis}, M. 1998, \apj, 500, 525

\bibitem[{{Schneider} {et~al.}(2010){Schneider}, {Richards}, {Hall}, {Strauss},
  {Anderson}, {Boroson}, {Ross}, {Shen}, {Brandt}, {Fan}, {Inada}, {Jester},
  {Knapp}, {Krawczyk}, {Thakar}, {Vanden Berk}, {Voges}, {Yanny}, {York},
  {Bahcall}, {Bizyaev}, {Blanton}, {Brewington}, {Brinkmann}, {Eisenstein},
  {Frieman}, {Fukugita}, {Gray}, {Gunn}, {Hibon}, {Ivezi{\'c}}, {Kent}, {Kron},
  {Lee}, {Lupton}, {Malanushenko}, {Malanushenko}, {Oravetz}, {Pan}, {Pier},
  {Price}, {Saxe}, {Schlegel}, {Simmons}, {Snedden}, {SubbaRao}, {Szalay}, \&
  {Weinberg}}]{schneider2010}
{Schneider}, D.~P., {Richards}, G.~T., {Hall}, P.~B., {et~al.} 2010, \aj, 139,
  2360

\bibitem[{{Serra} {et~al.}(2012){Serra}, {Oosterloo}, {Morganti}, {Alatalo},
  {Blitz}, {Bois}, {Bournaud}, {Bureau}, {Cappellari}, {Crocker}, {Davies},
  {Davis}, {de Zeeuw}, {Duc}, {Emsellem}, {Khochfar}, {Krajnovi{\'c}},
  {Kuntschner}, {Lablanche}, {McDermid}, {Naab}, {Sarzi}, {Scott}, {Trager},
  {Weijmans}, \& {Young}}]{serra2012}
{Serra}, P., {Oosterloo}, T., {Morganti}, R., {et~al.} 2012, \mnras, 422, 1835

\bibitem[{{Shan} {et~al.}(2017){Shan}, {Kneib}, {Li}, {Comparat}, {Erben},
  {Makler}, {Moraes}, {Van Waerbeke}, {Taylor}, {Charbonnier}, \&
  {Pereira}}]{shan2017}
{Shan}, H., {Kneib}, J.-P., {Li}, R., {et~al.} 2017, \apj, 840, 104

\bibitem[{{Sharma} \& {Nath}(2013)}]{sharma2013}
{Sharma}, M., \& {Nath}, B.~B. 2013, \apj, 763, 17

\bibitem[{{Sharma} {et~al.}(2012){Sharma}, {McCourt}, {Quataert}, \&
  {Parrish}}]{sharma2012}
{Sharma}, P., {McCourt}, M., {Quataert}, E., \& {Parrish}, I.~J. 2012, \mnras,
  420, 3174

\bibitem[{{Shull} {et~al.}(2017){Shull}, {Danforth}, {Tilton}, {Moloney}, \&
  {Stevans}}]{shull2017}
{Shull}, J.~M., {Danforth}, C.~W., {Tilton}, E.~M., {Moloney}, J., \&
  {Stevans}, M.~L. 2017, \apj, 849, 106

\bibitem[{{Singh} {et~al.}(2018){Singh}, {Majumdar}, {Nath}, \&
  {Silk}}]{singh2018}
{Singh}, P., {Majumdar}, S., {Nath}, B.~B., \& {Silk}, J. 2018, \mnras, 478,
  2909

\bibitem[{{Slepian} {et~al.}(2017){Slepian}, {Eisenstein}, {Brownstein},
  {Chuang}, {Gil-Mar{\'{\i}}n}, {Ho}, {Kitaura}, {Percival}, {Ross}, {Rossi},
  {Seo}, {Slosar}, \& {Vargas-Maga{\~n}a}}]{slepian2017}
{Slepian}, Z., {Eisenstein}, D.~J., {Brownstein}, J.~R., {et~al.} 2017, \mnras,
  469, 1738

\bibitem[{Smailagi{\'c} {et~al.}(2018)Smailagi{\'c}, Prochaska, Burchett, Zhu,
  \& M{\'e}nard}]{smailagic2018}
Smailagi{\'c}, M., Prochaska, J.~X., Burchett, J., Zhu, G., \& M{\'e}nard, B.
  2018, The Astrophysical Journal, 867, 106

\bibitem[{{Spitzer}(1978)}]{spitzer1978}
{Spitzer}, L. 1978, {Physical processes in the interstellar medium},
  doi:10.1002/9783527617722

\bibitem[{{Stern} {et~al.}(2018){Stern}, {Faucher-Gigu{\`e}re}, {Hennawi},
  {Hafen}, {Johnson}, \& {Fielding}}]{stern2018}
{Stern}, J., {Faucher-Gigu{\`e}re}, C.-A., {Hennawi}, J.~F., {et~al.} 2018,
  \apj, 865, 91

\bibitem[{{Stewart} {et~al.}(2011){Stewart}, {Kaufmann}, {Bullock}, {Barton},
  {Maller}, {Diemand}, \& {Wadsley}}]{stewart2011}
{Stewart}, K.~R., {Kaufmann}, T., {Bullock}, J.~S., {et~al.} 2011, \apj, 738,
  39

\bibitem[{{Stoehr} {et~al.}(2008){Stoehr}, {White}, {Smith}, {Kamp},
  {Thompson}, {Durand}, {Freudling}, {Fraquelli}, {Haase}, {Hook}, {Kimball},
  {K{\"u}mmel}, {Levay}, {Lombardi}, {Micol}, \& {Rogers}}]{stoehr2008}
{Stoehr}, F., {White}, R., {Smith}, M., {et~al.} 2008, in Astronomical Society
  of the Pacific Conference Series, Vol. 394, Astronomical Data Analysis
  Software and Systems XVII, ed. R.~W. {Argyle}, P.~S. {Bunclark}, \& J.~R.
  {Lewis}, 505

\bibitem[{{Telfer} {et~al.}(2002){Telfer}, {Zheng}, {Kriss}, \&
  {Davidsen}}]{telfer2002}
{Telfer}, R.~C., {Zheng}, W., {Kriss}, G.~A., \& {Davidsen}, A.~F. 2002, \apj,
  565, 773

\bibitem[{{The Astropy Collaboration} {et~al.}(2018){The Astropy
  Collaboration}, Price-Whelan, Sip{\H o}cz, G{\"u}nther, Lim, Crawford,
  Conseil, Shupe, Craig, Dencheva, Ginsburg, VanderPlas, Bradley,
  P{\'e}rez-Su{\'a}rez, de~Val-Borro, Contributors), Aldcroft, Cruz,
  Robitaille, Tollerud, Committee), Ardelean, Babej, Bach, Bachetti, Bakanov,
  Bamford, Barentsen, Barmby, Baumbach, Berry, Biscani, Boquien, Bostroem,
  Bouma, Brammer, Bray, Breytenbach, Buddelmeijer, Burke, Calderone,
  Rodr{\'\i}guez, Cara, Cardoso, Cheedella, Copin, Corrales, Crichton,
  D'Avella, Deil, Depagne, Dietrich, Donath, Droettboom, Earl, Erben, Fabbro,
  Ferreira, Finethy, Fox, Garrison, Gibbons, Goldstein, Gommers, Greco,
  Greenfield, Groener, Grollier, Hagen, Hirst, Homeier, Horton, Hosseinzadeh,
  Hu, Hunkeler, Ivezi{\'c}, Jain, Jenness, Kanarek, Kendrew, Kern, Kerzendorf,
  Khvalko, King, Kirkby, Kulkarni, Kumar, Lee, Lenz, Littlefair, Ma, Macleod,
  Mastropietro, McCully, Montagnac, Morris, Mueller, Mumford, Muna, Murphy,
  Nelson, Nguyen, Ninan, N{\"o}the, Ogaz, Oh, Parejko, Parley, Pascual, Patil,
  Patil, Plunkett, Prochaska, Rastogi, Janga, Sabater, Sakurikar, Seifert,
  Sherbert, Sherwood-Taylor, Shih, Sick, Silbiger, Singanamalla, Singer,
  Sladen, Sooley, Sornarajah, Streicher, Teuben, Thomas, Tremblay, Turner,
  Terr{\'o}n, van Kerkwijk, de~la Vega, Watkins, Weaver, Whitmore, Woillez,
  Zabalza, \& Contributors)}]{astropy2018}
{The Astropy Collaboration}, Price-Whelan, A.~M., Sip{\H o}cz, B.~M., {et~al.}
  2018, The Astronomical Journal, 156, 123

\bibitem[{{Thom} {et~al.}(2012){Thom}, {Tumlinson}, {Werk}, {Prochaska},
  {Oppenheimer}, {Peeples}, {Tripp}, {Katz}, {O'Meara}, {Ford}, {Dav{\'e}},
  {Sembach}, \& {Weinberg}}]{thom2012}
{Thom}, C., {Tumlinson}, J., {Werk}, J.~K., {et~al.} 2012, \apjl, 758, L41

\bibitem[{{Tinker} {et~al.}(2017){Tinker}, {Brownstein}, {Guo}, {Leauthaud},
  {Maraston}, {Masters}, {Montero-Dorta}, {Thomas}, {Tojeiro}, {Weiner},
  {Zehavi}, \& {Olmstead}}]{tinker2017}
{Tinker}, J.~L., {Brownstein}, J.~R., {Guo}, H., {et~al.} 2017, \apj, 839, 121

\bibitem[{{Tumlinson} {et~al.}(2017){Tumlinson}, {Peeples}, \&
  {Werk}}]{tpw2017}
{Tumlinson}, J., {Peeples}, M.~S., \& {Werk}, J.~K. 2017, \araa, 55, 389

\bibitem[{{Tumlinson} {et~al.}(2011){Tumlinson}, {Thom}, {Werk}, {Prochaska},
  {Tripp}, {Weinberg}, {Peeples}, {O'Meara}, {Oppenheimer}, {Meiring}, {Katz},
  {Dav{\'e}}, {Ford}, \& {Sembach}}]{tumlinson2011}
{Tumlinson}, J., {Thom}, C., {Werk}, J.~K., {et~al.} 2011, Science, 334, 948

\bibitem[{{Tumlinson} {et~al.}(2013){Tumlinson}, {Thom}, {Werk}, {Prochaska},
  {Tripp}, {Katz}, {Dav{\'e}}, {Oppenheimer}, {Meiring}, {Ford}, {O'Meara},
  {Peeples}, {Sembach}, \& {Weinberg}}]{tumlinson2013}
---. 2013, \apj, 777, 59

\bibitem[{{Velander} {et~al.}(2014){Velander}, {van Uitert}, {Hoekstra},
  {Coupon}, {Erben}, {Heymans}, {Hildebrandt}, {Kitching}, {Mellier}, {Miller},
  {Van Waerbeke}, {Bonnett}, {Fu}, {Giodini}, {Hudson}, {Kuijken}, {Rowe},
  {Schrabback}, \& {Semboloni}}]{velander2014}
{Velander}, M., {van Uitert}, E., {Hoekstra}, H., {et~al.} 2014, \mnras, 437,
  2111

\bibitem[{{Voit} {et~al.}(2015){Voit}, {Bryan}, {O'Shea}, \&
  {Donahue}}]{voit2015}
{Voit}, G.~M., {Bryan}, G.~L., {O'Shea}, B.~W., \& {Donahue}, M. 2015, \apjl,
  808, L30

\bibitem[{{Voit} {et~al.}(2017){Voit}, {Meece}, {Li}, {O'Shea}, {Bryan}, \&
  {Donahue}}]{voit2017}
{Voit}, G.~M., {Meece}, G., {Li}, Y., {et~al.} 2017, \apj, 845, 80

\bibitem[{{Wechsler} \& {Tinker}(2018)}]{wt2018}
{Wechsler}, R.~H., \& {Tinker}, J.~L. 2018, \araa, 56, 435

\bibitem[{{Werk} {et~al.}(2012){Werk}, {Prochaska}, {Thom}, {Tumlinson},
  {Tripp}, {O'Meara}, \& {Meiring}}]{werk2012}
{Werk}, J.~K., {Prochaska}, J.~X., {Thom}, C., {et~al.} 2012, \apjs, 198, 3

\bibitem[{{Wiersma} {et~al.}(2009){Wiersma}, {Schaye}, \&
  {Smith}}]{wiersma2009}
{Wiersma}, R.~P.~C., {Schaye}, J., \& {Smith}, B.~D. 2009, \mnras, 393, 99

\bibitem[{{Wotta} {et~al.}(2019){Wotta}, {Lehner}, {Howk}, {O'Meara},
  {Oppenheimer}, \& {Cooksey}}]{wotta2019}
{Wotta}, C.~B., {Lehner}, N., {Howk}, J.~C., {et~al.} 2019, \apj, 872, 81

\bibitem[{{Wotta} {et~al.}(2016){Wotta}, {Lehner}, {Howk}, {O'Meara}, \&
  {Prochaska}}]{wotta2016}
{Wotta}, C.~B., {Lehner}, N., {Howk}, J.~C., {O'Meara}, J.~M., \& {Prochaska},
  J.~X. 2016, \apj, 831, 95

\bibitem[{{Xu} {et~al.}(2013){Xu}, {Cuesta}, {Padmanabhan}, {Eisenstein}, \&
  {McBride}}]{xu2013}
{Xu}, X., {Cuesta}, A.~J., {Padmanabhan}, N., {Eisenstein}, D.~J., \&
  {McBride}, C.~K. 2013, \mnras, 431, 2834

\bibitem[{{Yan} \& {Blanton}(2012)}]{yan2012}
{Yan}, R., \& {Blanton}, M.~R. 2012, \apj, 747, 61

\bibitem[{{Yoon} \& {Putman}(2013)}]{yp2013}
{Yoon}, J.~H., \& {Putman}, M.~E. 2013, \apjl, 772, L29

\bibitem[{{Yoon} \& {Putman}(2017)}]{yp2017}
---. 2017, \apj, 839, 117

\bibitem[{{Yoon} {et~al.}(2012){Yoon}, {Putman}, {Thom}, {Chen}, \&
  {Bryan}}]{yoon2012}
{Yoon}, J.~H., {Putman}, M.~E., {Thom}, C., {Chen}, H.-W., \& {Bryan}, G.~L.
  2012, \apj, 754, 84

\bibitem[{{Zahedy} {et~al.}(2017){Zahedy}, {Chen}, {Gauthier}, \&
  {Rauch}}]{zahedy2017a}
{Zahedy}, F.~S., {Chen}, H.-W., {Gauthier}, J.-R., \& {Rauch}, M. 2017, \mnras,
  466, 1071

\bibitem[{{Zahedy} {et~al.}(2019){Zahedy}, {Chen}, {Johnson}, {Pierce},
  {Rauch}, {Huang}, {Weiner}, \& {Gauthier}}]{zahedy2019}
{Zahedy}, F.~S., {Chen}, H.-W., {Johnson}, S.~D., {et~al.} 2019, \mnras, 484,
  2257

\bibitem[{{Zhu} {et~al.}(2014){Zhu}, {M{\'e}nard}, {Bizyaev}, {Brewington},
  {Ebelke}, {Ho}, {Kinemuchi}, {Malanushenko}, {Malanushenko}, {Marchante},
  {More}, {Oravetz}, {Pan}, {Petitjean}, \& {Simmons}}]{zhu2014}
{Zhu}, G., {M{\'e}nard}, B., {Bizyaev}, D., {et~al.} 2014, \mnras, 439, 3139

\end{thebibliography}
